\documentclass[final]{siamonline250211}

\usepackage{graphicx}%
\usepackage{multirow}%
\usepackage{amsmath,amssymb,amsfonts}%
\usepackage{mathrsfs}%
\usepackage{xcolor}
\usepackage{booktabs}%
\usepackage{listings}%
\usepackage{bm}
\usepackage{mathtools}

\usepackage{subfig}
\usepackage{array}
\usepackage{cases}

\usepackage{makecell}
\usepackage{rotating}
\usepackage{diagbox}
\usepackage{url}

\usepackage{amsfonts}
\usepackage{lipsum}
\usepackage{epstopdf}
\usepackage{algorithmic}
\ifpdf
  \DeclareGraphicsExtensions{.eps,.pdf,.png,.jpg}
\else
  \DeclareGraphicsExtensions{.eps}
\fi

\usepackage{enumitem}
\setlist[enumerate]{leftmargin=.5in}
\setlist[itemize]{leftmargin=.5in}

\newcommand{\R}{\bm{\mathrm{R}}}

\newcommand{\I}{\bm{\mathrm{I}}}
\newcommand{\Etr}{\bm{\mathrm{E}}_{tr}}
\newcommand{\Eg}{\bm{\mathrm{E}}_g}
\newcolumntype{M}[1]{>{\centering\arraybackslash}m{#1}}
\newcommand{\tb}[1]{\textbf{#1}}
\newcommand{\ul}[1]{\underline{#1}}

\newcommand{\X}{\mathcal{X}}
\newcommand{\Y}{\mathcal{Y}}
\newcommand{\W}{\mathcal{W}}
\newcommand{\Z}{\mathcal{Z}}
\newcommand{\bX}{\mathbb{X}}
\newcommand{\bY}{\mathbb{Y}}
\newcommand{\D}{\mathrm{D}}
\newcommand{\Xgt}{\bar{\bm{\mathcal{X}}}}
\newcommand{\wline}{\Xhline{1pt}}

\newsiamremark{remark}{Remark}
\newsiamremark{hypothesis}{Hypothesis}
\crefname{hypothesis}{Hypothesis}{Hypotheses}
\newsiamthm{claim}{Claim}
\newsiamremark{fact}{Fact}
\crefname{fact}{Fact}{Facts}

\newsiamremark{assump}{Assumption}

\headers{A Data-driven Loss Weighting Scheme across Heterogeneous Tasks for Image Denoising}{}

\title{A Data-driven Loss Weighting Scheme across Heterogeneous Tasks for Image Denoising\thanks{Submitted to the editors DATE.
\funding{This work was supported in part by the National Key Research and Development Program of China under Grant 2022YFA1004100; Fundamental and Interdisciplinary Disciplines Breakthrough Plan of the Ministry of Education of China (No. JYB2025XDXM101); National Key Research and Development Program of China under Grant 2024YFE0202900; RGC GRF 12300125; Tianyuan Fund for Mathematics of the National Natural Science Foundation of China (Grant No. 12426105); the China NSFC projects under contracts 62476214 and 62272375; and the GDSTC: Guangdong and Hong Kong Universities “1+1+1” Joint Research Collaboration Scheme project No.: 2025A0505000007.}}}

\author{Xiangyu Rui
\and Xiangyong Cao
\and Xile Zhao
\and Deyu Meng
\and Michael K. NG}

\author{Xiangyu Rui\thanks{Department of Mathematics, Hong Kong Baptist University, Kowloon Tong, Hong Kong
		(\email{xyrui.aca@gmail.com}).}
	\and Xiangyong Cao\thanks{Corresponding author. School of Computer Science and Technology and the Ministry of Education Key Laboratory for Intelligent Networks and Network Security, Xi’an Jiaotong University, Xi’an, Shaanxi, China  
		(\email{caoxiangyong@xjtu.edu.cn}).}
	\and Xile Zhao\thanks{School of Mathematical Sciences, University of Electronic Science and Technology of
		China, Chengdu, Sichuan, China (\email{xlzhao122003@163.com}).}
	\and Deyu Meng\thanks{Corresponding author. School of Mathematics and Statistics and Ministry of  Education Key Lab of Intelligent Networks and Network Security, Xi’an Jiaotong University,  Xi’an, Shaanxi, China, (\email{dymeng@mail.xjtu.edu.cn}).}
	\and Michael K. NG\thanks{Department of Mathematics, Hong Kong Baptist University, Kowloon Tong, Hong Kong
		(\email{michael-ng@hkbu.edu.hk}).}
	}

\usepackage{amsopn}

\ifpdf
\hypersetup{
pdftitle={A Data-driven Loss Weighting Scheme across Heterogeneous Tasks for Image Denoising},
pdfauthor={Xiangyu Rui, Xiangyong Cao, Xile Zhao, Deyu Meng and Michael K. NG}
}
\fi

\begin{document}

\maketitle

\begin{abstract}
In a variational denoising model, the weight in the data fidelity term plays the role of enhancing the noise-removal capability. It is profoundly correlated with noise information and also serves to balance the data fidelity and regularization terms. However, assigning the weight is expected to be challenging when the noise pattern is beyond the case of independent identical Gaussian distribution, e.g., impulse noise, stripe noise, or a mixture of several patterns, etc. Furthermore, how to leverage the weight to balance the data fidelity and regularization terms is even less evident. In this work, we propose a data-driven loss weighting (DLW) scheme to address these issues. Specifically, DLW trains a parameterized weight function (i.e., a neural network) that maps the noisy image to the weight. The training is achieved by a bilevel optimization framework, where the lower-level problem solves several denoising models with the same weight predicted by the weight function, and the upper-level problem minimizes the distance between the restored image and the clean image. In this way, information from both the noise and the regularization can be efficiently extracted to determine the weight function. DLW also facilitates the easy implementation of a trained weight function on denoising models. Numerical results verify the remarkable performance of DLW on improving the ability of various variational denoising models to handle different complex noise. This implies that DLW has the ability to transfer the knowledge of noise at the model level to heterogeneous tasks beyond the training ones. Moreover, the generalization theory underlying DLW is studied, validating its intrinsic transferability.

\end{abstract}



\section{Introduction}

Image denoising aims to restore the clean image $\X\in\mathbb{R}^{M}$ from the observed noisy image $\Y\in\mathbb{R}^{M}$. Over the past few decades, variational optimization methods for image denoising have been extensively studied. Mathematically, these methods can be summarized as follows:
\begin{equation}\label{opt-problem}
	\hat{\X}=\arg\min\limits_{\X}~\ell(\Y,\X) + \lambda R(\X),
\end{equation}
where $\hat{\X}\in\mathbb{R}^{M}$ is the restored image. There are two terms in the objective function. The data fidelity term $\ell(\Y,\X)$ measures the discrepancy between $\Y$ and $\X$. The regularization term $R(\X)$ makes the optimization problem well-posed. The trade-off parameter $\lambda$ balances the data fidelity and regularization terms in the objective function. The regularization term $R(\cdot)$ encodes the structure of the clean image, and the data fidelity term $\ell(\cdot, \cdot)$ encodes the noise statistics.

The incorporation of weight into the data fidelity term can improve the noise-removal capability of a denoising model. To clarify the functional principles of weight, we can formulate the model \eqref{opt-problem} as follows:
\begin{align}\label{weight-problem}
	\hat{\X}=\arg\min\limits_{\X}~\underbrace{\dfrac{1}{2}\|\W\odot(\Y - \X)\|^2}_{\ell(\Y,\X;\W)} + R(\X),
\end{align}
where ``$\|\cdot\|$" denotes the $\ell_2$ norm, ``$\odot$" denotes the Hadamard product, and elements of the weight $\W$ are non-negative. In the data fidelity term $\ell(\Y,\X;\W)$, $\W$ weights $\Y$ elementwise. For example, if a region $\Y(\Omega)$ of the noisy image $\Y$ is more heavily polluted than other regions, it should be less ``trustworthy" for constructing $\X$. Thus, it is reasonable to set $\W(\Omega)$ to be smaller, implying that $\X(\Omega)$ need not be close to $\Y(\Omega)$. This toy example shows that a reasonable $\W$ is deeply tied to noise. However, determining $\W$ according to noise is rather challenging when the noise pattern is complex, e.g., impulse noise, stripe noise, or a mixture of several patterns. Besides noise, $\W$ also balances the data fidelity term and the regularization term. Unlike the scalar trade-off parameter $\lambda$ in \eqref{opt-problem}, $\W$ provides a finer-grained, elementwise coupling, which is more implicit and poses greater evaluation challenges. 

Keeping $\W$ constant is a trivial approach; however this often limits the model’s performance. Computing a non-trivial $\W$ in existing image denoising methods mainly involves the following forms. a) $\W$ is set by an empirical formula which roughly follows the rule that $\W$ should be inversely proportional to the estimated noise intensity ``$|\Y - \X|$". For example, Wang et al. \cite{wang2020weighted} set the weight as $\exp(-\xi (\Y - \X)^2)$ and update the weight in each iteration when solving the denoising model. However, this scheme fails to capture the intrinsic characteristics of the weight.
b) Taking model \eqref{weight-problem} as the maximum a posteriori (MAP) problem, $\W$ is obtained from the negative log-likelihood ``$-\log p(\Y|\X)$". Consequently, this method requires making assumptions on the noise distribution. For example, Liu et al. \cite{liu2012weighted} assume that the noise distribution is the mixture of several distributions. In practice, they use Gaussian mixture noise and the weight is derived from the corresponding log-likelihood. Xu et al. \cite{xu2017multi} present a multi-channel WNNM model. The noise in each channel follows a Gaussian distribution, but the variances differ across channels. The weights on a single channel are the same and are set as the noise precision. Chen et al. \cite{chen2022color} enhanced this method by setting the weight as the combination of noise precision and the identity matrix.  Liu et al. \cite{li2015reweighted} assume that the image is corrupted by Poisson noise and use additive Gaussian noise to approximate the Poisson noise. The weight is derived from the negative log of the Gaussian distribution. c) $\W$ is also considered to be a parameter of a hypothetical noise distribution. It is calculated and iteratively updated via inference algorithms such as Expectation-Maximization (EM) and variational inference. For example, Cao et al. \cite{PMOEP} model noise with the mixture of exponential power distributions and use EM to solve the denoising model. Jiang et al. \cite{Adaptive-Hyperspectral-Mixed-Noise-Removal} assume that the noise follows a Gaussian mixture distribution and use EM to solve the corresponding MAP problem. Chen et al. \cite{chen2017denoising} also assume that the noise follows a Gaussian mixture distribution but the problem is related to variational inference and the weight is updated by the corresponding rule. d) There are also special cases for computing the weight for image denoising models. Li et al. \cite{li2010multiplicative} consider the Aubert–Aujol model for removing multiplicative noise. The weight in their model is related to image structures. They are iteratively updated via the rule derived from Euler–Lagrange equation from the energy function.

In summary, non-trivial weighting strategies in existing methods mainly depend on empirical formulation or specific hypothetical noise distributions and are usually model-specific. As the noise pattern becomes more complex, the implementation difficulty of these methods increases accordingly. Besides, the regularization term is usually ignored and only the noise information is considered, which would also limit the performance.

To address the issues, we consider predicting the weight $\W$ in the data fidelity term $\frac{1}{2}\|\W\odot(\Y - \X)\|^2$ automatically by a parameterized ``weight function". In this work, we propose a data-driven loss weighting (DLW) scheme to train a ``weight function" in the form of a simple neural network ``$h_\theta$", which we called DLWnet for short. The DLWnet takes the noisy image, which contains all image and noise information, as input and directly generates the weight as output. Unlike all previous works, we consider letting the denoising models of the form \eqref{weight-problem} themselves, along with the data, determine this weight function. There are two basic but very important facts to help to realize such an idea:
\begin{itemize}
	\item Different weights $\W$ yield different solutions to one model \eqref{weight-problem}; therefore, the best solution should imply the most suitable weight.
	\item Different choices of regularization term $R(\cdot)$ with the same weight would also yield different restored images $\hat{\X}$.
\end{itemize}
To address such issues, in this study we describe the implicit relationship between the restored image and its proper weight function by a bilevel optimization framework based on noisy/clean image training pairs. Specifically, such a bilevel framework is composed of two problems. The lower-level problem leverages multiple denoising models of the form \eqref{weight-problem} to generate the restored images by optimizing these models. Each of them shares the same weight $\W$ predicted by a weighting function $h_\theta$ that maps an image to its corresponding weight. The upper-level problem aims to optimize the network parameters $\theta$ by minimizing the discrepancy between the restored and corresponding ground-truth images. The above bi-level optimization process completes the training of $h_\theta$. We see that $h_\theta$ is essentially guided to learn to extract noise and regularization features with very specific physical meaning from models and training data. Then, $h_\theta$ can be readily used for estimating the weight of other unseen noisy images as well as for other regularizations. Subsequently, the predicted weight helps enhance the performance of a denoising model \eqref{weight-problem}, enabling it to adapt to complex noise. 

Note that denoising models with remarkable performance often have complicated regularizations and may be incompatible with training $h_\theta$ in the bilevel framework. Thus, we consider using several denoising models with simpler regularizations (namely, source models) to train $h_\theta$. This raises another question. What effect does the choice of the source models have on the performance of a target denoising model? To analyze such model-level generalization of $h_\theta$, we conduct several experiments and analyze the results. Besides, we have also preliminarily conducted a theoretical analysis of the generalization error from the source models to the target model based on the proposed model divergence, which helps to understand the intrinsic nature of the generalization behavior of $h_\theta$ across heterogeneous image denoising models. The learning theory on generalization has been extensively developed in transfer learning \cite{Boosting-for-transfer-learning,Transfer-Learning-using-Kolmogorov-Complexity-Basic-Theory-and-Empirical-Evaluations}, domain adaptation \cite{a-theory-of-learning-from-different-domain,Domain-Adaptation-in-Regression,A-New-PAC-Bayesian-Perspective-on-Domain-Adaptation,Risk-Bounds-for-Transferring-Representations-With-and-Without-Fine-Tuning}, domain generalization \cite{Generalizing-from-Several-Related-Classification-Tasks-to-a-New-Unlabeled-Sample,Domain-Generalization-via-Invariant-Feature-Representation,Domain-Generalization-by-Marginal-Transfer-Learning,Domain-Adversarial-Neural-Networks-for-Domain-Generalization}, and so on. The theories of these works are developed mainly based on a presumed common environment in which all tasks are sampled. The main differences between our analysis and other generalization theories are as follows. Firstly, we measure the individual task divergence rather than the environment, which is more pertinent to our problem. Secondly, existing generalization theories mainly focus on classification problems, while our work is, to the best of our knowledge, the first theoretical work on generalization in the field of image denoising.

In summary, the contribution of this study is as follows:
\begin{enumerate}
	\item  We propose a data-driven loss weighting (DLW) scheme to learn the weight function $h_\theta$ that predicts the weight for the data fidelity term in a variational model for image denoising. The proposed scheme does not require any empirical or hypothetical assumptions about noise, and using denoising models to determine the weight function.
	
	\item The learned $h_\theta$ by the proposed DLW scheme can directly predict weights for new noisy images, and the weight helps improve the denoising model's ability to remove the complex noise. Extensive experiments demonstrate the capability of our DLWnet to handle various complex noises.
	
	\item The learned $h_\theta$ can be readily applied to a new image denoising model that has different regularization terms. Substantial experiments on the source model combinations and their effects on heterogeneous target models show that $h_\theta$ has good generalization ability at the model level.
	
	\item Some preliminary theories are established to analyze the generalization error of DLW at the model level. The theories help to understand what effect the choice of source models has on the target model.
\end{enumerate}

The rest of this work is organized as follows. Sec. \ref{sec-proposed-method} presents our DLW scheme for weight prediction, together with discussions on the weight and a theoretical analysis of the generalization error. Sec. \ref{sec-experiments} reports experiments verifying the effectiveness of our DLWnet and its generalization ability on noise patterns and regularizations beyond those seen during training. Finally, we conclude the paper.

\section{Proposed Method}\label{sec-proposed-method}
\begin{figure*}[t]
	\centering
	\includegraphics[width=15cm]{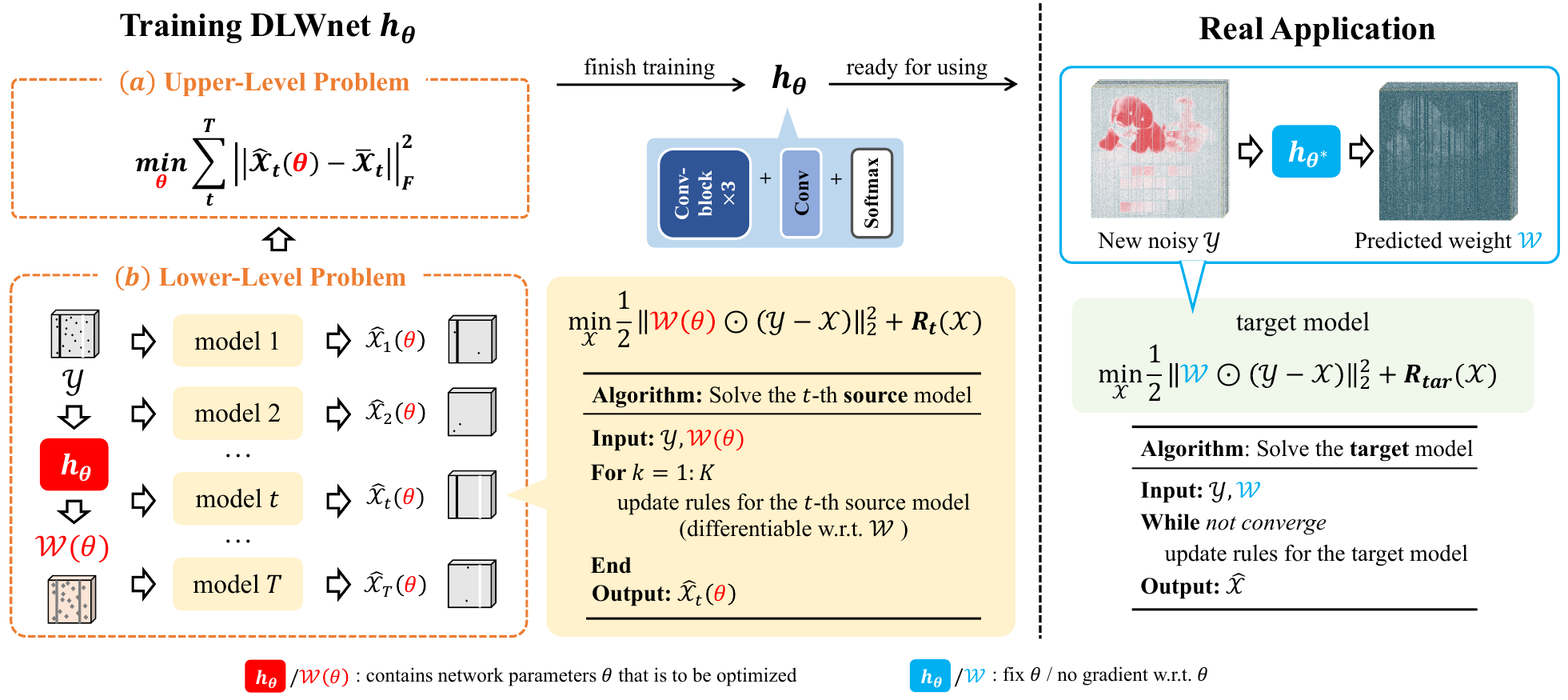}
	\caption{Overview of the proposed DLW scheme. The left part illustrates the training of the weight function $h_\theta$ (DLWnet). This process consists of a lower-level problem solving $T$ heterogeneous image denoising problems (all using the same $h_\theta$) and an upper-level problem minimizing the distance between the restored image $\hat{\X}(\theta)$ and the ground-truth image $\Xgt$. The right part shows the application of the trained $h_\theta$. In this stage, $h_\theta$ predicts the weight for a noisy image, which is then used in a target image denoising model, helping the model achieve better performance. }
	\label{fig-main}
\end{figure*}

As mentioned in the introduction, we want to move beyond empirical formulas or hypothetical noise distributions when designing the weight. To address the issue, we utilize a parameterized function to build a mapping from the noisy image $\Y$ to its weight $\W$, which is intended to reflect noise information:
\begin{align}\label{equ:W}
	\W = h_\theta(\Y).
\end{align}
Let $\theta$ denote the network parameters. The DLWnet $h_\theta$ takes the noisy image $\Y$ as input and outputs the weight $\W$ as the output. Correspondingly, the data fidelity term $\frac{1}{2}\|\W\odot(\Y - \X)\|^2$ can be rewritten as $\frac{1}{2}\|h_\theta(\Y)\odot(\Y - \X)\|^2$. In this section, we will introduce our DLW scheme to derive the weight function $h_\theta$ for image denoising models, including its training and practical application. A overview of this process is shown in \cref{fig-main}.

\subsection{Notations}
We use $\X/\Y$ and $\X \odot \Y$ to denote element-wise division and multiplication of $\X$ and $\Y$, respectively. Unless otherwise specified, the square operator $\X^2$ in this work denotes element-wise square. The $i$-th element of $\X$ is written as $\X(i)$. The norm $\|\X\|$ denotes the Frobenius norm of $\X$, i.e., $\|X\|=\sqrt{\sum_{i=1}^{M}\X(i)^2}$. The inner product of $\X$ and $\Y$ is denoted by $\langle \X, \Y \rangle$. The set $\{1,2, \dots ,N\}$ is denoted by $[N]$ for short.

\subsection{\texorpdfstring{Training DLWnet $h_\theta$ from multiple source models}{}}\label{sec-Auto-Wei-Sch}
\begin{figure*}[t]
	\centering
	\includegraphics[width=13cm]{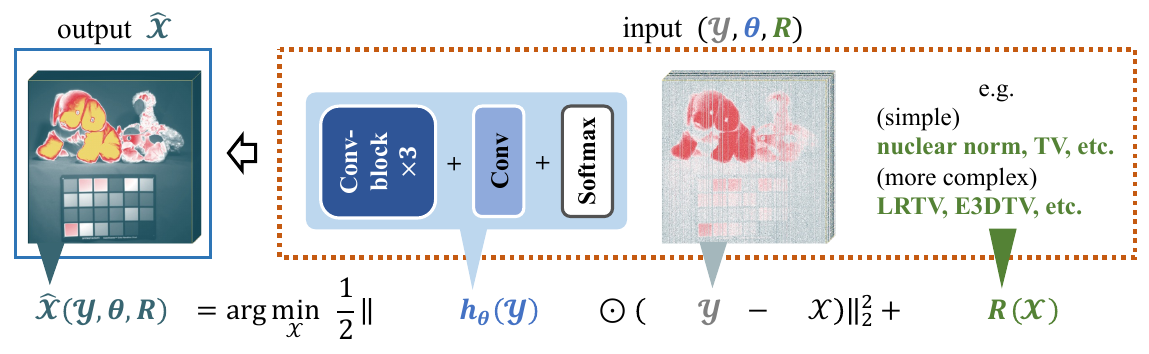}
	\caption{The solution $\hat{\X}$ to a denoising problem is implicitly related to three elements. They are the noisy image $\Y$, the network parameter $\theta$ (because $\W = h_\theta(\Y)$) and the regularization term $R(\X)$. This fact is the basis of the proposed DLW scheme.}
	\label{fig-mapping}
\end{figure*}

This section introduces the proposed DLW scheme. Combining \eqref{equ:W}, we can rewrite problem \eqref{weight-problem} as
\begin{align}\label{def-x}
	\hat{\X}(\Y,\theta,R)=\arg\min\limits_{\X}\dfrac{1}{2}\|h_\theta(\Y)\odot(\Y - \X)\|^2 + R(\X).
\end{align}
Here, the solution to the optimization problem \eqref{weight-problem}, i.e., the restored image $\hat{\X}$, is determined by three elements. Clearly, the noisy image $\Y$ is the fundamental element in determining $\hat{\X}$. The second element is the network parameter $\theta$. It first determines the weight $\W=h_\theta(\Y)$, which then influences the optimization results. The third element is the regularization term $R$ that distinguishes between different denoising models. Different denoising models also result in different restored images. Overall, the noisy image $\Y$, the network parameter $\theta$ and the model regularization $R$ collectively determine the final restored image $\hat{\X}$. Fig.~\ref{fig-mapping} briefly shows their relationship. Note that such mapping is implicit because it is obtained by solving an optimization problem. 

From \eqref{def-x}, it is clear that to train the network $h_\theta$, we need ``clean/noisy" training data and denoising models. Thus, a bi-level optimization framework is formulated as follows:

\begin{subequations}
	\begin{numcases}{}
		\min_\theta ~\dfrac{1}{T}\sum_{t=1}^T\dfrac{1}{N_t}\sum_{i_t=1}^{N_t}\ell_{up}\left( \hat{\X}_{t,i_t}(\theta), \Xgt_{t,i_t} \right), \label{multi-bilevel-up}\\
		\hat{\X}_{t,i_t}(\theta)=  \arg\min\limits_{\X}
		\dfrac{1}{2}\|h_\theta(\Y_{t,i_t})\odot(\Y_{t,i_t}-\X)\|^2+ R_t(\X), ~(t=1, \dots ,T). \label{multi-bilevel-down}
	\end{numcases}
\end{subequations}
Problem \eqref{multi-bilevel-up} and \eqref{multi-bilevel-down} are referred to as upper-level and lower-level problems, respectively.  There are $T$ different denoising models in the lower-level problem \eqref{multi-bilevel-down}. $R_t(\X)$ is the regularization term of the $t$-th denoising model. The training dataset is written as $\{(\Y_{t,i_t}, \Xgt_{t,i_t})\}_{i_t=1}^{N_t}{}_{t=1}^{T}$, where $\Y_{t,i_t}$ denotes the $i_t$-th noisy image for the $t$-th denoising model and $\Xgt_{t,i_t}$ is the corresponding ground-truth image. $N_t$ is the number of training pairs for the $t$-th model. For convenience, $\hat{\X}_{t,i_t}(\theta)$ denotes $\hat{\X}(\Y_{t,i_t}, \theta,  R_t)$, which means the solution to the $t$-th denoising model for the noisy image $\Y_{t,i_t}$. 

It should be noted that training $h_\theta$ using the proposed bi-level optimization framework \eqref{multi-bilevel-up}-\eqref{multi-bilevel-down} does not rely on prior knowledge of the weight $\W$. It completely allows the denoising models to decide the weight function $h_\theta$ because the output of $h_\theta$ plays the same role as the weights in the denoising model. Besides, the lower-level problem \eqref{multi-bilevel-down} produces the restored images from different denoising models using the same weight function $h_\theta$. Thus, $h_\theta$ is able to integrate noise information provided by multiple models. This should be important because $h_\theta$ trained by a single denoising model would inevitably be biased toward that model. Then, in the upper-level problem \eqref{multi-bilevel-up}, the network parameter $\theta$ is optimized by minimizing the distance between $\hat{\X}$ and $\Xgt$, which carries the ``weighting rule" from $h_\theta$.

To solve the bilevel optimization problem \eqref{multi-bilevel-up}-\eqref{multi-bilevel-down}, we utilize an efficient method called ``unrolling". The idea of unrolling is to apply an algorithm to solve the lower problem with a fixed number of iterations. The entire process of iteration replaces the lower-level problem. Consequently, the algorithm's output serves as an approximation of the true solution $\hat{\X}(\theta)$. An example using Alternating Direction Method of Multipliers (ADMM) \cite{ADMM} is summarized in \cref{algo-weight}. Details can be found in \cref{sec:append-algo}. There are two points of concern for the lower-level problem. First, we need to set an appropriate number of iterations because the algorithm typically requires many steps to converge, and the computational graph of each step must be stored. An excessive number of steps not only increases computational resource requirements but also compromises efficiency. Second, each updating rule in the algorithm should be differentiable with respect to (w.r.t.) its variables so that the final $\hat{\X}$ is differentiable w.r.t. $\W$ and gradient-descent-based methods can be used for the upper-level problem. Thus, we tend to choose simple denoising models (source models) when designing the lower-level problem for more efficient training. For example, denoising models with nuclear norm or total variation norm as regularization term usually have closed-form solutions for the corresponding $\X$-subproblem (i.e., the update for $\X$) and are relatively easy to optimize. For the upper-level problem \eqref{multi-bilevel-up}, general methods for optimizing neural networks can be used, e.g., Adam \cite{ADAM}.  By chain rule, the gradients w.r.t. $\theta$ backpropagate through the iteration process. Specifically, let $\X_{k} = g_{k}[\X_{k-1}; h_\theta(\Y)]$ denote the entire $k$-th step of \cref{algo-weight}\footnote{The complete expression for $g_{k}$ is $g_{k}[\X_{k-1};h_\theta(\Y)] = \arg\min\limits_{\X}\mu/2\left\|\X - (\Z_{k} - \Gamma_{k-1}/\mu)\right\|^2 + R(\X)$, where $\Z_{k} = (\W^2\odot\Y + \mu\X_{k-1} + \Gamma_{k-1})/(\W^2 + \mu)$ and $\W = h_\theta(\Y)$. The total number of iterations is $K$}. The lower-level problem then is replaced by
\begin{align}\label{equ:approx-x}
	\hat{\X}_{K}(\theta) = g_{K}[g_{K-1}[ \dots  g_1[\X_0;h_\theta(\Y)];h_\theta(\Y)]; h_\theta(\Y)]
\end{align} 
Let $H_k^{(\X)} = \frac{\nabla g_k}{\nabla \X_{k-1}}$ and $H_k^{(\W)} = \frac{\nabla g_k}{\nabla h_\theta(\Y)}\frac{\nabla h_\theta(\Y)}{\nabla \theta}$. The gradient w.r.t. network parameter $\theta$ can be roughly written as
\begin{align}\label{equ:gradient}
	\dfrac{\nabla \ell_{up}(\hat{\X}_{K}(\theta),\Xgt)}{\nabla \theta}=\dfrac{\nabla \ell_{up}(\hat{\X}_{K},\Xgt)}{\nabla \hat{\X}_{K}}\underbrace{\left[ H_K^{(\X)}\left[ H_{K-1}^{(\X)}\left[ \dots  H_1^{(\X)} + H_1^{(\W)} \right] + H_{K-1}^{(\W)} \right] + H_K^{(\W)}\right]}_{\text{through iterations}}
\end{align}
In practice, the above gradient is calculated using PyTorch's Autograd.

\begin{algorithm}[t]
	\renewcommand{\algorithmicrequire}{\textbf{Input:}}
	\renewcommand{\algorithmicensure}{\textbf{Initialization:}}
	\renewcommand{\algorithmicreturn}{\textbf{Output:}}
	\caption{An example of ADMM algorithm to solve denoising problem \eqref{weight-problem}}
	\label{algo-weight}
	\begin{algorithmic}[1]
		\REQUIRE noisy image $\Y$, weight $\W$.
		\ENSURE $\X_0, \Z_0, \Gamma_0, \mu$ 
		\FOR{$k = 1:K$}
		\STATE $\Z_{k} = \dfrac{\W^2\odot\Y + \mu\X_{k-1} + \Gamma_{k-1}}{\W^2 + \mu}$ 
		\STATE $\X_{k} = \arg\min\limits_{\X}\dfrac{\mu}{2}\left\|\X - \left(\Z_{k} - \dfrac{\Gamma_{k-1}}{\mu}\right)\right\|^2 + R(\X)$ 
		\STATE $\Gamma_k = \Gamma_{k-1} + \mu(\X_{k} - \Z_{k})$ 
		\ENDFOR \\
		\RETURN $\hat{\X}=\X_K$
	\end{algorithmic}
\end{algorithm}

\subsection{\texorpdfstring{Applications of DLWnet $h_\theta$}{}}\label{sec:appplication}
In the training phase, $h_\theta$ learns how to extract noise information from a noisy image and generate the corresponding weight. The weight is to be used for a specific image denoising model to better remove complex noise. The implementation of a trained $h_\theta$ is straightforward. Specifically, the data fidelity term $\|h_\theta(\Y)\odot(\Y - \X)\|^2$ is incorporated into a target image denoising model. The resulting general form for applying $h_\theta$ is as follows:
\begin{align}\label{target-model}
	\min\limits_{\X}~ & \dfrac{1}{2}\|\W\odot(\Y - \X)\|^2 + \tau R_{tar}(\X), \nonumber\\
	& \mathrm{where}~\W = h_\theta(\Y).
\end{align}
Note that the target model does not necessarily need to be one of the source models. If the goal is to achieve better performance, denoising models usually need to be carefully designed and often involve complex regularizations. However, training convenience and efficiency are the main criteria for selecting source models. The denoising problem \eqref{target-model} is solved using general optimization problems using suitable algorithms, e.g., proximal gradient (PG)~\cite{shi2015proximal}, half quadratic splitting (HQS)~\cite{sun2020learning}, and ADMM \cite{ADMM}. 

There are also some special cases of problem \eqref{target-model}. For example, when the deep image prior (DIP) \cite{DIP, Noise2noise, S2DIP} is implemented as an implicit image regularization, problem \eqref{target-model} can be modified into the following form:
\begin{align}\label{wdip}
	\min_\eta~ & \|\W\odot\left(g_\eta(\mathcal{Z}) - \Y\right)\|^2 + \tau R(g_\eta(\mathcal{Z})), \nonumber \\
	& \mathrm{where}~\W = h_\theta(\Y).
\end{align}
In problem \eqref{wdip}, $g_\eta(\cdot)$ is a deep neural network (DNN) that takes random noise $\mathcal{Z}$ as input and outputs the denoised image. Another example is the plug-and-play (PnP) framework \cite{PnP2}. Specifically, PnP also treats $R_{tar}(\cdot)$ as an implicit term and applies a suitable algorithm to solve problem \eqref{target-model}. During this process, a pre-trained Gaussian denoiser serves as the solver for the subproblem associated with the implicit regularization $R_{tar}$.

Compared with previous methods, our $h_\theta$ can directly generate the appropriate weight and learn the underlying weighting rules from data and models. It should also be noted that our $h_\theta$ is not designed for one specific model. It can be easily implemented in models with different regularization terms.

\subsection{Theoretical analysis}\label{sec-app-theory}
When the source model is the same as the target model, we obtain the ideal $h_\theta$ for the target model. On the contrary, there exists an error when we apply $h_\theta$ if the source and the target models are different. Specifically, such ``error" can be described as the difference between the average performance of the utilized $h_\theta$ and that of the ideal $h_\theta$ applied to the target model. We use ``$\Eg$" to denote this error, which is also called the generalization error. In this section, we present a preliminary estimate of the upper bound related to the generalization error.

More generally, suppose there are $S$ target models of the form in problem \eqref{weight-problem}. They are written as
\begin{align}\label{new-task}
	\min_{\X}\dfrac{1}{2}\|\W\!\odot\!(\Y\!- \!\X)\|^2\! +\!  R_t(\X),~t=T \!+ \!1, \dots ,T \! + \! S,
\end{align}
For convenience, the subscript character used in \eqref{multi-bilevel-down} for $T$ source models is also used for the $S$ target models, with the indices for target models starting from $T+1$ and ending at $T+S$. Then our main result concerning the error $\Eg$ resulting from transferring a learned $h_\theta$ to target models is presented in the following theorem. 

Let $\bY$ and $\bX$ denote the observed image space and the original image space, respectively. We have that $\Y\in\bY$ and $\Xgt\in\bX$. We assume that the observed image space $\bY$ and the original image space $\bX$ are bounded within a cube in $\mathbb{R}^M$, i.e., $\bY,\bX\subset [-B_d, B_d]^M$ for some positive constant $B_d$. Let $\D$ represent a joint distribution defined on $\bY\times\bX$. To be specific, the joint probability density function (PDF) $p(\Y,\X)$ is usually decomposed as
\begin{align*}
	p(\Y,\X) = p(\Y|\X)p(\X),
\end{align*}
where $p(\X)$ represents the PDF of the clean images $\X$, and $p(\Y|\X)$ describes the process of adding noise from $\X$ to $\Y$. The joint data distribution $\D$ is independent of the choice of source and target models, since images and the presence of noise in images are objective facts.\footnote{Note that the joint distribution of restored image and noisy image is relevant to the choice of source and target models, which is not considered in this work.}

We first present a lemma that guarantees the existence of an implicit mapping from $\Y$, $\W$ and $R$ to $\hat{\X}$, i.e., $\hat{\X}(\Y,\theta,R)$ as defined in \eqref{def-x}. The proof is provided in \cref{sec:appendix}. 
\begin{lemma}\label{theo:lemma1}
	Suppose that $\W\in[\varepsilon, B_w]^M$ for some positive constants $0<\varepsilon\leq B_w$, and that $R$ is proper, closed and convex. Then the solution to model \eqref{weight-problem} is unique. 
\end{lemma}

In the remainder of this work, we also use $f_{R}^{h}(\Y)$ to denote the restored image $\hat{\X}(\Y,\theta, R)$ and refer to ``$f_{R}^h$" as a source model. The $T$ source models $\{f_{ R_t}^h\}_{t=1}^T$ in the lower-level problem \eqref{multi-bilevel-down} can be aggregated into a set:
\begin{align*}
	\mathcal{F}_{\mathcal{H}}:=\left\{\left(f_{R_1}^h, \dots ,f_{R_T}^h\right)|h\in\mathcal{H}\right\}.
\end{align*}
Each element of $\mathcal{F}_{\mathcal{H}}$ is a $T$-tuple of source models $(f_{R_1}^h, \dots ,f_{R_T}^h)$. $\mathcal{H}$ is a collection of networks whose architectures are identical but whose parameter values differ. The network $h$ is shared among $T$ models within an element of $\mathcal{F}_{\mathcal{H}}$, and the differences across elements lie only in the network's parameter values.

The derivation of the generalization error upper bound consists of three parts. The first part evaluates the training error upper bound. The second part evaluates the difference between the source models and the target models. Finally, the generalization error upper bound can be decomposed into the sum of the training error and the model difference. In the remainder of this section, we present the derivation of each part in turn.

\subsubsection{Training Error Estimation}
The objective function of upper-level problem \eqref{multi-bilevel-up} can be equivalently written as 
\begin{align}\label{equ:append-ERM}
	\min_{h\in\mathcal{H}}~\dfrac{1}{T}\sum_{t=1}^T\dfrac{1}{N_t}\sum_{i_t=1}^{N_t}\ell_{up}\left( f_{ R_t}^h(\Y_{t,i_t}), {\Xgt}_{t,i_t} \right),
\end{align}
which is the \textit{empirical risk minimization} (ERM) problem \cite{FoML} of $h\in\mathcal{H}$ on training dataset $\mathbf{S} = \{(\Y_{t,i_t}, {\Xgt}_{t,i_t})\}_{i_t=1}^{N_t}{}_{t=1}^{T}$. Therefore, our training loss, or empirical risk, refers to
\begin{align*}
	\hat{\R}_{tr}(h) := \dfrac{1}{T}\sum_{t=1}^T\dfrac{1}{N_t}\sum_{i_t=1}^{N_t}\ell_{up}\left( f_{ R_t}^h(\Y_{t,i_t}), {\Xgt}_{t,i_t} \right).
\end{align*}
And the expected risk is
\begin{align*}
	\R_{tr}(h) := \dfrac{1}{T}\sum_{t=1}^T\mathbb{E}_{(\Y,\X)\sim\D}\left[\ell_{up}\left( f_{ R_t}^h(\Y), \X \right)\right].
\end{align*}
The expected risk takes expectations over joint distribution $\D$. Correspondingly, the ``best" DLWnets are defined as
\begin{align*}
	 \hat{h}_{tr} := \arg\min_{h\in\mathcal{H}}\hat{\R}_{tr}(h),~ h^*_{tr} := \arg\min_{h\in\mathcal{H}}\R_{tr}(h),
\end{align*}
where $\hat{h}_{tr}$ is the final trained DLWnet under the bi-level framework \eqref{multi-bilevel-up}-\eqref{multi-bilevel-down} with a finite number of training samples, and $h^*_{tr}$ is the ideal DLWnet derived from the underlying data distribution $\D$. Since the optimal $h^*_{tr}$ is intractable, the training error $\Etr$, which measures the ``closeness'' between $\hat{h}_{tr}$ and $h^*_{tr}$, is then defined as:
\begin{align}
	\Etr:=\R_{tr}(\hat{h}_{tr}) - \R_{tr}(h^*_{tr}).
\end{align}

Next, we estimate the upper bound of $\Etr$. To
achieve this goal, we first present some mild assumptions on $\mathcal{H}$ and $\ell_{up}$, as well as on the properties of $f_{R}^h$.
\begin{assump}\label{theo:assump-h}
	$\forall h\in\mathcal{H}, \Y\in\mathbb{Y}$, we have that $h(\Y)\in[\varepsilon, B_w]^M$, where $0<\varepsilon\leq B_w$.
\end{assump}
\begin{assump}\label{theo:assump-loss}
	$\ell_{up}$ satisfies $|\ell_{up}(\hat{\X},\X)|\leq B_l$ for all $\hat{\X}, \X\in[-B_d, B_d]^M$. And $\ell_{up}(\hat{\X}, \X)$ is $L_n$-Lipschitz w.r.t. $\hat{\X}$ for all $\X\in[-B_d, B_d]^M$.
\end{assump}
\begin{remark}
	The mean squared error (MSE) $\frac{1}{M}||\hat{\X} - \X||^2$ is bounded by $4B_d^2$ and is $\frac{4B_d}{\sqrt{M}}$-Lipschitz w.r.t. $\hat{\X}\in[-B_d, B_d]^M$ for all $\X\in[-B_d, B_d]^M$.
\end{remark}
\begin{lemma}\label{theo:lemma2}
	Suppose that $R$ is proper, closed and convex, and that \cref{theo:assump-h} holds. Then $\forall \Y\in\bY$ and $\forall h_1,h_2\in\mathcal{H}$, we have that
	\begin{align}
		\left\|f_{R}^{h_1}(\Y) - f_{R}^{h_2}(\Y) \right\| \leq L_H \left\| h_1(\Y) - h_2(\Y) \right\|.
	\end{align}
	The Lipschitz constant $L_H$ only depends on $B_d$, $\varepsilon$ and $B_w$.
\end{lemma}
\cref{theo:lemma2} shows the Lipschitz continuity of $f_{R}^h$ w.r.t. weight $h$, which means that if two weights $h_1(\Y)$ and $h_2(\Y)$ are sufficiently close, the corresponding restored images $f_{R}^{h_1}(\Y)$ and $f_{R}^{h_2}(\Y)$ are also close enough. 

We use Gaussian complexity to capture the richness of a set of functions by measuring how well the functions in this set can fit random noise. 
\begin{definition}[Empirical Gaussian complexity \cite{Rademacher-and-Gaussian-Complexities}]
	Let $\mathcal{F}$ be a set of functions mapping from $\mathcal{Z}$ to the inverval $[a,b]$, and let $S=(z_1, \dots ,z_m)$ be a fixed sample of size $m$ with elements in $\mathcal{Z}$. Then, the empirical Gaussian complexity of $\mathcal{F}$ w.r.t. the sample $S$ is defined as:
	\begin{align*}
		\hat{\mathfrak{G}}_{S} (\mathcal{F}):=\mathop{\mathbb{E}}\limits_{\bm{g}}\left[ \sup\limits_{f\in\mathcal{F}}\dfrac{1}{m}\sum_{i=1}^{m}g_{i} f(z_i) \right],~g_i\sim\mathcal{N}(0,1),
	\end{align*} 
	where $\bm{g}=(g_1, \dots ,g_m)^T$.
\end{definition}
The empirical Gaussian complexity of $\mathcal{F}_\mathcal{H}$ is then defined as
\begin{align*}
	\hat{\mathfrak{G}}_\mathbf{S}(\mathcal{F}_\mathcal{H}) & := \mathop{\mathbb{E}}\limits_{\bm{G}}\left[ \sup\limits_{h\in\mathcal{H}}\dfrac{1}{T}\sum_{t=1}^{T}\dfrac{1}{N_t}\sum_{i_t=1}^{N_t}\sum_{m=1}^{M}g_{t,i_t,m} f_{ R_t}^h(\Y_{t,i_t})_m \right],~ g_{t,i_t,m}\sim\mathcal{N}(0,1),\\
\end{align*}
where $\bm{G} = (g_{1,1,1}, \dots ,g_{t,i_t,m}, \dots ,g_{T,N_T,M})^T$ and $f_{ R_t}^h(\cdot)_m$ is the $m$-th element of $f_{ R_t}^h(\cdot)$.

Based on \cref{theo:assump-h} and \cref{theo:lemma2}, we can decouple $\hat{\mathfrak{G}}_\mathbf{S}(\mathcal{F}_\mathcal{H})$ in terms of $\hat{\mathfrak{G}}_\mathbf{S}(\mathcal{H})$ via an analogous chain rule \cite{Rademacher-and-Gaussian-Complexities}. 
We now present our main conclusion regarding the training error $\Etr$.

\begin{theorem}[Training error]\label{theo:train}
	Suppose that \cref{theo:assump-h} and \cref{theo:assump-loss} hold. $\{R_t\}_{t=1}^T$ are proper, closed and convex. Then, for any $\delta>0$, with probability at least $1-\delta$, the following inequality holds:
	\begin{align}\label{train-error-bound}
		\Etr \leq 6L_nL_H\hat{\mathfrak{G}}_S(\mathcal{H}) + \dfrac{6B_l}{T}\sqrt{\sum_{t=1}^T\dfrac{1}{N_t}}\sqrt{\dfrac{\log\frac{2}{\delta}}{2}}.
	\end{align}
\end{theorem}

The first term of \eqref{train-error-bound} is about the Gaussian complexity of $\mathcal{H}$. It implies that the range of the network’s outputs can affect training accuracy. For example, if the network only outputs a constant (i.e., the weight is uniform), the Gaussian complexity $\hat{\mathfrak{G}}_{S}(\mathcal{F})$ equals zero. The second term of \eqref{train-error-bound} is mainly about the sizes of the training data sets, i.e., the $N_t$ values. To lower the value of this term, we need to increase the $N_t$ values. This means that more training data should be used. Imagine that we have an infinite amount of training data; then each $N_t$ goes to infinity and the second term becomes zero. Besides, if one source model lacks sufficient number of training samples, i.e., $N_j$ is too small for some $j$, the value of the second part then becomes very large even if we enlarge $N_t$ for $t\neq j$. In practice, lacking sufficient training data for some models means that the information from these models is not well-learned by the network, which in turn increases the uncertainty of training.

\subsubsection{Generalization Error Estimation}
Similar to the training error, we establish the test error for $S$ target models \eqref{new-task} as:
\begin{align*}
	\R_{te}(h) := \dfrac{1}{S}\sum_{t=T+1}^{T+S}\mathbb{E}_{(\Y,\X)\sim\D}\left[ \ell_{up}\left( f_{ R_t}^h(\Y), \X \right) \right],
\end{align*}
where $\{R_t\}_{t=T+1}^{T+S}$ are usually different from $\{R_t\}_{t=1}^T$. The corresponding ``best" DLWnet for the target models can then be defined as:
\begin{align*}
	h^*_{0} = \arg\min_{h\in\mathcal{H}}\R_{te}(h).
\end{align*}
The generalization error is thus defined as:
\begin{align*}
	\Eg := \R_{te}(\hat{h}_{tr}) - \R_{te}(h^*_0).
\end{align*}
Specifically, $\R_{te}(\hat{h}_{tr})$ represents the test error of the trained DLWnet $\hat{h}_{tr}$, and $\R_{te}(h^*_0)$ is the infimum of the test error over the set $\mathcal{H}$. To estimate $\Eg$, our main challenge lies in measuring the divergence between the source and target tasks, i.e., in calculating $\R_{te}(h) - \R_{tr}(h)$. 

The major difficulty of calculating $\R_{te}(h)-\R_{tr}(h)$ comes from the implicit form of $f_{R}^h$. We thus seek an approximation of $f_{R}^h$ to ease the calculation using the modified gradient step \cite{proximal-algorithm}. Specifically, by taking the derivative of the objective function in \eqref{weight-problem} w.r.t. $\X$, we obtain that the solution $\hat{\X}$ should satisfy:
\begin{align*}
	 \W^2\odot(\hat{\X}-\Y) + \nabla R(\hat{\X}) & = 0 , \nonumber \\
	\Rightarrow  \hat{\X} + \dfrac{1}{\W^2}\odot \nabla R(\hat{\X}) & = \Y , \nonumber\\
	\Rightarrow  \left(\I + \dfrac{1}{\W^2}\odot\nabla R\right)(\hat{\X}) & = \Y, \nonumber \\
	\Rightarrow  \hat{\X} & = \left(\I + \dfrac{1}{\W^2}\odot\nabla R\right)^{-1}(\Y), \label{resolvent}
\end{align*}
where $\I$ represents the identity mapping. The operator $(\I + g)^{-1}$ is known as the resolvent operator \cite{equilibrium}. If $g$ is nonlinear, its resolvent form is still implicit and difficult to analyze. Therefore, we approximate Eq. \eqref{resolvent} by using a linear approximation of $\nabla R$ \cite{proximal-algorithm, equilibrium}:
\begin{align}\label{appro-prox}
	\left(\I + \dfrac{1}{\W^2}\odot\nabla R\right)^{-1} \approx \I - \dfrac{1}{\W^2}\odot\nabla R.
\end{align}
In the remainder of this section, we analyze the generalization error using the following approximated form of $f_{R}^h$:
\begin{align}\label{f-approx}
	\bar{f}_{R}^h(\Y) := \left(\I - \dfrac{1}{h^2}\odot\nabla R\right)(\Y) = \Y - \dfrac{1}{h(\Y)^2}\odot\nabla R(\Y).
\end{align}
Suppose that $\nabla R(\cdot)$ is bounded on $\bY$ for some constant $\bar{B}_y$, i.e., $\nabla R(\Y)\in [-\bar{B}_y, \bar{B}_y]^M$. We have that
\begin{align*}
	\left\|\bar{f}_{R}^{h_1}(\Y) - \bar{f}_{R}^{h_2}(\Y) \right\| & = \left\|\left( \dfrac{1}{h_1(\Y)^2} - \dfrac{1}{h_2(\Y)^2} \right)\odot \nabla R(\Y)\right\| \\
	& \leq \dfrac{2B_w\bar{B}_y}{\varepsilon^4}\|h_1(\Y) - h_2(\Y)\|
\end{align*}
Thus, $\bar{f}_{R}^h$ is also Lipschitz continuous w.r.t. $h$ with Lipschitz constant $\bar{L}_H := 2B_w\bar{B}_y/\varepsilon^4$. By utilizing Eq. \eqref{f-approx}, we first present a lemma to estimate $\vert \R_{te}(h)-\R_{tr}(h) \vert$.
\begin{lemma}\label{lemma3}
	Suppose that the mapping $\bar{f}_{R}^h$ is of the form $\I - \frac{1}{h^2}\odot\nabla R$ and $\ell_{up}$ is the MSE loss. For any $h\in\mathcal{H}$, we have
	\begin{align*}  
		\vert\R_{te}(h) - \R_{tr}(h)\vert \leq \mathbb{E}_{(\Y,\X)\sim\D}\left\{\dfrac{4B_d}{\sqrt{M}\varepsilon}A_1 + \dfrac{1}{M\varepsilon^2}A_2\right\},
	\end{align*}
	where
	\begin{align}
		A_1 & =  \left\| \left(\dfrac{1}{T}\sum_{t=1}^T\nabla R_t - \dfrac{1}{S}\sum_{t=T+1}^{T+S}\nabla R_t\right)(\Y) \right\|, \label{lemma3-A1} \\
		A_2 & = \left|\dfrac{1}{T}\sum_{t=1}^T\left\| \left( \nabla R_t \right)(\Y)\right\|_2^2 - \dfrac{1}{S}\sum_{t=T+1}^{T+S}\left\| \left( \nabla R_t \right)(\Y)\right\|_2^2\right|. \label{lemma3-A2}
	\end{align}
\end{lemma}

\cref{lemma3} shows that the divergence $\vert\R_{te}(h) - \R_{tr}(h)\vert$ is bounded by quantities involving the regularization gradients. The first term $A_1$ of the upper bound compares the average difference between the regularization gradients of the source and target tasks, while the second term $A_2$ compares the same average difference but in terms of the squared norms of the regularization gradients.

Based on \cref{lemma3}, we can directly estimate the generalization error $\Eg$ as follows:
\begin{theorem}[Generalization error]\label{theorem-ge}
	Suppose that \cref{theo:assump-h} and \cref{theo:assump-loss} hold,  $\{R_t\}_{t=1}^T$ and $\{R_t\}_{t=T+1}^{T+S}$ are proper, closed, and convex, $\bar{f}_{R}^h$ is of the form $\I - \frac{1}{h^2}\odot \nabla R$, and $\ell_{up}$ is the MSE loss. Then for any $\delta>0$, with probability at least $1-\delta$, we have
	\begin{align}\label{gene-bound}
		\Eg \leq 6L_n\bar{L}_H\hat{\mathfrak{G}}_S(\mathcal{H}) + \dfrac{6B_l}{T}\sqrt{\sum_{t=1}^T\dfrac{1}{N_t}}\sqrt{\dfrac{\log\frac{2}{\delta}}{2}} + \mathbb{E}_{(\Y,\X)\sim\D}\left\{\dfrac{8B_d}{\sqrt{M}\varepsilon}A_1 + \dfrac{2}{M\varepsilon^2}A_2 \right\},
	\end{align}
	where $A_1$ takes the form of \eqref{lemma3-A1} and $A_2$ takes the form of \eqref{lemma3-A2}.
\end{theorem}
The upper bound in \eqref{gene-bound} contains two parts. The leading two terms represent the aforementioned upper bound on the training error from \cref{theo:train}, which constrains the distance between $\hat{h}_{tr}$ and $h^*_{tr}$ and reveals how well $\hat{h}_{tr}$ is learned. The last term is the primary component in evaluating the generalization error, as it measures the diversity between the source and target models. Both $A_1$ and $A_2$ involve regularization gradient terms\footnote{based on the approximate form of the solution to denoising problem \eqref{weight-problem}, as the optimization form itself is intractable for analysis.} and each of them reveals one kind of ``model divergence" between source and target models. Specifically, $A_1$ directly calculates the difference between the averaged source regularization gradients and the averaged target regularization gradients. $A_2$ first calculates the norm of each regularization gradient at $\Y$, and then calculates the difference between the averaged norm values of the source models and those of the target models. We believe that $A_1$ mainly reflects the overall regularization content of the source and target models respectively, while $A_2$ provides a more detailed distinction among the regularizations within each model.

In the special case where there is only one source model and one target model ($T=1$ and $S=1$), an estimation of the generalization error without using the approximation form \eqref{appro-prox} can be derived as follows:
\begin{theorem}[Generalization error: single source and single target model]\label{theo:generalization-121}
	Suppose that \cref{theo:assump-h} and \cref{theo:assump-loss} hold, the source model regularization $R_1$ and the target model regularization $R_2$ are proper, closed, and convex, and $\ell_{up}$ is the MSE loss. Then for any $\delta>0$, with probability at least $1-\delta$, we have
	\begin{align}\label{gene-bound-single}
		\Eg \leq 6L_n L_H\hat{\mathfrak{G}}_S(\mathcal{H}) + \dfrac{6B_l}{\sqrt{N}}\sqrt{\dfrac{\log\frac{2}{\delta}}{2}} + \dfrac{8B_d\sqrt{2B_r}}{\varepsilon\sqrt{M}},
	\end{align}
	where $B_r = \max_{\X\in \mathbb{X}}|R_1(\X) -  R_2(\X)|$.
\end{theorem}

\begin{remark}
	When the target and source models are the same, the last terms of \eqref{gene-bound} and \eqref{gene-bound-single} reach their minimum value, i.e., zero, which means that there is no error introduced by the model discrepancy.
\end{remark}

In Sec.~\ref{sec-expe-simple-target}, we show some connections between Theorem \ref{theorem-ge} and simple experiments. However, it is also fair to say that the real applications are much more complex and include many other factors that may influence the generalization behavior of $h_\theta$, such as training efficiency. Theorem \ref{theorem-ge} provides a preliminary perspective for understanding the model-level generalization error of the proposed DLW.

\section{Experimental results}\label{sec-experiments}

We then conduct experiments to verify the effectiveness of our proposed DLW scheme for image denoising models. For convenience, the denoising models implemented with the DLWnet $h_\theta$ are prefixed with ``DLW-". 

\subsection{Setting of source models}
Three basic denoising models are used as source models to train DLWnet. Their regularization terms are nuclear norm (NN)~\cite{LRMR}, spatial total variation norm (TV)~\cite{SSTV}, and spectral total variation norm (TVS)~\cite{SSTV}. The NN can enforce the bands of a image to lie in a low-dimensional subspace. TV and TVS characterize the smoothness property in the spatial and spectral dimensions, respectively. These regularizations are widely used for image denoising. They are also relatively easy to optimize. ADMM (shown in Algorithm \ref{algo-weight}) is used to solve them due to its fast convergence and simple formulation. Algorithm details are included in \cref{sec-appendix-algodetails}.

In the training framework \eqref{multi-bilevel-up}-\eqref{multi-bilevel-down}, the above three denoising models can generate a total of seven different combinations for the lower-level problem \eqref{multi-bilevel-down} to train the DLWnet, as listed in \cref{tab-tlw-type}.
The abbreviations ``N", ``T" and ``TS" mean using DLW-NN, DLW-TV and DLW-TVS as single source models, respectively. Similarly, the abbreviation ``N+T" means that DLW-NN and DLW-TV are used to train the DLWnet, and thus the number of models $T$ in \eqref{multi-bilevel-down} is two. Consequently, we derive seven different DLWnets from the seven model combinations.

\begin{table}[t]
	\centering
	\fontsize{8.5}{9.5}\selectfont
	\renewcommand{\arraystretch}{1.15}
	\caption{Setting of source models: There are seven possible combinations. Each of them could train an individual DLWnet $h_\theta$.}
	\label{tab-tlw-type}
	\begin{tabular}{M{3cm} | c  c   c  c  c  c  c}
		\wline
		& \multicolumn{7}{c}{corresponding trained DLWnet names} \\
		\hline
		source models $\downarrow$ &   N   &    T   &   TS   &   N+T   &   N+TS   &   T+TS   & N+T+TS   \\
		\wline
		DLW-NN      & \checkmark &            &       & \checkmark & \checkmark &&\checkmark     \\
		
		DLW-TV      &            & \checkmark &        & \checkmark &&\checkmark& \checkmark      \\
		
		DLW-TVS     &            &            & \checkmark  & &\checkmark&\checkmark&\checkmark\\
		\Xhline{0.8pt}
	\end{tabular}
\end{table}

\subsection{Details of training and testing setting}
Two kinds of images are considered: hyperspectral images (HSI) and color images. For HSI, the CAVE dataset\footnote{\url{https://cave.cs.columbia.edu/repository/Multispectral}} \cite{CAVE} is used for training. Specifically, we randomly select 20 images to generate the training pairs. We crop 2500 overlapping patches of size $64\times64\times31$. After rotating and flipping, the total number of training patch pairs is increased to 20000. Additionally, six image datasets are used for testing, including the remaining 10 images in the CAVE dataset; 10 images of size $512\times512\times31$ from ICVL dataset \cite{ICVL}; the Washington DC Mall data\footnote{\url{https://engineering.purdue.edu/~biehl/MultiSpec/hyperspectral.html}} of size $200\times200\times152$; the PaviaU data\footnote{\url{https://www.ehu.eus/ccwintco/index.php/Hyperspectral_Remote_Sensing_Scenes}} of size $340\times340\times70$; and the Urban data \cite{Urban} and Indian Pines data. The first four testing datasets are used for synthetic image denoising experiments, and the last two datasets are used for real image denoising. For color images, the BSDS dataset \footnote{\url{https://www2.eecs.berkeley.edu/Research/Projects/CS/vision/bsds/}} is used for training and testing. The images used for training are cropped into small patches of size $64\times 64\times 3$ and the total number of training patches is 32000. Ten images from the rest of the dataset are used for testing.

First, we generate five kinds of complex noise for synthetic experiments. Note that only the noise from case 1 is used to generate the paired ``clean/noisy" training patches. In the testing stage, the pairs generated by all five types of noise are used. The details of the noisy image generation process are as follows:

\textbf{Case 1} (\textit{Gaussian + Impulse}): Each band in the image is corrupted with Gaussian noise, and the noise level is uniformly selected from the range [10,70]. The number of bands corrupted with additional impulse noise is 10, 10, 40, and 20 for CAVE, ICVL, DC, and PaviaU, respectively. The impulse noise ratio ranges from 0.1 to 0.5.

\textbf{Case 2} (\textit{Gaussian + Stripe}): All the settings are the same as in Case 1, except that stripe noise is used instead of impulse noise. The stripes noise ratio ranges from 0.05 to 0.2.

\textbf{Case 3} (\textit{Gaussian + Deadline}): All the settings are the same as in Case 1, except that the impulse noise is replaced by deadline noise, with a ratio ranging from 0.05 to 0.2.

\textbf{Case 4} (\textit{Spatial-Spectral Variant Gaussian}): Each band of the image is corrupted with the spatial-spectral variant Gaussian noise. The noise level for each band is randomly generated from a normal distribution, with values ranging between 10 and 70.

\textbf{Case 5} (\textit{Mixture}): The images are corrupted with all the noise types from Case 1 to 4.

In the training phase, the total number of iterations $K$ is set to 15 for DLW-NN and 20 for DLW-TV and DLW-TVS, respectively. Adam \cite{ADAM} is used to optimize the network parameters. The number of epochs is set to 10. The initial learning rate is 1e-3 and decays by a factor of 0.8 every epoch. The batch size is set to 10. All the experiments are conducted on a PC with Intel Core i7-8700K processor, and a GeForce RTX 2080 Ti with 11GB memory. It takes 4 to 6 hours to train each DLWnet for HSI, and 0.5 hour to train DLWnet for color images.

In this work, DLWnet $h_\theta(\cdot)$ is a four-layer CNN. Each of the first three layers is followed by the ``ReLU'' activation function. The output of the final layer is activated by the softmax function and is then multiplied by $M$, so that the elements of $\W$ averaged to one. For HSI, the 3-dimensional (3D) convolution kernel is used since HSIs usually have a diverse number of spectral bands. For color images, we simply use the 2D convolutional kernel.

Two quantitative measures are used to evaluate the denoising performance, namely, peak signal-to-noise ratio (PSNR) and structural similarity (SSIM). 

\subsection{Weight visualization}
Clearly, the principle of DLWnet is to extract a shared explicit weighting scheme across different noisy images and diverse source models. This enables it to capture a unified weighting policy for general noisy images (with diverse noise types) and different source models (with diverse regularization terms). This implies that the weight scheme extracted by the trained DLWnet for a test image should deliver both its noise information and its structural prior knowledge. 

\begin{figure}[t]
	\centering
	\includegraphics[width=14cm]{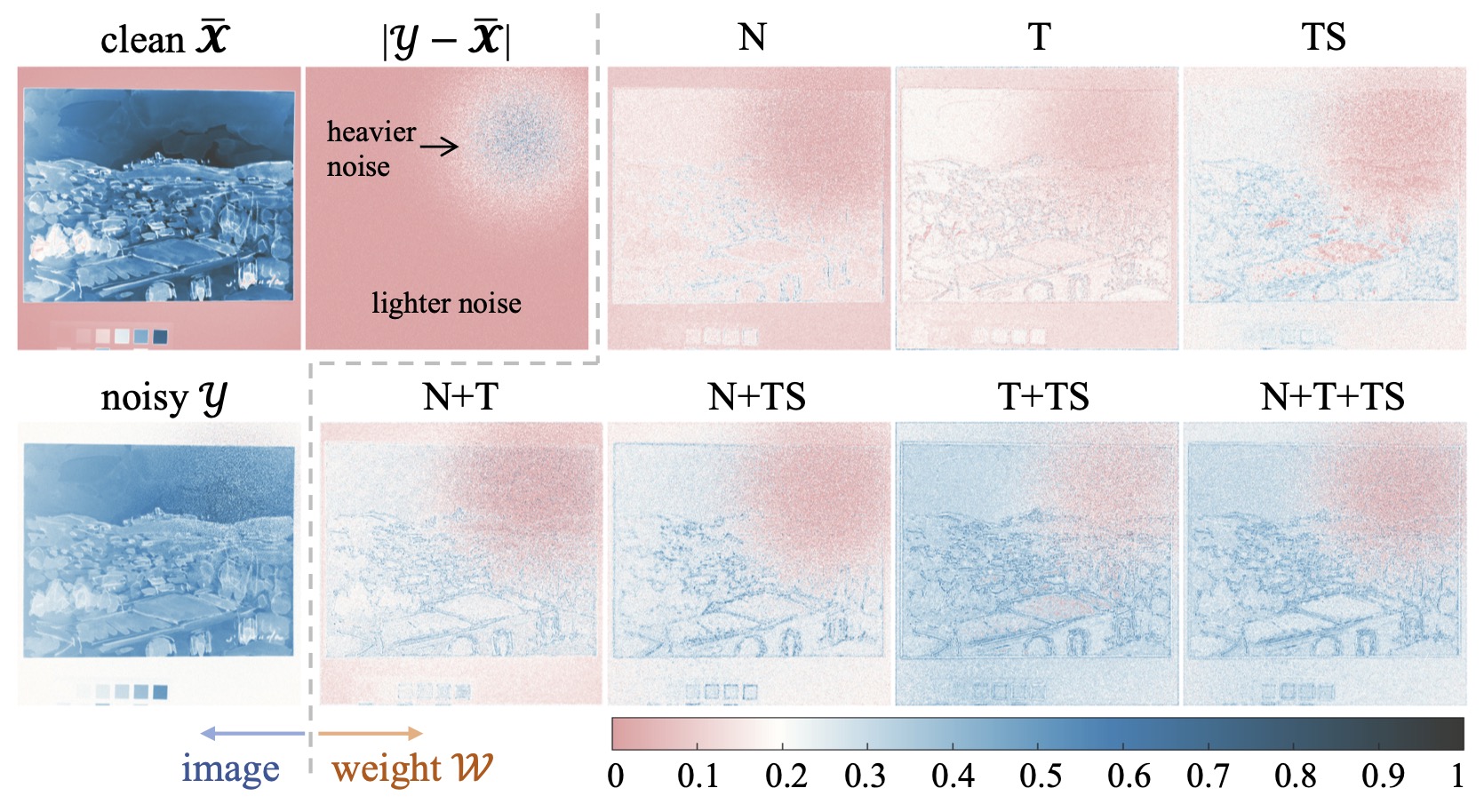}
	\caption{Visual comparison of the weights $\W$ predicted by different types of DLWnets. There are two observations: 1) All seven DLWnets can recognize the regions with heavier noise and assign smaller weights to them, clearly demonstrating the denoising assistance capability of DLWnet. 2) DLWnets additionally extract certain image structures, and the extracted structures vary across different DLWnets.}
	\label{fig-w}
\end{figure}

To intuitively illustrate this capability of DLWnet, we visualize the predicted weights calculated by the seven different DLWnets on a typical test image in \cref{fig-w}. Specifically, one band of the ``watercolor" image in the CAVE dataset is corrupted with spatially non-i.i.d Gaussian noise, and the top-right area of the image is more heavily corrupted than the remaining areas. From the figure, the aforementioned analysis is clearly validated. First, the heavy noise information located in the top-right area can be recognized by the weights predicted by all trained DLWnets. The weight values in this noisy area are very small, meaning that all DLWnets successfully recognize this polluted area and attempt to suppress its negative influence on the recovery by assigning relatively smaller weights to it. Second, all generated weights from the DLWnets show evident structural shapes consistent with those of the clean image. In particular, more important image structures for recovery, e.g., edges and textures, are emphasized by the DLWnet by assigning larger weights to them. Moreover, the emphasis on image structures varies across different weights, implying the underlying impact of different source models on the DLWnet.    

\subsection{Application of DLWnet}

In this section, we select three popular complex noise removal models for images as the target models to which we apply the trained DLWnet. These models are LRTV \cite{LRTV}, E3DTV \cite{E3DTV} and LRTFDFR \cite{LRTFDFR}. Specifically, LRTV \cite{LRTV} is a matrix-based method that utilizes both spatial smoothness property and spectral low-rankness. E3DTV \cite{E3DTV} is a tensor-based method that exploits the sparsity of the base matrix of the gradient map in each mode. LRTFDFR \cite{LRTFDFR} employs spectral low-rankness, group sparsity on the gradient map of the base matrix, and column continuity of the spectral factor matrix. The data fidelity terms of LRTV and E3DTV both use the $\ell_1$ norm. The data fidelity term of LRTFDFR includes both the weighted $\ell_1$ norm and the $\ell_2$ norm.

For each of the three target models, we modify it by simply replacing the data fidelity term with $\|h_\theta(\Y)\odot(\Y - \X)\|^2$. The modified models implemented with our DLWnet are called DLW-LRTV, DLW-E3DTV and DLW-LRTFDFR, respectively. They are also called \textit{our} methods for convenience. In this section, the DLWnet of type `N+T+TS' is applied to the three DLW-models. The algorithms for solving the DLW-models are inherited from the original models. More specifically, DLW-LRTV and DLW-E3DTV are optimized by ADMM which is also adopted in the original LRTV and E3DTV models. As for DLW-LRTFDFR, we also adopt ADMM because the proximal alternating minimization algorithm used by LRTFDFR can hardly be directly inherited. More details are presented in \cref{sec-appendix-expresultss}.

\begin{table}[t]
	\renewcommand{\arraystretch}{1.15}
	\newcommand{\mysize}{1.9cm}
	\newcommand{\firstc}{2.3cm}
	\fontsize{8.5}{9.5}\selectfont
	\caption{Average test performance of different denoising competing methods on the ICVL dataset. The best results in each \textbf{column} are in \textbf{bold}, and the second best results in each \textbf{column} are with \ul{underline}.}
	\label{tab-exist-com}
	\centering
	\begin{tabular}{M{\firstc} M{\mysize} M{\mysize}  M{\mysize} M{\mysize} M{\mysize}}
		\Xhline{0.8pt}
		noise $\rightarrow$ & Case 1 & Case 2 & Case 3 & Case 4 & Case 5 \\
		\cline{2-6}
		index $\rightarrow$ & PSNR/SSIM & PSNR/SSIM & PSNR/SSIM & PSNR/SSIM & PSNR/SSIM \\
		\hline
		LRMR &	24.35/0.7214	&	27.77/0.8198	&	26.45/0.7962	&	26.15/0.7794	&	23/0.6868 \\
		LRTV &	31.67/0.9059	&	32.91/0.9252	&	31.34/0.9135	&	35.49/0.9543	&	30.3/0.8971 \\
		NMoG &	28.9/0.8718	&	30.32/0.9088	&	29.09/0.9013	&	26.68/0.8422	&	24.7/0.7553 \\
		HyRes &	30/0.8829	&	35.09/0.9566	&	32.16/0.9308	&	32.25/0.9241	&	26.84/0.8357 \\
		FastHyMix &	31.09/0.8914	&	 \tb{37.30}/\tb{0.9680} 	&	 \ul{34.19}/\tb{0.9571} 	&	36.23/0.961	&	27.9/0.8498 \\
		CTV-RPCA &	31.09/0.8636	&	30.99/0.8579	&	30.37/0.8484	&	28.94/0.7944	&	28.45/0.7923 \\
		E3DTV &	 \ul{34.61}/\ul{0.9511} 	&	34.27/0.9473	&	33.52/0.9425	&	32.88/0.9271	&	31.63/0.9154 \\
		LRTFDFR &	29.69/0.8376	&	30.83/0.8479	&	27.92/0.7729	&	31.52/0.8553	&	28.42/0.8213 \\
		HSI-DeNet &	29.33/0.8588	&	28.95/0.8389	&	28.39/0.8308	&	30.31/0.8843	&	28.37/0.842 \\
		HSI-CNN &	34.36/0.932	&	 \ul{35.65}/0.9551	&	33.64/0.9456	&	 \tb{37.38}/\ul{0.9649} 	&	32.02/0.9206 \\
		\hline
		DLW-LRTV &	34.94/0.9495	&	35.62/\ul{0.9593} 	&	34.52/0.9523	&	 \ul{37.00}/\tb{0.9692} 	&	 \tb{33.90}/\tb{0.9480} \\
		DLW-E3DTV &	 \tb{35.41}/\tb{0.9529} 	&	35.46/0.959	&	 \tb{35.20}/\ul{0.9568} 	&	34.55/0.9476	&	33.26/0.9346 \\
		DLW-LRTFDFR &	34.09/0.9505	&	34.44/0.954	&	33.75/0.9464	&	35.42/0.964	&	 \ul{33.43}/\ul{0.9469} \\
		\Xhline{0.8pt}
	\end{tabular}
\end{table}

\begin{figure}[t]
	\newcommand{\mysize}{2.3cm}
	\newcommand{\minu}{0pt}
	\fontsize{8.5}{9.5}\selectfont
	\renewcommand{\arraystretch}{1.15}
	\centering
	\begin{minipage}[t]{\mysize}
		\centering
		\includegraphics[width=\mysize]{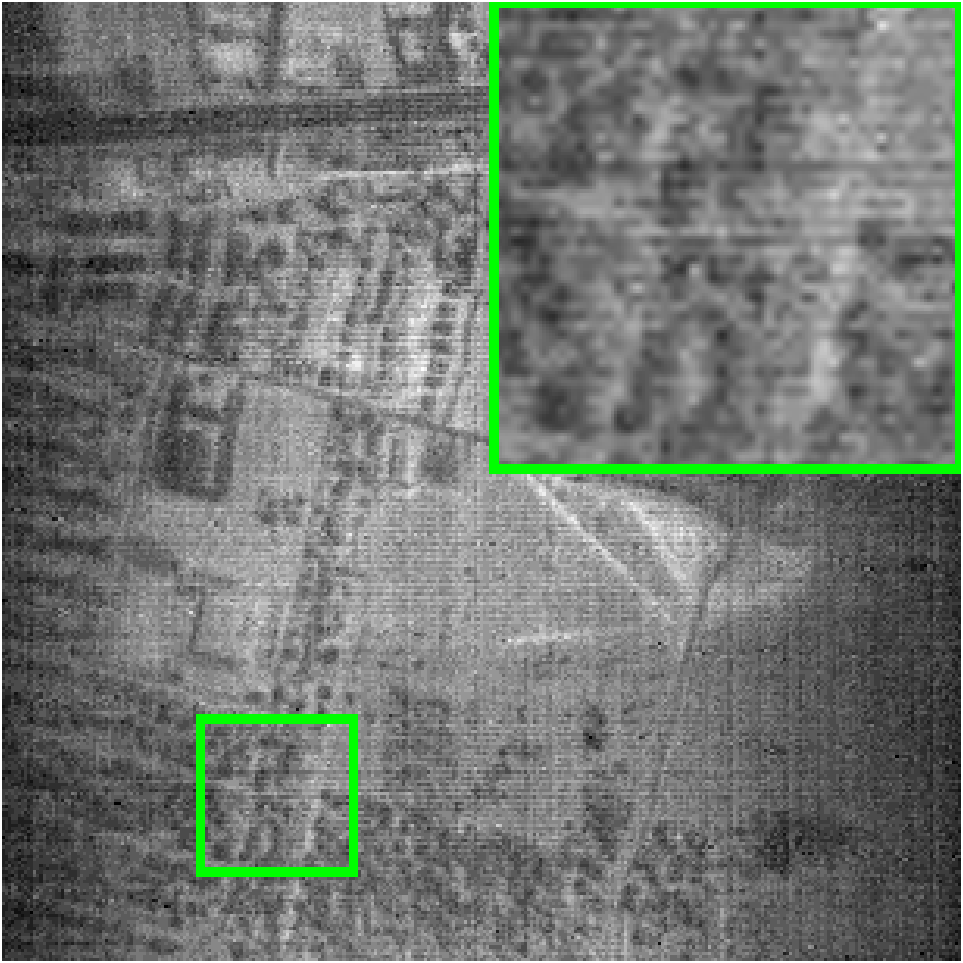}\\
		Urban
	\end{minipage}\hspace{\minu}
	\begin{minipage}[t]{\mysize}
		\centering
		\includegraphics[width=\mysize]{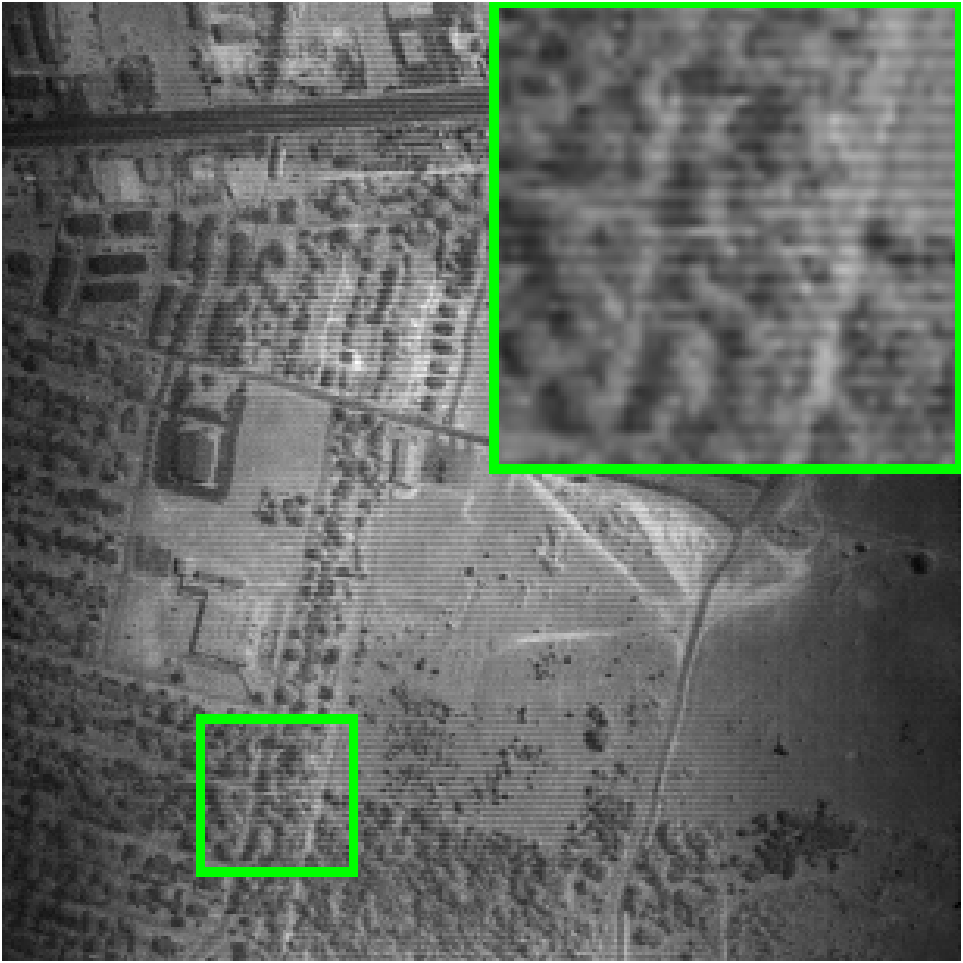}\\
		LRMR
	\end{minipage}\hspace{\minu}
	\begin{minipage}[t]{\mysize}
		\centering
		\includegraphics[width=\mysize]{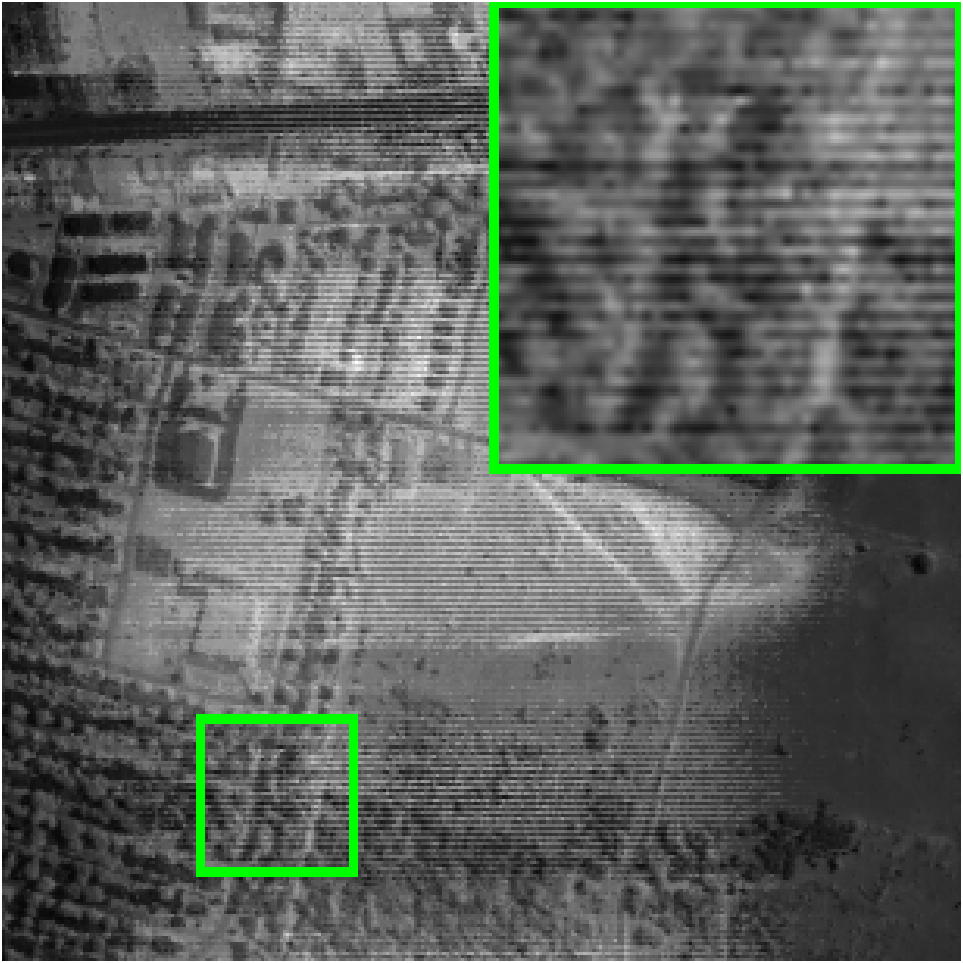}\\
		LRTV
	\end{minipage}\hspace{\minu}
	\begin{minipage}[t]{\mysize}
		\centering
		\includegraphics[width=\mysize]{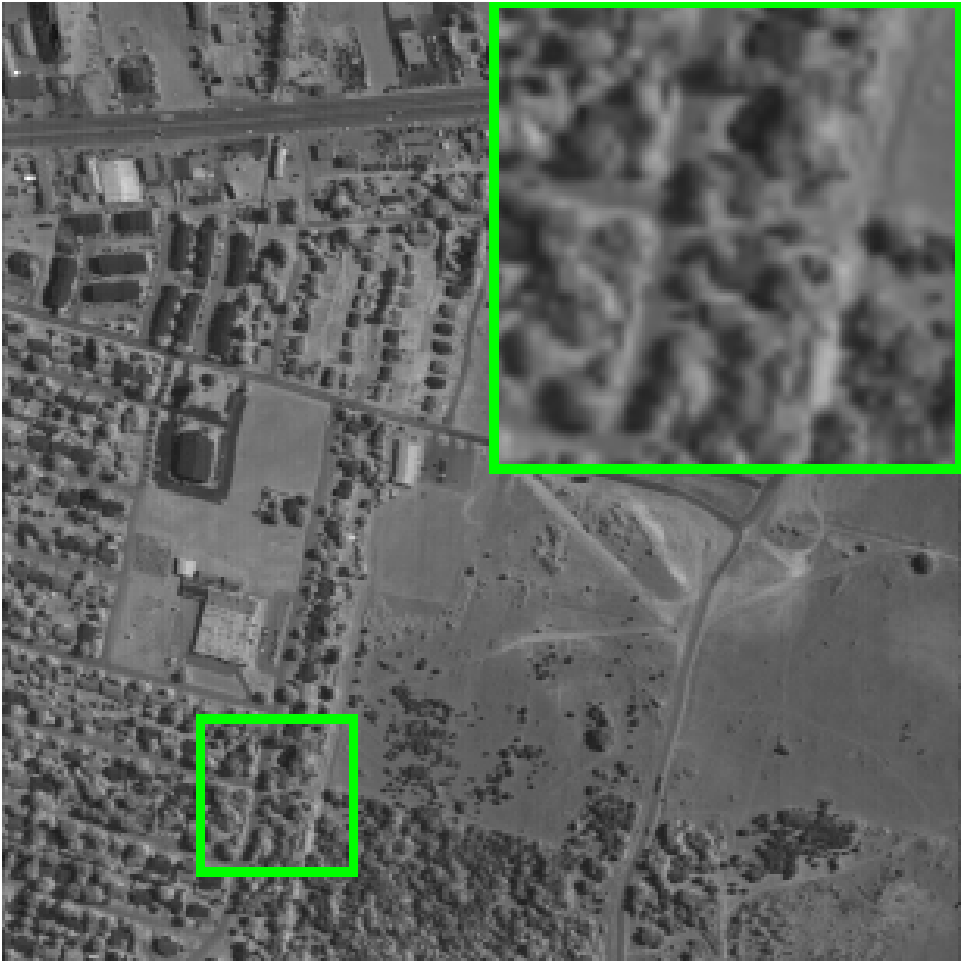}\\
		NMoG
	\end{minipage}
	\begin{minipage}[t]{\mysize}
		\centering
		\includegraphics[width=\mysize]{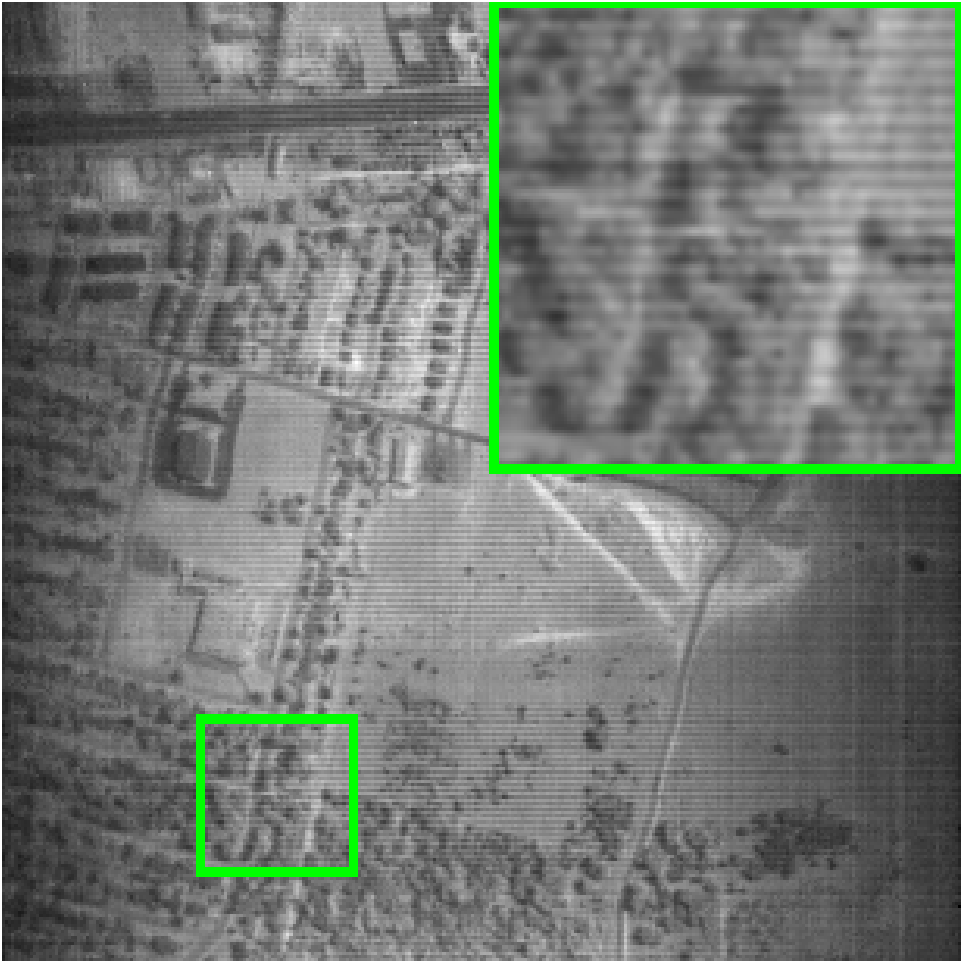}\\
		HyRes
	\end{minipage}\hspace{\minu}
	\begin{minipage}[t]{\mysize}
		\centering
		\includegraphics[width=\mysize]{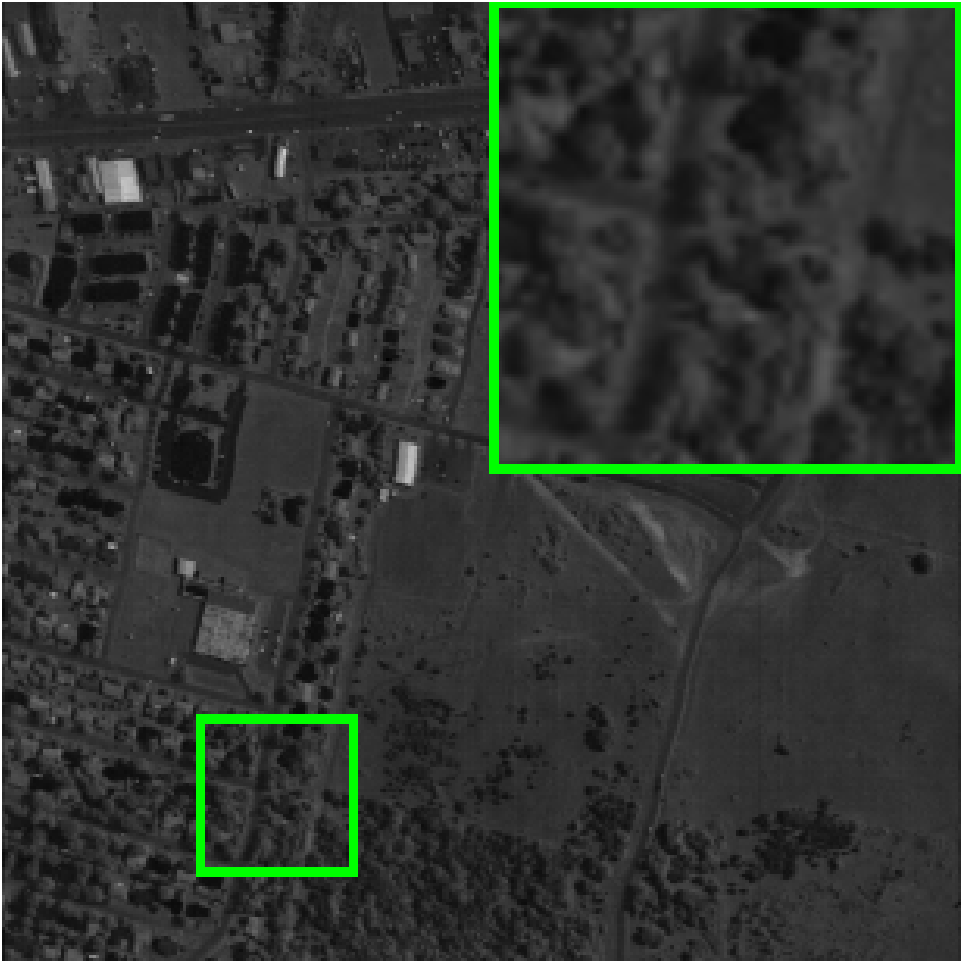}\\
		FastHyMix
	\end{minipage}\vspace{2pt} \\
	\begin{minipage}[t]{\mysize}
		\centering
		\includegraphics[width=\mysize]{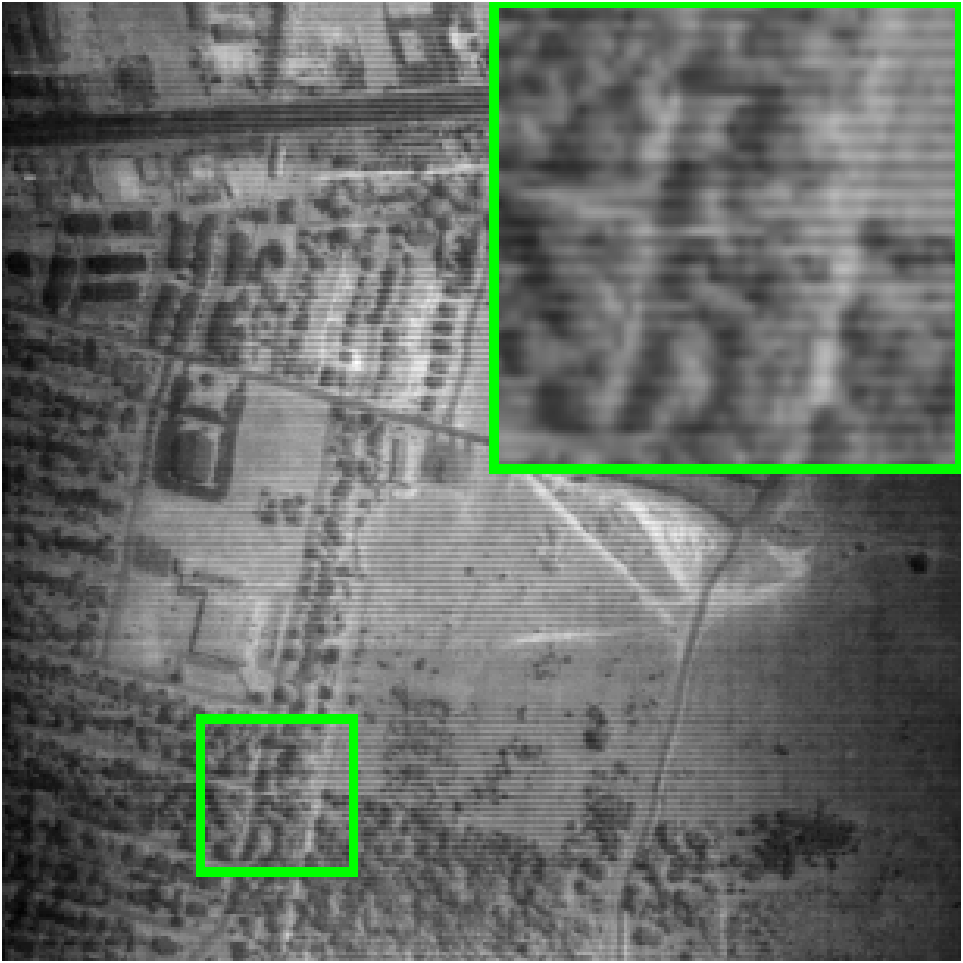}\\
		CTV-RPCA
	\end{minipage}\hspace{\minu}
	\begin{minipage}[t]{\mysize}
		\centering
		\includegraphics[width=\mysize]{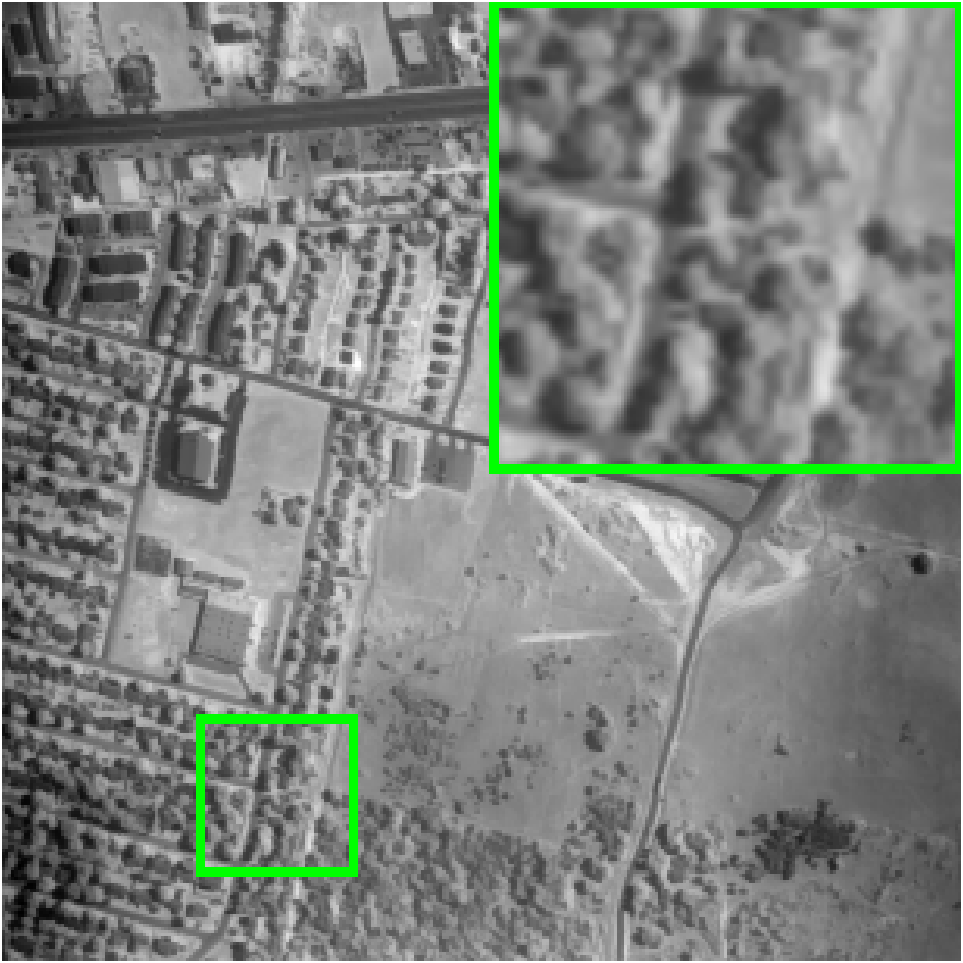}\\
		E3DTV
	\end{minipage} 
	\begin{minipage}[t]{\mysize}
		\centering
		\includegraphics[width=\mysize]{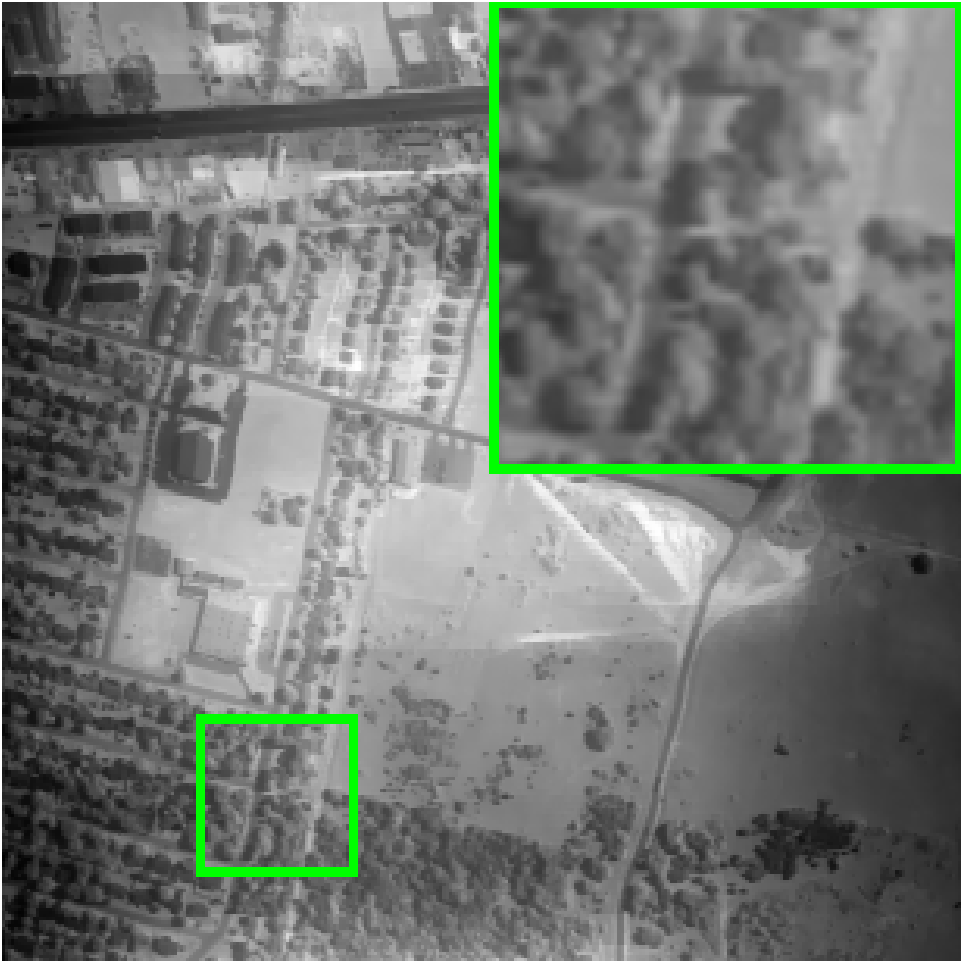}\\
		LRTFDFR
	\end{minipage}\hspace{\minu}
	\begin{minipage}[t]{\mysize}
		\centering
		\includegraphics[width=\mysize]{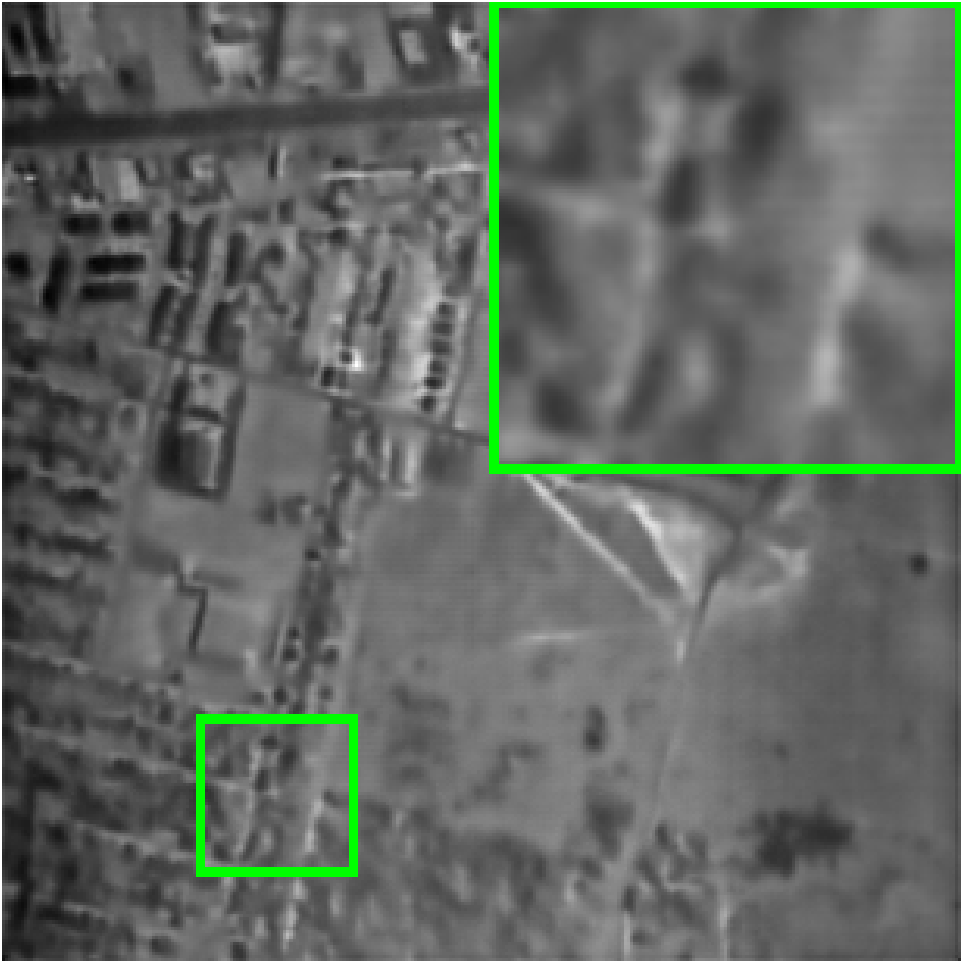}\\
		HSI-DeNet
	\end{minipage}\hspace{\minu}
	\begin{minipage}[t]{\mysize}
		\centering
		\includegraphics[width=\mysize]{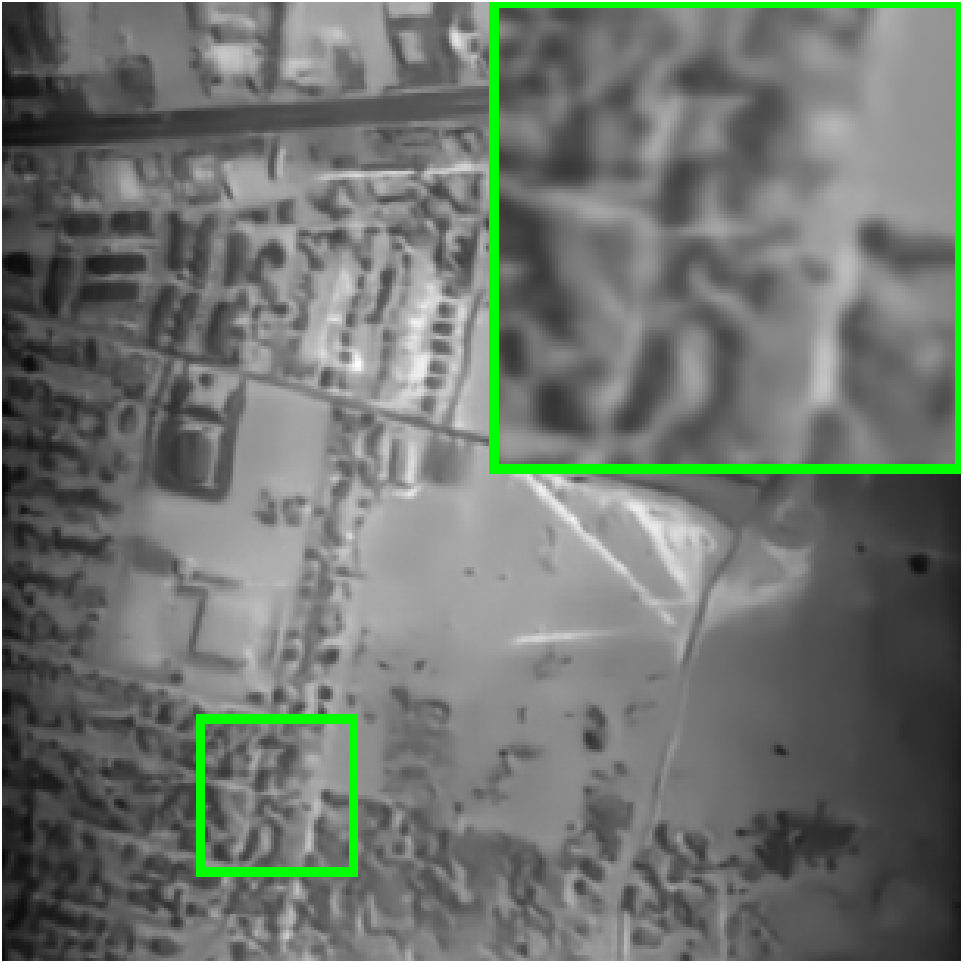}\\
		HSI-CNN
	\end{minipage}\hspace{\minu}
	\begin{minipage}[t]{\mysize}
		\centering
		\includegraphics[width=\mysize]{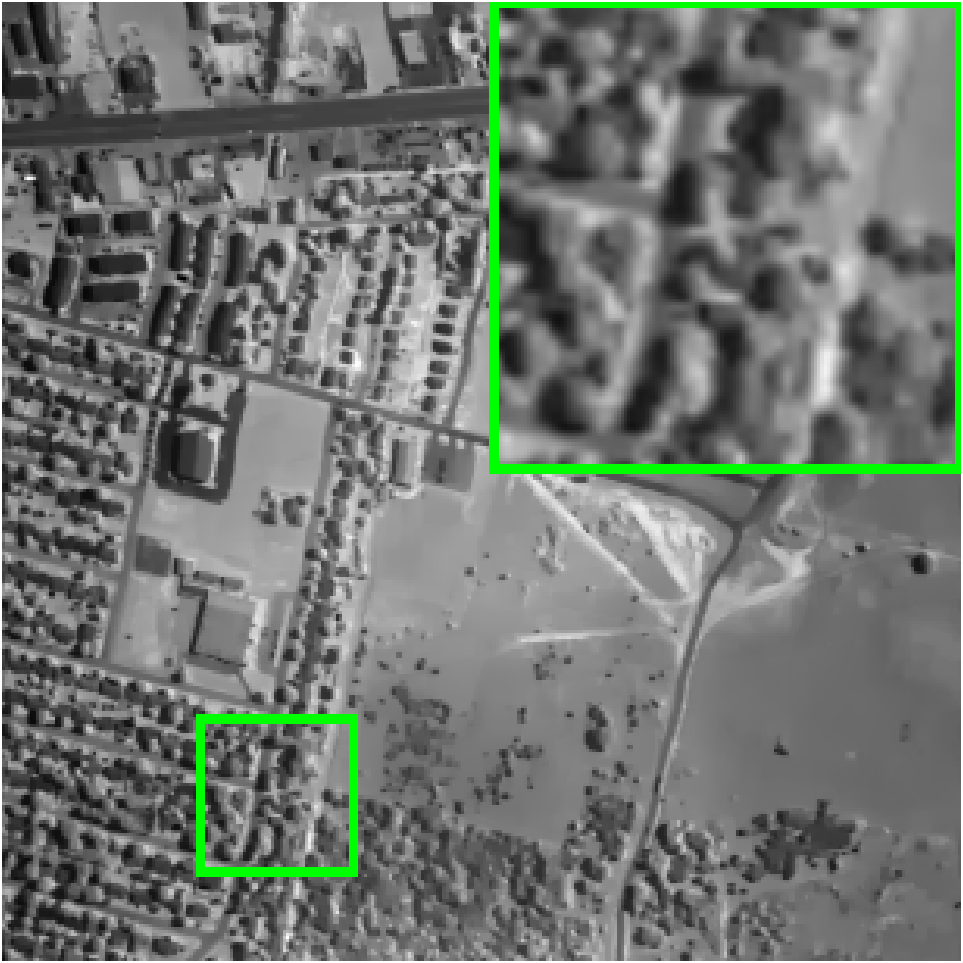}\\
		DLW-E3DTV
	\end{minipage}
	\caption{Denoising results by different compared methods on the real noisy image Urban (the 104th band is shown).}
	\label{fig-comall}
\end{figure}

We compare the three DLW-methods with several advanced complex noise removal methods for images, including LRMR \cite{LRMR}, LRTV \cite{LRTV}, NMoG \cite{chen2017denoising}, HyRes \cite{HyRes}, FastHyMix \cite{FastHyMix}, CTV-RPCA \cite{CTV}, E3DTV \cite{E3DTV}, LRTFDFR \cite{LRTFDFR}, HSI-DeNet \cite{HSI-DeNet} and HSI-CNN \cite{Hyperspectral-Image-Denoising-Employing-a-Spatial-Spectral-Deep-Residual-Convolutional-Neural-Network}. The quantitative results are reported in \cref{tab-exist-com}. From the table, we can clearly observe that our DLWnet dramatically improves the denoising results, as seen when comparing LRTV with DLW-LRTV or E3DTV with DLW-E3DTV. Additionally, it can also be seen that our methods generally perform best on different types of noise patterns, even though only Case 1 noise is used in the training process. This verifies that the DLWnet can learn a general weight prediction rule for a wide range of complex noise types and thus help the model adapt effectively to more noise types. Although FastHyMix achieves the best results in Case 2 (Gaussian+stripe), it does not perform well on the more complex noise in Case 5. We also compare our methods with all competing methods on a real noisy image dataset, Urban. The visual comparison is presented in \cref{fig-comall}. It can be observed that the image restored by our DLW-E3DTV method achieves the best visual quality, while other competing methods do not fully remove the noise, and their restored images often contain obvious blurring or stripe noise. More experimental results are presented in \cref{sec-appendix-expresultss}.

From the above synthetic and real image denoising experiments, the effectiveness of our DLWnet is reflected in two aspects. First, the trained DLWnet helps boost the denoising performance of traditional image denoising models. It should also be noted that the DLWnet is not limited to use on the LRTV, E3DTV and LRTFDFR models, but could be integrated into general image denoising models. Second, the DLWnet tends to learn a general weight prediction rule that helps a model effectively adapt to a wider range of noise types than those used in its training.

\subsection{Model generalization analysis}\label{sec-generalization-exp}
In this section, we further explore how DLWnet behaves across various source and target models.

\subsubsection{Simple target models}\label{sec-expe-simple-target}
In this experiment, we choose the target models from the source models (i.e., DLW-NN, DLW-TV and DLW-TVS). Since the divergence between the source and target models is relatively clear, we can examine the generalization ability of all seven DLWnets across different regularizations.

\begin{table}[t]
	\renewcommand{\arraystretch}{1.15}
	\newcommand{\mysize}{1cm}
	\fontsize{8.5}{9.5}\selectfont
	\caption{Average test performance of DLW-NN, DLW-TV and DLW-TVS on ICVL dataset. The best results in each \textbf{row} are in \textbf{bold}, and the second best results in each \textbf{row} are with \ul{underline}.}
	\label{tab-three-all}
	\centering
	\begin{tabular}{M{1.2cm} | M{\mysize} M{\mysize} M{\mysize} M{\mysize} M{\mysize} M{\mysize} M{\mysize} M{\mysize+0.3cm} M{\mysize}}
		\wline
		\multirow{2}{1.2cm}{\backslashbox[1.6cm]{noise}{source}}
		& noisy & N & T & TS & N+T & N+TS & T+TS & N+T+TS & Abl\\
		\cline{2-10}
		& \multicolumn{9}{c}{evaluation index: PSNR/SSIM} \\
		\wline
		\multicolumn{10}{c}{target model 1 : \tb{DLW-NN}} \cr
		\hline
		Case 1 & 14.86/ 0.3554 & \ul{33.16}/ \tb{0.9452} & 22.44/ 0.6523 & 27.38/ 0.8348 & 32.53/ 0.9413 & \tb{33.31}/ \ul{0.944} & 26.93/ 0.8292 & 32.93/ 0.9434 & 19.69/ 0.5787 \\
		Case 2 & 16.16/ 0.3998 & \ul{33.6}/ \ul{0.9493} & 22.31/ 0.6449 & 27.8/ 0.8421 & 32.95/ 0.9431 & \tb{33.71}/ \tb{0.9502} & 26.83/ 0.8257 & 33.35/ 0.9482 & 23.37/ 0.7198 \\
		Case 3 & 16.06/ 0.395 & \ul{32.54}/ \tb{0.9461} & 22.37/ 0.65 & 26.98/ 0.832 & 31.8/ 0.9352 & \tb{32.59}/ \ul{0.9433} & 26.51/ 0.8228 & 32.16/ 0.9406 & 22.7/ 0.7194 \\
		Case 4 & 13.13/ 0.3197 & \tb{32.39}/ \tb{0.9267} & 18.99/ 0.5332 & 25.09/ 0.7592 & 31.25/ 0.9138 & \ul{32.05}/ \ul{0.9253} & 23.71/ 0.7321 & 31.54/ 0.9215 & 22.0/ 0.7352 \\
		Case 5 & 12.94/ 0.2762 & \tb{30.43}/ \tb{0.9029} & 19.89/ 0.5492 & 23.9/ 0.7127 & 29.46/ 0.8907 & \ul{29.85}/ \ul{0.895} & 23.87/ 0.7287 & 29.61/ 0.8941 & 18.98/ 0.5719 \\
		\hline
		\multicolumn{10}{c}{target model 2 : \tb{DLW-TV}} \cr
		\hline
		Case 1 & 14.86/ 0.3554 & 29.27/ 0.8355 & \tb{32.84}/ \tb{0.9274} & 27.53/ 0.7872 & 31.52/ 0.8894 & 29.1/ 0.8343 & \ul{31.53}/ \ul{0.8898} & 30.97/ 0.8757 & 24.54/ 0.6534 \\
		Case 2 & 16.16/ 0.3998 & 29.4/ 0.8459 & \tb{32.95}/ \tb{0.9356} & 27.67/ 0.7935 & \ul{31.81}/ \ul{0.8987} & 29.23/ 0.8461 & 31.72/ 0.8974 & 31.19/ 0.8862 & 26.48/ 0.7492 \\
		Case 3 & 16.06/ 0.395 & 28.97/ 0.8385 & \tb{31.94}/ \tb{0.9275} & 26.75/ 0.7762 & \ul{31.03}/ \ul{0.8916} & 28.82/ 0.8372 & 30.76/ 0.8877 & 30.6/ 0.8793 & 25.62/ 0.7415 \\
		Case 4 & 13.13/ 0.3197 & 28.65/ 0.8272 & \tb{31.38}/ \tb{0.9087} & 26.73/ 0.7695 & \ul{30.89}/ \ul{0.8807} & 28.53/ 0.8283 & 30.8/ 0.8804 & 30.4/ 0.8697 & 27.52/ 0.7792 \\
		Case 5 & 12.94/ 0.2762 & 27.58/ 0.8003 & \tb{29.67}/ \tb{0.8774} & 25.19/ 0.7229 & \ul{29.16}/ \ul{0.8555} & 27.5/ 0.7985 & 28.97/ 0.8503 & 28.96/ 0.8427 & 22.5/ 0.5717 \\
		\hline
		\multicolumn{10}{c}{target model 3 : \tb{DLW-TVS}} \cr
		\hline
		Case 1 & 14.86/ 0.3554 & 31.26/ 0.9125 & 24.17/ 0.6991 & \tb{33.22}/ \tb{0.9249} & 30.49/ 0.9094 & \ul{32.35}/ 0.9238 & 32.03/ \ul{0.9241} & 31.77/ 0.924 & 23.02/ 0.7394 \\
		Case 2 & 16.16/ 0.3998 & 31.04/ 0.9135 & 23.97/ 0.6863 & \tb{33.48}/ \tb{0.9292} & 30.22/ 0.9065 & \ul{32.34}/ \ul{0.9267} & 32.08/ 0.9257 & 31.71/ 0.9253 & 25.92/ 0.8516 \\
		Case 3 & 16.06/ 0.395 & 31.3/ 0.9151 & 24.05/ 0.6942 & \tb{32.85}/ \ul{0.9243} & 30.31/ 0.906 & \ul{32.49}/ \tb{0.926} & 31.59/ 0.9193 & 31.73/ 0.9239 & 25.7/ 0.8499 \\
		Case 4 & 13.13/ 0.3197 & 29.52/ 0.8833 & 21.3/ 0.593 & \tb{32.0}/ \tb{0.9041} & 28.67/ 0.8733 & \ul{30.96}/ \ul{0.9018} & 30.54/ 0.8957 & 30.28/ 0.8981 & 25.05/ 0.815 \\
		Case 5 & 12.94/ 0.2762 & 29.59/ 0.8711 & 21.88/ 0.6039 & \ul{30.47}/ 0.8784 & 28.59/ 0.8599 & \tb{30.6}/ \tb{0.8858} & 29.49/ 0.8723 & 29.84/ \ul{0.8807} & 22.93/ 0.7202 \\
		\wline
	\end{tabular}
\end{table}

The average experimental results on the ICVL dataset are shown in \cref{tab-three-all}. The first row indexes seven types of DLWnets, and the first column indexes the different noise types. For example, the ``T" column means that the DLWnet applied to the three target models is trained using only the source model DLW-TV. The quantitative results for five noise types are reported. In \cref{fig-part1-nuclr} we present the visual results of DLW-NN using different DLWnet.\footnote{Unless otherwise specified, the pesudo-color image is constructed using the 23rd, 12th, and 4th spectral bands of a HSI.} Clearly, the DLWnet of type `N' achieves the best visual performance.

From \cref{tab-three-all}, we first observe that although the DLWnet is trained using only Case 1 noise, it can generalize well to other noise types (Cases 2-5), which have not been seen during training. Besides, it seems that some DLWnets always perform better on a specific target model. To explore the underlying rules, we compute the average PSNR for all five noise cases based on \cref{tab-three-all} and present the results in \cref{tab-three-com}. Combining the two tables, we have the following observations. First, the corresponding DLWnet tends to perform best when the source models and the target models match. For example, in \cref{tab-three-all}, if the DLWnet type is `N', the best denoising performance is obtained when the target model is DLW-NN (see second column). The phenomenon can also be seen in \cref{tab-three-com}, where the diagonal values are the highest. Second, the DLWnet of type `T' does not generalize well to the other two models (i.e., `N' and `TS'), while the `N'-type DLWnet generalizes better than the other two types of DLWnets. Third, although the inconsistency between source and target models degrades the performance, the performance on the target models is still good enough. Fourth, when the number of target models grows, the performance also rises as the number of source models increases (see the last column of \cref{tab-three-com}). This is because the DLWnet can capture more regularization information from multiple source models.

\begin{table*}[t]
	\renewcommand{\arraystretch}{1.15}
	\newcommand{\mysize}{1.55cm}
	\fontsize{8.5}{9.5}\selectfont
	\caption{Average PSNR/SSIM values of all five kinds of noise patterns obtained by DLW-NN, DLW-TV and DLW-TVS on ICVL dataset. The best results in each \tb{column} are in \textbf{bold}, and the second best results in each \tb{column} are with \ul{underline}.}
	\label{tab-three-com}
	\centering
	\begin{tabular}{M{1.3cm} M{\mysize} M{\mysize} M{\mysize} M{\mysize} M{\mysize} M{\mysize}  M{\mysize}}
		\Xhline{0.8pt}
		
		target $\rightarrow$ source $\downarrow$  & DLW-NN     &   DLW-TV    &   DLW-TVS   & DLW-NN \& DLW-TV & DLW-NN \& DLW-TVS & DLW-TV \& DLW-TVS & DLW-NN \& DLW-TV \& DLW-TVS  \\
		\Xhline{0.8pt}
		N        & \tb{32.4}/\tb{0.934}  & 28.8/0.830 & 30.5/0.899 & 30.6/0.882 & 31.5/0.917 & 29.7/0.864 & 30.6/0.888           \\
		
		T        & 21.2/0.606 & \tb{31.8}/\tb{0.915} & 23.1/0.655 & 26.5/0.761 & 22.1/0.631 & 27.4/0.785 & 25.3/0.726           \\
		
		TS       & 26.2/0.796 & 26.8/0.770 & \tb{32.4}/\ul{0.912} & 26.5/0.783 & 29.3/0.854 & 29.6/0.841 & 28.5/0.826           \\
		
		N+T        & 31.6/0.925 & \ul{30.9}/\ul{0.883} & 29.7/0.891 & \tb{31.2}/\tb{0.904} & 30.6/0.908 & 30.3/0.887 & 30.7/\ul{0.900} \\
		
		N+TS       & \ul{32.3}/\ul{0.932} & 28.6/0.829 & \ul{31.8}/\tb{0.913} & 30.5/0.880 & \tb{32.0}/\tb{0.922} & 30.2/0.871 & \ul{30.9}/0.891 \\
		
		T+TS       & 25.6/0.788 & 30.8/0.881 & 31.1/0.907 & 28.2/0.834 & 28.4/0.848 & \tb{31.0}/\tb{0.894} & 29.2/0.859\\
		
		N+T+TS       & 31.9/0.930 & 30.4/0.871 & 31.1/0.910 & \ul{31.2}/\ul{0.900} & \ul{31.5}/\ul{0.920} & \ul{30.8}/\ul{0.891} & \tb{31.1}/\tb{0.904}\\
		\Xhline{0.8pt}
	\end{tabular}
\end{table*}

We can also get further insight from \cref{tab-three-com} regarding \cref{theorem-ge}. For convenience, we assume that the training error (first two terms of the upper bound) can be ignored. Denote the sum of the last two terms of the upper bound as 
\begin{align}
	& U(x, y) := \mathbb{E}_{(\Y,\X)\sim\D}\left\{\dfrac{8B_d}{\sqrt{M}\varepsilon}A_1 \right\} + \mathbb{E}_{(\Y,\X)\sim\D}\left\{\dfrac{2}{M\varepsilon^2}A_2\right\}, \nonumber \\
	& \mathrm{where}~ x=\{\text{source models}\}, y=\{\text{target models}\}. 
\end{align}
Then, a smaller $U(x,y)$ implies that the generalization error is more likely to be small. When the source and target models are the same, we have $U(y, y) = 0$. Correspondingly, in \cref{tab-three-com}, the diagonal results are generally the best since the source and target models exactly match. When the source and target models are different, there are also some related observations. Let us first focus on the second column of \cref{tab-three-com}. The target model is DLW-NN. By the definition of $U$, we have $U(\text{N+T, DLW-NN})=\frac{1}{2}U(\text{T, DLW-NN})$ and $U(\text{N+TS,  DLW-NN})=\frac{1}{2}U(\text{TS, DLW-NN})$. \cref{tab-three-com} also shows that the performance of `N+T' is better than `T', and the performance of `N+TS' is better than that of `TS'. Besides, the performance of `TS' is better than that of `T', suggesting that $U(\text{TS, DLW-NN})<U(\text{T, DLW-NN})$ and thus $U(\text{N+TS, DLW-NN})<U(\text{N+T, DLW-NN})$. Correspondingly, \cref{tab-three-com} shows that the performance of `N+TS' is better than that of `N+T'. More similar observations can be found for other source and target combinations in \cref{tab-three-com}. It should be noted that in addition to the intrinsic aspects revealed by our learning theory---which help us understand the insightful working mechanism underlying our method, particularly its generalization ability---other practical factors, such as training effectiveness, may also affect the generalization behavior. This is similar to the role of conventional learning theory results.

\begin{figure}[t]
	\renewcommand{\arraystretch}{1.15}
	\newcommand{\mysize}{2.6cm}
	\fontsize{8.5}{9.5}\selectfont
	\newcommand{\minivs}{4pt}
	\newcommand{\vs}{2pt}
	\centering
	\begin{minipage}[t]{\mysize*5+1cm}
		\centering
		\begin{minipage}[b]{\mysize}
			\centering
			clean 
			\includegraphics[width=\mysize]{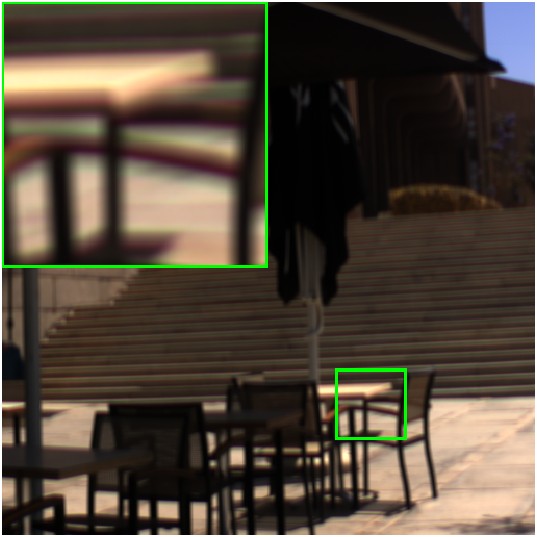} \\
			(PSNR, SSIM)
		\end{minipage}\hspace{\minivs}
		\begin{minipage}[b]{\mysize}
			\centering
			noisy 
			\includegraphics[width=\mysize]{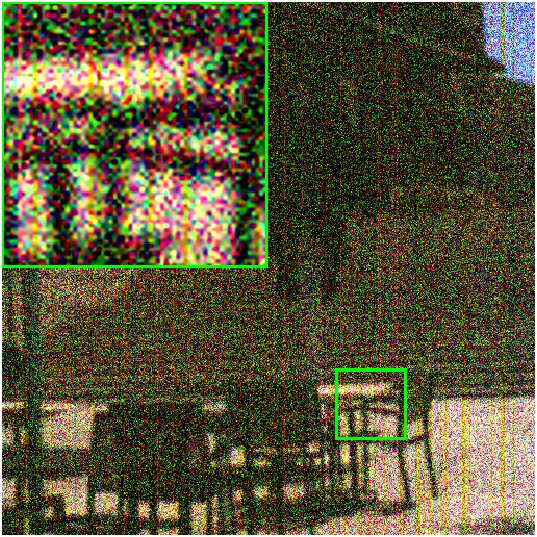} \\
			(14.75, 0.2441)
		\end{minipage}\hspace{\minivs}
		\begin{minipage}[b]{\mysize}
			\centering
			N 
			\includegraphics[width=\mysize]{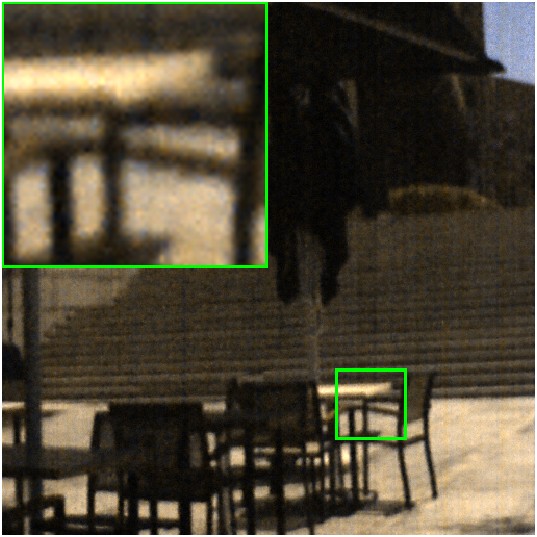} \\
			(\tb{33.26}, \tb{0.8872})
		\end{minipage}\hspace{\minivs}
		\begin{minipage}[b]{\mysize}
			\centering
			T 
			\includegraphics[width=\mysize]{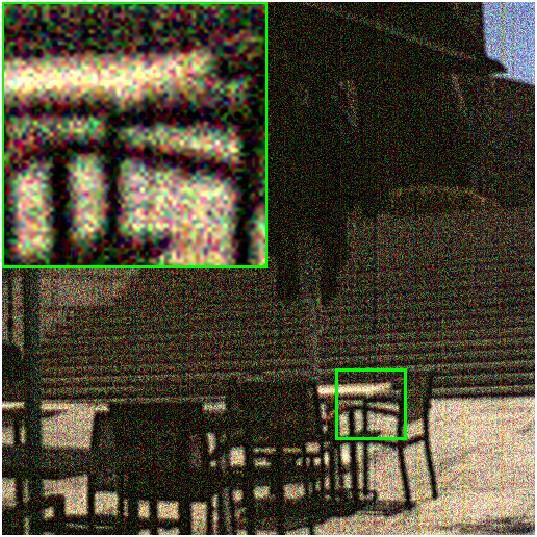} \\
			(20.6, 0.4874)
		\end{minipage} \vspace{1pt} \\
		\begin{minipage}[b]{\mysize}
			\centering
			TS 
			\includegraphics[width=\mysize]{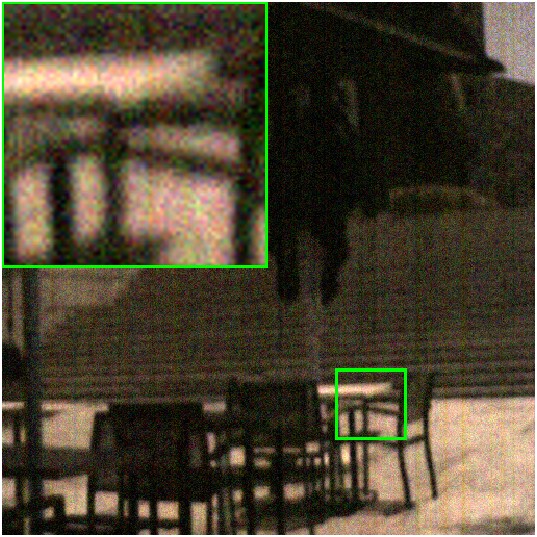} \\
			(25.81, 0.7159)
		\end{minipage}\hspace{\minivs}
		\begin{minipage}[b]{\mysize}
			\centering
			N+T 
			\includegraphics[width=\mysize]{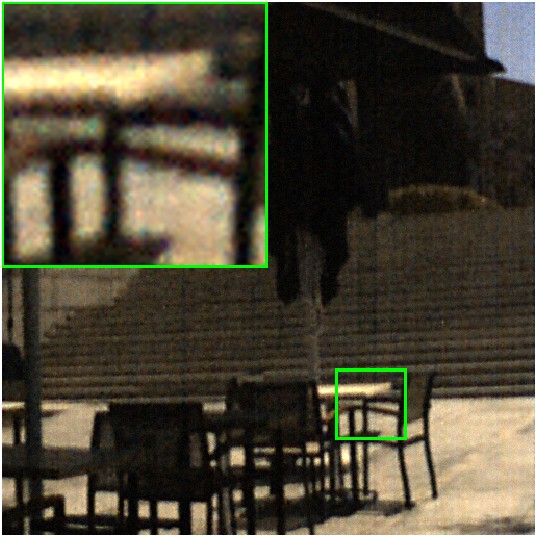} \\
			(31.94, 0.8845)
		\end{minipage}\hspace{\minivs}
		\begin{minipage}[b]{\mysize}
			\centering
			N+TS 
			\includegraphics[width=\mysize]{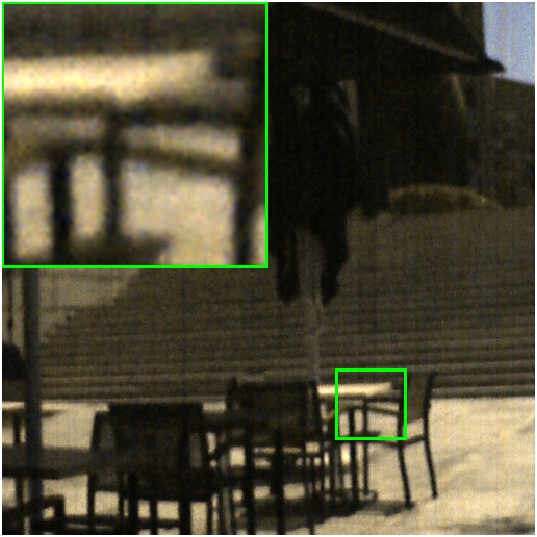} \\
			(\ul{32.25}, \ul{0.8856})
		\end{minipage}\hspace{\minivs}
		\begin{minipage}[b]{\mysize}
			\centering
			T+TS 
			\includegraphics[width=\mysize]{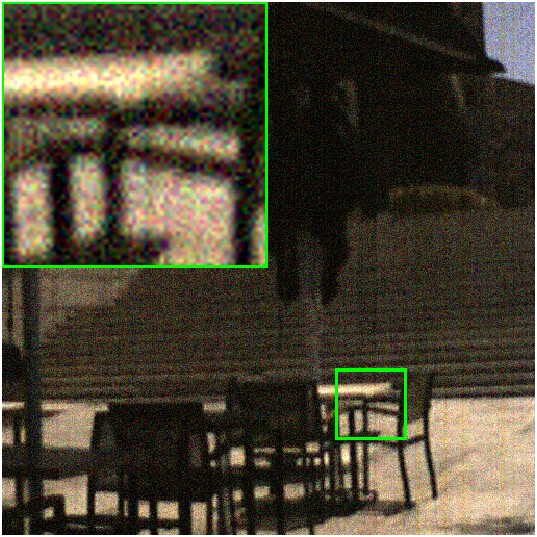} \\
			(24.88, 0.6978)
		\end{minipage}\hspace{\minivs}
		\begin{minipage}[b]{\mysize}
			\centering
			N+T+TS 
			\includegraphics[width=\mysize]{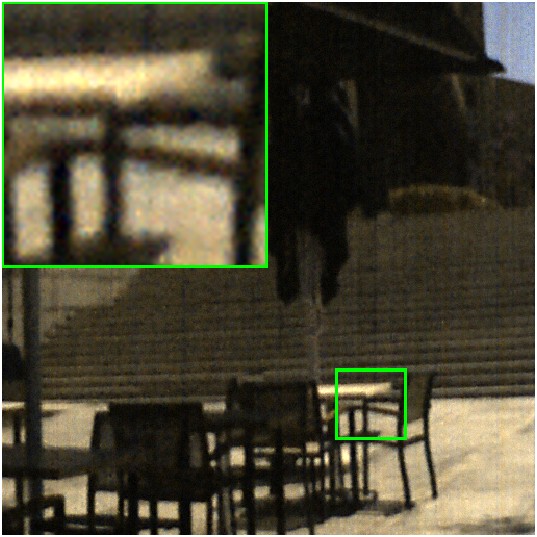} \\
			(32.05, 0.8829)
		\end{minipage}\hspace{\minivs}
	\end{minipage}
	\caption{Denoising results (pesudo-color image) of DLW-NN on image ``BGU\_0522-1136" of ICVL dataset. The noisy type is ``mixture" (Case 5).}
	\label{fig-part1-nuclr}
\end{figure}

\subsubsection{Complex target models}\label{sec-complex-target-model}
We apply the seven trained DLWnets to LRTV \cite{LRTV}, E3DTV \cite{E3DTV}, and LRTFDFR \cite{LRTFDFR}. The denoising results on the ICVL dataset are illustrated in \cref{tab-tran-all}. The second column of each sub-table (named ``\textit{*Original}") in \cref{tab-tran-all} records the performance of the original LRTV, E3DTV, and LRTFDFR models.\footnote{Note that the data fidelity terms of LRTV, E3DTV both use the $\ell_1$ norm. The data fidelity term of LRTFDFR includes the $\ell_2$ norm and the weighted $\ell_1$ norm. The ablation study using a uniform weight in our weighted models (i.e., the $\ell_2$ norm) is different from those ``Original" models.} Additionally, we have the following observations from \cref{tab-tran-all}. First, most models outperform their original counterparts. This can be attributed to the powerful ability of the data fidelity term $\|h_\theta(\Y)\odot(\Y - \X)\|^2$ to describe complex noise. Second, although only ``Gaussian+impulse" noise is used to train the DLWnet, the DLW-models, whose weights are predicted by the trained DLWnet, can also achieve superior results to the original models in other noise cases, which further proves the effectiveness of our DLWnet in handling complex noise. This can be explained by the fact that our DLWnet has learned the essential noise characteristics from the training data and thus can generalize well to other noise types.

\begin{table}[t]
	\renewcommand{\arraystretch}{1.15}
	\newcommand{\mysize}{0.92cm}
	\fontsize{8.5}{9.5}\selectfont
	\caption{Complex target models: average test performance of LRTV, E3DTV and LRTFDFR and their corresponding DLW-models on ICVL dataset. The best results in each \textbf{row} are in \textbf{bold}, and the second best results in each \textbf{row} are with \ul{underline}.}
	\label{tab-tran-all}
	\centering
	\begin{tabular}{M{1.2cm} | M{\mysize}  M{\mysize} M{\mysize}  M{\mysize} M{\mysize} M{\mysize} M{\mysize} M{\mysize} M{\mysize+0.36cm} M{\mysize}}
		\hline
		\multirow{2}{1.2cm}{\backslashbox[1.6cm]{noise}{source}} & noisy & *\textit{Ori} & N & T & TS & N+T & N+TS & T+TS & N+T+TS & Abl\\
		\cline{2-11}
		& \multicolumn{10}{c}{evaluation index: PSNR/SSIM}\\
		\Xhline{0.8pt}
		\multicolumn{11}{c}{target model 1 : \tb{LRTV}} \cr
		\hline
		Case 1 & 14.86/ 0.355 & 31.67/ 0.906 & 34.44/ 0.942 & 28.76/ 0.815 & 31.45/ 0.912 & \ul{34.54}/ \ul{0.95} & 34.3/ 0.942 & 34.31/ 0.942 & \tb{34.94}/ \tb{0.95} & 24.42/ 0.746 \\
		Case 2 & 16.16/ 0.4 & 32.91/ 0.925 & 35.23/ 0.954 & 29.23/ 0.835 & 32.73/ 0.928 & \ul{35.41}/ \ul{0.959} & 35.06/ 0.954 & 35.05/ 0.95 & \tb{35.62}/ \tb{0.959} & 28.66/ 0.857 \\
		Case 3 & 16.06/ 0.395 & 31.34/ 0.914 & 34.21/ 0.948 & 28.85/ 0.83 & 31.01/ 0.918 & \ul{34.22}/ \tb{0.953} & 34.08/ 0.946 & 33.57/ 0.944 & \tb{34.52}/ \ul{0.952} & 27.58/ 0.852 \\
		Case 4 & 13.13/ 0.32 & 35.49/ 0.954 & 36.61/ 0.965 & 32.25/ 0.915 & 34.44/ 0.952 & \ul{36.9}/ \tb{0.971} & 36.59/ 0.966 & 36.61/ 0.969 & \tb{37.0}/ \ul{0.969} & 27.74/ 0.822 \\
		Case 5 & 12.94/ 0.276 & 30.3/ 0.897 & \ul{33.57}/ 0.942 & 28.94/ 0.835 & 30.3/ 0.91 & 33.34/ \ul{0.947} & 33.55/ 0.943 & 33.07/ 0.943 & \tb{33.9}/ \tb{0.948} & 23.44/ 0.719 \\
		\hline
		\multicolumn{11}{c}{target model 2 : \tb{E3DTV}} \cr
		\hline
		Case 1 & 14.86/ 0.355 & 34.61/ 0.951 & 34.97/ 0.947 & 30.34/ 0.874 & 34.37/ 0.94 & 35.16/ \tb{0.956} & 34.95/ 0.947 & \tb{35.7}/ \ul{0.955} & \ul{35.41}/ 0.953 & 25.36/ 0.738 \\
		Case 2 & 16.16/ 0.4 & 34.27/ 0.947 & 35.04/ 0.953 & 29.76/ 0.857 & 34.77/ 0.947 & 35.18/ \ul{0.961} & 35.1/ 0.954 & \tb{35.81}/ \tb{0.961} & \ul{35.46}/ 0.959 & 30.6/ 0.869 \\
		Case 3 & 16.06/ 0.395 & 33.52/ 0.943 & 35.06/ 0.954 & 29.59/ 0.859 & 33.53/ 0.941 & 34.63/ \ul{0.957} & \ul{35.1}/ 0.953 & 34.83/ 0.956 & \tb{35.2}/ \tb{0.957} & 29.95/ 0.876 \\
		Case 4 & 13.13/ 0.32 & 32.88/ 0.927 & 34.12/ 0.942 & 27.81/ 0.792 & 33.53/ 0.933 & 34.13/ 0.948 & 34.19/ 0.943 & \tb{34.77}/ \tb{0.949} & \ul{34.55}/ \ul{0.948} & 28.91/ 0.823 \\
		Case 5 & 12.94/ 0.276 & 31.63/ 0.915 & 33.21/ 0.931 & 27.75/ 0.789 & 31.15/ 0.911 & 32.61/ \ul{0.934} & \ul{33.23}/ 0.93 & 32.77/ 0.932 & \tb{33.26}/ \tb{0.935} & 24.84/ 0.716 \\
		\hline
		\multicolumn{11}{c}{target model 3 : \tb{LRTFDFR}} \cr
		\hline
		Case 1 & 14.86/ 0.355 & 29.69/ 0.838 & 33.63/ 0.942 & 27.97/ 0.812 & 32.03/ 0.922 & 33.7/ 0.951 & 33.81/ 0.944 & \tb{34.16}/ \tb{0.953} & \ul{34.09}/ \ul{0.951} & 23.43/ 0.72 \\
		Case 2 & 16.16/ 0.4 & 30.83/ 0.848 & 34.21/ 0.95 & 28.66/ 0.837 & 32.82/ 0.929 & 34.17/ 0.952 & 34.03/ 0.949 & \tb{34.56}/ \ul{0.953} & \ul{34.44}/ \tb{0.954} & 29.34/ 0.852 \\
		Case 3 & 16.06/ 0.395 & 27.92/ 0.773 & 33.25/ 0.94 & 28.29/ 0.827 & 31.25/ 0.906 & 33.1/ 0.942 & 33.37/ 0.941 & \ul{33.43}/ \ul{0.943} & \tb{33.75}/ \tb{0.946} & 28.45/ 0.856 \\
		Case 4 & 13.13/ 0.32 & 31.52/ 0.855 & 35.41/ 0.962 & 31.12/ 0.91 & 34.7/ 0.953 & 35.23/ \ul{0.964} & 35.23/ 0.961 & \tb{35.51}/ 0.963 & \ul{35.42}/ \tb{0.964} & 26.56/ 0.784 \\
		Case 5 & 12.94/ 0.276 & 28.42/ 0.821 & 33.0/ 0.938 & 28.41/ 0.834 & 31.61/ 0.925 & 32.7/ 0.944 & 33.22/ 0.941 & \ul{33.22}/ \tb{0.948} & \tb{33.43}/ \ul{0.947} & 22.87/ 0.709 \\
		\Xhline{0.8pt}
	\end{tabular}
\end{table}

\cref{tab-tran-com} averages the PSNR/SSIM results of all five noise cases. From \cref{tab-tran-all} and \cref{tab-tran-com}, we can observe that for DLW-LRTV, the `N+T+TS'-type and `N+T'-type DLWnets obtain the best and second-best performance, which is reasonable since LRTV mainly considers low-rankness and spatial smoothness, corresponding to the 'N' and 'T' regularizations, respectively. As for DLW-E3DTV and DLW-LRTFDFR, the last two DLWnets achieve the best results, which can be attributed to the fact that a DLWnet learned from diverse models can capture more image priors and thus generalize well to a wide range of new target models. Additionally, we notice that the `T'-type DLWnet achieves the lowest performance among the original and other DLW-models. The reason might be that the TV regularization only captures spatial features while neglecting spectral correlation, which have been verified to be of great importance for image denoising.

Further, \cref{fig-part2-syn} shows the visual comparison of LRTV and its corresponding models with different DLWnets. It can be easily seen that except for the `T'-type DLWnet, the proposed models with other DLWnets can effectively remove complex noise.

We conducted a classification experiment on the Indian Pines dataset. The scene in this dataset contains 16 classes, and the number of samples per class ranges from 20 to 2455. The classifier for this experiment is the basic support vector machine (SVM). For each class, we randomly select 15 samples for training, and the remaining samples are used for testing. For each method, the experiment is conducted 10 times. The best result for each method is presented in \cref{fig-classification}. The overall accuracy is used to measure the classification results. From  \cref{fig-classification}, we can see that the denoised images produced by the model with the proposed weighting scheme achieve higher classification accuracy than those produced by its corresponding model, indicating that the proposed weighting scheme can help improve the denoising performance.

For color image denoising, we apply `T'-type DLWnet to the SVTV model \cite{SVTV}. The details of how to solve DLW-SVTV model are presented in the Appendix. The denoising results are summrized in \cref{tab-tran-svtv}. It is clearly seen that the DLWnet helps improve the SVTV model by a large margin.

\begin{table}[ht]
	\renewcommand{\arraystretch}{1.15}
	\fontsize{8.5}{9.5}\selectfont
	\caption{Average PSNR/SSIM values of all five kinds of noise patterns obtained by LRTV, E3DTV, LRTFDFR and their corresponding DLW-models on ICVL dataset.}
	\label{tab-tran-com}
	\centering
	\begin{tabular}{M{2cm} !{\vrule width1pt} M{2cm} M{2cm} M{2cm}}
		\Xhline{0.8pt}
		\backslashbox[2.4cm]{source}{target} &    LRTV   &   E3DTV   &   LRTFDFR  \\
		\hline
		*\textit{Original}  &  32.34/0.9192	&	33.38/0.9367	&	29.68/0.827 \\
		N                   & 34.81/0.9499	&	34.48/0.9455	&	33.9/0.9464\\
		T                   & 29.61/0.8459	&	29.05/0.8342	&	28.89/0.8439\\
		TS                  & 31.99/0.924	&	33.47/0.9344	&	32.48/0.9271\\
		N+T                 & \ul{34.88}/\tb{0.9560}	&	34.34/\tb{0.9511}	&	33.78/0.9505\\
		N+TS                & 34.72/0.9501	&	34.51/0.9453	&	33.93/0.9474\\
		T+TS                & 34.52/0.9493	&	\tb{34.78}/\ul{0.9507}	&	\ul{34.18}/\ul{0.9516}\\
		N+T+TS              & \tb{35.20}/\ul{0.9557}	&	\ul{34.78}/0.9502	&	\tb{34.23}/\tb{0.9524}\\
		\Xhline{0.8pt}
	\end{tabular}
\end{table}

\begin{table}[t]
	\renewcommand{\arraystretch}{1.15}
	\newcommand{\mysize}{2cm}
	\fontsize{8.5}{9.5}\selectfont
	\caption{Complex target models: average test performance of SVTV and the corresponding DLW-model on BSDS dataset. The best results in each \textbf{row} are in \textbf{bold}.}
	\label{tab-tran-svtv}
	\centering
	\begin{tabular}{M{1.2cm} | M{\mysize} M{\mysize}  M{\mysize} }
		\hline
		\multirow{2}{1.2cm}{\backslashbox[1.6cm]{noise}{source}} & noisy & *\textit{Ori} & T \\
		\cline{2-4}
		& \multicolumn{3}{c}{evaluation index: PSNR/SSIM}\\
		\Xhline{0.8pt}
		Case 1 & 10.35/0.0843 & 19.54/0.4183 & \tb{23.75}/\tb{0.5724}  \\
		Case 2 & 17.20/0.2813 & 24.39/0.6250 & \tb{25.54}/\tb{0.6665}  \\
		Case 3 & 12.65/0.1664 & 19.25/0.4983 & \tb{21.52}/\tb{0.5730}  \\
		Case 4 & 13.68/0.2116 & 23.17/0.5348 & \tb{23.91}/\tb{0.5786}  \\
		Case 5 & 8.91/0.0578 & 16.90/0.3671 & \tb{18.75}/\tb{0.4674}  \\
		\Xhline{0.8pt}
	\end{tabular}
\end{table}

\begin{figure}[t]
	\renewcommand{\arraystretch}{1.15}
	\newcommand{\mysize}{2.6cm}
	\fontsize{8.5}{9.5}\selectfont
	\newcommand{\minivs}{4pt}
	\newcommand{\vs}{2pt}
	\centering
	\begin{minipage}[t]{\mysize*5+1cm}
		\centering
		\begin{minipage}[b]{\mysize}
			\centering
			clean 
			\includegraphics[width=\mysize]{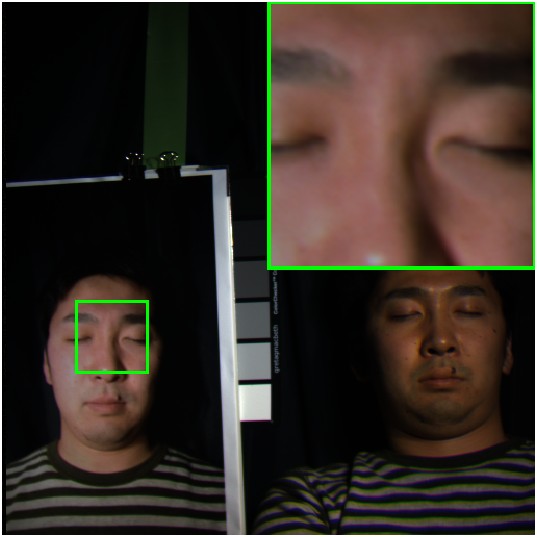} \
			(PSNR, SSIM)
		\end{minipage}\hspace{\minivs}
		\begin{minipage}[b]{\mysize}
			\centering
			noisy 
			\includegraphics[width=\mysize]{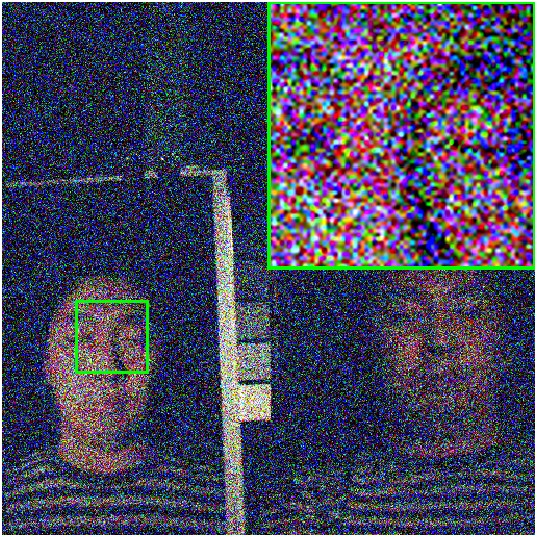} \
			(10.86, 0.1026)
		\end{minipage}\hspace{\minivs}
		\begin{minipage}[b]{\mysize}
			\centering
			\textit{*Original}  
			\includegraphics[width=\mysize]{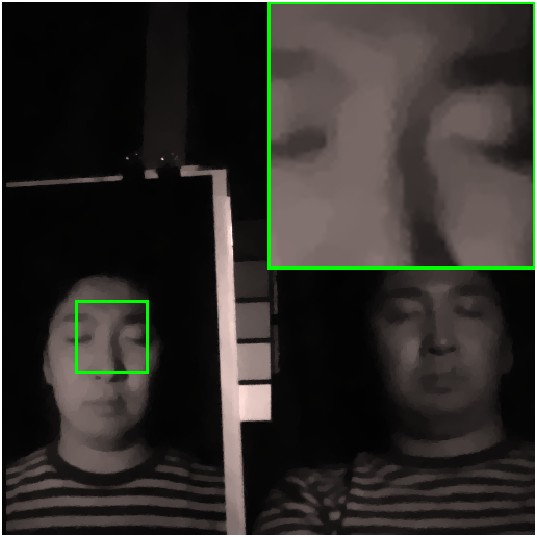} \
			(32.96, \tb{0.9087})
		\end{minipage}\hspace{\minivs}
		\begin{minipage}[b]{\mysize}
			\centering
			N 
			\includegraphics[width=\mysize]{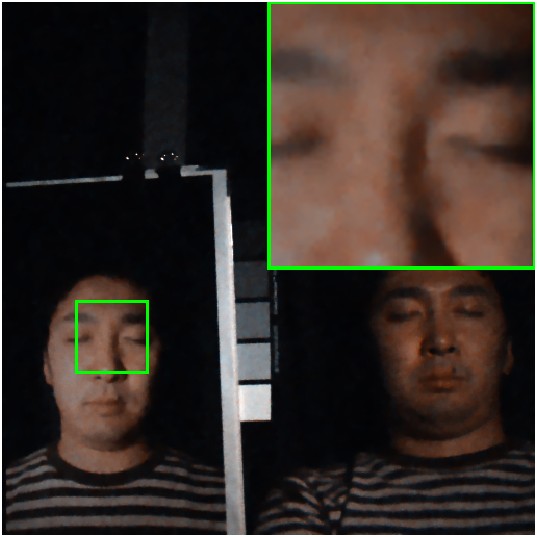} \
			(\ul{36.75}, 0.8761)
		\end{minipage}\hspace{\minivs}
		\begin{minipage}[b]{\mysize}
			\centering
			T 
			\includegraphics[width=\mysize]{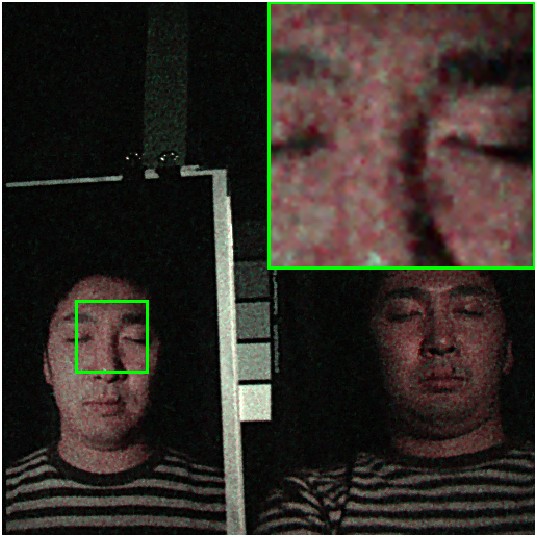} \
			(31.95, 0.7171)
		\end{minipage} \vspace{1pt} \\
		\begin{minipage}[b]{\mysize}
			\centering
			TS 
			\includegraphics[width=\mysize]{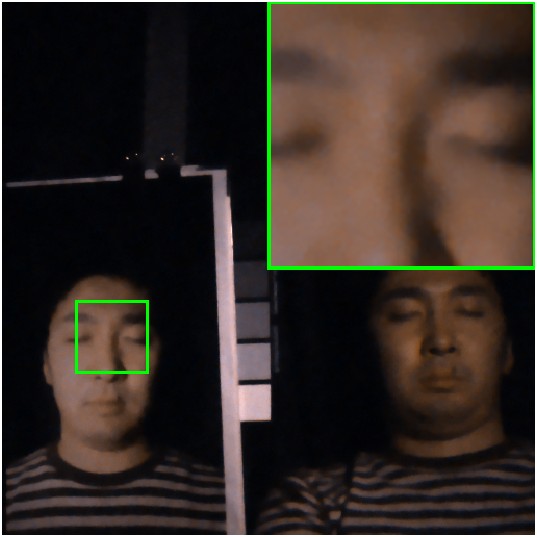} \
			(34.59, 0.8955)
		\end{minipage}\hspace{\minivs}
		\begin{minipage}[b]{\mysize}
			\centering
			N+T 
			\includegraphics[width=\mysize]{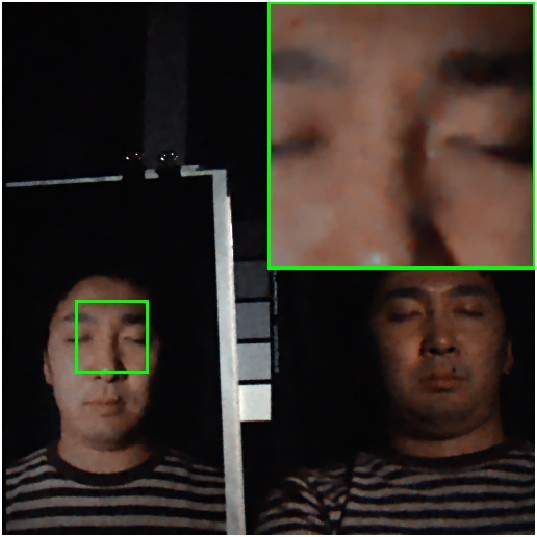} \
			(36.46, 0.8578)
		\end{minipage}\hspace{\minivs}
		\begin{minipage}[b]{\mysize}
			\centering
			N+TS 
			\includegraphics[width=\mysize]{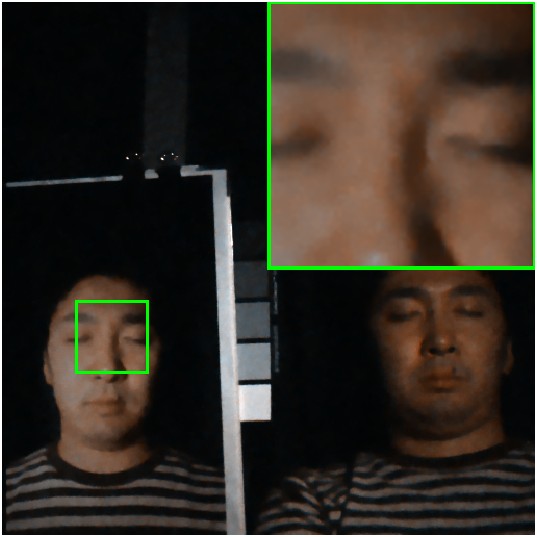} \
			(\tb{37.11}, \ul{0.9083})
		\end{minipage}\hspace{\minivs}
		\begin{minipage}[b]{\mysize}
			\centering
			T+TS 
			\includegraphics[width=\mysize]{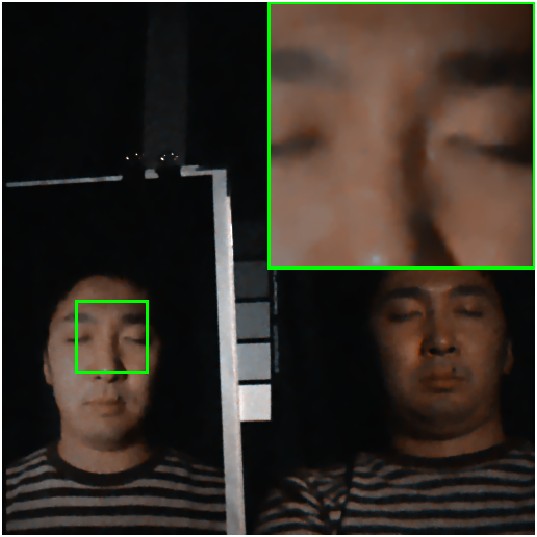} \
			(36.0, 0.8259)
		\end{minipage}\hspace{\minivs}
		\begin{minipage}[b]{\mysize}
			\centering
			N+T+TS 
			\includegraphics[width=\mysize]{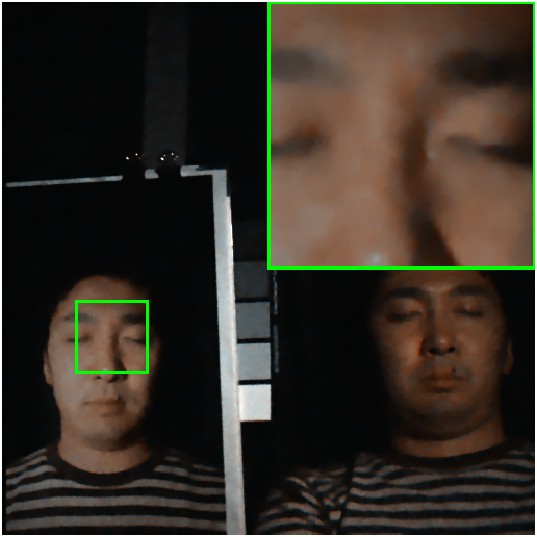} \
			(36.72, 0.8637)
		\end{minipage}\hspace{\minivs}
		
	\end{minipage}
	\caption{Denoising results (pesudo-color image) of LRTV and its corresponding DLW-models on image ``photo and face" image of CAVE dataset. The noisy type is ``Gaussian+impulse" (Case 1).}
	\label{fig-part2-syn}
\end{figure}

\begin{figure}[t]
	\newcommand{\mywid}{3cm}
	\newcommand{\myvs}{0pt}
	\newcommand{\mymi}{-0.2cm}
	\fontsize{8}{9}\selectfont
	\renewcommand{\arraystretch}{1.15}
	\centering
	\begin{tabular}{l  l  l l}
		\hspace{\mymi}
		\begin{minipage}[t]{\mywid}
			\centering
			\includegraphics[width=\mywid]{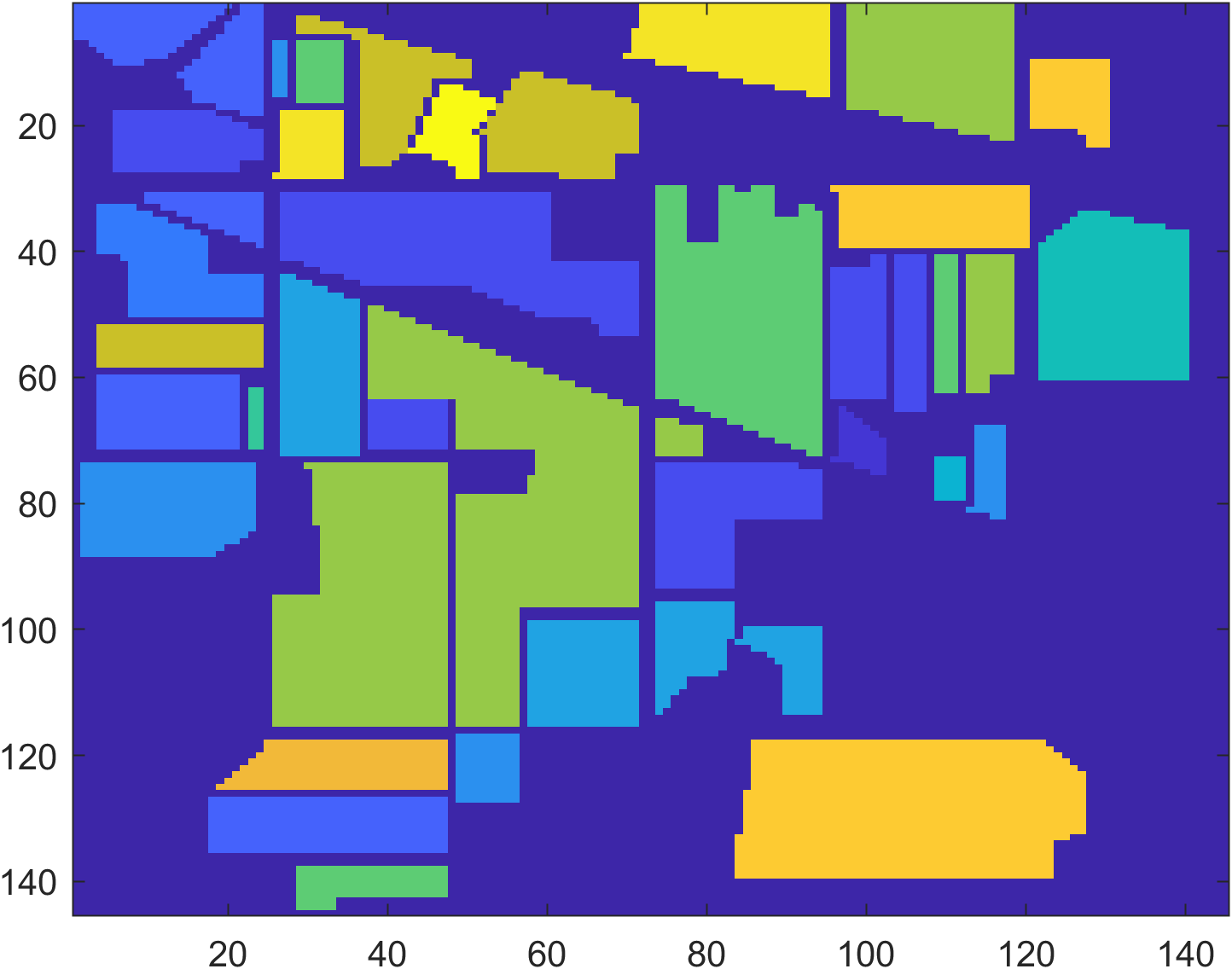}\\
			label \\
			accuracy
		\end{minipage} \hspace{\mymi} &
		\hspace{\mymi}
		\begin{minipage}[t]{\mywid}
			\centering
			\includegraphics[width=\mywid]{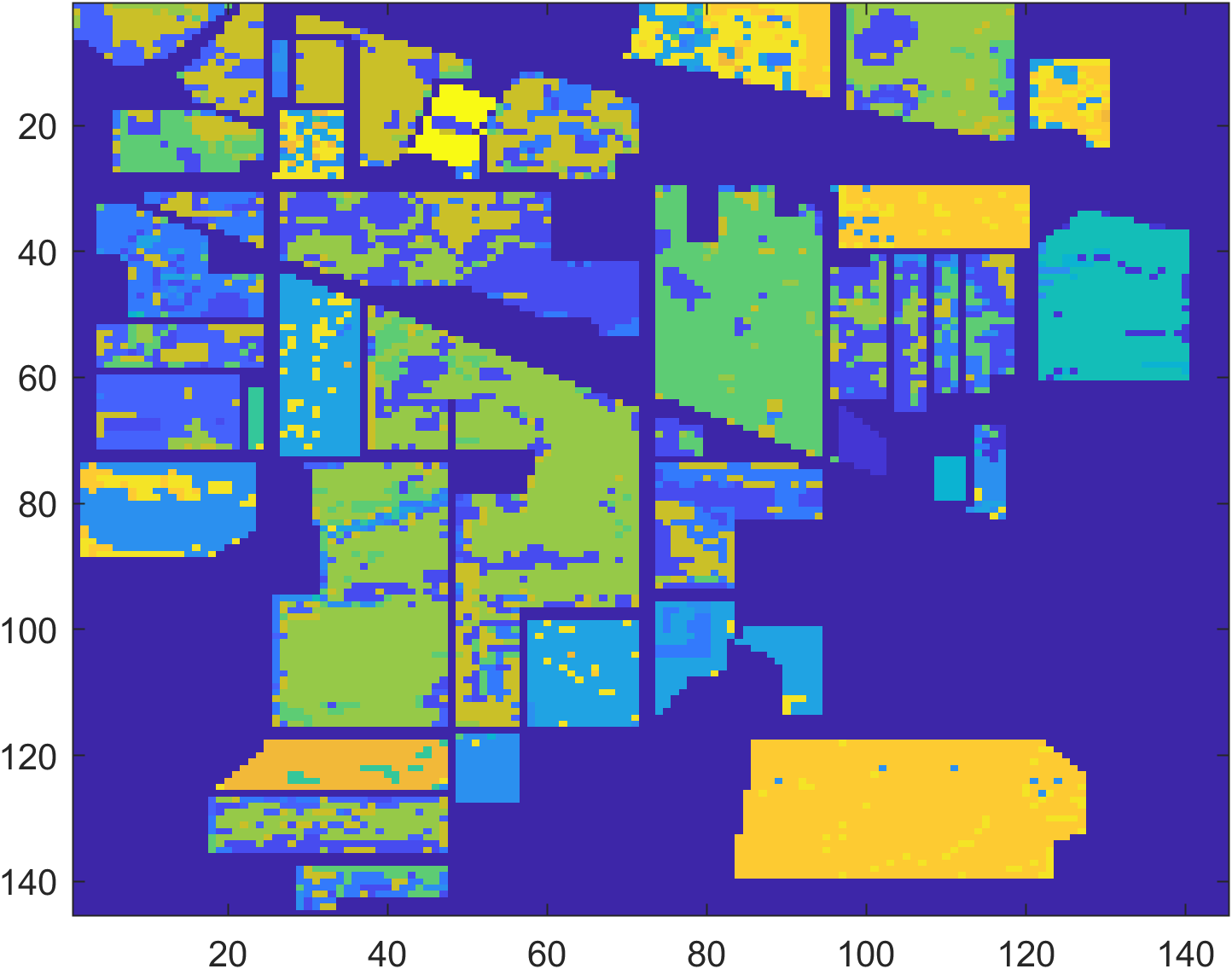}\\
			LRTV \\ 
			64.77\% \vspace{3pt}
		\end{minipage} \hspace{\mymi} &
		\hspace{\mymi}
		\begin{minipage}[t]{\mywid}
			\centering
			\includegraphics[width=\mywid]{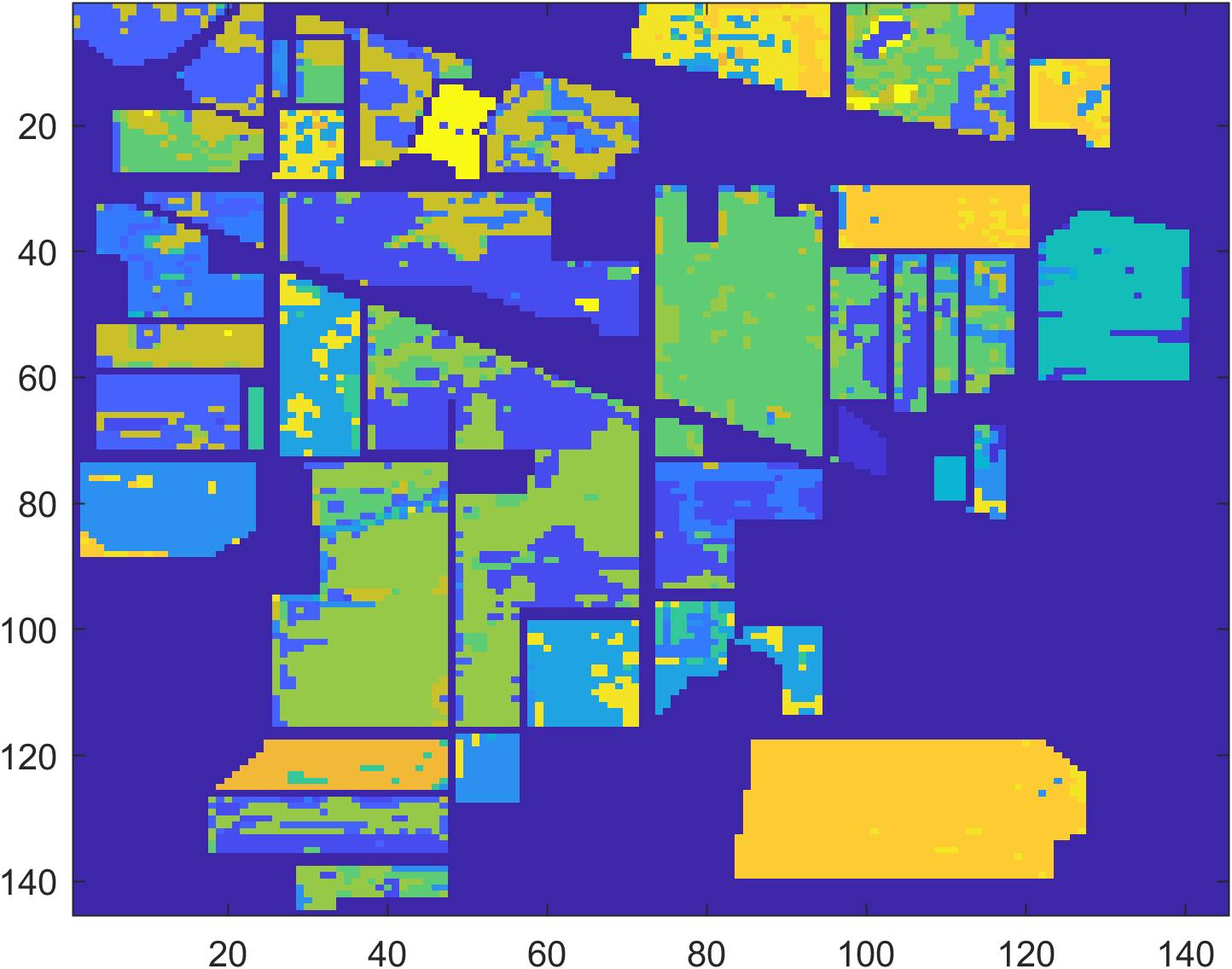}\\
			E3DTV \\
			66.47\%
		\end{minipage} \hspace{\mymi} & 
		\hspace{\mymi}
		\begin{minipage}[t]{\mywid}
			\centering
			\includegraphics[width=\mywid]{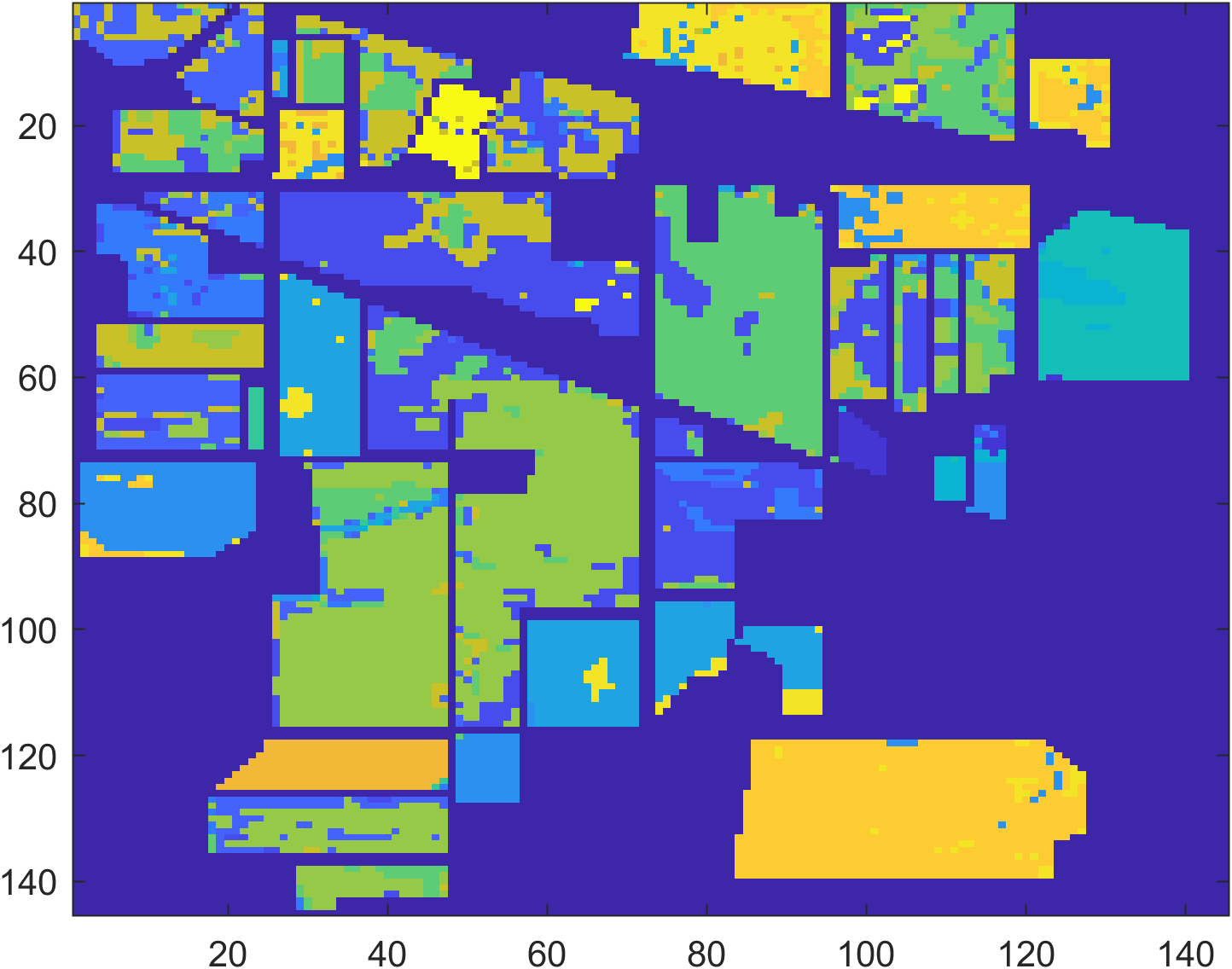}\\
			LRTFDFR \\
			71.45\%
		\end{minipage} \\ 
		\hspace{\mymi}
		\begin{minipage}[t]{\mywid}
			\centering
			\includegraphics[width=\mywid]{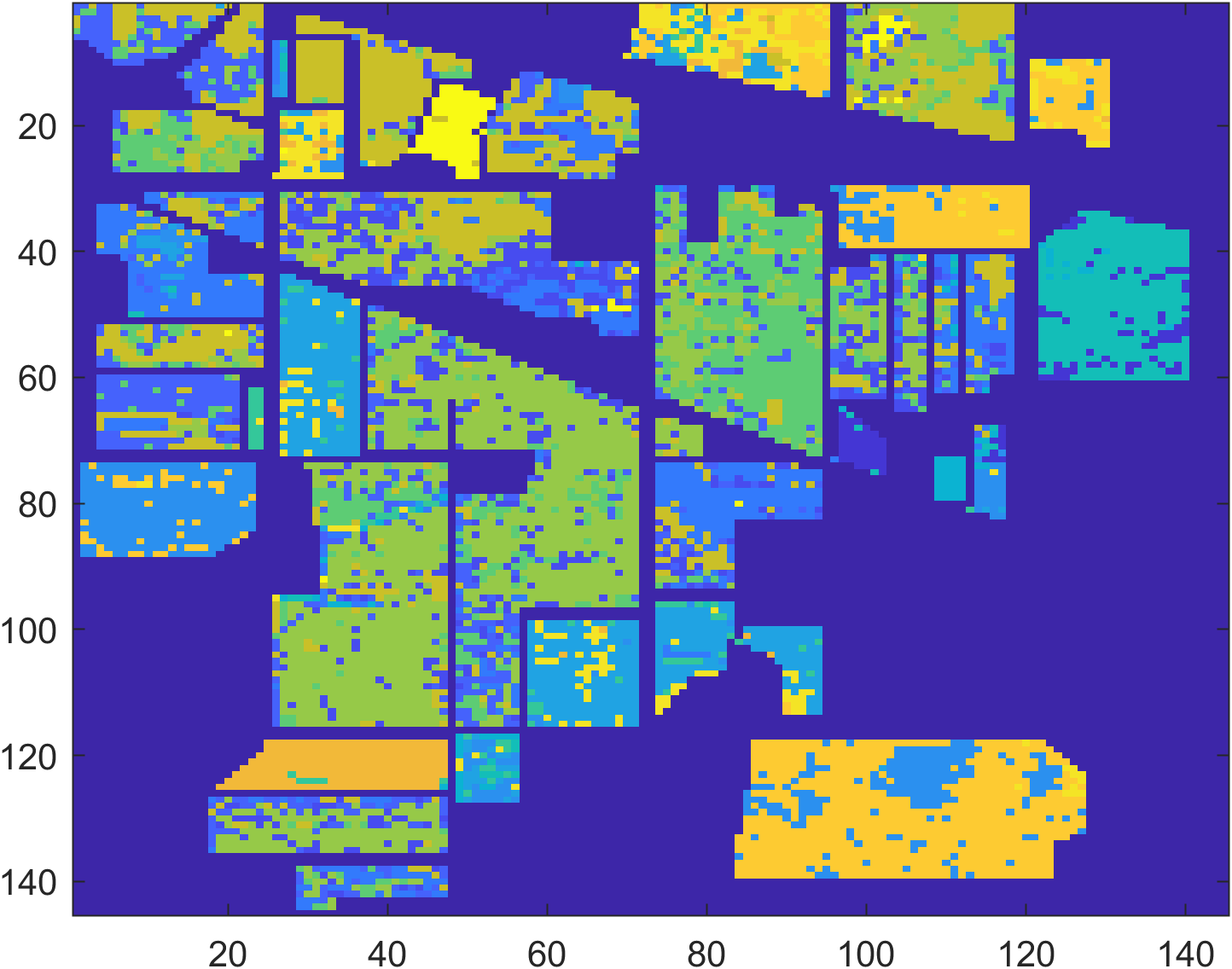}\\
			noisy \\
			55.43\%
		\end{minipage} \hspace{\mymi} &
		\hspace{\mymi}
		\begin{minipage}[t]{\mywid}
			\centering
			\includegraphics[width=\mywid]{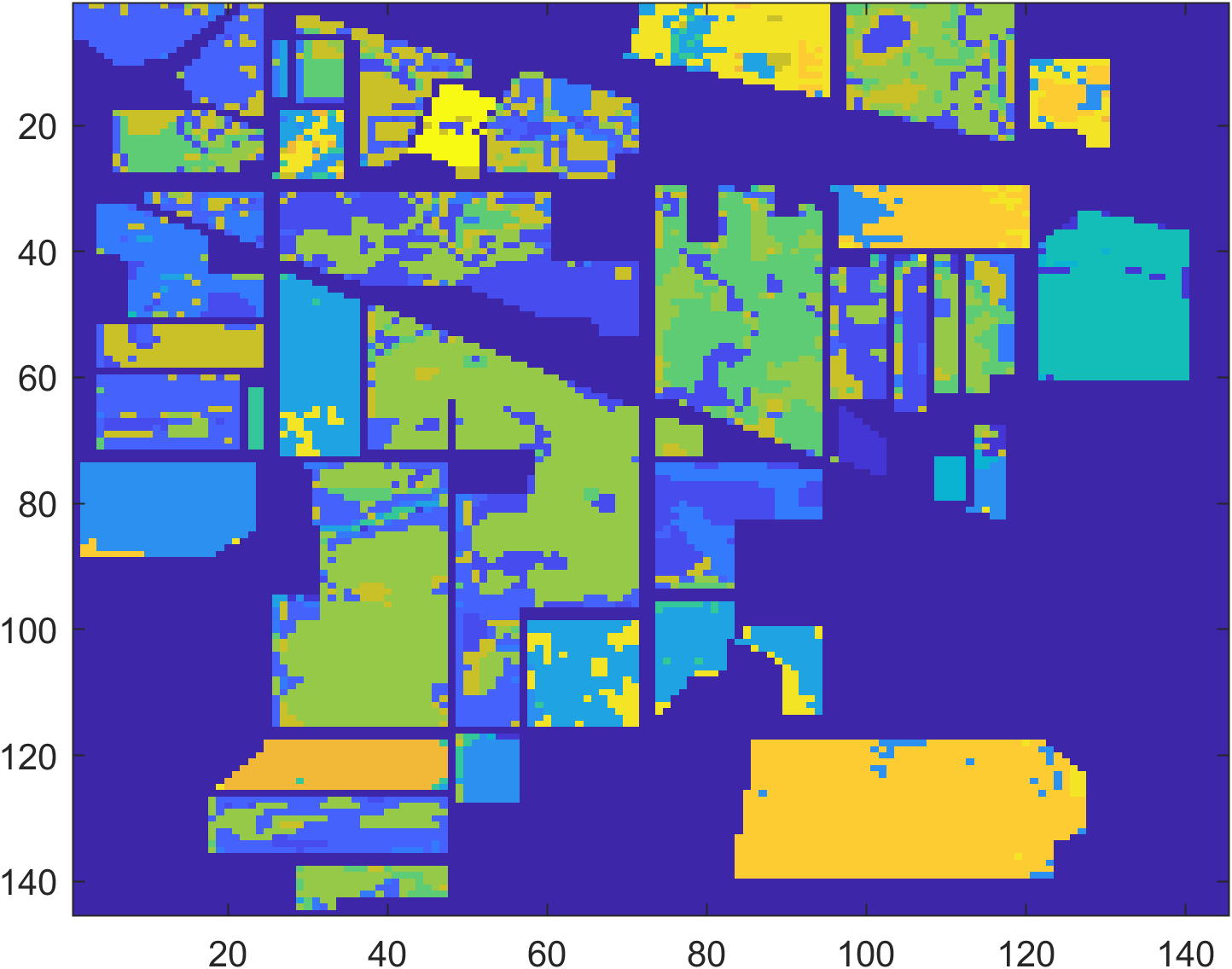}\\
			DLW-LRTV \\
			\tb{67.27\%}
		\end{minipage} \hspace{\mymi} &
		\hspace{\mymi}
		\begin{minipage}[t]{\mywid}
			\centering
			\includegraphics[width=\mywid]{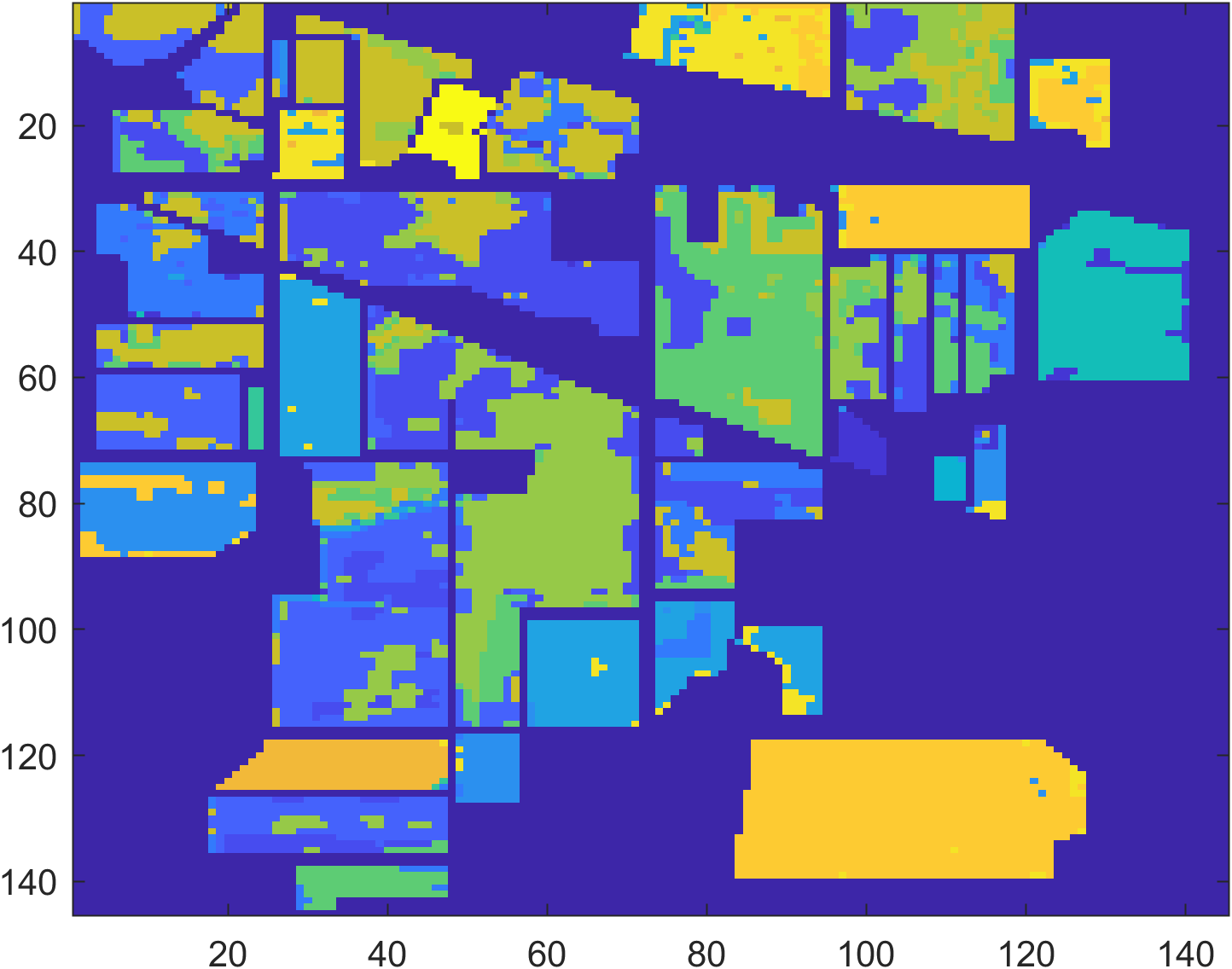}\\
			DLW-E3DTV \\
			\tb{71.61\%}
		\end{minipage} \hspace{\mymi} &
		\hspace{\mymi}
		\begin{minipage}[t]{\mywid}
			\centering
			\includegraphics[width=\mywid]{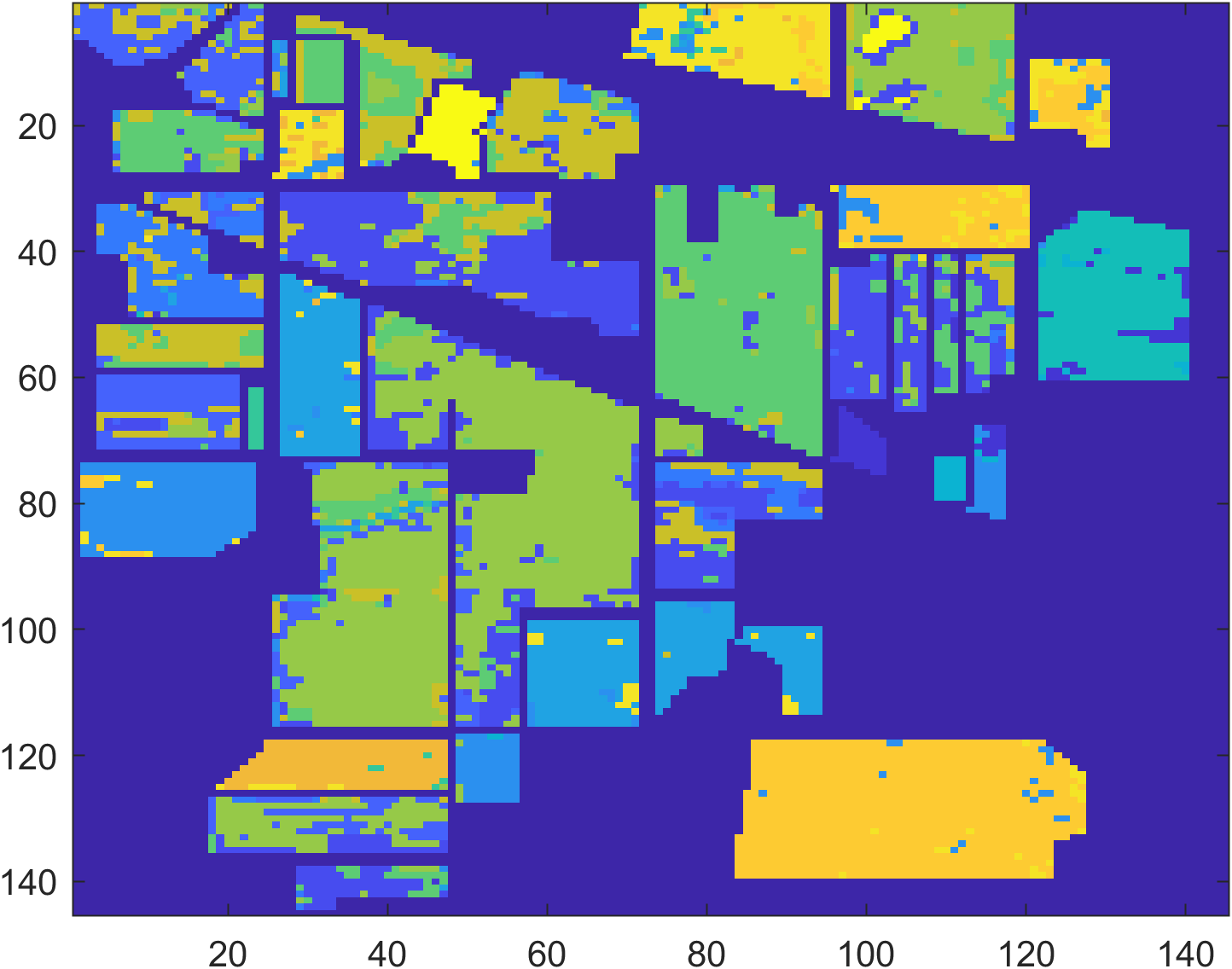}\\
			DLW-LRTFDFR \\
			\tb{73.69\%}
		\end{minipage}
	\end{tabular}
	\caption{Classification results of the compared methods on the Indian Pines dataset.}
	\label{fig-classification}
\end{figure}

\subsubsection{Special target models: DIP and PnP}\label{sec-DIP}
As mentioned in \cref{sec:appplication}, the DIP problem of the form \eqref{wdip} and the PnP framework can be seen as special kinds of target models. In this experiment, we use S2DIP \cite{S2DIP} as the backbone model for DIP and SERT \cite{SERT}\footnote{Pre-trained Gaussian model is provided in \url{https://github.com/MyuLi/SERT}} as the deep denoiser for the PnP framework.

\begin{figure*}[t]
	\newcommand{\mysize}{3cm}
	\newcommand{\vs}{1pt}
	\fontsize{8.5}{9.5}\selectfont
	\renewcommand{\arraystretch}{1.35}
	\centering
	\begin{minipage}[t]{\mysize}
		\centering
		\includegraphics[width=\mysize]{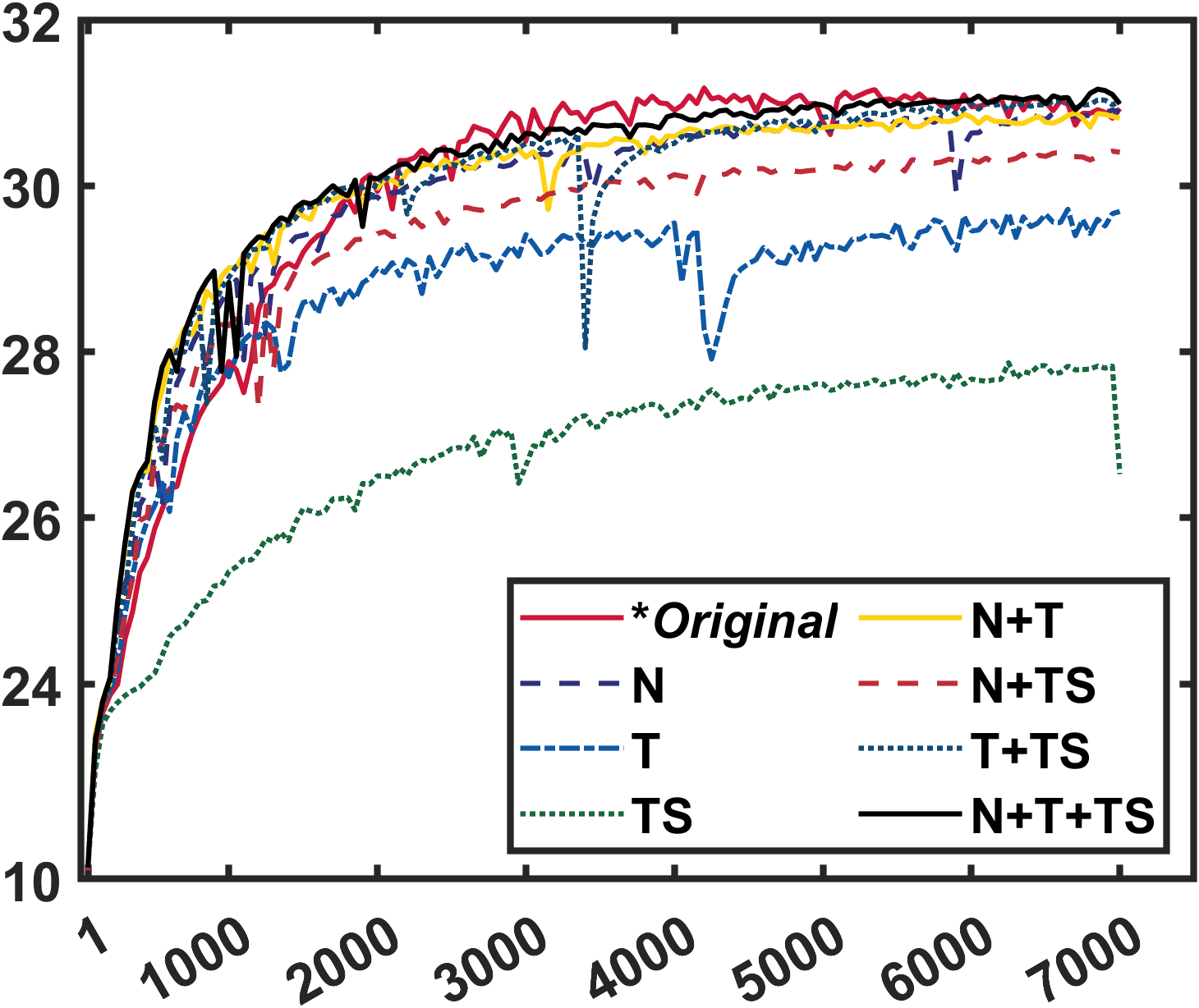}\\
		Case 1
	\end{minipage}\hspace{\vs}
	\begin{minipage}[t]{\mysize}
		\centering
		\includegraphics[width=\mysize]{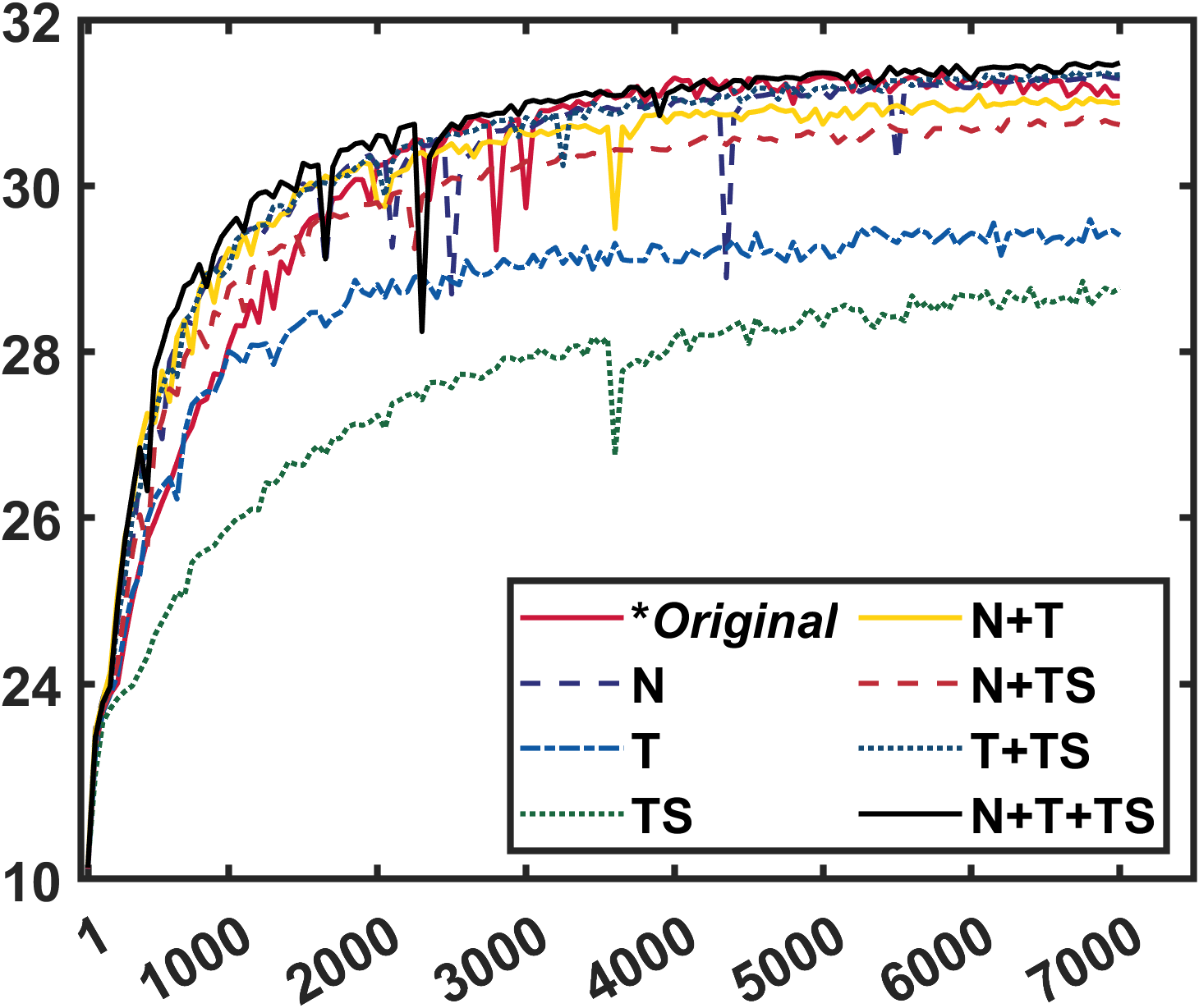}\\
		Case 2
	\end{minipage}\hspace{\vs}
	\begin{minipage}[t]{\mysize}
		\centering
		\includegraphics[width=\mysize]{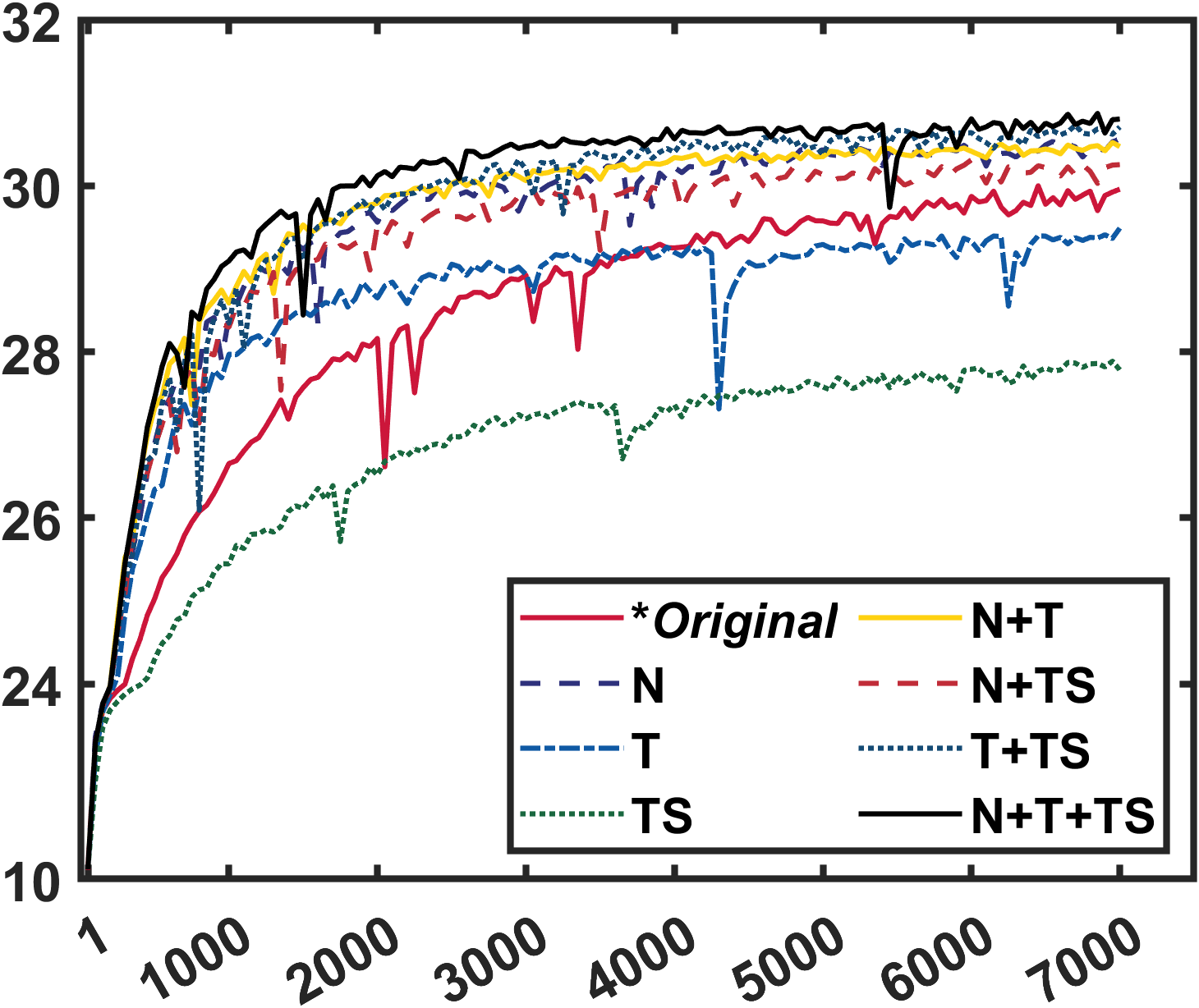}\\
		Case 3
	\end{minipage} 
	\begin{minipage}[t]{\mysize}
		\centering
		\includegraphics[width=\mysize]{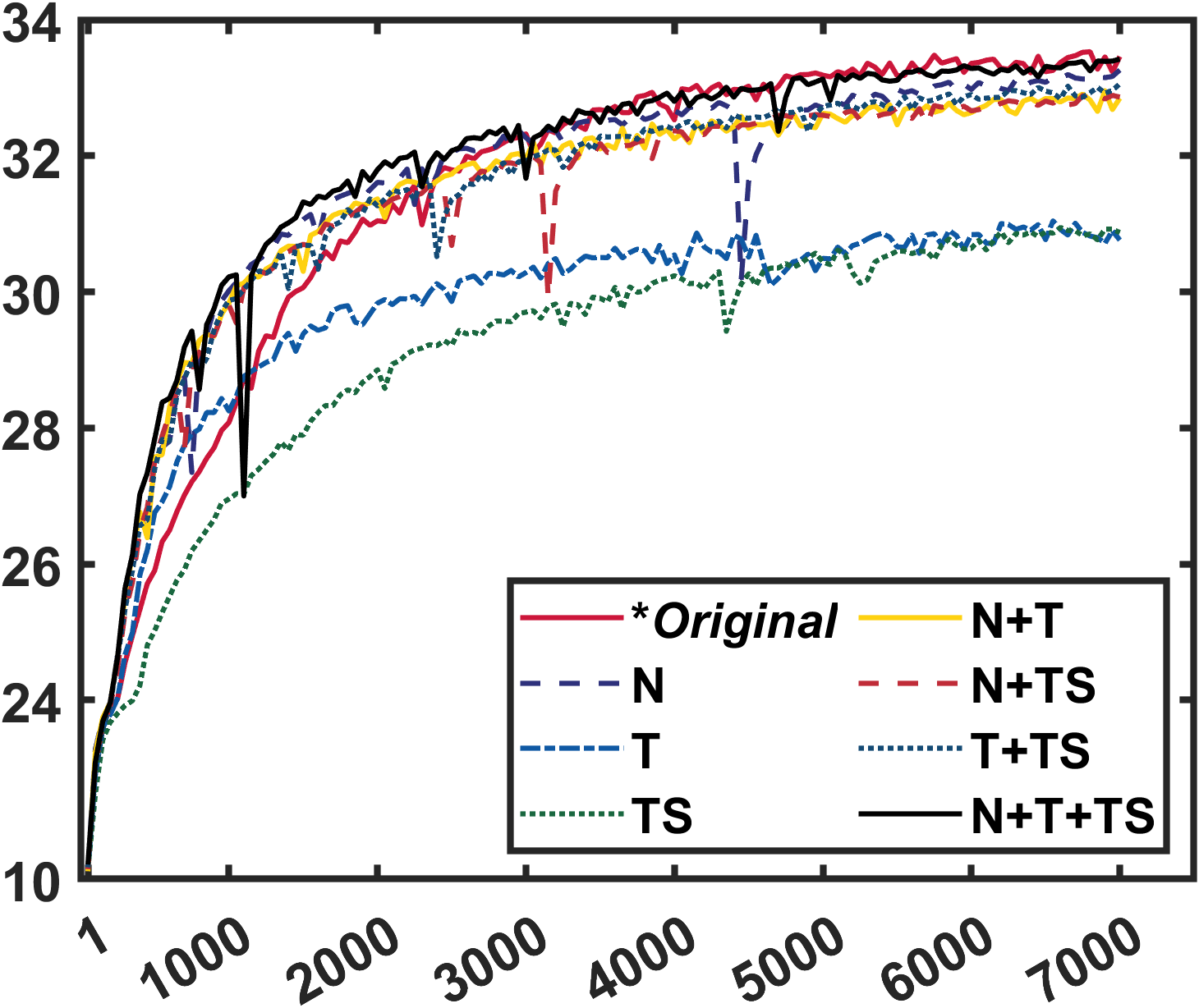}\\
		Case 4
	\end{minipage}\hspace{\vs}
	\begin{minipage}[t]{\mysize}
		\centering
		\includegraphics[width=\mysize]{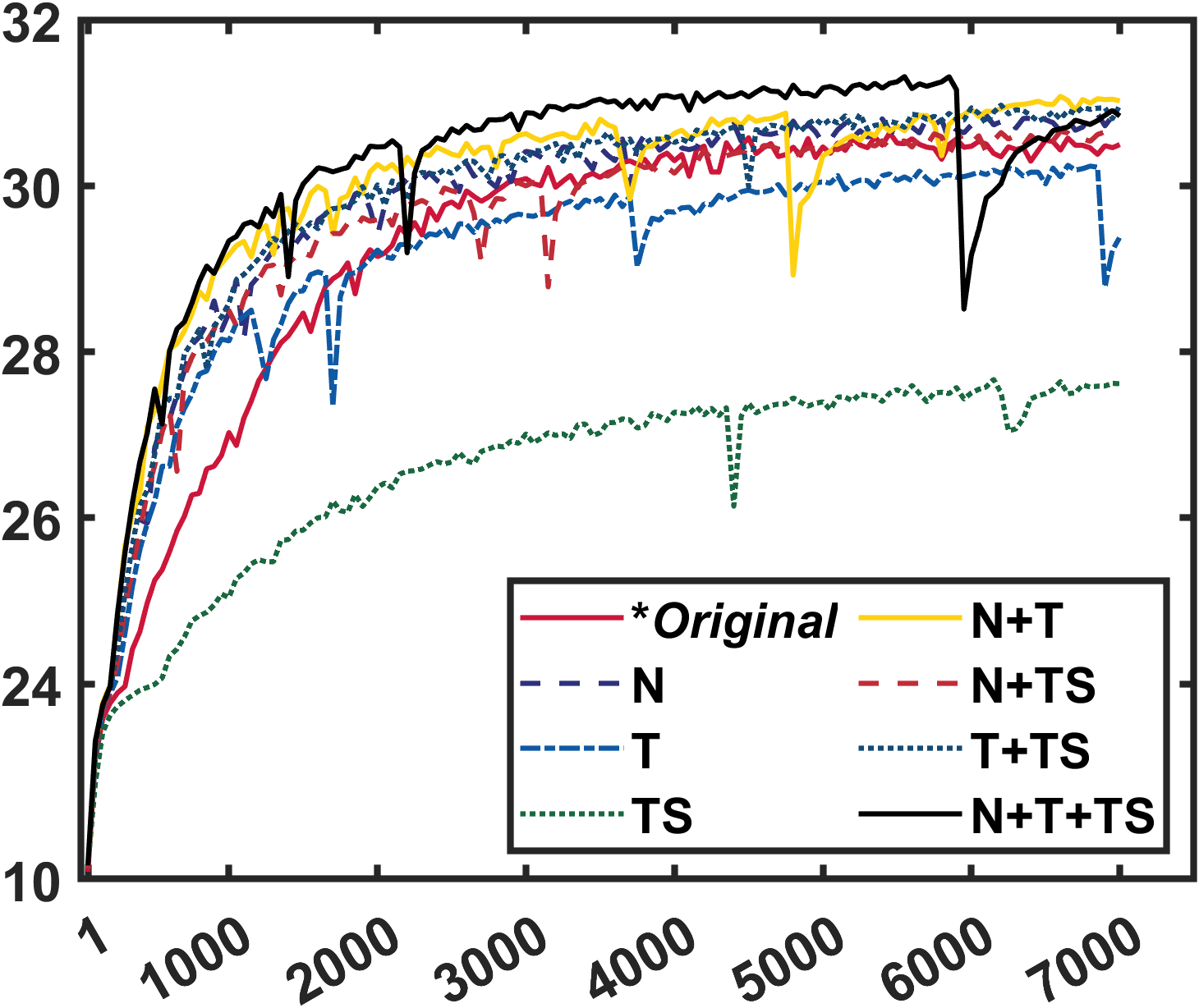}\\
		Case 5
	\end{minipage}
	\caption{The tendency of PSNR values during iterations for S2DIP and DLW-S2DIP with different DLWnets. }
	\label{fig-dip-converge}
\end{figure*}

\cref{tab-S2DIP} shows the experimental results of DIP problem on the CAVE dataset in five noise cases. Each experiment is repeated three times, and the average results are recorded. In \cref{tab-S2DIP}, the ``max" column presents the maximum PSNR during training. The ``final" column shows the average PSNR of the last 100 iterations and represents the converged results. The ``\textbar d.v\textbar" column shows the absolute difference between the maximum and the final values. From \cref{tab-S2DIP}, we can see that the DLW-S2DIP model outperforms the original model in most noise cases. For example, most DLW-S2DIP models perform better than the original model in the complex noise case (i.e., Case 5). Besides, the ``\textbar d.v\textbar" value of the DLW-S2DIP model is always smaller than that of S2DIP, which reveals that the DLW-S2DIP models are more stable than the original model. A different observation from previous experiments is that the `T'-type DLWnet does not perform the worst among all DLWnets. This is perhaps due to the extra implicit image prior brought by network $g_\eta$. To illustrate the convergence, we also plot the PSNR trend during the training process in \cref{fig-dip-converge}. Except for the `TS'-type DLWnet, DLW-S2DIP with the other DLWnets converges very quickly in the early stage. In Case 5, we see that the DLW-S2DIP with the `N+T+TS'-type DLWnet suffers from a large vibration at around the 6000th iteration, which causes a large ``\textbar d.v\textbar" value of 0.424 in \cref{tab-S2DIP}. However, its performance keeps rising after the 6000th iteration and finally reaches the third-best performance.

\cref{tab-sert-cave} shows the denoising results for the PnP framework on the CAVE dataset. We see that for ``stripe" and ``non i.i.d gaussian" noise (Cases 2 and 4, respectively), SERT achieves good performance, possibly because the noise characteristics of these two noise types are similar to those of Gaussian noise. However, for ``impulse", ``deadline" and ``mixture" noise (Case 1, 3 and 5, respectively), the effectiveness of SERT apparently declines. In most cases, the proposed DLWnet helps improve the denoising performance, especially in complex noise cases where the noise characteristics are not quite the same as those of Gaussian noise.

\begin{table}[t]
	\renewcommand{\arraystretch}{1.15}
	\newcommand{\mysize}{1.1cm}
	\fontsize{8.5}{9.5}\selectfont
	\caption{PSNR results of ``jelly\_beans" in CAVE dataset for S2DIP and DLW-S2DIP with different DLWnets. The best results in each \textbf{column} are in \textbf{bold}, and the second best results in each \textbf{column} are with \ul{underline}.}
	\label{tab-S2DIP}
	\centering
	\begin{tabular}{M{1.2cm}  M{0.7cm}  M{\mysize+0.3cm} M{\mysize}  M{\mysize} M{\mysize} M{\mysize}  M{\mysize} M{\mysize} M{\mysize}}
		\Xhline{0.8pt}
		\multicolumn{10}{c}{target model: S2DIP} \\
		\Xhline{0.8pt}
		\backslashbox[1.6cm]{noise}{source}& & *\textit{Original} & N & T & TS & N+T & N+TS & T+TS & N+T +TS \\
		\hline
		\multirow{3}{*}{Case 1} &	max & \tb{31.20} 	&	30.95	&	29.85	&	27.94	&	30.93	&	30.46	&	31.1	&	\ul{31.16} \\
		& final &	30.91	&	30.87	&	29.62	&	27.48	&	30.84	&	30.37	&	\ul{31.02} 	&	\tb{31.09} \\
		& \textbar d.v\textbar &	0.282	&	\tb{0.074} 	&	0.232	&	0.461	&	0.088	&	0.095	&	0.078	&	\ul{0.076} \\
		\hline
		\multirow{3}{*}{Case 2} &	max &	31.41	&	31.35	&	29.63	&	28.9	&	31.11	&	30.85	&	\ul{31.43} 	&	\tb{31.54} \\
		& final &	31.18	&	31.28	&	29.41	&	28.72	&	31.02	&	30.77	&	\ul{31.34} 	&	\tb{31.45} \\
		& \textbar d.v\textbar &	0.229	&	\tb{0.076} 	&	0.226	&	0.176	&	0.087	&	0.083	&	0.09	&	\ul{0.081} \\
		\hline
		\multirow{3}{*}{Case 3} &	max &	30.13	&	30.6	&	29.5	&	27.96	&	30.58	&	30.35	&	\ul{30.80} 	&	\tb{30.87} \\
		& final &	29.89	&	30.47	&	29.37	&	27.83	&	30.45	&	30.22	&	\ul{30.67} 	&	\tb{30.78} \\
		& \textbar d.v\textbar &	0.231	&	0.137	&	0.125	&	0.139	&	\ul{0.123} 	&	0.13	&	0.131	&	\tb{0.097} \\
		\hline
		\multirow{3}{*}{Case 4} &	max &	\tb{33.60} 	&	33.26	&	31.18	&	31.05	&	32.99	&	32.94	&	33.1	&	\ul{33.48} \\
		& final &	\tb{33.44} 	&	33.17	&	30.89	&	30.92	&	32.83	&	32.86	&	33	&	\ul{33.37} \\
		& \textbar d.v\textbar &	0.155	&	\ul{0.092} 	&	0.29	&	0.13	&	0.165	&	\tb{0.085} 	&	0.096	&	0.117\\
		\hline
		\multirow{3}{*}{Case 5} &	max &	30.67	&	30.94	&	30.31	&	27.72	&	 \ul{31.12} 	&	30.74	&	31.02	&	\tb{31.33} \\
		& final &	30.43	&	30.8	&	29.48	&	27.62	&	\tb{31.03} 	&	30.62	&	30.9	&	\ul{30.91} \\
		& \textbar d.v\textbar &	0.242	&	0.145	&	0.831	&	 \ul{0.100}	&	\tb{0.092}	&	0.125	&	0.113	&	0.424\\
		\Xhline{0.8pt}
	\end{tabular}
\end{table} 

\begin{table}[t]
	\renewcommand{\arraystretch}{1.15}
	\newcommand{\mysize}{1cm}
	\fontsize{8.5}{9.5}\selectfont
	\caption{Average test performance of deep-plug-and-play using SERT on CAVE dataset. The best results in each \textbf{row} are in \textbf{bold}, and the second best results in each \textbf{row} are with \ul{underline}.}
	\label{tab-sert-cave}
	\centering
	\begin{tabular}{M{1.2cm} | M{\mysize} M{\mysize} M{\mysize} M{\mysize} M{\mysize} M{\mysize} M{\mysize} M{\mysize} M{\mysize+0.3cm}}
		\wline
		\multirow{2}{1.2cm}{\backslashbox[1.6cm]{noise}{source}}
		& noisy & SERT & N & T & TS & N+T & N+TS & T+TS & N+T+TS \\
		\cline{2-10}
		& \multicolumn{9}{c}{evaluation index: PSNR/SSIM} \\
		\wline
		Case 1 & 14.71/ 0.2733 & 25.77/ 0.7037 & 34.98/ 0.9137 & 31.59/ 0.8336 & 34.28/ 0.9143 & 35.12/ \tb{0.9264} & 35.01/ 0.9161 & \ul{35.23}/ \ul{0.9232} & \tb{35.23}/ 0.9217 \\
		Case 2 & 16.16/ 0.3014 & 35.13/ \ul{0.9494} & 35.54/ 0.9461 & 31.78/ 0.8806 & 34.52/ 0.9290 & 35.27/ 0.9470 & 35.40/ 0.9440 & \tb{35.58}/ \tb{0.9501} & \ul{35.56}/ 0.9490 \\
		Case 3 & 16.19/ 0.3119 & 33.38/ 0.9415 & \tb{35.22}/ \tb{0.9449} & 31.56/ 0.8785 & 33.78/ 0.9206 & 34.82/ 0.9408 & 35.14/ 0.9398 & 34.94/ 0.9424 & \ul{35.20}/ \ul{0.9447} \\
		Case 4 & 13.13/ 0.2365 & \ul{34.63}/ \tb{0.9451} & 34.45/ 0.9297 & 30.12/ 0.8404 & 33.56/ 0.9202 & 34.22/ 0.9298 & 34.48/ 0.9316 & \tb{34.68}/ \ul{0.9383} & 34.61/ 0.9341 \\
		Case 5 & 12.91/ 0.2068 & 25.71/ 0.7000 & 33.08/ 0.8904 & 29.14/ 0.7648 & 31.85/ 0.8794 & 32.86/ \tb{0.8986} & \tb{33.20}/ 0.8920 & 32.89/ 0.8942 & \ul{33.19}/ \ul{0.8950} \\
		\wline
	\end{tabular}
\end{table}

\subsection{Discussions}
In \cref{tab-numparams}, we present the number of parameters of all networks used in this work. Compared with other networks that are directly used for denoising, the DLWnet is more lightweight. We think that learning the noise information should be easier than learning to directly denoise, and a relatively simpler network can help improve the denoising performance in a variational denoising model.

In \cref{tab-time}, we present the running time of each model for reference. Note that in the testing phase, the trained network is fixed, and inferencing the network output takes very little time on GPU devices ($<$0.4s). Thus, the network inference time can be omitted. Overall, the incorporation of DLWnet into the denoising model hardly increases the model's computational complexity. This is mainly because when solving problem \eqref{weight-problem}, the only subproblem related to $\W$ has a simple closed-form solution (see \cref{algo-weight}).

\begin{table}[ht]
	\renewcommand{\arraystretch}{1.15}
	\newcommand{\mysize}{1.5cm}
	\fontsize{8.5}{9.5}\selectfont
	\caption{Number of parameters of networks. Units are in millions (M).}
	\label{tab-numparams}
	\centering
	\begin{tabular}{M{1.5cm} !{\vrule width1pt} M{\mysize} M{\mysize} M{\mysize} M{\mysize} M{\mysize} M{\mysize+0.3cm}}
		\Xhline{0.8pt}
		& HSI-DeNet & HSI-CNN & S2DIP & SERT & DLWnet (HSI) & DLWnet (\text{color image})  \\
		\Xhline{0.8pt}
        \#Params & 2.76M & 0.46M & 1.69M & 1.90M & 0.11M & 0.02M\\
		\Xhline{0.8pt}
	\end{tabular}
\end{table}

\begin{table}[ht]
	\renewcommand{\arraystretch}{1.15}
	\newcommand{\mysize}{1cm}
	\fontsize{8.5}{9.5}\selectfont
	\caption{Running time of each model. The unit is seconds.}
	\label{tab-time}
	\centering
	\begin{tabular}{ M{2cm} | M{\mysize} M{\mysize} M{\mysize} M{\mysize} M{\mysize} M{\mysize} M{\mysize} M{\mysize+0.3cm}}
		\wline
		& *\textit{Ori} & N & T & TS & N+T & N+TS & T+TS & N+T+TS \\
		\wline
		LRTV & 81.31 & 55.08 & 48.84 & 54.02 & 51.51 & 52.12 & 51.20 & 51.14\\
		E3DTV & 55.71 & 59.11 & 57.53 & 66.60 & 60.27 & 60.80 & 63.16 &60.72 \\
		LRTFDFR & 43.02 & 40.89 & 41.13 & 40.59 & 40.77 & 40.63 & 41.14 & 41.22 \\
		SVTV & 10.30 & - & 51.72 & - & - & - & - & -  \\
		\wline
	\end{tabular}
\end{table}

\section{Conclusions}
\label{sec:conclusions}
In this work, we propose an automatic weighting scheme for the weighted image denoising models. The weight is predicted by an explicit neural network mapping called DLWnet, which can be learned under a bi-level optimization framework. The trained DLWnet has been shown to generalize to other weighted image denoising models in a plug-and-play manner. Experimental results substantiate that the proposed DLWnet can help a new weighted image denoising model finely adapt to complex noise types and thus can generalize well to different image datasets. Furthermore, experimental results show that the DLWnet trained with diverse regularizers (i.e., models) can be used for new models to further improve their denoising performance. Additionally, we theoretically prove a generalization error upper bound of the DLWnet when it is plugged into a new weighted denoising model, revealing its insightful generalization capability.

The proposed automatic weighting scheme still has several limitations. Firstly, the learning of the DLWnet requires that the iterative optimization algorithm used to solve the lower-level problem to be differentiable w.r.t. weight $\W$. However, some common operators only have sub-gradients, such as nuclear norm and total variation (TV) norm, which may render the training of DLWnet unstable. 
Secondly, the theoretical analysis of generalization error for multiple source and target models is based on an approximation of the solution to the source models. Moreover, the source and target models are symmetric in $U(x,y)$, i.e., $U(x,y)=U(y,x)$. However, in practice the situation could be different when the source and target models are exchanged. 

Additionally, the question of how to further improve the generalization ability of DLWnet still needs to be more deeply and comprehensively investigated. Some possible strategies may help, such as constraining specific structures of the predicted parameters \cite{learning-an-explicit-hyper-pred}. Furthermore, we will consider ways to address and incorporate more physical insights underlying image noise to further improve the quality of our extracted weighting scheme and enhance its denoising effect. We will investigate these issues in future research.

\appendix
\section{Details about \cref{algo-weight}}\label{sec:append-algo}
Algorithm \ref{algo-weight} shows how to apply ADMM to solve the denoising problem \eqref{weight-problem}. Specifically, we first introduce an auxiliary variable $\Z$, and transform \eqref{weight-problem} into the following consensus form
\begin{align*}
	\min\limits_{\X,\Z} & ~\dfrac{1}{2}\|\W\odot(\Y - \Z)\|^2 + R(\X), \nonumber \\
	& \mathrm{s.t.}~\X = \Z.
\end{align*}
The corresponding augmented Lagrangian is
\begin{align*}
	L(\X,\Z,\Gamma)=\dfrac{1}{2}\|\W\odot(\Y - \Z)\|^2 + R(\X) + \langle \Gamma, \X-\Z\rangle + \dfrac{\mu}{2}\|\X - \Z\|^2,
\end{align*}
where $\Gamma$ is the Lagrange multiplier and $\mu>0$ is a parameter. The updating rules of ADMM is written as
\begin{subequations}
	\begin{numcases}{}
		\X\leftarrow \arg\min_{\X}~L(\X,\Z,\Gamma), \label{append-admm-x}\\
		\Z\leftarrow \arg\min_{\Z}~L(\X,\Z,\Gamma), \label{append-admm-z}\\
		\Gamma\leftarrow \Gamma + \mu(\X - \Z). 
	\end{numcases}
\end{subequations}
\eqref{append-admm-x} and \eqref{append-admm-z} are called ``$\X$-" and ``$\Z$-" subproblems, respectively. The $\X$-subproblem has the specific form of
\begin{align*}
	\arg\min_\X~\dfrac{\mu}{2}\left\|\X - (\Z - \dfrac{\Gamma}{\mu})\right\|^2 + R(\X),
\end{align*}
This problem may not have closed-form solution, which depends on the choice of $R(\cdot)$. The $\Z$-subproblem has the specific form of
\begin{align*}
	& \arg\min_\Z~\dfrac{1}{2}\|\W\odot(\Y - \Z)\|^2 + \dfrac{\mu}{2}\left\| \Z - (\X + \dfrac{\Gamma}{\mu}) \right\|^2, \\
	= & \dfrac{\W^2\odot\Y + \mu\X + \Gamma}{\W^2 + \mu}. \nonumber
\end{align*}

\section{Proof of Lemmas and Theories}\label{sec:appendix}
\begin{proof}[Proof of \cref{theo:lemma1}]
	Let $\ell(\X):=\dfrac{1}{2}\|\W\odot(\Y-\X)\|^2$. For any $\eta\in[0,1]$, we have that
	\begin{align*}
		& \eta \ell(\X) + (1-\eta)\ell(\Z) - \ell(\eta\X + (1-\eta)\Z) \\
		= & \dfrac{\eta}{2}\|\W\odot\X\|^2 + \dfrac{1-\eta}{2}\|\W\odot\Z\|^2 - \dfrac{1}{2}\|\W\odot(\eta\X + (1-\eta)\Z)\|^2  \\
		= & \dfrac{1}{2}\eta(1-\eta)\|\W\odot\X\|^2 + \dfrac{1}{2}\eta(1-\eta)\|\W\odot\Z\|^2 -\eta(1-\eta)\langle \W\odot\X,\W\odot\Z \rangle \\
		= & \dfrac{1}{2}\eta(1-\eta)\|\W\odot(\X - \Z)\|^2 \\
		\geq  & \dfrac{\varepsilon^2}{2}\eta(1-\eta)\|\X - \Z\|^2.
	\end{align*}
	Thus, the objective function of \eqref{weight-problem} is a closed and strongly convex function as a sum of the closed and $\varepsilon^2$-strongly convex function $\ell(\X)$ and the closed and convex function $R(\X)$. According to Theorem 5.25 in \cite{First-order-Methods-in-Optimization}, the solution to \eqref{weight-problem} is unique.
\end{proof}

\begin{proof}[Proof of \cref{theo:lemma2}]
1)	Let $\hat{\X_1}= f_R^{h_1}(\Y)$ and $\hat{\X_2}= f_R^{h_2}(\Y)$ for short, and $\W_1 = h_1(\Y)$,  $W_2 = h_2(\Y)$. By definition, 
	\begin{align*}
		\hat{\X_1} & =\arg\min\limits_{\X} ~ \dfrac{1}{2}\left\|\W_1\odot(\Y-\X) \right\|^2 +  R(\X), \\
		\hat{\X_2} & =\arg\min\limits_{\X} ~ \dfrac{1}{2}\left\|\W_2\odot(\Y-\X) \right\|^2  + R(\X). 
	\end{align*}
	Applying Fermat's optimality condition, we have that there exists $\mathcal{G}_1\in\partial R(\hat{\X_1})$ and $\mathcal{G}_2\in\partial R(\hat{\X_2})$, such that
	\begin{equation}\label{proof-lemm2-1}
		\begin{aligned}
			\W_1^2\odot(\hat{\X_1} - \Y) + \mathcal{G}_1 & = 0, \\
			\W_2^2\odot(\hat{\X_2} - \Y) + \mathcal{G}_2 & = 0.
		\end{aligned}
	\end{equation}
	By the definition of subgradient, we also have 
	\begin{align*}
		R(\hat{\X_2}) - R(\hat{\X_1}) \geq  & \langle \mathcal{G}_1, \hat{\X_2} - \hat{\X_1} \rangle, \\
		R(\hat{\X_1}) - R(\hat{\X_2}) \geq  & \langle \mathcal{G}_2, \hat{\X_1} - \hat{\X_2} \rangle.
	\end{align*}
	By adding the two inequalities above, we obtain:
	\begin{align*}
		0 \geq \langle \mathcal{G}_2 - \mathcal{G}_1, \hat{\X_1} - \hat{\X_2} \rangle 
	\end{align*}
	Further combining with \eqref{proof-lemm2-1}, we have that
	\begin{align*}
		0 \geq \langle \mathcal{G}_2 - \mathcal{G}_1, \hat{\X_1} - \hat{\X_2} \rangle  = \left\langle \W_2^2\odot(\Y - \hat{\X_2}) - \W_1^2\odot(\Y - \hat{\X_1}), \hat{\X_1} - \hat{\X_2} \right\rangle. 
	\end{align*}
	And thus
	\begin{align}
		\left\langle \Y\odot(\W_1^2 - \W_2^2), \hat{\X_1} - \hat{\X_2} \right\rangle & \geq \left\langle \W_1^2\odot\hat{\X_1} - \W_2^2\odot\hat{\X_2}, \hat{\X_1} - \hat{\X_2} \right\rangle, \nonumber\\
		& = \langle \W_1^2\odot(\hat{\X_1} -\hat{\X_2}) + (\W_1^2 - \W_2^2)\odot\hat{\X_2}, \hat{\X_1} - \hat{\X_2} \rangle, \nonumber \\
		\Rightarrow \langle (\Y - \hat{\X_2})\odot(\W_1^2 - \W_2^2), & \hat{\X_1} - \hat{\X_2} \rangle \geq \langle \W_1^2\odot(\hat{\X_1} - \hat{\X_2}), \hat{\X_1} - \hat{\X_2} \rangle. \label{equ:lemma2-1}
	\end{align}
	For the right side of \eqref{equ:lemma2-1}, we have
	\begin{align*}
		\langle \W_1^2\odot(\hat{\X_1} - \hat{\X_2}), \hat{\X_1} - \hat{\X_2} \rangle = \|\W_1\odot(\hat{\X_1} - \hat{\X_2})\|^2 \geq \varepsilon^2\|\hat{\X_1} - \hat{\X_2}\|^2 \geq 0.
	\end{align*}
	Thus,
	\begin{align*}
		\varepsilon^2\|\hat{\X_1} - \hat{\X_2}\|^2  & \leq  \langle (\Y - \hat{\X_2})\odot(\W_1^2 - \W_2^2), \hat{\X_1} - \hat{\X_2} \rangle \\
		& \leq \|(\Y - \hat{\X_2})\odot(\W_1^2 - \W_2^2)\| \|\hat{\X_1} - \hat{\X_2}\| \\
		\Rightarrow  \left\|f_R^{h_1}(\Y) - f_R^{h_2}(\Y)\right\| = \|\hat{\X_1} - \hat{\X_2}\| & \leq \dfrac{1}{\varepsilon^2}\left\|\left(\Y - f_R^{h_2}(\Y)\right)\odot\left(h_1(\Y)^2 - h_2(\Y)^2\right)\right\|.
		\stepcounter{equation}\tag{\theequation}\label{equ:proof-lemma2-add1}
	\end{align*} \\
2) Select $h'\in\mathcal{H}$ such that $h'(\Y)\equiv w$ for a constant $w \in [\varepsilon, B_w]$. Then $f_R^{h'}$ becomes a general proximal operator. By the nonexpansivity of proximal operator (Theorem 6.42 in \cite{First-order-Methods-in-Optimization}), we directly have that 
\begin{align}\label{equ:proof-lemma2-add2}
	     \|f_R^{h'}(\Y_1) - f_R^{h'}(\Y_2)\| \leq \|\Y_1 - \Y_2\|
\end{align}
3) Select $h'\in\mathcal{H}$ such that $h'(\Y)\equiv w$ for a constant $w \in [\varepsilon, B_w]$. For any $\Y\in\bY$ and any $h\in\mathcal{H}$, we have that
\begin{align*}
    \|f_R^h(\Y)\| & \leq \|f_R^h(\Y) - f_R^{h'}(0)\| + \|f_R^{h'}(0)\| \\
    & = \|f_R^h(\Y) - f_R^{h'}(\Y) + f_R^{h'}(\Y) - f_R^{h'}(0)\| + \|f_R^{h'}(0)\| \\
    & \leq \|f_R^h(\Y) - f_R^{h'}(\Y)\| + \|f_R^{h'}(\Y) - f_R^{h'}(0)\| + \|f_R^{h'}(0)\| \\
    & \leq \dfrac{1}{\varepsilon^2}\left\|\left(\Y - f_R^{h'}(\Y)\right)\odot\left(h(\Y)^2 - h'(\Y)^2\right)\right\| + \|\Y\| + \|f_R^{h'}(0)\| \\
    & \leq \dfrac{(B_w - \varepsilon)^2}{\varepsilon^2}\left(\|\Y\| + \left\|f_R^{h'}(\Y)\right\|\right) + \|\Y\| + \|f_R^{h'}(0)\| \\
    & \leq \dfrac{B_w^2}{\varepsilon^2}\left(\|\Y\| + \left\|f_R^{h'}(\Y) - f_R^{h'}(0) + f_R^{h'}(0)\right\|\right) + \|\Y\| + \|f_R^{h'}(0)\| \\
    & \leq \dfrac{B_w^2}{\varepsilon^2}\left(\|\Y\| + \|\Y\| + \|f_R^{h'}(0)\|\right) + \|\Y\| + \|f_R^{h'}(0)\| \\
    & \leq \left(\dfrac{2B_w^2}{\varepsilon^2} + 1\right)\|\Y\| + \left( \dfrac{B_w^2}{\varepsilon^2} + 1\right)\|f_R^{h'}(0)\|.
\end{align*}
The third inequality uses \eqref{equ:proof-lemma2-add1} and \eqref{equ:proof-lemma2-add2}. Since $\Y$ is bounded and $f_R^{h'}(0)$ is a constant, we have that $f_R^h(\Y)$ is uniformly bounded for any $\Y\in\bY$ and any $h\in\mathcal{H}$. \\
4) From \eqref{equ:proof-lemma2-add1}, we have that
\begin{align*}
	\left\|f_R^{h_1}(\Y) - f_R^{h_2}(\Y)\right\| & \leq \dfrac{1}{\varepsilon^2}\left\|\left(\Y - f_R^{h_2}(\Y)\right)\odot\left(h_1(\Y)^2 - h_2(\Y)^2\right)\right\| \\
	& = \dfrac{1}{\varepsilon^2}\left\|\left(\Y - f_R^{h_2}(\Y)\right)\odot(h_1(\Y) + h_2(\Y))\odot\left(h_1(\Y) - h_2(\Y)\right)\right\| \\
	& \leq \dfrac{2B_w}{\varepsilon^2}\left\|\left(\Y - f_R^{h_2}(\Y)\right)\odot\left(h_1(\Y) - h_2(\Y)\right)\right\| \\
	& \leq L_H\|h_1(\Y) - h_2(\Y)\|,
\end{align*}
for some constant $L_H$. The last inequality is derived from the boundness of $\Y$ and $f_R^h(\Y)$. The proof is then completed.
\end{proof}

\begin{proof}[Proof of \cref{theo:train}]
	The proof consists of several parts. We will firstly present an error bound \cite{FoML}\cite{Learning-an-explicit-hyper-parameter-prediction-function-conditioned-on-tasks} and then decouple the complexity of $\mathcal{F}_\mathcal{H}$ using \cref{theo:lemma2}.
	
	1) Let $\mathcal{Q}:=Q_1\otimes  \dots \otimes Q_T$ be Cartesian product of $T$ function classes $Q_t$s, with element $\bm{q}=(q_1, \dots ,q_T)$ and $q_t\in Q_t$ for all $t\in [T]$. The function $q_t$ in $Q_t$ satisfies $0\leq q_t \leq B_l$ for all $t\in[T]$. The data set $\mathbf{S}=\{z_{t,i_t}\}_{i_t=1}^{N_t}{}_{t=1}^{T}$ consists of $\sum_{t=1}^T N_t$ data points i.i.d sampled from distribution $\D$. For a $\bm{q}\in\mathcal{Q}$, let $\hat{\mathbb{E}}_\mathbf{S}[\bm{q}] = \frac{1}{T}\sum_{t=1}^T\frac{1}{N_t}\sum_{i_t=1}^{N_t}q_t(z_{t,i_t})$, and $\mathbb{E}[\bm{q}] = \mathbb{E}[\hat{\mathbb{E}}_\mathbf{S}[\bm{q}]]=\frac{1}{T}\sum_{t=1}^T\mathbb{E}_{z\sim\D}[q_t(z)]$.
	
	We then define
	\begin{align*}
		\Phi(\mathbf{S}) := \sup\limits_{\bm{q}\in\mathcal{Q}}\left( \mathbb{E}[\bm{q}] - \hat{\mathbb{E}}_\mathbf{S}[\bm{q}] \right).
	\end{align*}
	Let $\mathbf{S}'$ be another dataset that is identical to $\mathbf{S}$ except for the only element at the $(t,i_t)$-th position, i.e., $z_{t,i_t}'\neq z_{t,i_t}$ where $z_{t,i_t}'\in\mathbf{S}'$ and $z_{t,i_t}\in\mathbf{S}$. Then we have
	\begin{align*}
		\Phi(\mathbf{S}') - \Phi(\mathbf{S}) & \leq\sup\limits_{\bm{q}\in\mathcal{Q}}\left( \hat{\mathbb{E}}_\mathbf{S}[\bm{q}] - \hat{\mathbb{E}}_{\mathbf{S}'}[\bm{q}] \right), \\
		& = \sup\limits_{\bm{q}\in\mathcal{Q}}\left( \dfrac{q_t(z_{t,i_t}) - q_t(z_{t,i_t}')}{TN_t} \right) ,\\
		& \leq \dfrac{B_l}{TN_t}.
	\end{align*}
	According to McDiarmid's inequality, we have
	\begin{align*}
		\mathbb{P}\left[ \Phi(\mathbf{S}) - \mathbb{E}_\mathbf{S}[\Phi(\mathbf{S})]\geq\epsilon \right] \leq \exp{\left( \dfrac{T^2}{B_l^2}\dfrac{-2\epsilon^2}{\sum_{t=1}^T\frac{1}{N_t} } \right)}.
	\end{align*}
	Let $\frac{\delta}{2}=\exp{\left( \frac{T^2}{B_l^2}\frac{-2\epsilon^2}{\sum_{t=1}^T\frac{1}{N_t} } \right)}$. The above inequality is equal to say that with probability at least $1-\frac{\delta}{2}$, the following inequality holds:
	\begin{align}\label{equ:theo1-1-1}
		\Phi(\mathbf{S}) \leq \mathbb{E}_\mathbf{S}[\Phi(\mathbf{S})] + \dfrac{B_l}{T}\sqrt{\sum_{t=1}^T\dfrac{1}{N_t}}\sqrt{\dfrac{1}{2}\log\dfrac{2}{\delta}}.
	\end{align}
	Next, we estimate $\mathbb{E}_\mathbf{S}[\Phi(\mathbf{S})]$ as follows:
	\begin{align*}
		\mathbb{E}_\mathbf{S}[\Phi(\mathbf{S})] & = \mathop{\mathbb{E}}\limits_\mathbf{S}\left[ \sup_{\bm{q}\in\mathcal{Q}}\left( \mathbb{E}[\bm{q}] - \hat{\mathbb{E}}_\mathbf{S}[\bm{q}] \right) \right]\\
		& = \mathop{\mathbb{E}}\limits_\mathbf{S}\left[ \sup_{\bm{q}\in\mathcal{Q}}\mathbb{E}_{\mathbf{S}'}\left( \hat{\mathbb{E}}_{\mathbf{S}'}[\bm{q}] - \hat{\mathbb{E}}_\mathbf{S}[\bm{q}] \right) \right] \\
		& \leq \mathop{\mathbb{E}}\limits_{\mathbf{S},\mathbf{S}'}\left[ \sup_{\bm{q}\in\mathcal{Q}}\left( \hat{\mathbb{E}}_{\mathbf{S}'}[\bm{q}] - \hat{\mathbb{E}}_{\mathbf{S}}[\bm{q}] \right) \right] \\
		& = \mathop{\mathbb{E}}\limits_{\mathbf{S},\mathbf{S}'}\left[ \sup_{\bm{q}\in\mathcal{Q}} \dfrac{1}{T}\sum_{t=1}^T\dfrac{1}{N_t}\sum_{i_t=1}^{N_t}\left(q_t(z'_{t,i_t})-q_t(z_{t,i_t}) \right) \right] \\
		& = \mathop{\mathbb{E}}\limits_{\Sigma,\mathbf{S},\mathbf{S}'}\left[ \sup_{\bm{q}\in\mathcal{Q}} \dfrac{1}{T}\sum_{t=1}^T\dfrac{1}{N_t}\sum_{i_t=1}^{N_t}\sigma_{t,i_t}\left(q_t(z'_{t,i_t})-q_t(z_{t,i_t}) \right) \right] \\
		& \leq \mathop{\mathbb{E}}\limits_{\Sigma,\mathbf{S}'}\left[ \sup_{\bm{q}\in\mathcal{Q}} \dfrac{1}{T}\sum_{t=1}^T\dfrac{1}{N_t}\sum_{i_t=1}^{N_t}\sigma_{t,i_t}q_t(z'_{t,i_t})\right] + \mathop{\mathbb{E}}\limits_{\Sigma,\mathbf{S}}\left[ \sup_{\bm{q}\in\mathcal{Q}} \dfrac{1}{T}\sum_{t=1}^T\dfrac{1}{N_t}\sum_{i_t=1}^{N_t}-\sigma_{t,i_t}q_t(z_{t,i_t})\right] \\
		& = 2\mathop{\mathbb{E}}\limits_{\Sigma,\mathbf{S}}\left[ \sup_{\bm{q}\in\mathcal{Q}} \dfrac{1}{T}\sum_{t=1}^T\dfrac{1}{N_t}\sum_{i_t=1}^{N_t}\sigma_{t,i_t}q_t(z_{t,i_t})\right] \\
		& = 2\mathfrak{R}(\mathcal{Q}).
	\end{align*}
	
	Checking the conditions of McDiarmid's inequality again to $\hat{\mathfrak{R}}_\mathbf{S}(\mathcal{Q})$, that is
	\begin{align*}
		\hat{\mathfrak{R}}_\mathbf{S}'(\mathcal{Q}) - \hat{\mathfrak{R}}_\mathbf{S}(\mathcal{Q}) \leq  \mathop{\mathbb{E}}\limits_{\Sigma} \left[ \sup\limits_{\bm{q}\in\mathcal{Q}} \dfrac{\sigma_{t,i_t}\left( q(z_{t,i_t}') - q(z_{t,i_t}) \right)}{TN_t} \right] \leq \dfrac{B_l}{TN_t}.
	\end{align*}
	Thus, with probability at least $1-\frac{\delta}{2}$, the following inequality holds:
	\begin{align}\label{equ:theo1-1-2}
		\mathfrak{R}(\mathcal{Q}) \leq \hat{\mathfrak{R}}_\mathbf{S}(\mathcal{Q}) + \dfrac{B_l}{T}\sqrt{\sum_{t=1}^T\dfrac{1}{N_t}}\sqrt{\dfrac{1}{2}\log\dfrac{2}{\delta}}.
	\end{align}
	
	Combining \eqref{equ:theo1-1-1} and \eqref{equ:theo1-1-2}, we have that with probability at least $1-\delta$, the following inequality holds:
	\begin{align}\label{proof-train-1}
		\mathbb{E}[\bm{q}] \leq \hat{\mathbb{E}}_\mathbf{S}[\bm{q}] + 2\hat{\mathfrak{R}}_\mathbf{S}(\mathcal{Q}) + 3\dfrac{B_l}{T}\sqrt{\sum_{t=1}^T\dfrac{1}{N_t}}\sqrt{\dfrac{1}{2}\log\dfrac{2}{\delta}}.
	\end{align}
	
	The first part of the proof is then completed. Just let $z_{t,i_t}=(y_{t,i_t},x_{t,i_t})$ and  $q_{t}(z_{t,i_t})=\ell_{up}(f_t^h(y_{t,i_t}),x_{t,i_t})$, we will get very close to the final training error estimation. But let us first decouple $\hat{\mathfrak{R}}_\mathbf{S}(\mathcal{Q})$ by some so-called contraction inequalities and \cref{theo:lemma2}, which leads to the second part of the proof.
	
	2) Let $Q_t = \{ q_t(y,x)=\ell_{up}(f_t^h(y),x) | h\in\mathcal{H}, y\in\bY, x\in\bX\}$. 
	\begin{align*}
		\hat{\mathfrak{R}}_\mathbf{S}(\mathcal{Q}) & = \mathop{\mathbb{E}}\limits_{\Sigma}\left[ \sup_{\bm{q}\in\mathcal{Q}} \dfrac{1}{T}\sum_{t=1}^T\dfrac{1}{N_t}\sum_{i_t=1}^{N_t}\sigma_{t,i_t}q_t(z_{t,i_t})\right]\\
		& = \mathop{\mathbb{E}}_{\Sigma}\left[ \sup_{h\in\mathcal{H}} \dfrac{1}{T}\sum_{t=1}^T\dfrac{1}{N_t}\sum_{i_t=1}^{N_t}\sigma_{t,i_t}\ell_{up}\left(f_t^h(y_{t,i_t}),x_{t,i_t}\right)\right].
	\end{align*}
	According to \cite{A-vector-contraction-inequality-for-rademacher-complexities} and \cref{theo:assump-loss}, we apply the vector-contraction inequality and get
	\begin{align*}
		& \mathbb{E}_{\Sigma}\left[ \sup_{h\in\mathcal{H}} \dfrac{1}{T}\sum_{t=1}^T\dfrac{1}{N_t}\sum_{i_t=1}^{N_t}\sigma_{t,i_t}\ell_{up}(f_t^h(y_{t,i_t}),x_{t,i_t})\right] \leq \\
		& \sqrt{\dfrac{\pi}{2}}L_n\mathbb{E}_{G}\left[ \sup_{h\in\mathcal{H}} \dfrac{1}{T}\sum_{t=1}^T\dfrac{1}{N_t}\sum_{i_t=1}^{N_t}\sum_{m=1}^Mg_{ti_tm}f_t^h(y_{t,i_t})_m \right],
	\end{align*}
	where $g_{ti_tm}\overset{i.i.d}{\sim}\mathcal{N}(0,1)$, $G$ is the set of all $g_{ti_tms}$s and $f_t^h(y_{t,i_t})_m$ is the $m$-th element of $f_t^h(y_{t,i_t})$. The above inequality implies that
	\begin{align*}
		\hat{\mathfrak{R}}_\mathbf{S}(\mathcal{G}) \leq \sqrt{\dfrac{\pi}{2}}L_n\hat{\mathfrak{G}}_\mathbf{S}(\mathcal{F}_\mathcal{H}) \leq \dfrac{3}{2}L_n\hat{\mathfrak{G}}_\mathbf{S}(\mathcal{F}_\mathcal{H}).
	\end{align*}
	
	We could further decouple $\hat{\mathfrak{G}}_\mathbf{S}(\mathcal{F}_\mathcal{H})$ using \cref {theo:lemma2}.
	
	Similar to \cite{Rademacher-and-Gaussian-Complexities}, we define two zero mean Gaussian processes
	\begin{align*}
		X_h & = \dfrac{1}{T}\sum_{t=1}^T\dfrac{1}{N_t}\sum_{i_t=1}^{N_t}\sum_{m=1}^Mg_{ti_tm}f_t^h(y_{t,i_t})_m, \\
		Y_h & = \dfrac{1}{T}\sum_{t=1}^T\dfrac{1}{N_t}\sum_{i_t=1}^{N_t}\sum_{m=1}^M\bar{g}_{t i_t m}L_H h(y_{t,i_t})_m.
	\end{align*}
	Then it can be deduced that:
	\begin{align*}
		& \mathbb{E}_{G}\left[ (X_h - X_{h'})^2 \right] \\
		= & \mathbb{E}_{G}\left[ \left[ \dfrac{1}{T}\sum_{t=1}^T\dfrac{1}{N_t}\sum_{i_t=1}^{N_t}\sum_{m=1}^M\left( f_t^h(y_{t,i_t})_m - f_t^{h'}(y_{t,i_t})_m \right) g_{t i_t m} \right]^2 \right] \\
		= & \dfrac{1}{T^2}\sum_{t=1}^T\dfrac{1}{N_t^2}\sum_{i_t=1}^{N_t}\sum_{m=1}^M\left( f_t^h(y_{t,i_t})_m - f_t^{h'}(y_{t,i_t})_m \right)^2 \\
		= & \dfrac{1}{T^2}\sum_{t=1}^T\dfrac{1}{N_t^2}\sum_{i_t=1}^{N_t}\|f_t^h(y_{t,i_t}) - f_t^{h'}(y_{t,i_t}) \|^2 \\
		\leq & \dfrac{1}{T^2}\sum_{t=1}^T\dfrac{1}{N_t^2}\sum_{i_t=1}^{N_t}L_H^2 \|h(y_{t,i_t}) - h'(y_{t,i_t}) \|^2 \\
		= & \dfrac{1}{T^2}\sum_{t=1}^T\dfrac{1}{N_t^2}\sum_{i_t=1}^{N_t}\sum_{m=1}^ML_H^2 \left( h(y_{t,i_t})_m - h'(y_{t,i_t})_m
		\right)^2 \\
		= & \mathbb{E}_{\bar{G}}\left[ \left[ \dfrac{1}{T}\sum_{t=1}^T\dfrac{1}{N_t}\sum_{i_t=1}^{N_t}\sum_{m=1}^M L_H \left( h(y_{t,i_t})_m  - h'(y_{t,i_t})_m \right) \bar{g}_{t i_t m} \right]^2 \right] \\
		= & \mathbb{E}_{\bar{G}} \left[ (Y_h - Y_{h'})^2 \right]. 
	\end{align*}
	The inequality above is derived by \cref{theo:lemma2}. Note that $\mathcal{H}$ is a function class parameterized by a finite number of real-valued $\theta$. Then applying the Sudakov-Fernique inequality to the general Gaussian process, we get
	\begin{align*}
		\mathbb{E}_G\left[ \sup\limits_{h\in\mathcal{H}}X_h \right] \leq \mathbb{E}_G\left[ \sup\limits_{h\in\mathcal{H}}Y_h \right],
	\end{align*}
	which is equivalent to
	\begin{align*}
		\hat{\mathfrak{G}}_\mathbf{S}(\mathcal{F}_\mathcal{H}) \leq L_H \hat{\mathfrak{G}}_\mathbf{S}(\mathcal{H}).
	\end{align*}
	
	The second part of the proof is then completed. Next, we will finally prove the proposed training error bound.
	
	3) \begin{align*}
		\Etr & = \R_{tr}(\hat{h}_{tr}) - \R_{tr}(h^*_{tr}) \\
		& = \underbrace{\R_{tr}(\hat{h}_{tr}) - \hat{\R}_{tr}(\hat{h}_{tr})}_{(a)} + \underbrace{\hat{\R}_{tr}(\hat{h}_{tr}) - \hat{\R}_{tr}(h^*_{tr})}_{(b)} + \underbrace{\hat{\R}_{tr}(h^*_{tr}) - \R_{tr}(h^*_{tr})}_{(c)}.
	\end{align*}
	According to the definition of $\hat{h}_{tr}$, $(b)\leq 0$. For $(a) + (c)$, we directly estimate it by the result \eqref{proof-train-1} in the first part of this proof. With probability at least $1-\delta$, the following inequality holds
	\begin{align*}
		(a)+(c) & \leq 4\hat{\mathfrak{R}}_\mathbf{S}(\mathcal{Q}) + 6\dfrac{B}{T}\sqrt{\sum_{t=1}^T\dfrac{1}{N_t}}\sqrt{\dfrac{1}{2}\log\dfrac{\delta}{2}} \\
		& \leq  6L_n\hat{\mathfrak{G}}_\mathbf{S}(\mathcal{F}_\mathcal{H}) + 6\dfrac{B}{T}\sqrt{\sum_{t=1}^T\dfrac{1}{N_t}}\sqrt{\dfrac{1}{2}\log\dfrac{\delta}{2}} \\
		& \leq 6L_n L_H \hat{\mathfrak{G}}_\mathbf{S}(\mathcal{H}) + 6\dfrac{B}{T}\sqrt{\sum_{t=1}^T\dfrac{1}{N_t}}\sqrt{\dfrac{1}{2}\log\dfrac{\delta}{2}}.
	\end{align*}
	
	The whole proof is then completed.
\end{proof}

\begin{proof}[Proof of \cref{lemma3}]
	For ease of notation, let $u_t^h = \frac{1}{h^2}\odot\nabla R_t$.
	\begin{align*}
		\R_{te}(h) - \R_{tr}(h) 
		& = \dfrac{1}{M}\mathbb{E}_{(\Y,\X)\sim\D}\left[ \dfrac{1}{S}\sum_{t=T+1}^{T+S} \|\Y - u_t^h(\Y) - \X\|_2^2 - \dfrac{1}{T}\sum_{t=1}^T\|\Y - u_t^h(\Y) - \X\|_2^2 \right] \\
		 & = \dfrac{1}{M}\mathbb{E}_{(\Y,\X)\sim\D}\left[ \dfrac{1}{S}\sum_{t=T+1}^{T+S}\|u_t^h(\Y)\|_2^2 -  \dfrac{1}{T}\sum_{t=1}^T\|u_t^h(\Y)\|_2^2 \right.  \\
		& ~~~~+ \left. 2\left\langle \Y-\X, \dfrac{1}{T}\sum_{t=1}^Tu_t^h(\Y) - \dfrac{1}{S}\sum_{t=T+1}^{T+S}u_t^h(\Y) \right\rangle \right] \\
		\Rightarrow \vert\R_{te}(h) - \R_{tr}(h)\vert &
		\leq \dfrac{1}{M}\mathbb{E}_{(\Y,\X)\sim\D}\left[ \left\vert \dfrac{1}{S}\sum_{t=T+1}^{T+S}\|u_t^h(\Y)\|_2^2 -  \dfrac{1}{T}\sum_{t=1}^T\|u_t^h(\Y)\|_2^2 \right\vert \right. \\
		& ~~~~ + \left. 2\| \Y-\X\| \left\|\dfrac{1}{T}\sum_{t=1}^Tu_t^h(\Y) - \dfrac{1}{S}\sum_{t=T+1}^{T+S}u_t^h(\Y) \right\|_2 \right]. \\
	\end{align*}
	Note that
	\begin{align*}
		& \left\vert \dfrac{1}{S}\sum_{t=T+1}^{T+S}\|u_t^h(\Y)\|^2 -  \dfrac{1}{T}\sum_{t=1}^T\|u_t^h(\Y)\|^2 \right\vert \\
		\leq & \dfrac{1}{\varepsilon^2}\underbrace{\left\vert \dfrac{1}{S}\sum_{t=T+1}^{T+S}\|\nabla R_t(\Y)\|^2 -  \dfrac{1}{T}\sum_{t=1}^T\|\nabla R_t(\Y)\|^2 \right\vert}_{A_2}
	\end{align*}
	and
	\begin{align*}
		& \| \Y-\X\| \left\|\dfrac{1}{T}\sum_{t=1}^Tu_t^h(\Y) - \dfrac{1}{S}\sum_{t=T+1}^{T+S}u_t^h(\Y) \right\| \\
		\leq & \dfrac{2B_d\sqrt{M}}{\varepsilon}\underbrace{\left\|\dfrac{1}{T}\sum_{t=1}^T\nabla R_t(\Y) - \dfrac{1}{S}\sum_{t=T+1}^{T+S}\nabla R_t(\Y) \right\|}_{A_1}.
	\end{align*}
	Then we have
	\begin{align*}
		\vert\R_{te}(h) - \R_{tr}(h)\vert 
		\leq & \dfrac{1}{M}\mathbb{E}_{(\Y,\X)\sim\D}\left[ \dfrac{4B_d\sqrt{M}}{\varepsilon}A_1 + \dfrac{1}{\varepsilon^2}A_2\right]\\
		=    & \mathbb{E}_{(\Y,\X)\sim\D}\left[ \dfrac{4B_d}{\sqrt{M}\varepsilon}A_1 + \dfrac{1}{M\varepsilon^2}A_2\right].
	\end{align*}
	
	The proof is then completed.
\end{proof}

\begin{proof}[Proof of \cref{theorem-ge}]
	\begin{align*}
		\Eg = & \R_{te}(\hat{h}_{tr}) - \R_{te}(h^*_0) \\
		= & \underbrace{\R_{te}(\hat{h}_{tr}) - \R_{tr}(\hat{h}_{tr})}_{(a)} + \underbrace{\R_{tr}(\hat{h}_{tr}) - \R_{tr}(h_{tr}^*)}_{(b)} +  \underbrace{\R_{tr}(h_{tr}^*) - \R_{tr}(h_0^*)}_{(c)} + \underbrace{\R_{tr}(h_0^*) - \R_{te}(h^*_0)}_{(d)}.
	\end{align*}
	According to the definition of $h_{tr}^*$, we get $(c)\leq 0$. (b) is the training error that we have estimated as in Theorem \ref{theo:train}. And by Lemma \ref{lemma3}, we can directly get
	\begin{align*}
		(a)+(c) \leq \mathbb{E}_{(\Y,\X)\sim\D}\left[ \dfrac{8B_d}{\sqrt{M}\varepsilon}A_1 + \dfrac{2}{M\varepsilon^2}A_2\right].
	\end{align*}
	Then, the final result is achieved.
\end{proof}

\begin{proof}[Proof of \cref{theo:generalization-121}]	
	For any $h\in\mathcal{H}$, we have
	\begin{align*}
		& \R_{te}(h) - \R_{tr}(h) \\
		= & \dfrac{1}{M}\mathbb{E}_{(\Y,\X)\sim\D}\left[ \|f_1^h(\Y) - \X\|^2  - \|f_2^h(\Y) - \X\|^2 \right],\\
		= & \dfrac{1}{M}\mathbb{E}_{(\Y,\X)\sim\D}\left[ \left(\|f_1^h(\Y) - \X\|+\|f_2^h(\Y) - \X\|\right)\left(\|f_1^h(\Y) - \X\|-\|f_2^h(\Y) - \X\| \right)  \right],\\
		\leq & \dfrac{1}{M}\mathbb{E}_{(\Y,\X)\sim\D}\left[ 4B_d\sqrt{M}\left(\|f_1^h(\Y) - \X\|-\|f_2^h(\Y) - \X\| \right)  \right], \\
		\leq & \dfrac{1}{M}\mathbb{E}_{(\Y,\X)\sim\D}\left[ 4B_d\sqrt{M}\left(\|f_1^h(\Y) - f_2^h(\Y)\| \right)  \right] \stepcounter{equation}\tag{\theequation}\label{equ:proof-l3-1},  
	\end{align*}
	Let $\mathcal{L}_1(\X)$ and $\mathcal{L}_2(\X)$ denote the objective functions of source and target models, respectively (indicating that $\Y$ and $\W$ are temporarily held constant):
	\begin{align*}
		\mathcal{L}_1(\X) := \dfrac{1}{2}\|\W\odot(\Y - \X)\|^2 + R_1(\X), \\
		\mathcal{L}_2(\X) := \dfrac{1}{2}\|\W\odot(\Y - \X)\|^2 + R_2(\X),
	\end{align*} 
	Recall that $\mathcal{L}_1(\X)$ and $\mathcal{L}_2(\X)$ are proper, closed and $\varepsilon^2$-strongly convex function w.r.t. $\X$. And $f_1^h(\Y)$ and $f_2^h(\Y)$ are their unique minimizers respectively. Thus, we have following two equalities (Theorem 5.25 in \cite{First-order-Methods-in-Optimization}):
	\begin{align}
		\mathcal{L}_2(f_1^h(\Y)) - \mathcal{L}_2(f_2^h(\Y)) & \geq \dfrac{\varepsilon^2}{2}\|f_1^h(\Y) - f_2^h(\Y)\|^2, \label{equ:proof-l3-2}\\
		\mathcal{L}_1(f_2^h(\Y)) - \mathcal{L}_1(f_1^h(\Y)) & \geq \dfrac{\varepsilon^2}{2}\|f_2^h(\Y) - f_1^h(\Y) \|^2
	\end{align} 
	Then, we have that
	\begin{align*}
		  & \left[\mathcal{L}_2(f_1^h(\Y)) - \mathcal{L}_1(f_1^h(\Y))\right] + \left[\mathcal{L}_1(f_2^h(\Y)) - \mathcal{L}_2(f_2^h(\Y))\right] \\
		  = & \left[\mathcal{L}_2(f_1^h(\Y)) - \mathcal{L}_2(f_2^h(\Y))\right] + \left[ \mathcal{L}_1(f_2^h(\Y)) - \mathcal{L}_1(f_1^h(\Y))\right] \\
		  \geq & \varepsilon^2\|f_1^h(\Y) - f_2^h(\Y)\|^2
	\end{align*}
	Note that $\mathcal{L}_1(\X) - \mathcal{L}_2(\X)= R_1(\X) - R_2(\X)$. Let $B_r$ denote the maximum distance between $R_1$ and $R_2$ over $\mathbb{X}$, i.e., $B_r := \max_{\X\in \mathbb{X}}|R_1(\X) -  R_2(\X)|$. Then, we have that
	\begin{align}
		\varepsilon^2\|f_1^h(\Y) - f_2^h(\Y)\|^2 \leq & R_2(f_1^h(\Y)) - R_1(f_1^h(\Y)) + R_1(f_2^h(\Y)) - R_2(f_2^h(\Y)),\\
		& \leq 2B_r \nonumber\\
		\Rightarrow \|f_1^h(\Y) - f_2^h(\Y)\| & \leq \dfrac{\sqrt{2B_r}}{\varepsilon}, \label{equ:proof-l3-3}
	\end{align}
	Combining \cref{equ:proof-l3-1} and \cref{equ:proof-l3-3}, we have that
	\begin{align}
		\R_{te}(h) - \R_{tr}(h) \leq \dfrac{4B_d\sqrt{2B_r}}{\varepsilon\sqrt{M}}
	\end{align}
	The proof is then completed.
\end{proof}

\section{Details on solving DLW-models}\label{sec-appendix-algodetails}
We present the details of ADMM to solve the three basic models, i.e. DLW-NUCLR, DLW-TV, DLW-TVS, and three different target models, i.e. DLW-LRTV, DLW-E3DTV, and DLW-LRTFDFR.

\subsection{DLW-NN}
The DLW-NN aims to solve the following problem
\begin{align}\label{hw-nnm}
	\min\limits_{X}~\dfrac{1}{2}\|W\odot(Y-X)\|^2 + \lambda \|X\|_*,
\end{align}
where $W,X,Y\in\mathbb{R}^{hw\times b}$. We use the ADMM algorithm to solve it. Firstly, the problem can be equivalently written as
\begin{align*}
	\min\limits_{X,Z} & ~\dfrac{1}{2}\|W\odot(Y-Z)\|^2 + \lambda \|X\|_*, \\
	\mathrm{s.t.} & ~ X = Z.
\end{align*}
The corresponding augmented Lagrange function is
\begin{align*}
	\mathcal{L} = \dfrac{1}{2}\|W\odot(Y-Z)\|^2 + \lambda \|X\|_* + \dfrac{\mu}{2}\|X - Z \|^2 +\langle \Gamma, X-Z\rangle,
\end{align*}
where $\Gamma$ is the Lagrange multiplier. Then we iteratively solve the problem \eqref{hw-nnm}. Specifically, in the $k$-th step, the update rules are as follows.
\\
\noindent\textbf{1) update $Z$:} The $Z$-subproblem is
\begin{align*}
	\min\limits_{Z}~\dfrac{1}{2}\|W\odot(Y-Z)\|^2 + \dfrac{\mu}{2}\left\|X^{(k-1)} - Z + \dfrac{\Gamma^{(k-1)}}{\mu}\right\|^2.
\end{align*}
This subproblem has a closed-form solution
\begin{align*}
	Z^{(k)} = \dfrac{W^2\odot Y + \mu X^{(k-1)} + \Gamma^{(k-1)}}{W^2 + \mu}.
\end{align*}
\\
\noindent\textbf{2) update $X$:} The $X$-subproblem is
\begin{align*}
	\min\limits_{X}~\lambda \|X\|_* + \dfrac{\mu}{2}\left\|X - Z^{(k)} + \dfrac{\Gamma^{(k-1)}}{\mu}\right\|^2.
\end{align*}
This subproblem also has a closed-form solution
\begin{align*}
	X^{(k)} = U^{(k)}\mathrm{Shrink}\left(\Sigma^{(k)},\frac{\lambda}{\mu}\right)V^{(k)}{}^{T}.
\end{align*}
$U^{(k)}\Sigma^{(k)}V^{(k)}{}^{T}$ is the singular value decomposition (SVD) of $Z^{(k)} - \frac{\Gamma^{(k-1)}}{\mu}$. The shrinkage operator $\mathrm{Shrink}(\cdot,\cdot)$ is defined as
$$
{\textrm{Shrink}(X,\eta )=}\left\{
\begin{array}{cc}
	X-\eta, &  X\geq\eta;\\
	0,     &    -\eta\leq X\leq \eta;\\
	X+\eta,&   X\leq-\eta.
\end{array}%
\right. $$

\noindent\textbf{3) update $\Gamma$:}
\begin{align*}
	\Gamma^{(k)} & = \Gamma^{(k-1)} + \mu(X^{(k)} - Z^{(k)}).
\end{align*}

\subsection{DLW-TV}
The DLW-TV aims to solve the following problem
\begin{align}\label{hw-tv}
	\min\limits_{\X}~\dfrac{1}{2}\|\W\odot(\Y-\X)\|^2 + \lambda \|\X\|_{TV},
\end{align}
where $\W,\Y,\X\in\mathbb{R}^{h\times w\times b}$. We use the ADMM algorithm to solve it. Firstly, the problem can be equivalently written as
\begin{align*}
	\min\limits_{\X,\Z} & ~\dfrac{1}{2}\|\W\odot(\Y-\Z)\|^2 + \lambda \left( \|\X_1\|_{1} + \|\X_2\|_{1} \right),\\
	\mathrm{s.t.} & ~ \X_1 = \nabla_1\Z, \X_2 = \nabla_2\Z,
\end{align*}
where $\nabla_i$ is the difference operator along the $i$-th dimension. Specifically,
\begin{align*}
	\|\nabla_1\Z\|_{1} := & \sum_{i,j,n}|\Z(i+1,j,n) - \Z(i,j,n)|, \\
	\|\nabla_2\Z\|_{1} := & \sum_{i,j,n}|\Z(i,j+1,n) - \Z(i,j,n)|.
\end{align*}
The corresponding augmented Lagrange function is
\begin{align*}
	\mathcal{L} = & \dfrac{1}{2}\|\W\odot(\Y-\Z)\|^2 + \lambda \left( \|\X_1\|_{1} + \|\X_2\|_{1} \right) + \\
	& \dfrac{\mu}{2}\left( \|\X_1 - \nabla_1\Z \|^2 + \langle \Gamma_1, \X_1 - \nabla_1\Z\rangle + \|\X_2 - \nabla_2\Z \|^2 + \langle \Gamma_2, \X_2 - \nabla_2\Z\rangle\right),
\end{align*}
where $\Gamma_1$ and $\Gamma_2$ are the Lagrange multipliers. Then we iteratively solve the problem \eqref{hw-tv}. Specifically, in the $k$-th step, the updating rules are as follows.
\\
\noindent\textbf{1) update $\Z$:} The $\Z$-subproblem is
\begin{align}\label{hw-tv-z}
	\min\limits_{\Z}~ \dfrac{1}{2}\|\W\odot(\Y-\Z)\|^2 + \dfrac{\mu}{2}\left( \left\|\X_1^{(k-1)} - \nabla_1\Z + \dfrac{\Gamma_1^{(k-1)}}{\mu}\right\|^2 + \left\|\X_2^{(k-1)} - \nabla_2\Z + \dfrac{\Gamma_2^{(k-1)}}{\mu}\right\|^2\right).
\end{align}
Let $Z_n$ denote the $n$-th slice, $\Z(:,:,n)$, of $\Z$, and then \eqref{hw-tv-z} can be split into $b$ equivalent problems, and each
has closed-form solution 
\begin{align*}
	Z^{(k)}_n & = \mathbf{F}^{-1}\left( \dfrac{A_{1n} + (W_n^2/\mu)\odot\mathbf{F}(Y_n)}{A_{2n} + W^2_n/\mu}\right), n=1, \dots ,b,\\
	\mathrm{where} & ~A_{1n} = \mathbf{F}(\nabla_1)^*\odot\mathbf{F}(w_{1n})+\mathbf{F}(\nabla_2)^*\odot\mathbf{F}(w_{2n}),\\
	& ~A_{2n} = \mathbf{F}(\nabla_1)^*\odot\mathbf{F}(\nabla_1)+\mathbf{F}(\nabla_2)^*\odot\mathbf{F}(\nabla_2),\\
	& ~w_{1n} = X_{1n}^{(k-1)} + \dfrac{\Gamma_{1n}^{(k-1)}}{\mu},\\
	& ~w_{2n} = X_{2n}^{(k-1)} + \dfrac{\Gamma_{2n}^{(k-1)}}{\mu}.
\end{align*}
$\mathbf{F}$ and $\mathbf{F}^{-1}$ denote the discrete Fourier transform and its reverse, respectively. Finally, $\Z^{(k)}$ can be derived by aggregating its slices $Z_n^{(k)}$ for $n\in[b]$.
\\
\noindent\textbf{2) update $\X$:} The $\X_l$-subproblem is
\begin{align*}
	\min\limits_{\X}~\lambda \|\X\|_{1} + \dfrac{\mu}{2}\left\|\X - \nabla_l\Z + \dfrac{\Gamma^{(k-1)}_l}{\mu}\right\|^2,~l=1,2.
\end{align*}
Each has a closed-form solution as
\begin{align*}
	\X_l^{(k)} = \mathrm{Shrink}\left( \nabla_l\Z - \dfrac{\Gamma^{(k-1)}_l}{\mu}, \dfrac{\lambda}{\mu} \right),~l=1,2.
\end{align*}
\\
\noindent\textbf{3) update $\Gamma$:}
\begin{align*}
	\Gamma^{(k)}_1 & = \Gamma^{(k-1)}_1 + \mu(\X^{(k)}_1 - \nabla_1\Z^{(k)}), \\
	\Gamma^{(k)}_2 & = \Gamma^{(k-1)}_2 + \mu(\X^{(k)}_2 - \nabla_2\Z^{(k)}).
\end{align*}

\subsection{DLW-TVS}
The DLW-TVS aims to solve the following problem
\begin{align}\label{hw-tvs}
	\min\limits_{\X}~\dfrac{1}{2}\|\W\odot(\Y-\X)\|^2 + \lambda \|\X\|_{TVS},
\end{align}
where $\W,\Y,\X\in\mathbb{R}^{h\times w\times b}$. We use the ADMM algorithm to solve it. Firstly, the problem can be equivalently written as
\begin{align*}
	\min\limits_{\X,\Z} & ~\dfrac{1}{2}\|\W\odot(\Y-\Z)\|^2 + \lambda \|\X\|_{1},\\
	\mathrm{s.t.} & ~ \X = \nabla_3\Z,
\end{align*}
where $\nabla_3$ is the difference operator along the third dimension. Specifically,
\begin{align*}
	\|\nabla_3\Z\|_{1} := & \sum_{i,j,n}|\Z(i,j,n+1) - \Z(i,j,n)|.
\end{align*}
The corresponding augmented Lagrange function is
\begin{align*}
	\mathcal{L} = \dfrac{1}{2}\|\W\odot(\Y-\Z)\|^2 + \lambda \|\X\|_{1} + \dfrac{\mu}{2}\|\X - \nabla_3\Z\|^2  + \langle \Gamma, \X - \nabla_3\Z\rangle.
\end{align*}
Then we iteratively solve the problem \eqref{hw-tvs}. Specifically, in the $k$-th step, the updating rules are as follows.
\\
\noindent\textbf{1) update $\Z$:} The $\Z$-subproblem is
\begin{align*}
	\min\limits_{\Z}~ \dfrac{1}{2}\|\W\odot(\Y-\Z)\|^2 + \dfrac{\mu}{2}\left( \left\|\X^{(k-1)} - \nabla_3\Z + \dfrac{\Gamma^{(k-1)}}{\mu}\right\|^2 \right).
\end{align*}
Let $Z_{(3)}\in\mathbb{R}^{hw\times b}$ denote the third-mode unfolding of $\Z$, that is
\begin{align*}
	\Z(i,j,:) = Z_{(3)}((i-1)w+j,:).
\end{align*}
Then the subproblem has a closed-form solution as
\begin{align*}
	Z^{(k)}_{(3)} & = \mathbf{F}^{-1}\left( \dfrac{A_{1} + (W_{(3)}^2/\mu)\odot\mathbf{F}(Y_{(3)})}{A_{2} + W^2_{(3)}/\mu}\right),\\
	\mathrm{where} & ~A_{1} = \mathbf{F}(\nabla_2)^*\odot\mathbf{F}(w),\\
	& ~A_{2} = \mathbf{F}(\nabla_2)^*\odot\mathbf{F}(\nabla_2),\\
	& ~w = X_{(3)}^{(k-1)} + \dfrac{\Gamma_{(3)}^{(k-1)}}{\mu}.
\end{align*}
Finally, $\Z^{(k)}$ can be derived by folding $Z_{(3)}^{(k)}$ along the third mode.
\\
\noindent\textbf{2) update $\X$:} The $\X$-subproblem is
\begin{align*}
	\min\limits_{\X}~\lambda \|\X\|_{1} + \dfrac{\mu}{2}\left\|\X - \nabla_3\Z + \dfrac{\Gamma^{(k-1)}}{\mu}\right\|^2.
\end{align*}
Its solution is
\begin{align*}
	\X^{(k)} = \mathrm{Shrink}\left( \nabla_3\Z - \dfrac{\Gamma^{(k-1)}}{\mu}, \dfrac{\lambda}{\mu} \right).
\end{align*}
\\
\noindent\textbf{3) update $\Gamma$:}
\begin{align*}
	\Gamma^{(k)} & = \Gamma^{(k-1)} + \mu(\X^{(k)} - \nabla_3\Z^{(k)}).
\end{align*}

\subsection{DLW-LRTV}
The LRTV \cite{LRTV} method solves the following problem
\begin{align*}
	\min_{X,S\in\mathbb{R}^{hw\times b}} & ~||X||_* + \tau\|X\|_{HTV} + \lambda\|S\|_1, \\
	\mathrm{s.t.} & ~\|Y-X-S\|^2\leq\varepsilon, \mathrm{rank}(X)\leq r.
\end{align*}

We derive our DLW-LRTV by replacing $\|S\|_1$ with our weighted loss term, and the problem is formulated as
\begin{align}\label{hw-lrtv}
	\min_{X\in\mathbb{R}^{hw\times b}} & ~\dfrac{\kappa}{2}\|W\odot(Y-X)\|^2 + ||X||_* + \tau\|X\|_{HTV}, \nonumber\\
	\mathrm{s.t.} & ~\mathrm{rank}(X)\leq r.
\end{align}
Similar to LRTV, we also use the ADMM algorithm to solve our DLW-LRTV. Firstly, \eqref{hw-lrtv} can be equivalently written as
\begin{align*}
	\min_{X,Z,L} & ~\dfrac{\kappa}{2}\|W\odot(Y-Z)\|^2 + ||L||_* + \tau\|X\|_{HTV}, \\
	\mathrm{s.t.} & ~\mathrm{rank}(L)\leq r,L=Z,X=Z.
\end{align*}
Its augmented Lagrange function is
\begin{align*}
	\mathcal{L} = & \dfrac{\kappa}{2}\|W\odot(Y-Z)\|^2 + \|L\|_* + \tau\|X\|_{HTV}  + \dfrac{\mu_1}{2}\|L - Z \|^2 + \langle \Gamma_1, L-Z\rangle  + \\
	& \dfrac{\mu_2}{2}\|X - Z \|^2 + \langle \Gamma_2, X-Z\rangle.
\end{align*}
Then the iterative updating rules are:
\\
\noindent\textbf{1) update $L$:} The $L$-subproblem is
\begin{align*}
	\min\limits_L & ~\|L\|_* + \dfrac{\mu_1}{2}\left\|L - Z^{(k-1)} + \dfrac{\Gamma_1^{(k-1)}}{\mu_1}\right\|^2, \\
	\mathrm{s.t.} & ~\mathrm{rank}(L)\leq r.
\end{align*}
Its solution is
\begin{align*}
	L^{(k-1)} = U^{(k-1)}\mathrm{Shrink}\left(\Sigma^{(k-1)}_r, \dfrac{1}{\mu_1} \right)V^{(k-1)}{}^{T},
\end{align*}
where $U^{(k-1)}\Sigma^{(k-1)}V^{(k-1)}{}^{T}$ is SVD of $Z^{(k-1)} - \frac{\Gamma_1^{(k-1)}}{\mu_1}$. The first $r$
diagonal elements of $\Sigma_r$ are the same as $\Sigma$ while the remaining elements are all 0.
\\
\noindent\textbf{2) update $X$:}The $X$-subproblem is
\begin{align*}
	\min_X~\tau\|X\|_{HTV} + \dfrac{\mu_2}{2}\left\|X - Z^{(k-1)} + \dfrac{\Gamma_2^{(k-1)}}{\mu_2}\right\|^2.
\end{align*}
We inherit the fast gradient-based algorithm for solving this subproblem from the original LRTV. Details can be seen in \cite{LRTV}.
\\
\noindent\textbf{3) update $Z$:} The $Z$-subproblem is
\begin{align*}
	\min_Z~ & \dfrac{\kappa}{2}\|W\odot(Y-Z)\|^2 + \dfrac{\mu_1}{2}\left\|L^{(k)} - Z + \dfrac{\Gamma_1^{(k-1)}}{\mu_1}\right\|^2  \\
	& + \dfrac{\mu_2}{2}\left\|X^{(k)} - Z + \dfrac{\Gamma_2^{(k-1)}}{\mu_2}\right\|^2.
\end{align*}
It has a closed-form solution
\begin{align*}
	Z^{(k)}  & = \dfrac{\kappa W^2\odot Y + A}{\kappa W^2 + \mu_1 + \mu_2},\\
	\mathrm{where} & ~ A = \mu_1L^{(k)} + \mu_2X^{(k)} + \Gamma_1^{(k-1)} + \Gamma_2^{(k-1)}.
\end{align*}
\\
\noindent\textbf{4) update $\Gamma$:}
\begin{align*}
	\Gamma^{(k)}_1 & = \Gamma^{(k-1)}_1 + \mu(L^{(k)} - Z^{(k)}), \\
	\Gamma^{(k)}_2 & = \Gamma^{(k-1)}_2 + \mu(X^{(k)} - Z^{(k)}).
\end{align*}

The parameter $\tau$ and $\kappa$ are set as $0.001$ and $0.14$, respectively, for all experiments. The rank $r$ for CAVE and ICVL is set as 3, and 5 for PaviaU, DCmall, Indian Pine and Urban datasets.

\subsection{DLW-E3DTV}
The E3DTV method \cite{E3DTV} solves the following problem
\begin{align*}
	\min_{\X,U_n,V_n,\mathcal{E}} & ~ \sum_{n=1}^{3}\tau\|U_n\|_1 + \|\mathcal{E}\|_1, \\
	\mathrm{s.t.} & ~ \Y = \X + \mathcal{E}, ~ \nabla_n \X = U_n V_n^T, V_n^T V_n=I,\\
	& ~ V_n\in\mathbb{R}^{b\times r}, U_n\in\mathbb{R}^{hw\times r},n=1,2,3.
\end{align*}

And we formulate our DLW-E3DTV as
\begin{equation}\label{hw-e3dtv}
	\begin{aligned}
		\min_{\X,U_n,V_n,\mathcal{E}} & ~ \dfrac{1}{2}\|\W\odot(\Y - \Z)\|^2 + \sum_{n=1}^{3}\tau\|U_n\|_1, \\
		\mathrm{s.t.} & ~ \X = \Z,  ~ \nabla_n \X = U_n V_n^T, V_n^T V_n=I, \\
		& ~ V_n\in\mathbb{R}^{b\times r}, U_n\in\mathbb{R}^{hw\times r},n=1,2,3.
	\end{aligned}
\end{equation}
Similar to E3DTV, we also apply the ADMM algorithm to solve \eqref{hw-e3dtv}. The corresponding augmented Lagrange function is
\begin{align*}
	\mathcal{L} & = \dfrac{1}{2}\|\W\odot(\Y - \Z)\|^2 + \sum_{n=1}^{3}\tau\|U_n\|_1 +\dfrac{\mu}{2}\|\X - \Z \|^2 +\langle \Gamma_4, \X - \Z\rangle \\
	& + \dfrac{\mu}{2}\sum_{n=1}^3\left\{\|\nabla_n \X - U_n V_n^T\|^2 +\langle \Gamma_n, \nabla_n \X - U_n V_n^T\rangle\right\}.
\end{align*}
We need to iteratively solve the subproblems about $\X,\Z,U_n,V_n$. Note that for $\X,U_n,V_n$ subproblems are exactly the same type as in the original E3DTV algorithm, and we only have to change the corresponding inputs. Please refer to \cite{E3DTV} for details. In the $k$-th iteration, the $\Z$-subproblem is
\begin{align*}
	\min_{\Z}~ & \dfrac{1}{2}\|\W\odot(\Y - \Z)\|^2 + \dfrac{\mu}{2}\left\|\X^{(k-1)} - \Z + \dfrac{\Gamma_4^{(k-1)}}{\mu}\right\|^2.
\end{align*}
It has a closed-form solution
\begin{align*}
	\Z^{(k)} = \dfrac{\W^2\odot\Y + \mu\X^{(k-1)} + \Gamma_4^{(k-1)}}{\W^2 + \mu}.
\end{align*}

The trade-off parameter $\tau$ is set as $10^{-4}*\sqrt{h*w}$ for CAVE, ICVL, PaviaU and DCmall dataset, and $10^{-4}*5*\sqrt{h*w}$ for Indian Pine and Urban datasets. The tensor rank for CAVE and ICVL datasets is set as [4,4,4] and [5,5,5] for PaviaU, DCmall, Indian Pines and Urban datasets.

\subsection{DLW-LRTFDFR}
The LRTFDFR method \cite{LRTFDFR} solves the following problem
\begin{align*}
	\min_{A,\mathcal{B},\mathcal{S}} ~\dfrac{1}{2}\|\Y - \mathcal{B}\times_3 A - \mathcal{S}\|^2 + \tau\sum_{k=1}^2\|\W_k\odot(\mathcal{B}\times_k D_k)\|_{2,1}  + \lambda\|D_3 A\|^2 + \mu\|\W_s\odot\mathcal{S}\|_1.
\end{align*}

And we derive our DLW-LRTFDFR as
\begin{align}\label{hw-lrtfdfr}
	\min_{A,\mathcal{B}}~\dfrac{1}{2}\|\W\odot(\Y - \mathcal{B}\times_3 A)\|^2 + \tau\sum_{k=1}^2\|\W_k\odot(\mathcal{B}\times_k D_k)\|_{2,1} + \lambda\|D_3 A\|^2,
\end{align}
where $\W = h_\theta(\Y)$ is the predicted weight. The original PAM-based algorithm can not conveniently solve DLW-LRTFDFR. Thus, we instead use the ADMM algorithm. Again, the problem \eqref{hw-lrtfdfr} can be equivalently written as
\begin{align*}
	\min_{A,\mathcal{B},\Z} &~\dfrac{1}{2}\|\W\odot(\Y - \Z)\|^2 + \tau\sum_{k=1}^2\|\W_k\odot(\mathcal{B}\times_k D_k)\|_{2,1} + \lambda\|D_3 A\|^2, \\
	\mathrm{s.t.} & ~\Z = \mathcal{B}\times_3 A.
\end{align*}
The corresponding augmented Lagrange function is
\begin{align*}
	\mathcal{L} = & \dfrac{1}{2}\|\W\odot(\Y - \Z)\|^2 + \tau\sum_{k=1}^2\|\W_k\odot(\mathcal{B}\times_k D_k)\|_{2,1}  + \lambda\|D_3 A\|^2 + \\
	& \dfrac{\mu}{2}\|\Z - \mathcal{B}\times_3 A \|^2 +\langle \Gamma, \Z - \mathcal{B}\times_3 A\rangle.
\end{align*}
Note that the subproblems about $\mathcal{B},A$ are the same type as in the original PAM-based algorithm when solving LRTFDFR. Thus we also directly inherit the corresponding parts from the original algorithm. In the $k$-th iteration, the $\Z$-subproblem is
\begin{align*}
	\min_\Z~ \dfrac{1}{2}\|\W\odot(\Y - \Z)\|^2 +  \dfrac{\mu}{2}\left\|\Z - \mathcal{B}^{(k-1)}\times_3 A^{(k-1)} + \dfrac{\Gamma^{(k-1)}}{\mu}\right\|^2.
\end{align*}
Its closed-form solution is
\begin{align*}
	\Z^{(k)} = \dfrac{\W^2\odot\Y + \mu\mathcal{B}^{(k-1)}\times_3 A^{(k-1)} - \Gamma^{(k-1)}}{\W^2 + \mu}.
\end{align*}

The trade-off parameters $\tau$ and $\lambda$ are set as 0.01 and 0.1, respectively, for all experiments. The rank $r$ is set as 3 for the CAVE and ICVL datasets, 5 for the PaviaU and DCmall datasets, and 6 for the Indian Pine and Urban datasets.

\subsection{DLW-SVTV}
The original SVTV model for denoising is briefly written as
\begin{align*}
	\min_X~\dfrac{\lambda}{2}\|X - Y\|^2 + \|X\|_{SVTV}
\end{align*}  
And the corresponding DLW-SVTV model is
\begin{align}\label{hw-svtv}
	\min_X~\dfrac{\lambda}{2}\|W\odot(X - Y)\|^2 + \|X\|_{SVTV}
\end{align} \\
We solve the DLW-SVTV \eqref{hw-svtv} in the following way. The ADMM algorithm is applied, which is also suggested by the original SVTV work \cite{SVTV}. One auxiliary variable $Z$ is introduced and problem \eqref{hw-svtv} can be equivalently written as
\begin{align*}
	& \min_X~\dfrac{\lambda}{2}\|W\odot(Z - Y)\|^2 + \|X\|_{SVTV}, \\
	& \mathrm{s.t.}~Z=X.
\end{align*}
The corresponding augmented Lagrangian is
\begin{align*}
	L(Z,X,\Gamma) := \dfrac{\lambda}{2}\|W\odot(Z - Y)\|^2 + \|X\|_{SVTV} + \langle \Gamma, Z-X\rangle + \dfrac{\mu}{2}\|Z - X\|^2
\end{align*}
Consequently, the $Z$-subproblem is
\begin{align*}
	\min_Z~\dfrac{\lambda}{2}\|W\odot(Z - Y)\|^2 + \dfrac{\mu}{2}\left\|Z - \left(X - \dfrac{\Gamma}{\mu}\right) \right\|^2,
\end{align*}
which has a closed-form solution
\begin{align*}
	Z = \dfrac{\lambda W^2\odot Y + \mu X - \Gamma}{\lambda W^2 + \mu}.
\end{align*}
The $X$-subproblem is 
\begin{align*}
	\min_X~ \dfrac{\mu}{2}\left\|X -\left(Z + \dfrac{\Gamma}{\mu}\right) \right\|^2 + \|X\|_{SVTV},
\end{align*}
which is exactly the original SVTV model and can be solved by the released code\footnote{\url{https://github.com/weiwamng/SVTV_Image_restoration}}.

\section{Additional denoising Results}\label{sec-appendix-expresultss}
Six datasets are used for testing the performance of the proposed weighting scheme. We extract part of the original DCmall dataset to fine-tune the DLWnet trained on the CAVE dataset. The rest of DCmall with size $200\times 200\times 152$ is used for testing. The fine-tuned DLWnet is used for PaviaU, DCmall, Indian Pine, and Urban datasets. \cref{tab-exist-com-cave}, \cref{tab-exist-com-pavia} and
\cref{tab-exist-com-dc} show the quantitative denoising comparison on CAVE, PaviaU and DCmall datasets, respectively, for all the advanced complex noise removal methods and \textit{our} methods. We find that
some methods may achieve very high performance on some images and noise patterns. However, it can be observed that our method performs more robust across different images with diverse noise patterns, and can also gain superior denoising results in most cases.

\cref{tab-three-cave}, \cref{tab-three-pavia} and \cref{tab-three-dc} show the quantitative denoising comparison for the three basic models, respectively, with different DLWnets. Basically, we can see that when the source and target models match, the denoising performances are the best. For some images, it is also seen that DLWnet trained with combined source models gain the best denoising results.
For example, in \cref{tab-three-cave}, for Case 1 part of target model 1, type-`N+TS' DLWnet generalizes better than type-`N' DLWnet. However, it can be explored that the TVS regularization that describes the spectral smoothness property
has a relationship with the spectral low-rank property, and thus the corresponding DLWnets may perform in a relatively similar way. For some images, type-`N+TS' DLWnet is possible to surpass type-`N' DLWnet for target model DLW-NN. This is not contradictory to their overall generalization performance.

\cref{tab-tran-cave}, \cref{tab-tran-pavia} and \cref{tab-tran-dc} show the quantitative denoising results on CAVE, PaviaU and DCmall datasets, respectively, obtained by LRTV, E3DTV, LRTFDFR and our corresponding weighted models with different DLWnets.
Generally, our weighted models outperform the original models for most images. Generally, the type-`N+T+TS' DLWnet performs robustly and thus we advise first applying this DLWnet to a new noisy image.

\cref{tab-sert-icvl} show the quantitative denoising results of PnP framework using SERT on ICVL dataset. We see that for ``Gaussian+stripe" noise and ``Spatial-Spectral Variant Gaussian" noise, SERT performs better. And our DLWnet demonstrates more improvements under other complex noise cases.

\cref{fig-comall-cave-high} and \cref{fig-comall-indian} show the visual comparison of the CAVE and Indian Pines datasets, respectively, calculated by advanced denoising methods and our methods. For the Indian Pine datasets, our DLW-E3DTV can obtain better visual results although the noisy band is severely polluted. From \cref{fig-append-add-tv1} to \cref{fig-tran-urban}, we present the visual comparisons between each DLWnet on three simple target models and three complex target models (LRTV, E3DTV and LRTFDFR). \cref{fig-append-svtv} shows the visual comparisons between SVTV and DLW-SVTV. It can be easily observed that most DLWnets can help the denoising models to remove complex noise better.

\begin{table}[t]
	\renewcommand{\arraystretch}{1.15}
	\newcommand{\mysize}{1.9cm}
	\newcommand{\firstc}{2.3cm}
	\fontsize{8.5}{9.5}\selectfont
	\caption{Average test performance of different denoising competing methods on the CAVE dataset. The best results in each \textbf{column} are in \textbf{bold}, and the second best results in each \textbf{column} are with \ul{underline}.}
	\label{tab-exist-com-cave}
	\centering
	\begin{tabular}{M{\firstc} M{\mysize} M{\mysize}  M{\mysize} M{\mysize} M{\mysize}}
		\Xhline{0.8pt}
		noise $\rightarrow$ & Case 1 & Case 2 & Case 3 & Case 4 & Case 5 \\
		\cline{2-6}
		index $\rightarrow$ & PSNR/SSIM & PSNR/SSIM & PSNR/SSIM & PSNR/SSIM & PSNR/SSIM \\
		\hline
		LRMR 	&	24.48/0.6488	&	28.15/0.7314	&	26.92/0.713	&	26.45/0.6873	&	23.2/0.6083	\\
		LRTV 	&	30.88/0.8887	&	31.75/0.9046	&	30.86/0.8907	&	32.68/0.9246	&	30.49/0.8832	\\
		NMoG 	&	23.63/0.6713	&	28.1/0.817	&	26.79/0.8009	&	24.97/0.718	&	20.21/0.5506	\\
		HyRes 	&	28.39/0.7901	&	34.04/0.9247	&	31.86/0.907	&	31.99/0.899	&	26.02/0.7512	\\
		FastHyMix 	&	27.95/0.7949	&	33.39/0.9111	&	31.96/0.9208	&	33.13/0.9135	&	26.37/0.7657	\\
		CTV-RPCA 	&	30.87/0.7898	&	30.38/0.7648	&	30.42/0.7815	&	28.37/0.6851	&	29.06/0.736	\\
		E3DTV 	&	33.71/ \tb{0.9218} 	&	33.21/0.9099	&	32.98/0.9061	&	31.91/0.8822	&	31.6/ \ul{0.8835} 	\\
		LRTFDFR 	&	26.88/0.7848	&	26.28/0.752	&	25.93/0.7521	&	26.39/0.7597	&	25.21/0.7204	\\
		HSI-DeNet 	&	27.59/0.7028	&	26.84/0.6311	&	26.9/0.6329	&	27.88/0.7182	&	28.4/0.6949	\\
		HSI-CNN 	&	 \tb{34.04}/0.8607	&	 \tb{37.13}/\tb{0.9552} 	&	 \tb{35.44}/\tb{0.9521} 	&	 \tb{38.88}/\tb{0.9661} 	&	 \ul{32.19}/0.8604	\\
		\hline
		DLW-LRTV 	&	31.94/0.8851	&	32.78/0.9145	&	32.2/0.9045	&	33.11/0.9233	&	31.55/ \tb{0.8930} 	\\
		DLW-E3DTV 	&	 \ul{33.98}/\ul{0.9056} 	&	 \ul{34.66}/\ul{0.9392} 	&	 \ul{34.51}/\ul{0.9369} 	&	 \ul{33.96}/\ul{0.9257} 	&	 \tb{32.55}/0.8823	\\
		DLW-LRTFDFR 	&	31.32/0.8831	&	31.82/0.9042	&	31.57/0.8936	&	31.99/0.9039	&	30.96/0.8787	\\
		\Xhline{0.8pt}
	\end{tabular}
\end{table}

\begin{table}[t]
	\renewcommand{\arraystretch}{1.15}
	\newcommand{\mysize}{1.9cm}
	\newcommand{\firstc}{2.3cm}
	\fontsize{8.5}{9.5}\selectfont
	\caption{Average test performance of different denoising competing methods on the PaviaU dataset. The best results in each \textbf{column} are in \textbf{bold}, and the second best results in each \textbf{column} are with \ul{underline}.}
	\label{tab-exist-com-pavia}
	\centering
	\begin{tabular}{M{\firstc} M{\mysize} M{\mysize}  M{\mysize} M{\mysize} M{\mysize}}
		\Xhline{0.8pt}
		noise $\rightarrow$ & Case 1 & Case 2 & Case 3 & Case 4 & Case 5 \\
		\cline{2-6}
		index $\rightarrow$ & PSNR/SSIM & PSNR/SSIM & PSNR/SSIM & PSNR/SSIM & PSNR/SSIM \\
		\hline
		LRMR & 25.05/0.5965	&	28.43/0.7108	&	27.99/0.7091	&	27.01/0.654	&	24.24/0.5628	\\
		LRTV & 32.61/0.8793	&	33.1/0.8847	&	31.81/0.8687	&	34.91/0.9159	&	31.69/0.8792	\\
		NMoG & 30.53/0.8077	&	32.99/0.8636	&	31.41/0.8407	&	28.97/0.7502	&	14.9/0.1078	\\
		HyRes & 30.37/0.8523	&	 \ul{34.40}/\ul{0.9214} & 33.32/\ul{0.9166} & 31.19/0.8469	&	28.32/0.8065	\\
		FastHyMix & 31.89/0.8899	&	 \tb{35.55}/\tb{0.9349} &  \tb{33.68}/0.898	&	33.24/0.9046	&	29.09/0.8258	\\
		CTV-RPCA & 30.33/0.7887	&	30.52/0.7954	&	29.75/0.7774	&	28.56/0.7247	&	28.05/0.7224	\\
		E3DTV & 32.78/0.8656	&	33.06/0.876	&	32.33/0.8663	&	31.2/0.8345	&	30.89/0.8383	\\
		LRTFDFR & 20.54/0.4496	&	20.69/0.4503	&	20.5/0.4488	&	20.68/0.4505	&	20.23/0.4464	\\
		HSI-DeNet & 29.44/0.8194	&	30.78/0.846	&	30.48/0.8388	&	32.22/0.8932	&	29.94/0.8322	\\
		HSI-CNN & 32.72/0.8972	&	33.84/0.9158	&	32.79/0.9046	&	35.24/0.934	&	31.4/0.8904	\\
		\hline
		DLW-LRTV &  \ul{33.67}/0.9076	&	34.11/0.9143	&	33.2/0.9081	&	 \tb{35.84}/\ul{0.9348} &  \ul{33.21}/\ul{0.9139} 	\\
		DLW-E3DTV &  \tb{33.77}/\tb{0.9189} & 34/0.9244	&	 \ul{33.60}/\tb{0.9203} & 32.01/0.8933	&	32.13/0.8939	\\
		DLW-LRTFDFR & 33.63/\ul{0.9170}	&	33.64/0.9169	&	33.16/0.9109	&	 \ul{35.79}/\tb{0.9491}	&	 \tb{33.84}/\tb{0.9289}	\\
		\Xhline{0.8pt}
	\end{tabular}
\end{table}

\begin{table}[t]
	\renewcommand{\arraystretch}{1.15}
	\newcommand{\mysize}{1.9cm}
	\newcommand{\firstc}{2.3cm}
	\fontsize{8.5}{9.5}\selectfont
	\caption{Average test performance of different denoising competing methods on the DC dataset. The best results in each \textbf{column} are in \textbf{bold}, and the second best results in each \textbf{column} are with \ul{underline}.}
	\label{tab-exist-com-dc}
	\centering
	\begin{tabular}{M{\firstc} M{\mysize} M{\mysize}  M{\mysize} M{\mysize} M{\mysize}}
		\Xhline{0.8pt}
		noise $\rightarrow$ & Case 1 & Case 2 & Case 3 & Case 4 & Case 5 \\
		\cline{2-6}
		index $\rightarrow$ & PSNR/SSIM & PSNR/SSIM & PSNR/SSIM & PSNR/SSIM & PSNR/SSIM \\
		\hline
		LRMR & 26.07/0.7463	&	30.42/0.8676	&	29.78/0.8597	&	29.18/0.8372	&	25.59/0.734	\\
		LRTV & 28.1/0.855	&	31.87/0.9093	&	29.75/0.8911	&	34.15/0.942	&	28.56/0.8663	\\
		NMoG & 24.23/0.72	&	35.81/0.953	&	34.87/0.9549	&	31.61/0.9105	&	19.5/0.4763	\\
		HyRes & 32.3/0.8779	&	 \tb{37.11}/\tb{0.9666} &  \ul{35.53}/\ul{0.9549} & 33.16/0.9223	&	30/0.858	\\
		FastHyMix & 33.22/0.8942	&	 \ul{37.04}/\ul{0.9598} &  \tb{36.13}/\tb{0.9571} & 34.61/0.9466	&	31.12/0.8831	\\
		CTV-RPCA & 31.49/0.8781	&	31.57/0.8779	&	30.76/0.8627	&	29.33/0.8239	&	28.86/0.8175	\\
		E3DTV & 32.94/0.9115	&	34.21/0.9293	&	32.97/0.9138	&	32.09/0.8961	&	30.82/0.8822	\\
		LRTFDFR & 33.31/0.9238	&	35.2/0.9516	&	33.44/0.9333	&	 \tb{36.37}/\tb{0.9625} & 19.82/0.2832	\\
		HSI-DeNet & 27.64/0.8047	&	28.62/0.8449	&	28.56/0.8384	&	29.65/0.8798	&	28.31/0.8237	\\
		HSI-CNN & 31.12/0.8865	&	32.42/0.9145	&	31.66/0.9063	&	33.81/0.9326	&	30.48/0.8832	\\
		\hline
		DLW-LRTV & 30.43/0.8875	&	31.57/0.9122	&	29.98/0.896	&	33.69/0.9434	&	29/0.8794	\\
		DLW-E3DTV &  \tb{33.74}/\tb{0.9417} & 34.21/0.9497	&	33.68/0.9443	&	32.7/0.9224	&	 \ul{32.50}/\ul{0.9244} 	\\
		DLW-LRTFDFR &  \ul{33.67}/\ul{0.9399}	&	33.87/0.9434	&	33.28/0.9358	&	 \ul{35.44}/\ul{0.9608}	&	 \tb{33.73}/\tb{0.9423}	\\
		\Xhline{0.8pt}
	\end{tabular}
\end{table}

\begin{table}[ht]
	\renewcommand{\arraystretch}{1.15}
	\newcommand{\mysize}{1cm}
	\fontsize{8.5}{9.5}\selectfont
	\caption{Average test performance of DLW-NN, DLW-TV and DLW-TVS on CAVE dataset. The best results in each \textbf{row} are in \textbf{bold}, and the second best results in each \textbf{row} are with \ul{underline}.}
	\label{tab-three-cave}
	\centering
	\begin{tabular}{M{1.2cm} | M{\mysize} M{\mysize} M{\mysize} M{\mysize} M{\mysize} M{\mysize} M{\mysize} M{\mysize+0.3cm} M{\mysize}}
		\wline
		\multirow{2}{1.2cm}{\backslashbox[1.6cm]{noise}{source}}
		& noisy & N & T & TS & N+T & N+TS & T+TS & N+T+TS & Abl\\
		\cline{2-10}
		& \multicolumn{9}{c}{evaluation index: PSNR/SSIM} \\
		\wline
		\multicolumn{10}{c}{target model 1 : \tb{DLW-NN}} \cr
		\hline
		Case 1 & 14.71/ 0.2733 & 31.53/ 0.8639 & 22.83/ 0.5529 & 26.86/ 0.7545 & 31.45/ 0.8697 & \tb{31.72}/ \ul{0.8704} & 27.24/ 0.7529 & \ul{31.63}/ \tb{0.8705} & 19.69/ 0.5787 \\
		Case 2 & 16.16/ 0.3014 & 31.96/ 0.8984 & 22.06/ 0.5425 & 26.62/ 0.7411 & 31.87/ 0.899 & \tb{32.57}/ \tb{0.9108} & 26.64/ 0.7586 & \ul{32.26}/ \ul{0.909} & 23.37/ 0.7198 \\
		Case 3 & 16.19/ 0.3119 & \ul{31.58}/ 0.9026 & 22.42/ 0.566 & 26.36/ 0.7545 & 31.38/ 0.9003 & \tb{31.68}/ \ul{0.9045} & 26.72/ 0.7711 & 31.56/ \tb{0.9053} & 22.7/ 0.7194 \\
		Case 4 & 13.13/ 0.2365 & \ul{30.78}/ 0.8675 & 18.69/ 0.416 & 23.8/ 0.6227 & 30.42/ 0.8638 & \tb{30.99}/ \tb{0.8816} & 23.46/ 0.6383 & 30.69/ \ul{0.8774} & 22.0/ 0.7352 \\
		Case 5 & 12.91/ 0.2068 & \tb{29.48}/ 0.8249 & 20.33/ 0.4519 & 24.0/ 0.637 & \ul{29.29}/ \tb{0.8311} & 29.17/ 0.82 & 24.57/ 0.6596 & 29.22/ \ul{0.8253} & 18.98/ 0.5719 \\
		\hline
		\multicolumn{10}{c}{target model 2 : \tb{DLW-TV}} \cr
		\hline
		Case 1 & 14.71/ 0.2733 & 31.59/ 0.8472 & \tb{33.87}/ 0.8785 & 29.57/ 0.7788 & \ul{33.74}/ \tb{0.8928} & 31.32/ 0.8434 & 33.45/ \ul{0.8831} & 33.0/ 0.8749 & 25.79/ 0.6267 \\
		Case 2 & 16.16/ 0.3014 & 32.14/ 0.8849 & \tb{34.99}/ \tb{0.9369} & 29.91/ 0.7941 & \ul{34.41}/ \ul{0.9273} & 31.78/ 0.8745 & 33.99/ 0.9184 & 33.74/ 0.9155 & 29.78/ 0.828 \\
		Case 3 & 16.19/ 0.3119 & 31.65/ 0.8816 & \tb{34.08}/ \tb{0.9322} & 29.01/ 0.7873 & \ul{33.57}/ \ul{0.9208} & 31.28/ 0.8699 & 32.97/ 0.9094 & 33.03/ 0.9085 & 28.51/ 0.8245 \\
		Case 4 & 13.13/ 0.2365 & 31.25/ 0.8695 & \tb{33.43}/ \tb{0.9141} & 28.93/ 0.7776 & \ul{33.34}/ \ul{0.9115} & 30.97/ 0.8615 & 32.91/ 0.9013 & 32.79/ 0.9001 & 29.66/ 0.7815 \\
		Case 5 & 12.91/ 0.2068 & 29.67/ 0.8138 & \ul{31.0}/ 0.8385 & 27.06/ 0.7088 & \tb{31.29}/ \tb{0.8667} & 29.36/ 0.8029 & 30.66/ 0.8429 & 30.82/ \ul{0.8429} & 22.68/ 0.5067 \\
		\hline
		\multicolumn{10}{c}{target model 3 : \tb{DLW-TVS}} \cr
		\hline
		Case 1 & 14.71/ 0.2733 & 29.76/ 0.8032 & 24.74/ 0.6146 & \tb{32.28}/ \tb{0.8505} & 29.55/ 0.8094 & 31.32/ 0.8366 & \ul{31.48}/ \ul{0.843} & 30.97/ 0.8344 & 19.88/ 0.5081 \\
		Case 2 & 16.16/ 0.3014 & 29.87/ 0.8332 & 24.35/ 0.6363 & \tb{33.1}/ \tb{0.895} & 29.43/ 0.8339 & 31.81/ 0.8762 & \ul{31.98}/ \ul{0.8853} & 31.3/ 0.8702 & 21.96/ 0.7003 \\
		Case 3 & 16.19/ 0.3119 & 30.31/ 0.8497 & 24.75/ 0.6655 & \tb{32.67}/ \tb{0.8935} & 29.81/ 0.8479 & \ul{31.99}/ 0.8821 & 31.79/ \ul{0.8847} & 31.45/ 0.8766 & 21.92/ 0.7017 \\
		Case 4 & 13.13/ 0.2365 & 28.29/ 0.7944 & 21.62/ 0.531 & \tb{31.53}/ \tb{0.8647} & 27.78/ 0.7918 & 30.33/ 0.8453 & \ul{30.5}/ \ul{0.8541} & 29.83/ 0.8393 & 21.73/ 0.6641 \\
		Case 5 & 12.91/ 0.2068 & 28.49/ 0.769 & 22.63/ 0.5313 & \tb{30.19}/ \tb{0.8102} & 28.2/ 0.7744 & \ul{29.83}/ 0.7992 & 29.55/ \ul{0.8} & 29.46/ 0.7972 & 19.82/ 0.4935 \\
		\wline
	\end{tabular}
\end{table}

\begin{table}[ht]
	\renewcommand{\arraystretch}{1.15}
	\newcommand{\mysize}{1cm}
	\fontsize{8.5}{9.5}\selectfont
	\caption{Average test performance of DLW-NN, DLW-TV and DLW-TVS on PaviaU dataset. The best results in each \textbf{row} are in \textbf{bold}, and the second best results in each \textbf{row} are with \ul{underline}.}
	\label{tab-three-pavia}
	\centering
	\begin{tabular}{M{1.2cm} | M{\mysize} M{\mysize} M{\mysize} M{\mysize} M{\mysize} M{\mysize} M{\mysize} M{\mysize+0.3cm} M{\mysize}}
		\wline
		\multirow{2}{1.2cm}{\backslashbox[1.6cm]{noise}{source}}
		& noisy & N & T & TS & N+T & N+TS & T+TS & N+T+TS & Abl\\
		\cline{2-10}
		& \multicolumn{9}{c}{evaluation index: PSNR/SSIM} \\
		\wline
		\multicolumn{10}{c}{target model 1 : \tb{DLW-NN}} \cr
		\hline
		Case 1 & 14.94/ 0.178 & \tb{33.57}/ \tb{0.9013} & 21.52/ 0.4248 & 28.1/ 0.6735 & 31.58/ 0.8516 & \ul{33.47}/ \ul{0.8993} & 26.05/ 0.5763 & 32.61/ 0.8763 & 21.62/ 0.5983 \\
		Case 2 & 16.55/ 0.2108 & \tb{34.36}/ \tb{0.9141} & 22.28/ 0.4514 & 31.03/ 0.8416 & 32.03/ 0.8676 & \ul{34.19}/ \ul{0.9135} & 26.71/ 0.6134 & 33.05/ 0.8894 & 25.86/ 0.7773 \\
		Case 3 & 16.58/ 0.2084 & \tb{33.5}/ \tb{0.9081} & 22.25/ 0.4501 & 29.57/ 0.7978 & 31.19/ 0.8586 & \ul{33.2}/ \ul{0.9061} & 26.34/ 0.6061 & 32.13/ 0.8811 & 25.14/ 0.7693 \\
		Case 4 & 13.25/ 0.1433 & \tb{32.02}/ \tb{0.8735} & 18.99/ 0.3267 & 28.52/ 0.7289 & 29.75/ 0.8035 & \ul{31.7}/ \ul{0.8644} & 23.09/ 0.4403 & 30.7/ 0.8286 & 22.84/ 0.7005 \\
		Case 5 & 13.34/ 0.1398 & \tb{31.35}/ \tb{0.8563} & 19.65/ 0.3489 & 27.59/ 0.7239 & 29.36/ 0.7899 & \ul{30.91}/ \ul{0.8501} & 23.4/ 0.4548 & 30.17/ 0.8199 & 20.8/ 0.5699 \\
		\hline
		\multicolumn{10}{c}{target model 2 : \tb{DLW-TV}} \cr
		\hline
		Case 1 & 14.94/ 0.178 & 24.77/ 0.6083 & \tb{29.63}/ \tb{0.8201} & 24.91/ 0.6035 & 27.39/ 0.7102 & 24.68/ 0.6102 & \ul{28.51}/ \ul{0.7599} & 27.85/ 0.7295 & 23.0/ 0.514 \\
		Case 2 & 16.55/ 0.2108 & 24.9/ 0.6114 & \tb{29.82}/ \tb{0.829} & 24.56/ 0.585 & 27.63/ 0.7188 & 24.74/ 0.6127 & \ul{28.83}/ \ul{0.7728} & 28.14/ 0.7405 & 23.78/ 0.5375 \\
		Case 3 & 16.58/ 0.2084 & 24.87/ 0.6097 & \tb{29.38}/ \tb{0.8213} & 24.32/ 0.5845 & 27.17/ 0.7104 & 24.74/ 0.6099 & \ul{28.31}/ \ul{0.7641} & 27.74/ 0.7339 & 23.51/ 0.5316 \\
		Case 4 & 13.25/ 0.1433 & 23.97/ 0.5742 & \tb{28.22}/ \tb{0.7765} & 23.92/ 0.5608 & 26.71/ 0.6896 & 24.18/ 0.5895 & \ul{27.67}/ \ul{0.7366} & 26.98/ 0.7033 & 25.04/ 0.5972 \\
		Case 5 & 13.34/ 0.1398 & 24.31/ 0.5876 & \tb{28.04}/ \tb{0.7723} & 23.58/ 0.5571 & 26.47/ 0.6861 & 24.35/ 0.5933 & \ul{27.29}/ \ul{0.7301} & 26.85/ 0.7024 & 22.48/ 0.4799 \\
		\hline
		\multicolumn{10}{c}{target model 3 : \tb{DLW-TVS}} \cr
		\hline
		Case 1 & 14.94/ 0.178 & 30.43/ 0.8083 & 22.26/ 0.464 & \tb{32.01}/ \ul{0.8344} & 29.38/ 0.7799 & \ul{31.75}/ \tb{0.8393} & 30.52/ 0.8235 & 30.95/ 0.8312 & 19.49/ 0.5242 \\
		Case 2 & 16.55/ 0.2108 & 31.08/ 0.8264 & 22.69/ 0.4719 & \tb{32.9}/ \tb{0.8631} & 29.72/ 0.7969 & \ul{32.27}/ \ul{0.8529} & 30.7/ 0.8341 & 31.25/ 0.8413 & 19.76/ 0.5714 \\
		Case 3 & 16.58/ 0.2084 & 31.34/ 0.823 & 22.69/ 0.4698 & \tb{32.35}/ \tb{0.8514} & 29.7/ 0.792 & \ul{32.22}/ \ul{0.8483} & 30.34/ 0.8256 & 31.11/ 0.8357 & 19.8/ 0.5715 \\
		Case 4 & 13.25/ 0.1433 & 28.61/ 0.7642 & 20.13/ 0.3725 & \tb{30.63}/ \tb{0.8152} & 27.27/ 0.7187 & \ul{29.87}/ \ul{0.7991} & 28.48/ 0.7693 & 29.0/ 0.7825 & 19.8/ 0.5435 \\
		Case 5 & 13.34/ 0.1398 & 29.14/ 0.764 & 20.68/ 0.3927 & \tb{30.44}/ \tb{0.8127} & 27.74/ 0.7226 & \ul{30.1}/ \ul{0.799} & 28.41/ 0.7615 & 29.23/ 0.7842 & 19.74/ 0.5142 \\
		\wline
	\end{tabular}
\end{table}

\begin{table}[ht]
	\renewcommand{\arraystretch}{1.15}
	\newcommand{\mysize}{1cm}
	\fontsize{8.5}{9.5}\selectfont
	\caption{Average test performance of DLW-NN, DLW-TV and DLW-TVS on DCmall dataset. The best results in each \textbf{row} are in \textbf{bold}, and the second best results in each \textbf{row} are with \ul{underline}.}
	\label{tab-three-dc}
	\centering
	\begin{tabular}{M{1.2cm} | M{\mysize} M{\mysize} M{\mysize} M{\mysize} M{\mysize} M{\mysize} M{\mysize} M{\mysize+0.3cm} M{\mysize}}
		\wline
		\multirow{2}{1.2cm}{\backslashbox[1.6cm]{noise}{source}}
		& noisy & N & T & TS & N+T & N+TS & T+TS & N+T+TS & Abl\\
		\cline{2-10}
		& \multicolumn{9}{c}{evaluation index: PSNR/SSIM} \\
		\wline
		\multicolumn{10}{c}{target model 1 : \tb{DLW-NN}} \cr
		\hline
		Case 1 & 15.28/ 0.2639 & \tb{33.47}/ \tb{0.9318} & 22.51/ 0.5795 & 29.49/ 0.8376 & 31.15/ 0.8927 & \ul{32.8}/ \ul{0.9192} & 27.18/ 0.7302 & 32.23/ 0.9115 & 22.61/ 0.6991 \\
		Case 2 & 16.82/ 0.3034 & \tb{34.38}/ \tb{0.9432} & 23.06/ 0.5957 & 30.69/ 0.8816 & 31.39/ 0.8983 & \ul{33.49}/ \ul{0.9313} & 27.63/ 0.7537 & 32.58/ 0.9193 & 25.79/ 0.834 \\
		Case 3 & 16.86/ 0.3002 & \tb{33.59}/ \tb{0.9322} & 23.15/ 0.598 & 29.73/ 0.8584 & 30.69/ 0.8861 & \ul{32.72}/ \ul{0.9196} & 27.42/ 0.7499 & 31.9/ 0.9084 & 25.22/ 0.8188 \\
		Case 4 & 13.21/ 0.2058 & \tb{32.19}/ \tb{0.9061} & 19.43/ 0.4546 & 27.63/ 0.752 & 29.31/ 0.8467 & \ul{31.32}/ \ul{0.8813} & 23.82/ 0.5884 & 30.51/ 0.869 & 23.34/ 0.7418 \\
		Case 5 & 13.52/ 0.1972 & \tb{31.43}/ \tb{0.8972} & 20.53/ 0.4948 & 26.46/ 0.7095 & 29.16/ 0.8439 & \ul{30.93}/ \ul{0.8831} & 24.56/ 0.6184 & 30.4/ 0.8738 & 21.88/ 0.6651 \\
		\hline
		\multicolumn{10}{c}{target model 2 : \tb{DLW-TV}} \cr
		\hline
		Case 1 & 15.28/ 0.2639 & 23.72/ 0.4979 & \tb{28.18}/ \tb{0.7895} & 23.25/ 0.4668 & 25.82/ 0.6159 & 23.48/ 0.4932 & \ul{26.87}/ \ul{0.6865} & 26.34/ 0.6525 & 22.0/ 0.4337 \\
		Case 2 & 16.82/ 0.3034 & 23.93/ 0.5066 & \tb{28.56}/ \tb{0.8082} & 23.32/ 0.4681 & 26.01/ 0.6277 & 23.58/ 0.4965 & \ul{27.15}/ \ul{0.705} & 26.58/ 0.6684 & 22.97/ 0.4118 \\
		Case 3 & 16.86/ 0.3002 & 23.84/ 0.5039 & \tb{28.17}/ \tb{0.7973} & 23.06/ 0.4626 & 25.66/ 0.6178 & 23.54/ 0.4923 & \ul{26.72}/ \ul{0.6922} & 26.17/ 0.6566 & 22.76/ 0.4035 \\
		Case 4 & 13.21/ 0.2058 & 23.2/ 0.4621 & \tb{27.36}/ \tb{0.7728} & 22.94/ 0.4516 & 25.47/ 0.6065 & 23.25/ 0.4742 & \ul{26.33}/ \ul{0.6662} & 25.82/ 0.63 & 23.99/ 0.4998 \\
		Case 5 & 13.52/ 0.1972 & 23.35/ 0.4738 & \tb{27.17}/ \tb{0.7596} & 22.91/ 0.4539 & 25.1/ 0.5915 & 23.26/ 0.4718 & \ul{25.98}/ \ul{0.6516} & 25.55/ 0.62 & 21.71/ 0.4269 \\
		\hline
		\multicolumn{10}{c}{target model 3 : \tb{DLW-TVS}} \cr
		\hline
		Case 1 & 15.28/ 0.2639 & 28.18/ 0.8172 & 22.24/ 0.5945 & \tb{29.93}/ \tb{0.848} & 27.54/ 0.7958 & \ul{29.37}/ \ul{0.8415} & 28.71/ 0.8344 & 28.79/ 0.8356 & 20.86/ 0.6715 \\
		Case 2 & 16.82/ 0.3034 & 28.66/ 0.8321 & 22.47/ 0.5901 & \tb{30.4}/ \tb{0.8681} & 27.77/ 0.8056 & \ul{29.84}/ \ul{0.8576} & 28.9/ 0.8436 & 29.09/ 0.8463 & 22.04/ 0.7254 \\
		Case 3 & 16.86/ 0.3002 & 28.72/ 0.8297 & 22.57/ 0.5914 & \tb{30.11}/ \tb{0.8602} & 27.69/ 0.8006 & \ul{29.7}/ \ul{0.8531} & 28.73/ 0.8378 & 29.0/ 0.8418 & 22.11/ 0.7268 \\
		Case 4 & 13.21/ 0.2058 & 26.77/ 0.7695 & 20.11/ 0.487 & \tb{28.49}/ \tb{0.8051} & 25.85/ 0.7337 & \ul{28.03}/ \ul{0.7966} & 27.29/ 0.7826 & 27.38/ 0.7851 & 21.99/ 0.7106 \\
		Case 5 & 13.52/ 0.1972 & 26.95/ 0.7709 & 20.88/ 0.5265 & \tb{28.4}/ \ul{0.7996} & 26.17/ 0.7417 & \ul{28.15}/ \tb{0.8016} & 27.47/ 0.791 & 27.61/ 0.7946 & 21.01/ 0.665 \\
		\wline
	\end{tabular}
\end{table}

\begin{table}[t]
	\renewcommand{\arraystretch}{1.15}
	\newcommand{\mysize}{0.92cm}
	\fontsize{8.5}{9.5}\selectfont
	\caption{Complex target models: average test performance of LRTV, E3DTV and LRTFDFR and their corresponding DLW-models on CAVE dataset. The best results in each \textbf{row} are in \textbf{bold}, and the second best results in each \textbf{row} are with \ul{underline}.}
	\label{tab-tran-cave}
	\centering
	\begin{tabular}{M{1.2cm} | M{\mysize}  M{\mysize} M{\mysize}  M{\mysize} M{\mysize} M{\mysize} M{\mysize} M{\mysize} M{\mysize+0.36cm} M{\mysize}}
		\hline
		\multirow{2}{1.2cm}{\backslashbox[1.6cm]{noise}{source}} & noisy & *\textit{Ori} & N & T & TS & N+T & N+TS & T+TS & N+T+TS & Abl\\
		\cline{2-11}
		& \multicolumn{10}{c}{evaluation index: PSNR/SSIM}\\
		\Xhline{0.8pt}
		\multicolumn{11}{c}{target model 1 : \tb{LRTV}} \cr
		\hline
		Case 1 & 14.71/ 0.273 & 30.88/ \ul{0.889} & 31.73/ 0.876 & 29.03/ 0.753 & 29.83/ 0.862 & \tb{32.09}/ \tb{0.893} & 31.55/ 0.873 & 31.41/ 0.876 & \ul{31.94}/ 0.885 & 22.01/ 0.571 \\
		Case 2 & 16.16/ 0.301 & 31.75/ 0.905 & 32.70/ 0.911 & 29.90/ 0.816 & 30.89/ 0.882 & \ul{32.74}/ \tb{0.916} & 32.54/ 0.907 & 32.23/ 0.907 & \tb{32.78}/ \ul{0.915} & 26.88/ 0.735 \\
		Case 3 & 16.19/ 0.312 & 30.86/ 0.891 & 32.17/ 0.902 & 29.68/ 0.820 & 29.94/ 0.872 & \ul{32.18}/ \ul{0.905} & 31.99/ 0.896 & 31.51/ 0.897 & \tb{32.20}/ \tb{0.905} & 26.18/ 0.735 \\
		Case 4 & 13.13/ 0.237 & 32.68/ \ul{0.925} & \ul{33.15}/ 0.924 & 31.52/ 0.886 & 31.48/ 0.902 & \tb{33.16}/ \tb{0.926} & 32.94/ 0.918 & 32.63/ 0.918 & 33.11/ 0.923 & 26.10/ 0.695 \\
		Case 5 & 12.91/ 0.207 & 30.49/ 0.883 & 31.33/ 0.881 & 29.18/ 0.790 & 28.96/ 0.856 & \tb{31.61}/ \tb{0.898} & 31.26/ 0.883 & 30.85/ 0.884 & \ul{31.55}/ \ul{0.893} & 21.38/ 0.549 \\
		\hline
		\multicolumn{9}{c}{target model 2 : \tb{E3DTV}} \cr
		\hline
		Case 1 & 14.71/ 0.273 & 33.71/ \tb{0.922} & 33.46/ 0.893 & 31.09/ 0.822 & 33.11/ 0.896 & 33.97/ \ul{0.91} & 33.61/ 0.899 & \tb{34.17}/ 0.908 & \ul{33.98}/ 0.906 & 23.05/ 0.583 \\
		Case 2 & 16.16/ 0.301 & 33.21/ 0.91 & 34.32/ 0.931 & 31.38/ 0.865 & 33.96/ 0.924 & 34.49/ \ul{0.940} & 34.48/ 0.935 & \tb{34.95}/ \tb{0.942} & \ul{34.66}/ 0.939 & 29.57/ 0.783 \\
		Case 3 & 16.19/ 0.312 & 32.98/ 0.906 & 34.28/ 0.933 & 31.40/ 0.867 & 33.06/ 0.912 & 34.30/ \tb{0.939} & 34.36/ 0.933 & \ul{34.44}/ 0.936 & \tb{34.51}/ \ul{0.937} & 29.16/ 0.789 \\
		Case 4 & 13.13/ 0.237 & 31.91/ 0.882 & 33.51/ 0.917 & 29.33/ 0.799 & 33.13/ 0.912 & 33.64/ 0.926 & 33.67/ 0.920 & \tb{34.13}/ \tb{0.928} & \ul{33.96}/ \ul{0.926} & 28.24/ 0.727 \\
		Case 5 & 12.91/ 0.207 & 31.60/ \ul{0.884} & 32.31/ 0.876 & 28.86/ 0.749 & 30.99/ 0.862 & \ul{32.44}/ \tb{0.891} & 32.37/ 0.876 & 32.22/ 0.879 & \tb{32.55}/ 0.882 & 22.95/ 0.565 \\
		\hline
		\multicolumn{9}{c}{target model 3 : \tb{LRTFDFR}} \cr
		\hline
		Case 1 & 14.71/ 0.273 & 26.88/ 0.785 & 30.73/ 0.861 & 28.75/ 0.788 & 29.47/ 0.825 & \tb{31.39}/ \tb{0.888} & 30.91/ 0.872 & \ul{31.35}/ \ul{0.887} & 31.32/ 0.883 & 22.54/ 0.599 \\
		Case 2 & 16.16/ 0.301 & 26.28/ 0.752 & 31.54/ 0.897 & 29.60/ 0.846 & 30.37/ 0.859 & \tb{31.83}/ \ul{0.903} & 31.43/ 0.896 & 31.75/ 0.900 & \ul{31.82}/ \tb{0.904} & 27.64/ 0.760 \\
		Case 3 & 16.19/ 0.312 & 25.93/ 0.752 & 31.11/ 0.880 & 29.51/ 0.845 & 29.65/ 0.839 & \ul{31.41}/ \ul{0.890} & 31.16/ 0.882 & 31.35/ 0.889 & \tb{31.57}/ \tb{0.894} & 27.66/ 0.779 \\
		Case 4 & 13.13/ 0.237 & 26.39/ 0.760 & 31.98/ 0.899 & 30.70/ 0.892 & 30.75/ 0.871 & \tb{32.19}/ \tb{0.905} & 31.65/ 0.894 & 31.89/ 0.896 & \ul{31.99}/ \ul{0.904} & 26.09/ 0.700 \\
		Case 5 & 12.91/ 0.207 & 25.21/ 0.720 & 30.53/ 0.864 & 28.93/ 0.807 & 29.35/ 0.842 & 30.89/ 0.879 & 30.63/ 0.871 & \ul{30.89}/ \tb{0.882} & \tb{30.96}/ \ul{0.879} & 21.58/ 0.558 \\
		\Xhline{0.8pt}
	\end{tabular}
\end{table}

\begin{table}[t]
	\renewcommand{\arraystretch}{1.15}
	\newcommand{\mysize}{0.92cm}
	\fontsize{8.5}{9.5}\selectfont
	\caption{Complex target models: average test performance of LRTV, E3DTV and LRTFDFR and their corresponding DLW-models on PaviaU dataset. The best results in each \textbf{row} are in \textbf{bold}, and the second best results in each \textbf{row} are with \ul{underline}.}
	\label{tab-tran-pavia}
	\centering
	\begin{tabular}{M{1.2cm} | M{\mysize}  M{\mysize} M{\mysize}  M{\mysize} M{\mysize} M{\mysize} M{\mysize} M{\mysize} M{\mysize+0.36cm} M{\mysize}}
		\hline
		\multirow{2}{1.2cm}{\backslashbox[1.6cm]{noise}{source}} & noisy & *\textit{Ori} & N & T & TS & N+T & N+TS & T+TS & N+T+TS & Abl\\
		\cline{2-11}
		& \multicolumn{10}{c}{evaluation index: PSNR/SSIM}\\
		\Xhline{0.8pt}
		\multicolumn{11}{c}{target model 1 : \tb{LRTV}} \cr
		\hline
		Case 1 & 14.94/ 0.178 & 32.3/ 0.867 & 32.23/ 0.887 & 24.93/ 0.595 & 32.26/ 0.872 & 32.04/ 0.862 & \ul{32.63}/ \ul{0.89} & 32.43/ 0.87 & \tb{33.33}/ \tb{0.895} & 26.77/ 0.71 \\
		Case 2 & 16.55/ 0.211 & 32.85/ 0.873 & 33.05/ 0.899 & 24.95/ 0.583 & 32.91/ 0.888 & 32.11/ 0.868 & \ul{33.22}/ \ul{0.902} & 32.51/ 0.878 & \tb{33.73}/ \tb{0.908} & 31.5/ 0.827 \\
		Case 3 & 16.58/ 0.208 & 31.4/ 0.846 & \tb{33.1}/ 0.896 & 24.92/ 0.582 & 31.85/ 0.883 & 31.95/ 0.869 & 32.88/ \ul{0.899} & 31.84/ 0.878 & \ul{33.09}/ \tb{0.901} & 30.71/ 0.829 \\
		Case 4 & 13.25/ 0.143 & 30.97/ 0.836 & 30.01/ 0.847 & 23.88/ 0.536 & 31.17/ 0.857 & 30.36/ 0.825 & 30.86/ \ul{0.858} & \ul{31.25}/ 0.854 & \tb{31.66}/ \tb{0.869} & 30.07/ 0.767 \\
		Case 5 & 13.34/ 0.14 & 29.97/ 0.834 & 30.97/ 0.855 & 24.09/ 0.554 & 29.71/ 0.834 & 30.68/ 0.829 & \ul{31.05}/ \ul{0.861} & 30.64/ 0.837 & \tb{31.5}/ \tb{0.869} & 25.89/ 0.676 \\
		\hline
		\multicolumn{9}{c}{target model 2 : \tb{E3DTV}} \cr
		\hline
		Case 1 & 14.94/ 0.178 & 32.98/ 0.875 & 32.08/ 0.901 & 26.67/ 0.684 & 32.33/ 0.893 & 32.59/ 0.894 & 32.24/ 0.899 & \tb{33.47}/ \tb{0.913} & \ul{33.41}/ \ul{0.911} & 26.83/ 0.774 \\
		Case 2 & 16.55/ 0.211 & 33.24/ 0.881 & 32.8/ 0.911 & 26.5/ 0.672 & 32.58/ 0.898 & 33.05/ 0.905 & 32.71/ 0.906 & \ul{33.69}/ \tb{0.921} & \tb{33.86}/ \ul{0.92} & 31.54/ 0.869 \\
		Case 3 & 16.58/ 0.208 & 32.42/ 0.869 & 33.03/ 0.909 & 26.62/ 0.674 & 31.8/ 0.891 & 32.67/ 0.898 & 32.8/ 0.905 & \ul{33.17}/ \tb{0.915} & \tb{33.47}/ \ul{0.914} & 30.9/ 0.868 \\
		Case 4 & 13.25/ 0.143 & 31.38/ 0.841 & 30.24/ 0.874 & 25.05/ 0.605 & 30.65/ 0.862 & 31.35/ 0.877 & 30.46/ 0.871 & \tb{32.15}/ \ul{0.896} & \ul{32.02}/ \tb{0.896} & 31.29/ 0.866 \\
		Case 5 & 13.34/ 0.14 & 30.88/ 0.839 & 30.97/ 0.873 & 25.72/ 0.641 & 29.64/ 0.839 & 31.47/ 0.879 & 30.79/ 0.869 & \ul{31.78}/ \ul{0.89} & \tb{32.13}/ \tb{0.897} & 27.09/ 0.792 \\
		\hline
		\multicolumn{9}{c}{target model 3 : \tb{LRTFDFR}} \cr
		\hline
		Case 1 & 14.94/ 0.178 & 30.4/ 0.872 & 32.52/ 0.895 & 25.0/ 0.593 & \tb{33.39}/ \ul{0.911} & 32.4/ 0.885 & 33.03/ 0.908 & 33.14/ 0.909 & \ul{33.38}/ \tb{0.912} & 25.29/ 0.608 \\
		Case 2 & 16.55/ 0.211 & 31.29/ 0.89 & 33.31/ 0.905 & 24.93/ 0.582 & \tb{33.9}/ \tb{0.918} & 32.61/ 0.889 & \ul{33.61}/ \ul{0.916} & 33.11/ 0.908 & 33.55/ 0.913 & 28.81/ 0.705 \\
		Case 3 & 16.58/ 0.208 & 29.93/ 0.858 & \ul{33.09}/ 0.899 & 24.92/ 0.579 & 32.69/ 0.907 & 31.97/ 0.878 & \tb{33.38}/ \tb{0.911} & 32.47/ 0.9 & 33.05/ \ul{0.907} & 28.25/ 0.697 \\
		Case 4 & 13.25/ 0.143 & 28.37/ 0.815 & 30.55/ 0.849 & 23.8/ 0.534 & \tb{31.82}/ \tb{0.885} & 30.58/ 0.843 & 31.16/ 0.872 & \ul{31.39}/ 0.873 & 31.36/ \ul{0.874} & 26.46/ 0.599 \\
		Case 5 & 13.34/ 0.14 & 28.26/ 0.829 & 31.15/ 0.859 & 24.0/ 0.547 & 30.85/ \tb{0.885} & 30.53/ 0.842 & \ul{31.41}/ 0.877 & 30.98/ 0.866 & \tb{31.42}/ \ul{0.877} & 24.82/ 0.597 \\
		\Xhline{0.8pt}
	\end{tabular}
\end{table}

\begin{table}[t]
	\renewcommand{\arraystretch}{1.15}
	\newcommand{\mysize}{0.92cm}
	\fontsize{8.5}{9.5}\selectfont
	\caption{Complex target models: average test performance of LRTV, E3DTV and LRTFDFR and their corresponding DLW-models on PaviaU dataset. The best results in each \textbf{row} are in \textbf{bold}, and the second best results in each \textbf{row} are with \ul{underline}.}
	\label{tab-tran-dc}
	\centering
	\begin{tabular}{M{1.2cm} | M{\mysize}  M{\mysize} M{\mysize}  M{\mysize} M{\mysize} M{\mysize} M{\mysize} M{\mysize} M{\mysize+0.36cm} M{\mysize}}
		\hline
		\multirow{2}{1.2cm}{\backslashbox[1.6cm]{noise}{source}} & noisy & *\textit{Ori} & N & T & TS & N+T & N+TS & T+TS & N+T+TS & Abl\\
		\cline{2-11}
		& \multicolumn{10}{c}{evaluation index: PSNR/SSIM}\\
		\Xhline{0.8pt}
		\multicolumn{11}{c}{target model 1 : \tb{LRTV}} \cr
		\hline
	    Case 1 & 15.28/ 0.264 & 32.79/ 0.929 & 32.66/ 0.93 & 26.71/ 0.787 & 30.54/ 0.892 & 32.66/ 0.922 & 31.69/ 0.91 & \ul{33.03}/ \ul{0.93} & \tb{33.45}/ \tb{0.937} & 28.39/ 0.81 \\
	    Case 2 & 16.82/ 0.303 & 33.1/ 0.933 & \ul{33.4}/ \ul{0.94} & 26.34/ 0.748 & 31.3/ 0.912 & 32.79/ 0.922 & 31.89/ 0.915 & 33.1/ 0.931 & \tb{33.8}/ \tb{0.944} & 32.22/ 0.922 \\
	    Case 3 & 16.86/ 0.3 & 32.09/ 0.919 & \tb{33.15}/ \tb{0.936} & 26.44/ 0.75 & 30.45/ 0.897 & 32.16/ 0.916 & 31.54/ 0.909 & 32.46/ 0.925 & \ul{33.06}/ \ul{0.935} & 31.33/ 0.909 \\
	    Case 4 & 13.21/ 0.206 & 31.81/ 0.894 & 30.81/ 0.899 & 25.89/ 0.75 & 29.99/ 0.874 & 31.25/ 0.895 & 30.42/ 0.878 & 31.53/ 0.901 & \tb{32.25}/ \tb{0.914} & \ul{32.21}/ \ul{0.904} \\
	    Case 5 & 13.52/ 0.197 & 30.29/ 0.884 & 31.38/ 0.911 & 26.44/ 0.786 & 29.26/ 0.865 & 31.17/ 0.897 & 30.56/ 0.887 & \ul{31.72}/ \ul{0.914} & \tb{32.08}/ \tb{0.919} & 27.75/ 0.802 \\
	    \hline
	    \multicolumn{9}{c}{target model 2 : \tb{E3DTV}} \cr
	    \hline
	    Case 1 & 15.28/ 0.264 & 32.94/ 0.911 & 33.06/ \ul{0.939} & 26.52/ 0.787 & 31.21/ 0.904 & 32.91/ 0.929 & 32.56/ 0.927 & \ul{33.09}/ 0.936 & \tb{33.43}/ \tb{0.94} & 28.17/ 0.832 \\
	    Case 2 & 16.82/ 0.303 & \ul{33.99}/ 0.923 & \tb{34.15}/ \tb{0.951} & 26.29/ 0.757 & 32.32/ 0.924 & 33.04/ 0.928 & 33.43/ 0.939 & 33.28/ 0.939 & 33.8/ \ul{0.945} & 33.79/ 0.94 \\
	    Case 3 & 16.86/ 0.3 & \ul{33.34}/ 0.919 & \tb{33.81}/ \tb{0.945} & 26.43/ 0.756 & 31.27/ 0.91 & 32.28/ 0.913 & 32.98/ 0.931 & 32.75/ 0.932 & 33.26/ \ul{0.938} & 32.83/ 0.93 \\
	    Case 4 & 13.21/ 0.206 & 32.1/ 0.896 & 31.86/ 0.917 & 25.71/ 0.747 & 30.68/ 0.884 & 31.3/ 0.899 & 31.48/ 0.899 & 32.18/ 0.917 & \ul{32.34}/ \ul{0.919} & \tb{32.94}/ \tb{0.925} \\
	    Case 5 & 13.52/ 0.197 & 30.82/ 0.882 & \ul{31.92}/ \ul{0.92} & 26.22/ 0.779 & 30.06/ 0.88 & 31.35/ 0.902 & 31.32/ 0.901 & 31.91/ 0.917 & \tb{32.19}/ \tb{0.921} & 28.13/ 0.837 \\
	    \hline
	    \multicolumn{9}{c}{target model 3 : \tb{LRTFDFR}} \cr
	    \hline
	    Case 1 & 15.28/ 0.264 & \tb{33.8}/ 0.928 & 32.74/ 0.927 & 26.25/ 0.773 & 31.32/ 0.904 & 32.37/ 0.917 & 32.31/ 0.92 & 32.86/ \ul{0.935} & \ul{33.24}/ \tb{0.938} & 27.77/ 0.81 \\
	    Case 2 & 16.82/ 0.303 & \tb{35.29}/ \tb{0.95} & \ul{33.7}/ 0.937 & 25.88/ 0.737 & 31.97/ 0.913 & 32.42/ 0.913 & 32.88/ 0.926 & 32.82/ 0.932 & 33.4/ \ul{0.938} & 31.36/ 0.871 \\
	    Case 3 & 16.86/ 0.3 & \tb{34.07}/ \tb{0.942} & \ul{33.24}/ 0.93 & 26.03/ 0.737 & 30.97/ 0.897 & 31.69/ 0.897 & 32.28/ 0.918 & 32.37/ 0.924 & 32.88/ \ul{0.93} & 31.32/ 0.887 \\
	    Case 4 & 13.21/ 0.206 & \tb{33.15}/ \tb{0.92} & 31.37/ 0.895 & 25.37/ 0.73 & 30.3/ 0.875 & 30.65/ 0.879 & 30.78/ 0.882 & 31.75/ 0.911 & \ul{31.84}/ \ul{0.911} & 29.83/ 0.835 \\
	    Case 5 & 13.52/ 0.197 & 31.43/ 0.898 & 31.6/ 0.905 & 25.84/ 0.759 & 30.23/ 0.88 & 30.55/ 0.879 & 31.28/ 0.897 & \ul{31.81}/ \ul{0.917} & \tb{32.06}/ \tb{0.919} & 27.09/ 0.792 \\
		\Xhline{0.8pt}
	\end{tabular}
\end{table}

\begin{table}[t]
	\renewcommand{\arraystretch}{1.15}
	\newcommand{\mysize}{1cm}
	\fontsize{8.5}{9.5}\selectfont
	\caption{Average test performance of deep-plug-and-play using SERT on ICVL dataset. The best results in each \textbf{row} are in \textbf{bold}, and the second best results in each \textbf{row} are with \ul{underline}.}
	\label{tab-sert-icvl}
	\centering
	\begin{tabular}{M{1.2cm} | M{\mysize} M{\mysize} M{\mysize} M{\mysize} M{\mysize} M{\mysize} M{\mysize} M{\mysize} M{\mysize+0.3cm}}
		\wline
		\multirow{2}{1.2cm}{\backslashbox[1.6cm]{noise}{source}}
		& noisy & SERT & N & T & TS & N+T & N+TS & T+TS & N+T+TS \\
		\cline{2-10}
		& \multicolumn{9}{c}{evaluation index: PSNR/SSIM} \\
		\wline
		Case 1 & 14.86/ 0.3554 & 27.99/ 0.8695 & \ul{36.14}/ 0.9553 & 29.92/ 0.8435 & 35.39/ 0.9542 & 35.35/ 0.9564 & 36.03/ 0.9563 & \tb{36.25}/ \tb{0.9631} & 36.13/ \ul{0.96} \\
		Case 2 & 16.16/ 0.3998 & \tb{36.39}/ \tb{0.9679} & 35.94/ 0.9584 & 29.32/ 0.8295 & 35.31/ 0.9554 & 35.17/ 0.9549 & 35.75/ 0.9595 & \ul{36.02}/ \ul{0.9645} & 35.82/ 0.9613 \\
		Case 3 & 16.06/ 0.395 & 33.85/ \tb{0.9593} & \ul{35.29}/ 0.9557 & 29.02/ 0.8291 & 33.93/ 0.9466 & 34.35/ 0.9498 & \tb{35.35}/ 0.9568 & 34.82/ \ul{0.9579} & 35.2/ 0.9577 \\
		Case 4 & 13.13/ 0.3197 & \tb{35.87}/ \tb{0.9635} & 34.48/ 0.9377 & 27.84/ 0.784 & 33.95/ 0.9405 & 33.81/ 0.9319 & 34.47/ 0.9417 & \ul{34.91}/ \ul{0.9512} & 34.54/ 0.9424 \\
		Case 5 & 12.94/ 0.2762 & 27.66/ 0.8593 & \ul{33.18}/ 0.9248 & 27.44/ 0.7663 & 31.77/ 0.9212 & 32.08/ 0.9159 & \tb{33.41}/ \ul{0.9305} & 32.79/ \tb{0.9323} & 33.06/ 0.9281 \\
		\wline
	\end{tabular}
\end{table}

\begin{figure*}[ht]
	\newcommand{\mysize}{2.6cm}
	\newcommand{\minu}{0pt}
	\fontsize{8.5}{9.5}\selectfont
	\renewcommand{\arraystretch}{1.25}
	\centering
	\begin{minipage}[t]{\mysize}
		\centering
		\includegraphics[width=\mysize]{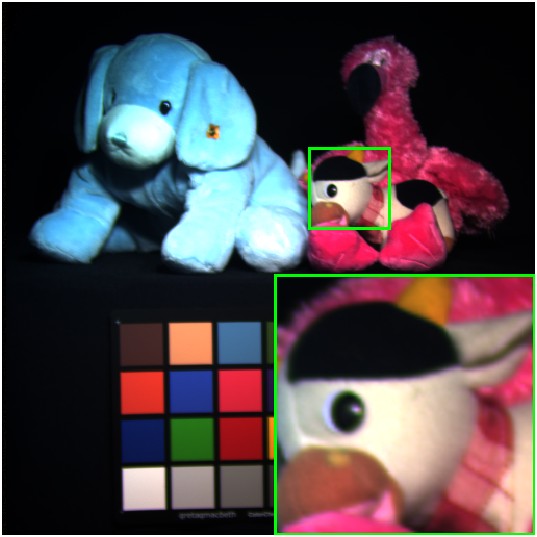}\\
		clean
		(PSNR, SSIM)
	\end{minipage}\hspace{\minu}
	\begin{minipage}[t]{\mysize}
		\centering
		\includegraphics[width=\mysize]{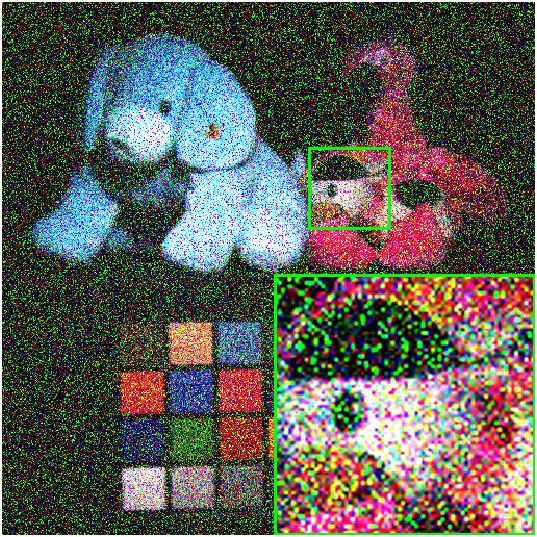}\\
		noisy \\(11.85, 0.1674)
	\end{minipage}\hspace{\minu}
	\begin{minipage}[t]{\mysize}
		\centering
		\includegraphics[width=\mysize]{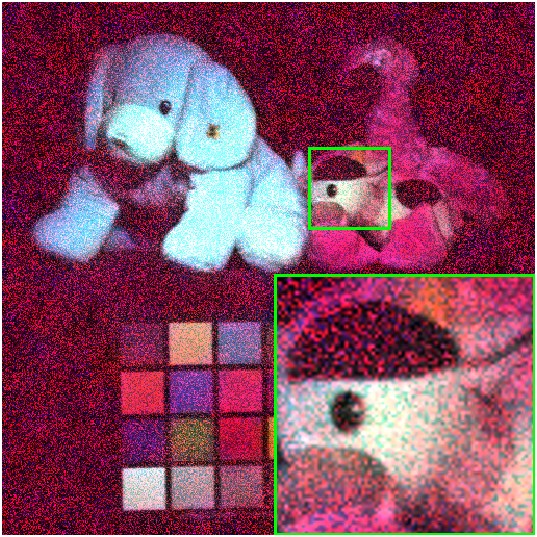}\\
		LRMR \\ (20.02,	0.4760)
	\end{minipage}\hspace{\minu}
	\begin{minipage}[t]{\mysize}
		\centering
		\includegraphics[width=\mysize]{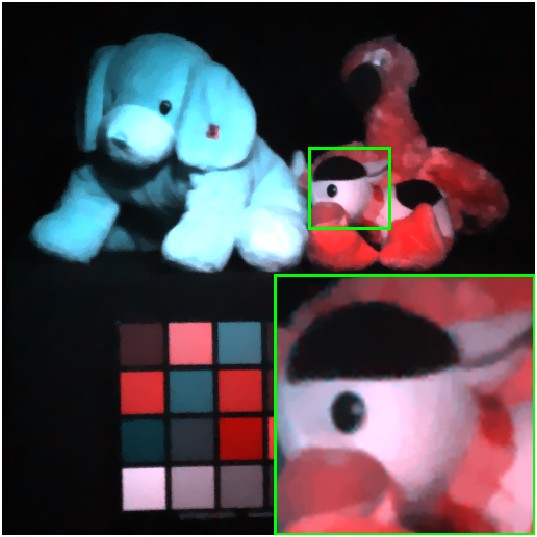}\\
		LRTV \\(29.92, 0.8950)
	\end{minipage}\hspace{\minu}
	\begin{minipage}[t]{\mysize}
		\centering
		\includegraphics[width=\mysize]{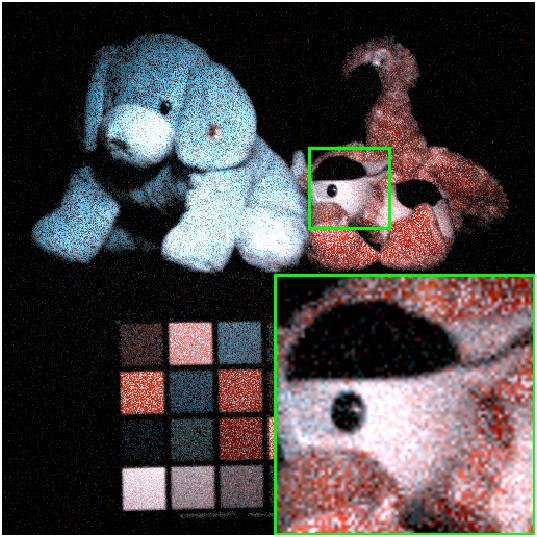}\\
		NMoG \\(21.71, 0.6735)
	\end{minipage} \vspace{4pt}\\
	\begin{minipage}[t]{\mysize}
		\centering
		\includegraphics[width=\mysize]{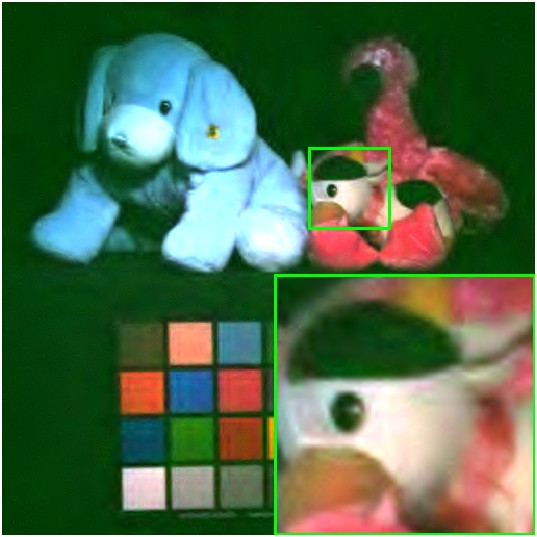}\\
		HyRes \\ (28.12, 0.7571)
	\end{minipage}\hspace{\minu}
	\begin{minipage}[t]{\mysize}
		\centering
		\includegraphics[width=\mysize]{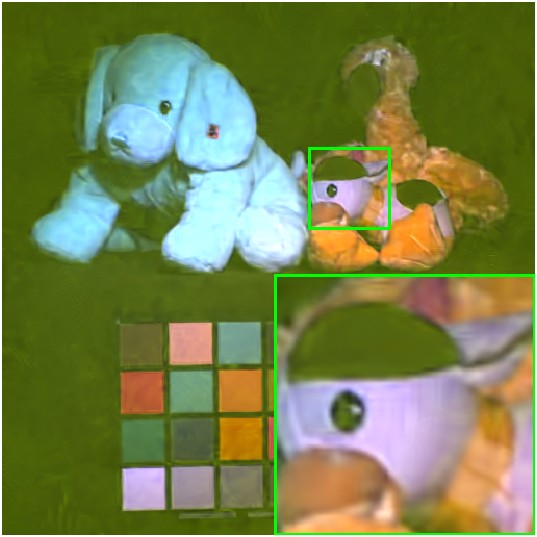}\\
		FastHyMix\\ (19.20, 0.5520)
	\end{minipage}\hspace{\minu}
	\begin{minipage}[t]{\mysize}
		\centering
		\includegraphics[width=\mysize]{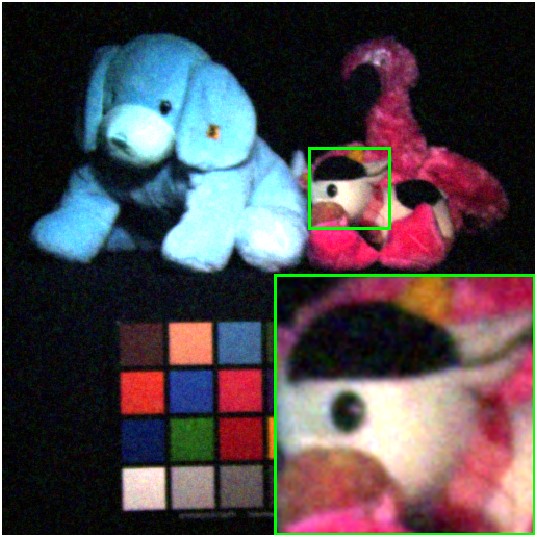}\\
		CTV \\(31.98, 0.7888)
	\end{minipage}\hspace{\minu}
	\begin{minipage}[t]{\mysize}
		\centering
		\includegraphics[width=\mysize]{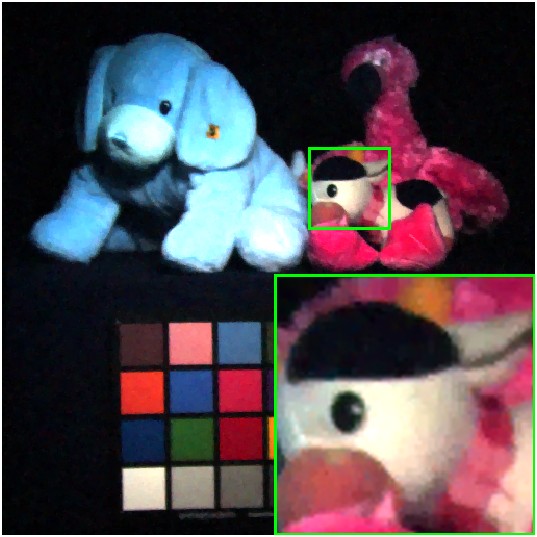}\\
		E3DTV \\(\ul{34.57}, \tb{0.9291})
	\end{minipage}\hspace{\minu}
	\begin{minipage}[t]{\mysize}
		\centering
		\includegraphics[width=\mysize]{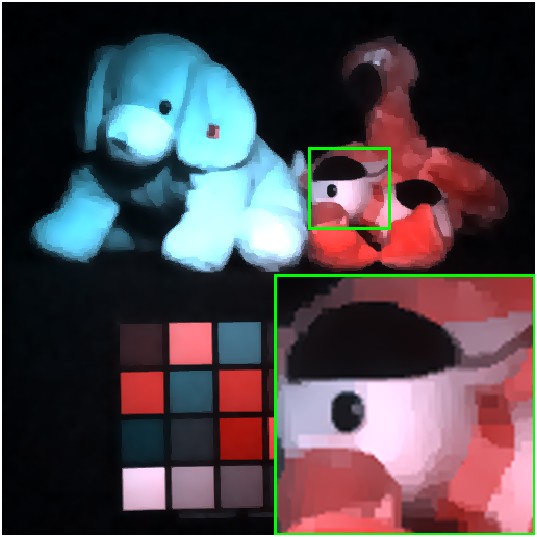}\\
		LRTFDFR \\(29.53, \ul{0.9278})
	\end{minipage} \vspace{4pt}\\
	\begin{minipage}[t]{\mysize}
		\centering
		\includegraphics[width=\mysize]{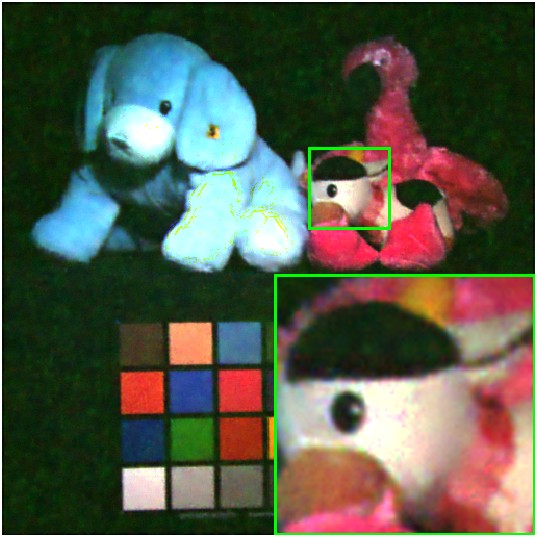}\\
		HSIDeNet\\ (23.33, 0.6969)
	\end{minipage}\hspace{\minu}
	\begin{minipage}[t]{\mysize}
		\centering
		\includegraphics[width=\mysize]{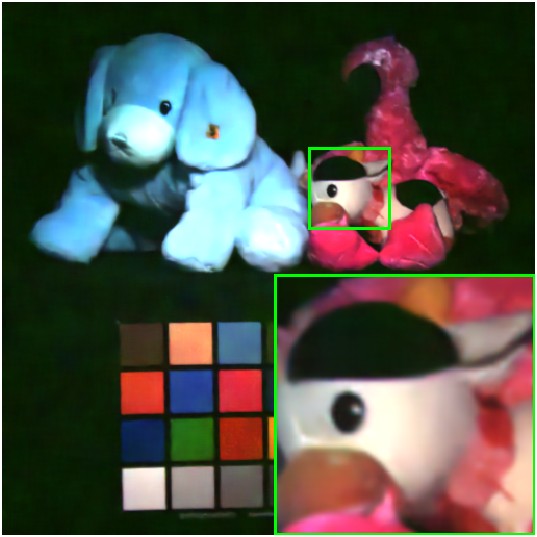}\\
		HSICNN\\ (31.98, 0.7696)
	\end{minipage}\hspace{\minu}
	\begin{minipage}[t]{\mysize}
		\centering
		\includegraphics[width=\mysize]{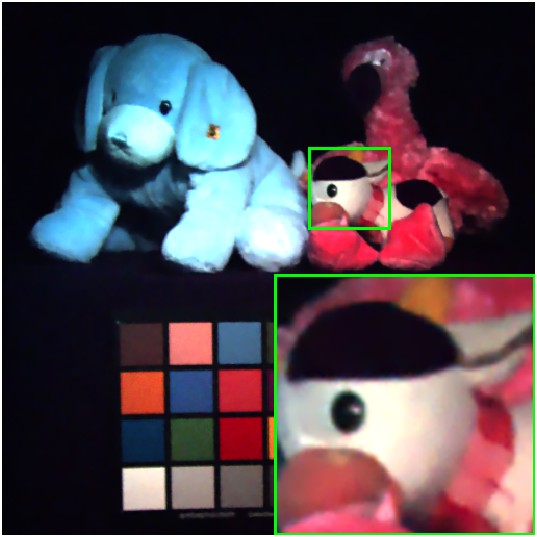}\\
		\textit{DLW-LRTV}\\ (32.71, 0.8949)
	\end{minipage}\hspace{\minu}
	\begin{minipage}[t]{\mysize}
		\centering
		\includegraphics[width=\mysize]{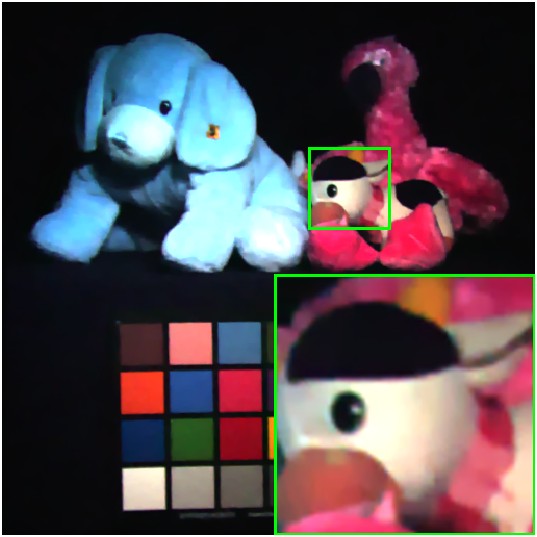}\\
		\textit{DLW-E3DTV} \\(\tb{34.62}, 0.8813)
	\end{minipage}\hspace{\minu}
	\begin{minipage}[t]{\mysize}
		\centering
		\includegraphics[width=\mysize]{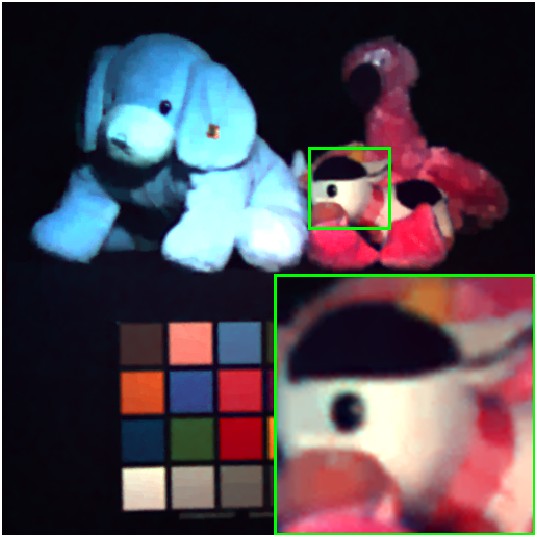}\\
		\textit{DLW-LRTFDFR}\\ (31.75, 0.8510)
	\end{minipage}
	\caption{Visual comparison (pesudo-color image) of the denoising performance of image ``stuffed toys" of CAVE dataset obtained by different competing denoising methods.}
	\label{fig-comall-cave-high}
\end{figure*}

\begin{figure*}[ht]
	\newcommand{\mysize}{2.6cm}
	\newcommand{\minu}{0pt}
	\fontsize{8.5}{9.5}\selectfont
	\renewcommand{\arraystretch}{1.25}
	\centering
	\begin{minipage}[t]{\mysize}
		\centering
		\includegraphics[width=\mysize]{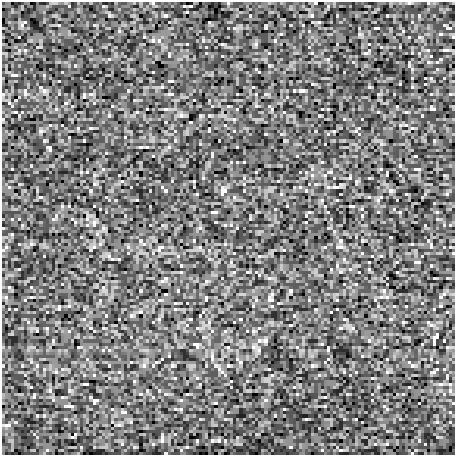}\\
		Band 151
	\end{minipage}\hspace{\minu}
	\begin{minipage}[t]{\mysize}
		\centering
		\includegraphics[width=\mysize]{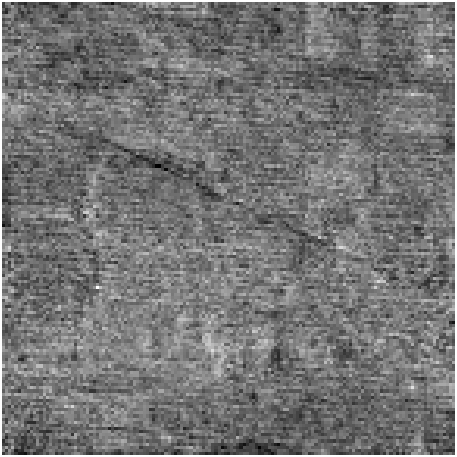}\\
		LRMR
	\end{minipage}\hspace{\minu}
	\begin{minipage}[t]{\mysize}
		\centering
		\includegraphics[width=\mysize]{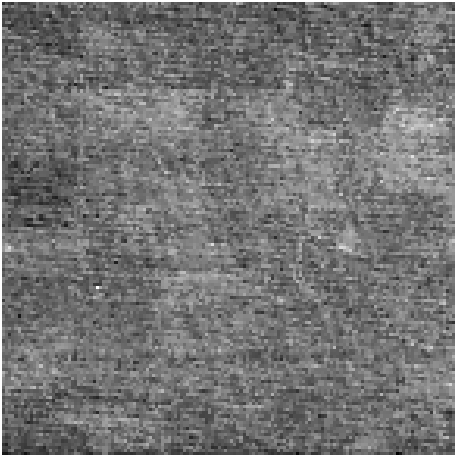}\\
		LRTV
	\end{minipage}\hspace{\minu}
	\begin{minipage}[t]{\mysize}
		\centering
		\includegraphics[width=\mysize]{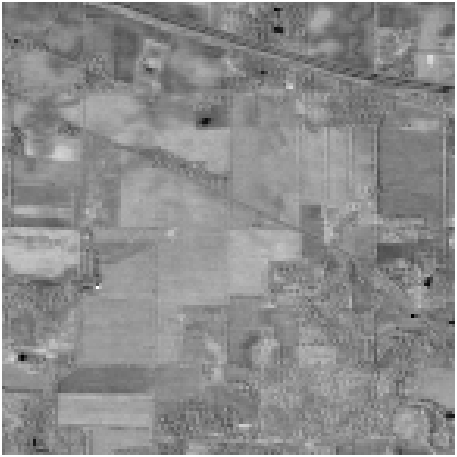}\\
		NMoG
	\end{minipage}
	\begin{minipage}[t]{\mysize}
		\centering
		\includegraphics[width=\mysize]{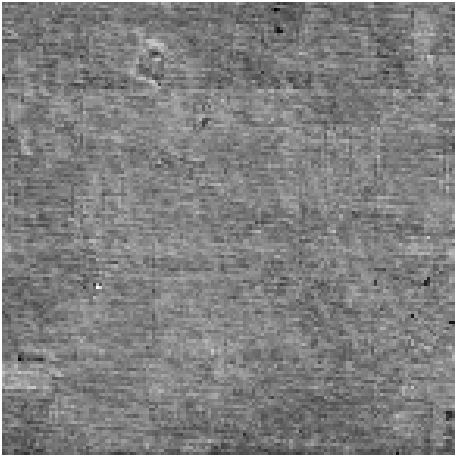}\\
		HyRes
	\end{minipage}\vspace{4pt}\\
	\begin{minipage}[t]{\mysize}
		\centering
		\includegraphics[width=\mysize]{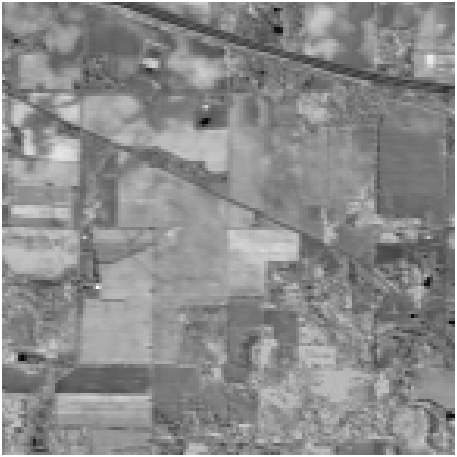}\\
		FastHyMix
	\end{minipage}\hspace{\minu}
	\begin{minipage}[t]{\mysize}
		\centering
		\includegraphics[width=\mysize]{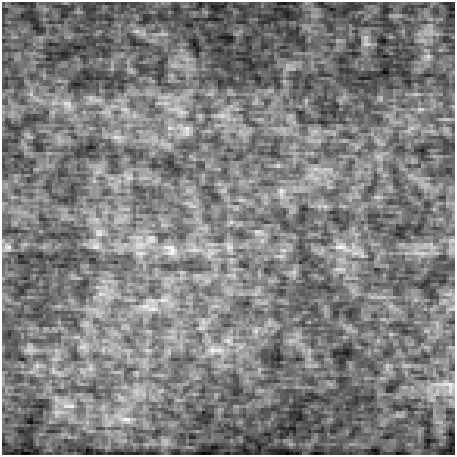}\\
		CTV
	\end{minipage}\hspace{\minu}
	\begin{minipage}[t]{\mysize}
		\centering
		\includegraphics[width=\mysize]{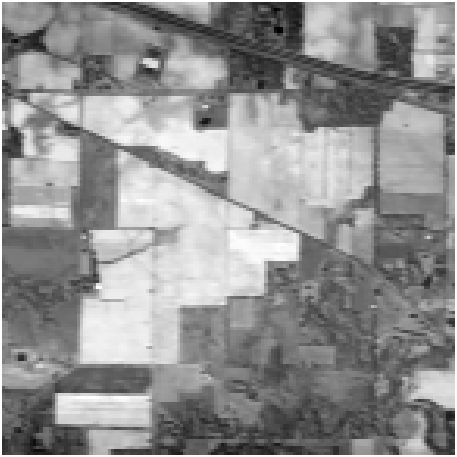}\\
		E3DTV
	\end{minipage}\hspace{\minu}
	\begin{minipage}[t]{\mysize}
		\centering
		\includegraphics[width=\mysize]{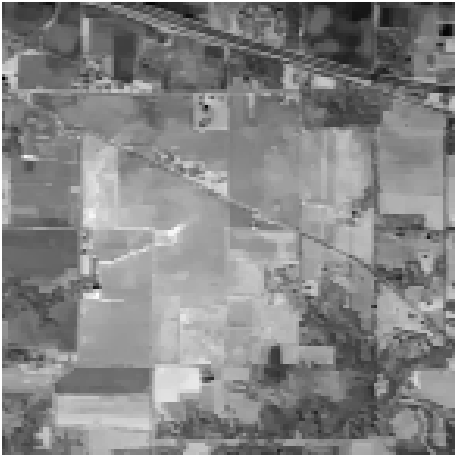}\\
		LRTFDFR
	\end{minipage}
	\begin{minipage}[t]{\mysize}
		\centering
		\includegraphics[width=\mysize]{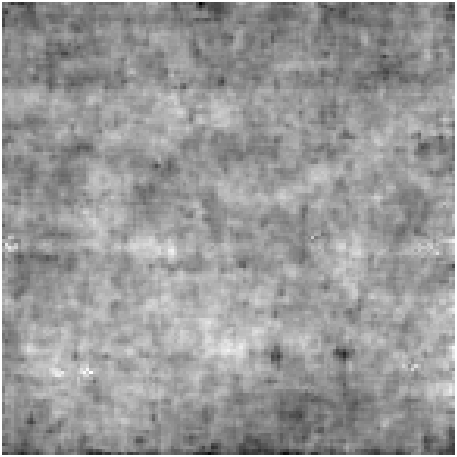}\\
		HSIDeNet
	\end{minipage}\vspace{4pt}\\
	\begin{minipage}[t]{\mysize}
		\centering
		\includegraphics[width=\mysize]{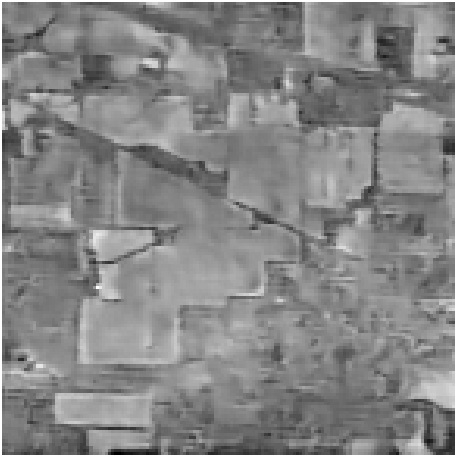}\\
		HSICNN
	\end{minipage}\hspace{\minu}
	\begin{minipage}[t]{\mysize}
		\centering
		\includegraphics[width=\mysize]{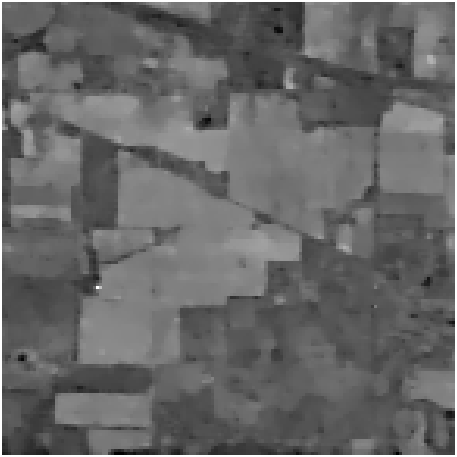}\\
		\textit{DLW-LRTV}
	\end{minipage}\hspace{\minu}
	\begin{minipage}[t]{\mysize}
		\centering
		\includegraphics[width=\mysize]{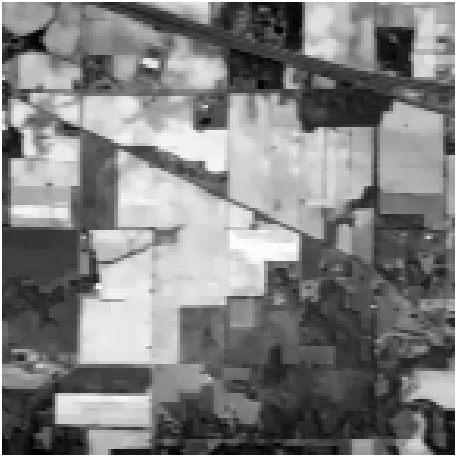}\\
		\textit{DLW-E3DTV}
	\end{minipage}\hspace{\minu}
	\begin{minipage}[t]{\mysize}
		\centering
		\includegraphics[width=\mysize]{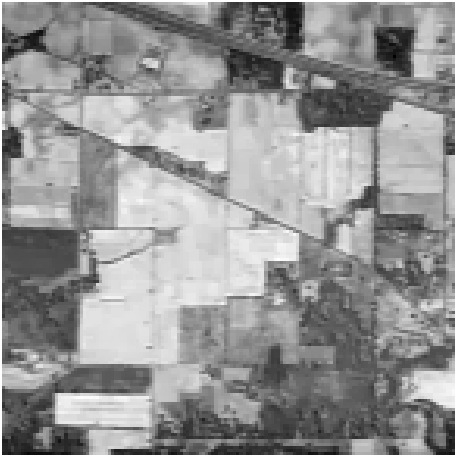}\\
		\textit{DLW-LRTFDFR}
	\end{minipage}\hspace{\minu}
	
	\caption{Visual comparison of the denoising performance on Indian Pine dataset obtained by different competing denoising methods.}
	\label{fig-comall-indian}
\end{figure*}

\begin{figure}[t]
	\renewcommand{\arraystretch}{1.15}
	\newcommand{\mysize}{2.6cm}
	\fontsize{8.5}{9.5}\selectfont
	\newcommand{\minivs}{5pt}
	\newcommand{\vs}{2pt}
	\centering
	\begin{minipage}[t]{\mysize*5+2cm}
		\centering
		\begin{minipage}[b]{\mysize}
			\centering
			clean 
			\includegraphics[width=\mysize]{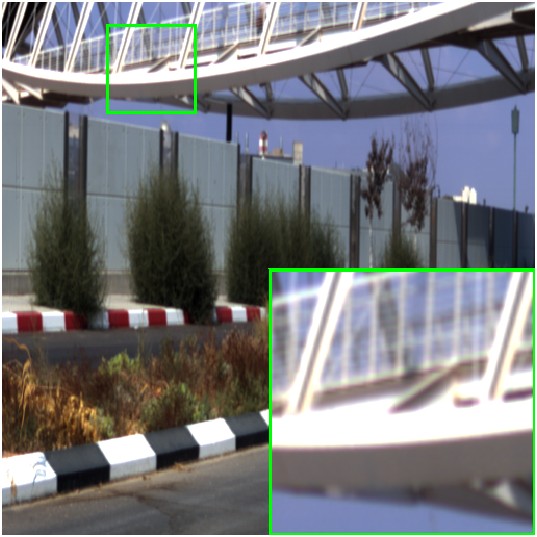} \\
			(PSNR, SSIM)
		\end{minipage}\hspace{\minivs}
		\begin{minipage}[b]{\mysize}
			\centering
			noisy 
			\includegraphics[width=\mysize]{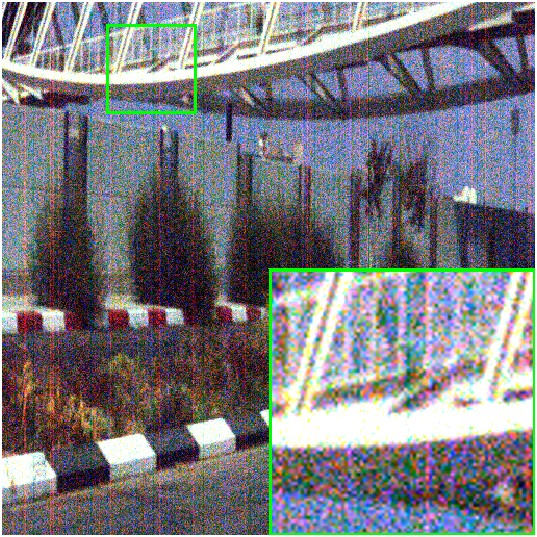} \\
			(17.67, 0.4624)
		\end{minipage}\hspace{\minivs}
		\begin{minipage}[b]{\mysize}
			\centering
			N 
			\includegraphics[width=\mysize]{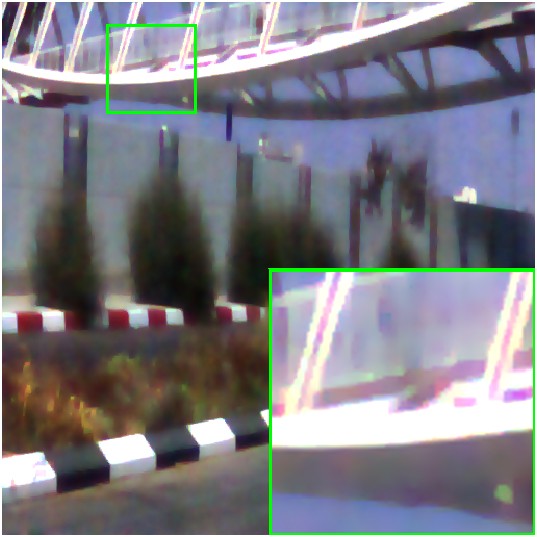} \\
			(29.2, 0.8601)
		\end{minipage}\hspace{\minivs}
		\begin{minipage}[b]{\mysize}
			\centering
			T 
			\includegraphics[width=\mysize]{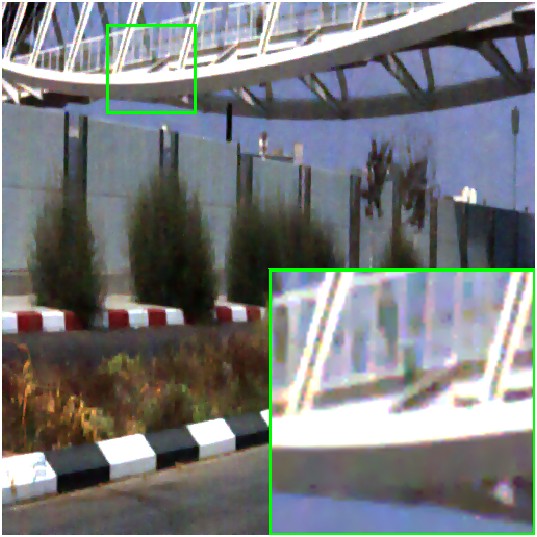} \\
			(\tb{32.48}, \tb{0.9357})
		\end{minipage} \vspace{6pt} \\
		\begin{minipage}[b]{\mysize}
			\centering
			TS 
			\includegraphics[width=\mysize]{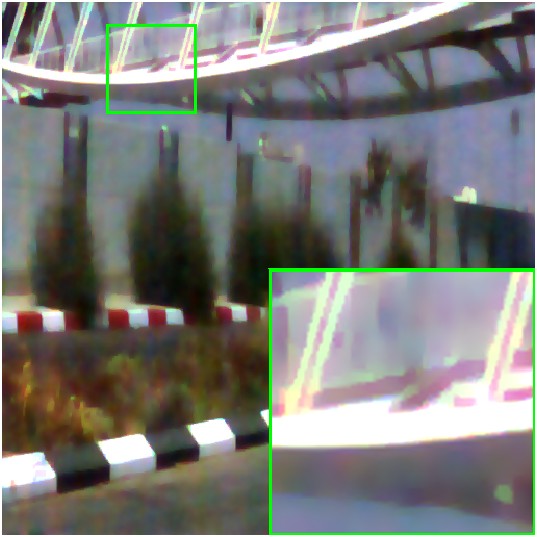} \\
			(27.87, 0.8275)
		\end{minipage}\hspace{\minivs}
		\begin{minipage}[b]{\mysize}
			\centering
			N+T 
			\includegraphics[width=\mysize]{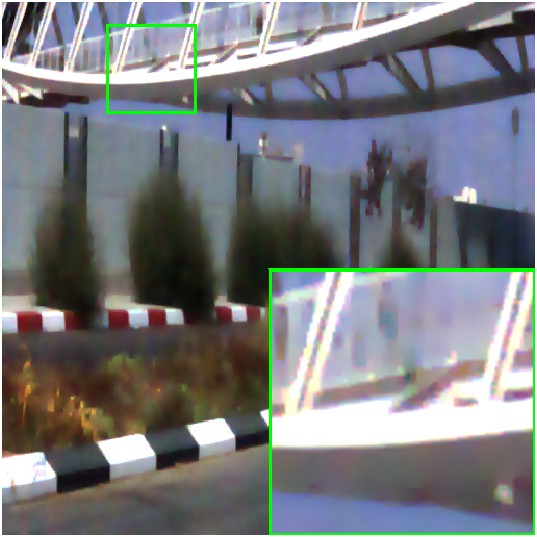} \\
			(\ul{31.65}, \ul{0.9034})
		\end{minipage}\hspace{\minivs}
		\begin{minipage}[b]{\mysize}
			\centering
			N+TS 
			\includegraphics[width=\mysize]{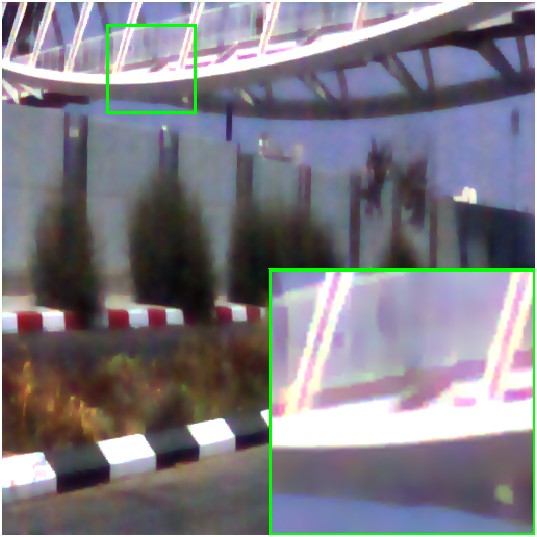} \\
			(29.18, 0.8595)
		\end{minipage}\hspace{\minivs}
		\begin{minipage}[b]{\mysize}
			\centering
			T+TS 
			\includegraphics[width=\mysize]{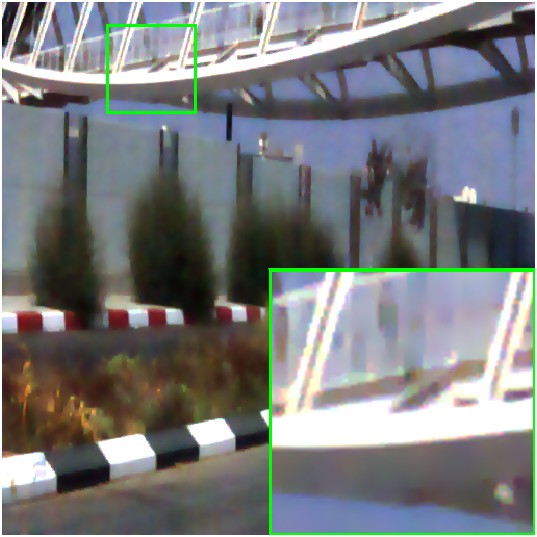} \\
			(31.48, 0.9022)
		\end{minipage}\hspace{\minivs}
		\begin{minipage}[b]{\mysize}
			\centering
			N+T+TS 
			\includegraphics[width=\mysize]{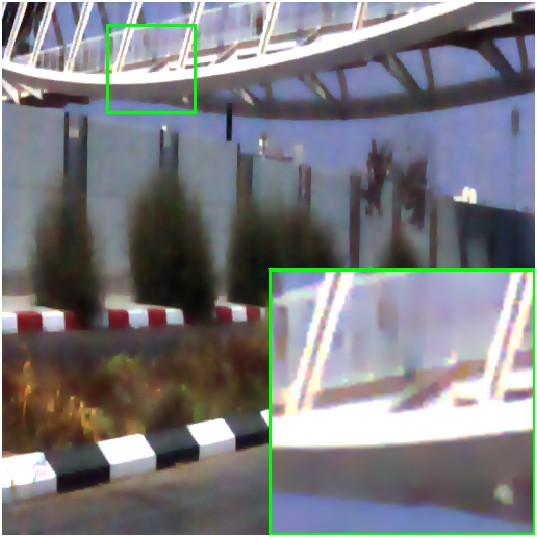} \\
			(31.13, 0.8941)
		\end{minipage}\hspace{\minivs}
	\end{minipage}
	\caption{Denoising results (pesudo-color image) of DLW-TV on image ``gavyam 0823-0933" of ICVL dataset. The noisy type is ``Gaussian+stripe" (Case 2).}
	\label{fig-append-add-tv1}
\end{figure}

\begin{figure}[t]
	\renewcommand{\arraystretch}{1.15}
	\newcommand{\mysize}{2.6cm}
	\fontsize{8.5}{9.5}\selectfont
	\newcommand{\minivs}{5pt}
	\newcommand{\vs}{2pt}
	\centering
	\begin{minipage}[t]{\mysize*5+2cm}
		\centering
		\begin{minipage}[b]{\mysize}
			\centering
			clean 
			\includegraphics[width=\mysize]{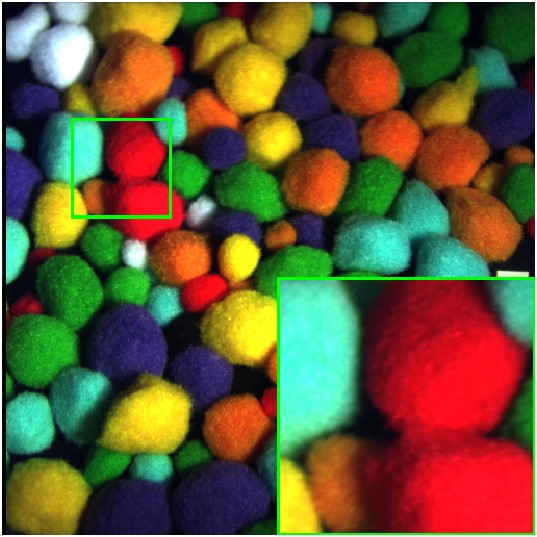} \\
			(PSNR, SSIM)
		\end{minipage}\hspace{\minivs}
		\begin{minipage}[b]{\mysize}
			\centering
			noisy 
			\includegraphics[width=\mysize]{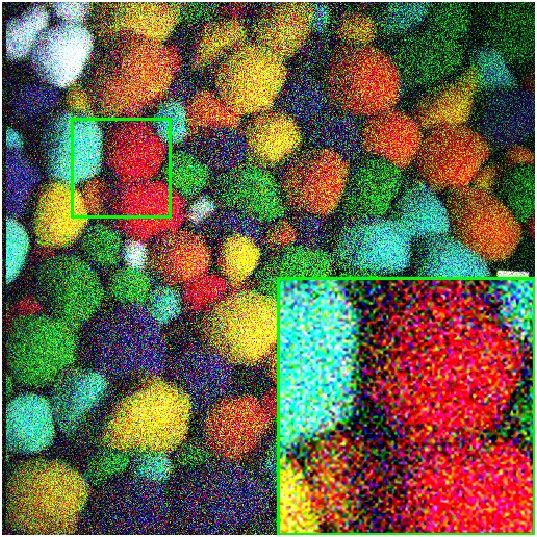} \\
			(13.19, 0.274)
		\end{minipage}\hspace{\minivs}
		\begin{minipage}[b]{\mysize}
			\centering
			N 
			\includegraphics[width=\mysize]{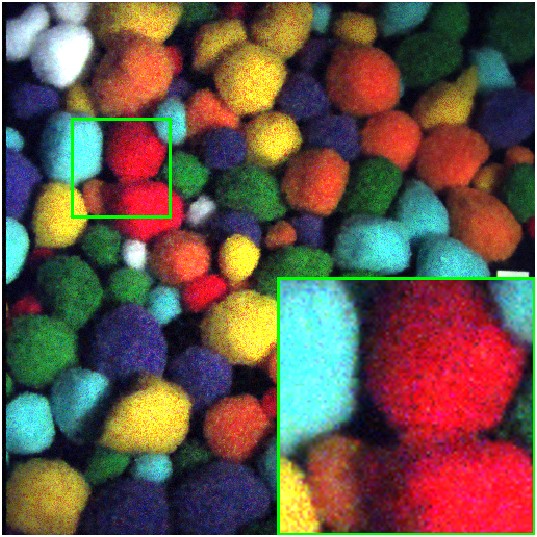} \\
			(25.55, 0.7706)
		\end{minipage}\hspace{\minivs}
		\begin{minipage}[b]{\mysize}
			\centering
			T 
			\includegraphics[width=\mysize]{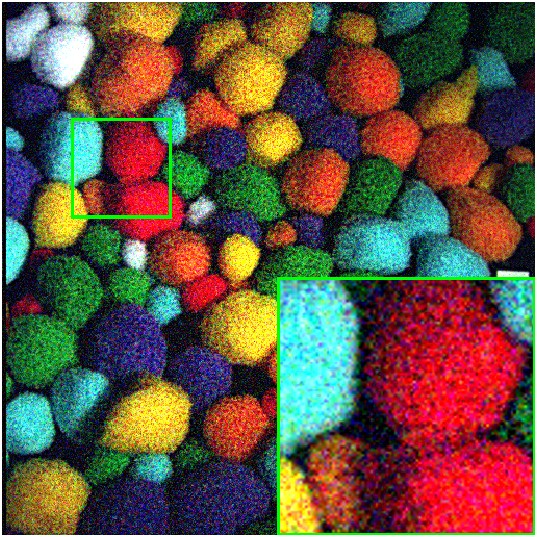} \\
			(20.66, 0.5331)
		\end{minipage} \vspace{6pt} \\
		\begin{minipage}[b]{\mysize}
			\centering
			TS 
			\includegraphics[width=\mysize]{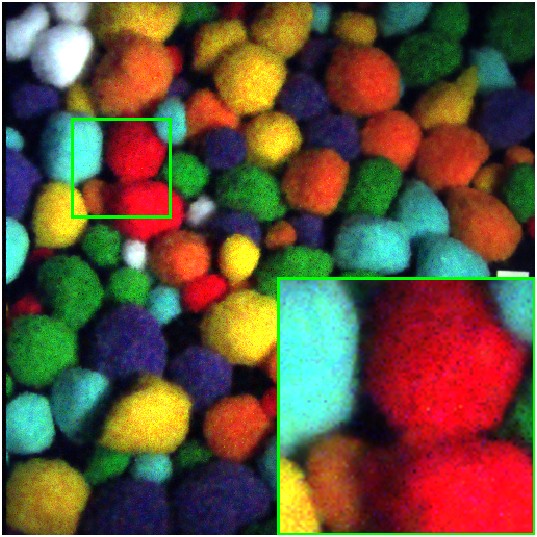} \\
			(\tb{29.25}, \tb{0.854})
		\end{minipage}\hspace{\minivs}
		\begin{minipage}[b]{\mysize}
			\centering
			N+T 
			\includegraphics[width=\mysize]{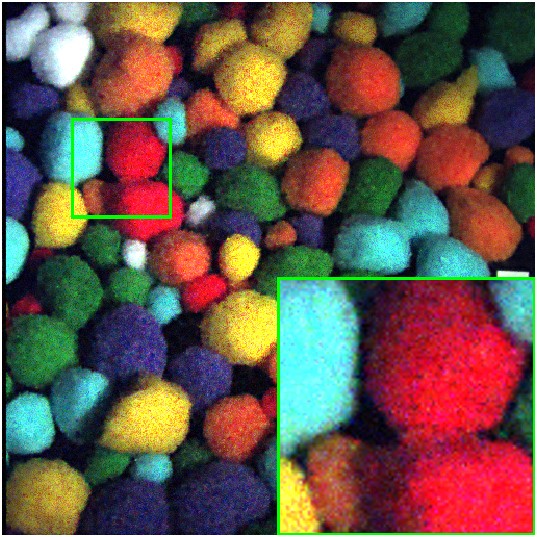} \\
			(25.15, 0.756)
		\end{minipage}\hspace{\minivs}
		\begin{minipage}[b]{\mysize}
			\centering
			N+TS 
			\includegraphics[width=\mysize]{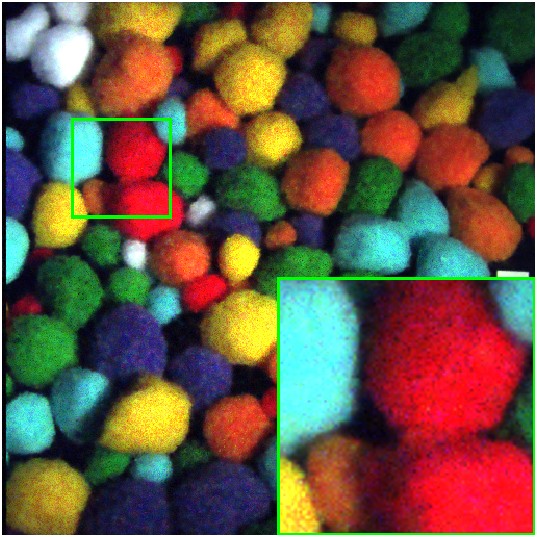} \\
			(27.93, 0.8369)
		\end{minipage}\hspace{\minivs}
		\begin{minipage}[b]{\mysize}
			\centering
			T+TS 
			\includegraphics[width=\mysize]{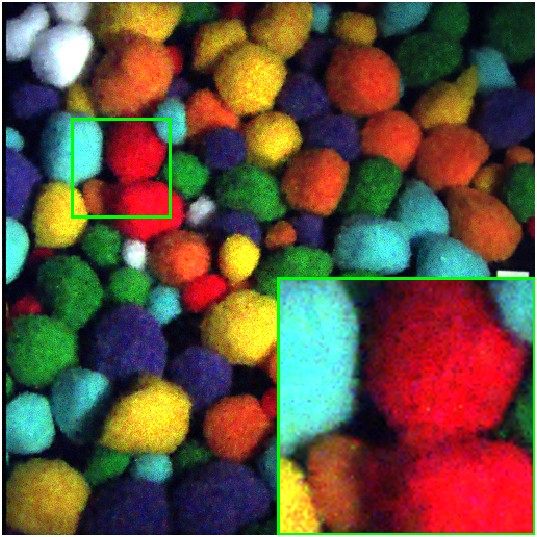} \\
			(\ul{28.11}, \ul{0.8415})
		\end{minipage}\hspace{\minivs}
		\begin{minipage}[b]{\mysize}
			\centering
			N+T+TS 
			\includegraphics[width=\mysize]{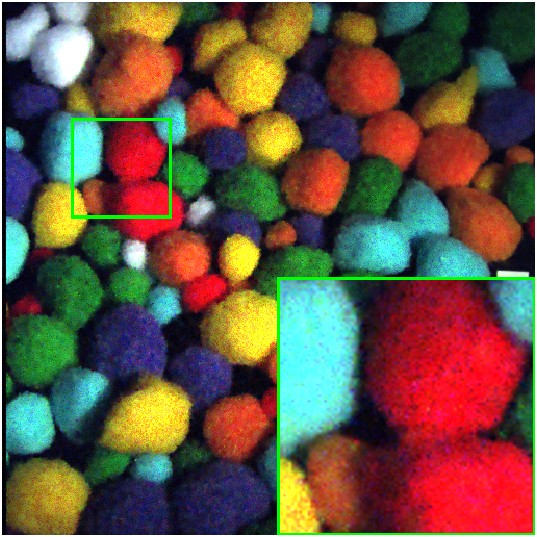} \\
			(27.41, 0.8266)
		\end{minipage}\hspace{\minivs}
	\end{minipage}
	\caption{Denoising results (pesudo-color image) of DLW-TVS on image ``pompoms" of CAVE dataset. The noisy type is ``Spatial-Spectral Variant Gaussian" (Case 4).}
	\label{fig-append-add-tvs1}
\end{figure}

\begin{figure}[t]
	\renewcommand{\arraystretch}{1.15}
	\newcommand{\mysize}{2.6cm}
	\fontsize{8.5}{9.5}\selectfont
	\newcommand{\minivs}{5pt}
	\newcommand{\vs}{2pt}
	\centering
	\begin{minipage}[t]{\mysize*5+2cm}
		\centering
		\begin{minipage}[b]{\mysize}
			\centering
			clean 
			\includegraphics[width=\mysize]{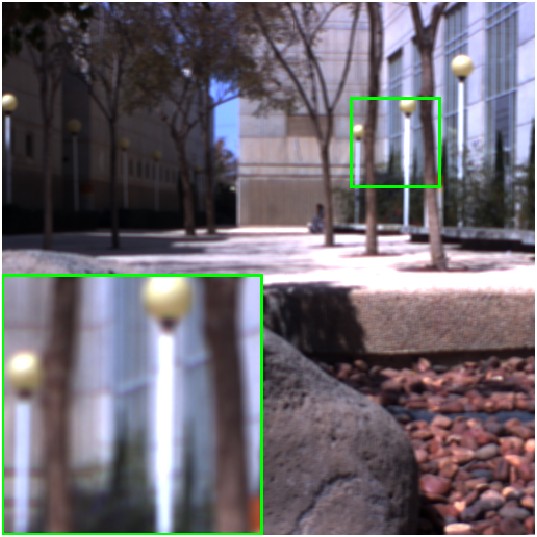} \\
			(PSNR, SSIM)
		\end{minipage}\hspace{\minivs}
		\begin{minipage}[b]{\mysize}
			\centering
			noisy 
			\includegraphics[width=\mysize]{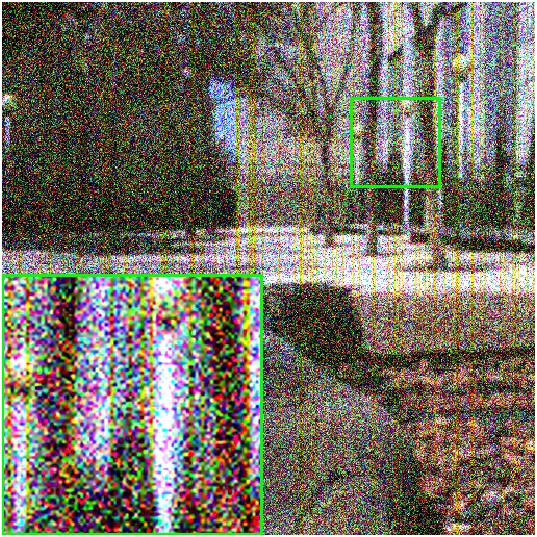} \\
			(12.11, 0.2338)
		\end{minipage}\hspace{\minivs}
		\begin{minipage}[b]{\mysize}
			\centering
			\textit{*Original}  
			\includegraphics[width=\mysize]{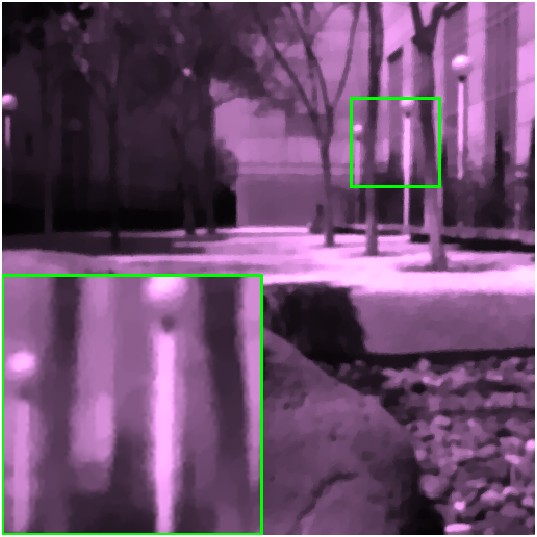} \\
			(26.75, 0.8858)
		\end{minipage}\hspace{\minivs}
		\begin{minipage}[b]{\mysize}
			\centering
			N 
			\includegraphics[width=\mysize]{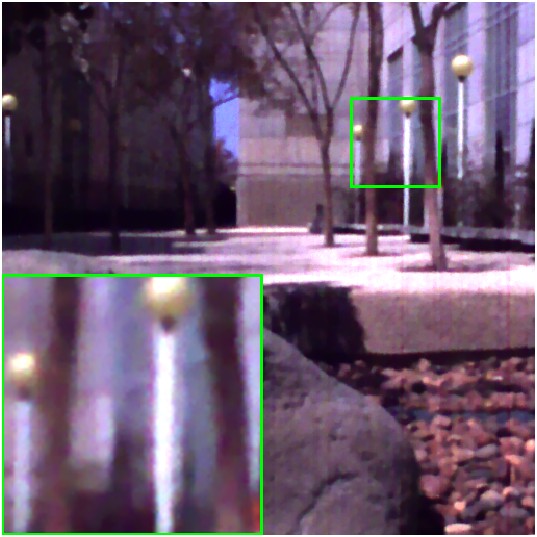} \\
			(\ul{31.45}, 0.9204)
		\end{minipage}\hspace{\minivs}
		\begin{minipage}[b]{\mysize}
			\centering
			T 
			\includegraphics[width=\mysize]{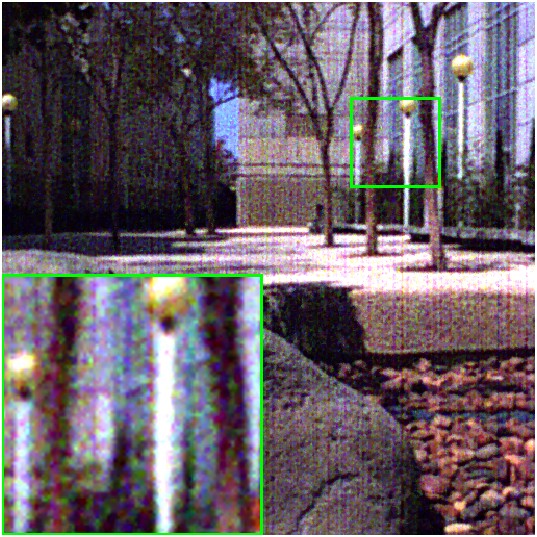} \\
			(25.54, 0.7384)
		\end{minipage} \vspace{6pt} \\
		\begin{minipage}[b]{\mysize}
			\centering
			TS 
			\includegraphics[width=\mysize]{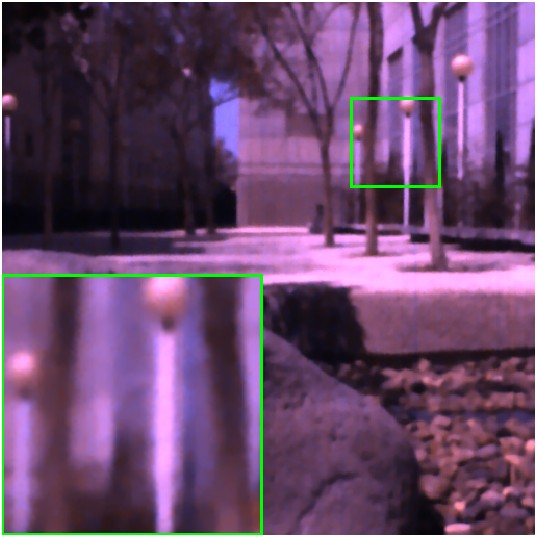} \\
			(27.77, 0.8933)
		\end{minipage}\hspace{\minivs}
		\begin{minipage}[b]{\mysize}
			\centering
			N+T 
			\includegraphics[width=\mysize]{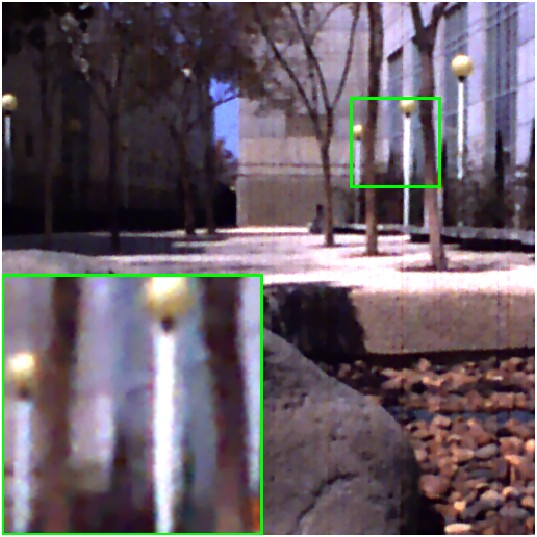} \\
			(30.54, 0.934)
		\end{minipage}\hspace{\minivs}
		\begin{minipage}[b]{\mysize}
			\centering
			N+TS 
			\includegraphics[width=\mysize]{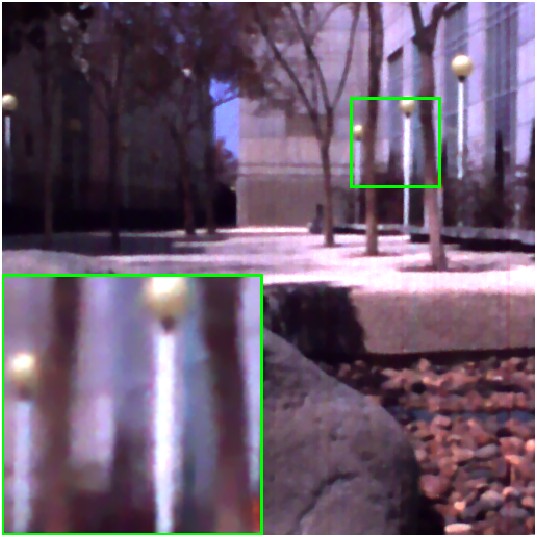} \\
			(\tb{31.52}, 0.9275)
		\end{minipage}\hspace{\minivs}
		\begin{minipage}[b]{\mysize}
			\centering
			T+TS 
			\includegraphics[width=\mysize]{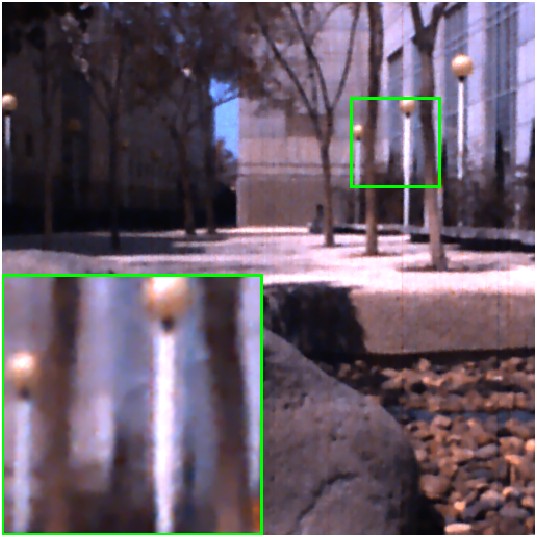} \\
			(30.22, \tb{0.9361})
		\end{minipage}\hspace{\minivs}
		\begin{minipage}[b]{\mysize}
			\centering
			N+T+TS 
			\includegraphics[width=\mysize]{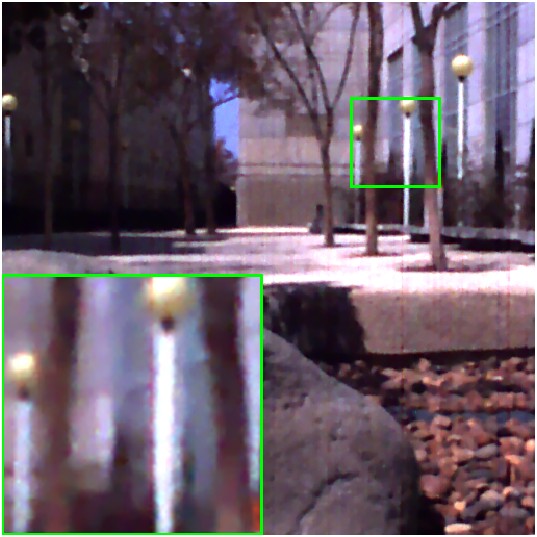} \\
			(31.21, \ul{0.9348})
		\end{minipage}\hspace{\minivs}
	\end{minipage}
	\caption{Denoising results (pesudo-color image) of LRTV and their corresponding ``DLW-" models on image ``photo and face" of CAVE dataset. The noisy type is ``mixture" (Case 5).}
	\label{fig-append-add-lrtv1}
\end{figure}

\begin{figure}[t]
	\renewcommand{\arraystretch}{1.15}
	\newcommand{\mysize}{2.6cm}
	\fontsize{8.5}{9.5}\selectfont
	\newcommand{\minivs}{5pt}
	\newcommand{\vs}{2pt}
	\centering
	\begin{minipage}[t]{\mysize*5+2cm}
		\centering
		\begin{minipage}[b]{\mysize}
			\centering
			clean 
			\includegraphics[width=\mysize]{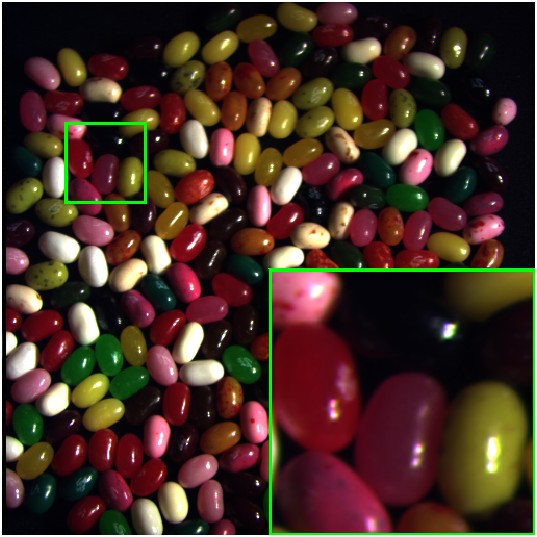} \\
			(PSNR, SSIM)
		\end{minipage}\hspace{\minivs}
		\begin{minipage}[b]{\mysize}
			\centering
			noisy 
			\includegraphics[width=\mysize]{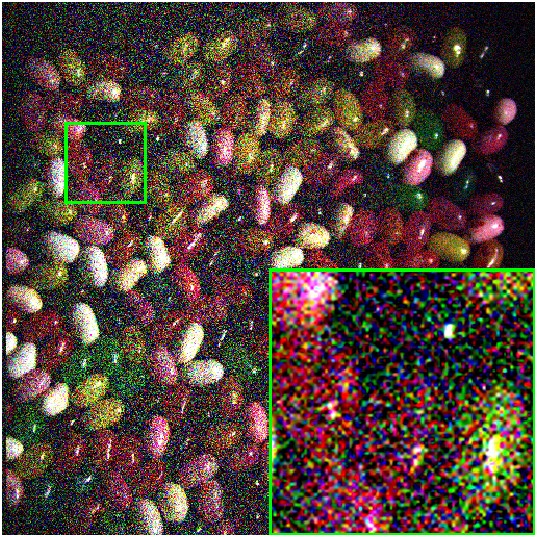} \\
			(13.78, 0.4417)
		\end{minipage}\hspace{\minivs}
		\begin{minipage}[b]{\mysize}
			\centering
			\textit{*Original}  
			\includegraphics[width=\mysize]{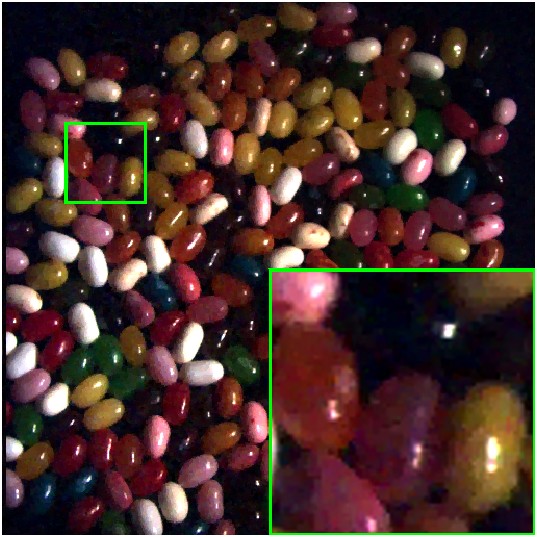} \\
			(30.58, 0.9132)
		\end{minipage}\hspace{\minivs}
		\begin{minipage}[b]{\mysize}
			\centering
			N 
			\includegraphics[width=\mysize]{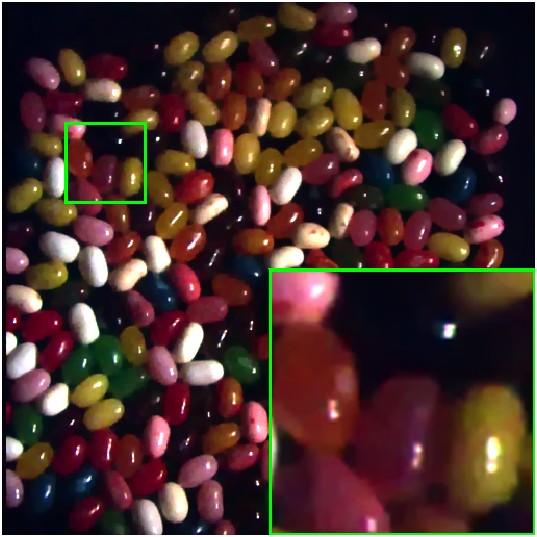} \\
			(32.07, 0.9275)
		\end{minipage}\hspace{\minivs}
		\begin{minipage}[b]{\mysize}
			\centering
			T 
			\includegraphics[width=\mysize]{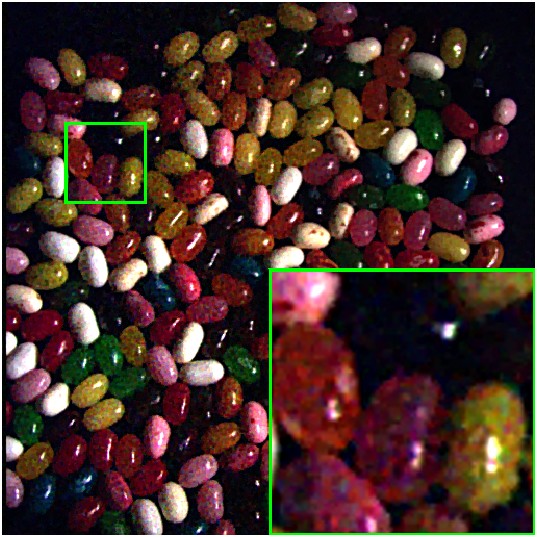} \\
			(28.62, 0.8831)
		\end{minipage} \vspace{6pt} \\
		\begin{minipage}[b]{\mysize}
			\centering
			TS 
			\includegraphics[width=\mysize]{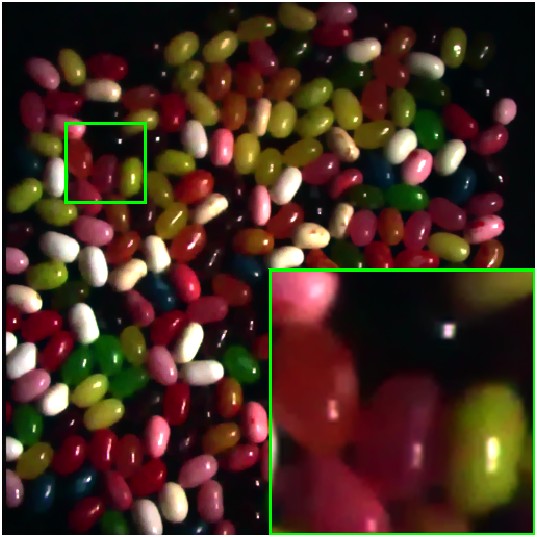} \\
			(31.39, 0.9244)
		\end{minipage}\hspace{\minivs}
		\begin{minipage}[b]{\mysize}
			\centering
			N+T 
			\includegraphics[width=\mysize]{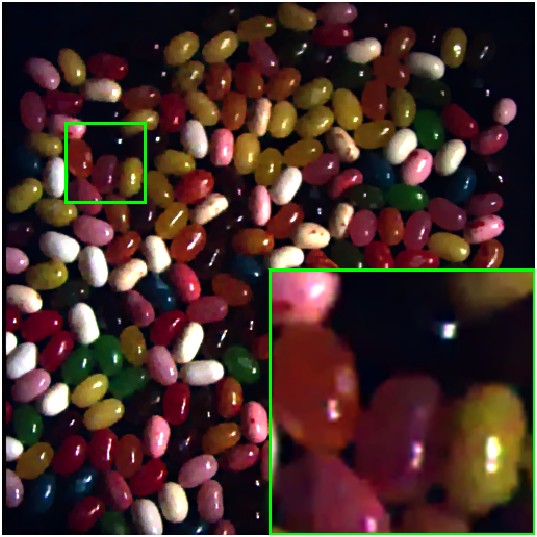} \\
			(32.32, 0.9405)
		\end{minipage}\hspace{\minivs}
		\begin{minipage}[b]{\mysize}
			\centering
			N+TS 
			\includegraphics[width=\mysize]{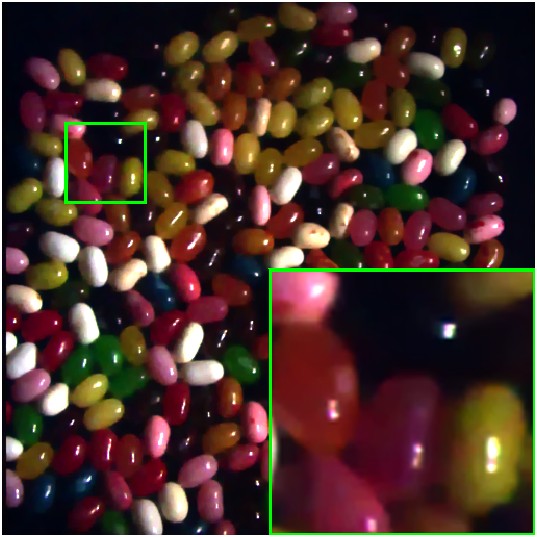} \\
			(32.76, 0.9415)
		\end{minipage}\hspace{\minivs}
		\begin{minipage}[b]{\mysize}
			\centering
			T+TS 
			\includegraphics[width=\mysize]{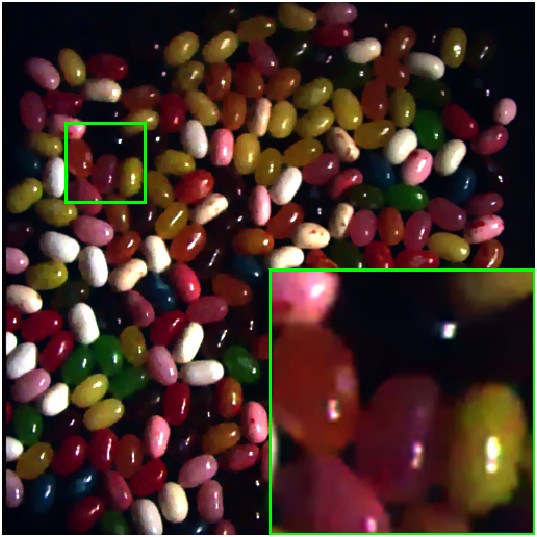} \\
			(\tb{32.87}, \tb{0.9464})
		\end{minipage}\hspace{\minivs}
		\begin{minipage}[b]{\mysize}
			\centering
			N+T+TS 
			\includegraphics[width=\mysize]{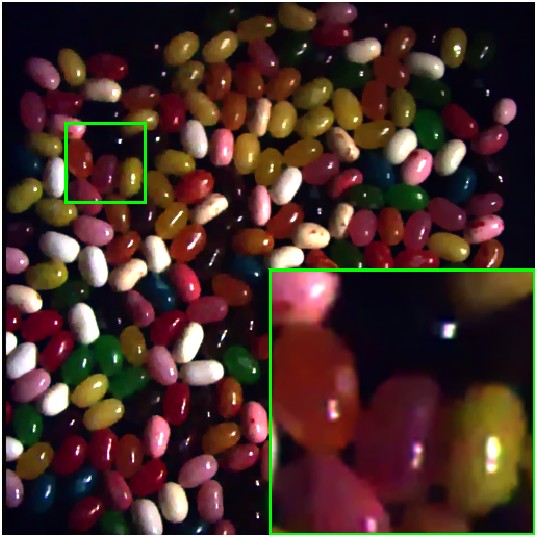} \\
			(\ul{32.79}, \ul{0.9436})
		\end{minipage}\hspace{\minivs}
	\end{minipage}
	\caption{Denoising results (pesudo-color image) of E3DTV and their corresponding ``DLW-" models on image ``jelly beans" of CAVE dataset. The noisy type is ``Spatial-Spectral Variant Gaussian" (Case 4).}
	\label{fig-append-add-e3dtv1}
\end{figure}

\begin{figure}[t]
	\renewcommand{\arraystretch}{1.15}
	\newcommand{\mysize}{2.6cm}
	\fontsize{8.5}{9.5}\selectfont
	\newcommand{\minivs}{5pt}
	\newcommand{\vs}{2pt}
	\centering
	\begin{minipage}[t]{\mysize*5+2cm}
		\centering
		\begin{minipage}[b]{\mysize}
			\centering
			clean 
			\includegraphics[width=\mysize]{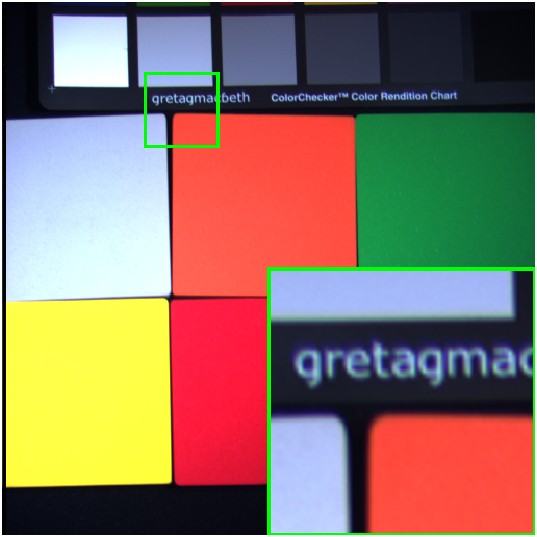} \\
			(PSNR, SSIM)
		\end{minipage}\hspace{\minivs}
		\begin{minipage}[b]{\mysize}
			\centering
			noisy 
			\includegraphics[width=\mysize]{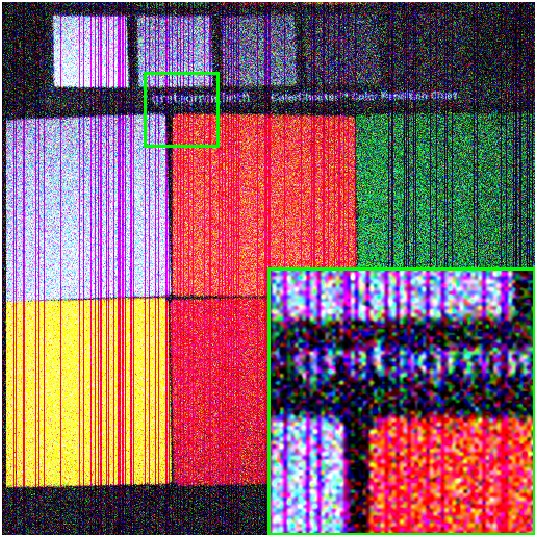} \\
			(13.48, 0.1731)
		\end{minipage}\hspace{\minivs}
		\begin{minipage}[b]{\mysize}
			\centering
			\textit{*Original}  
			\includegraphics[width=\mysize]{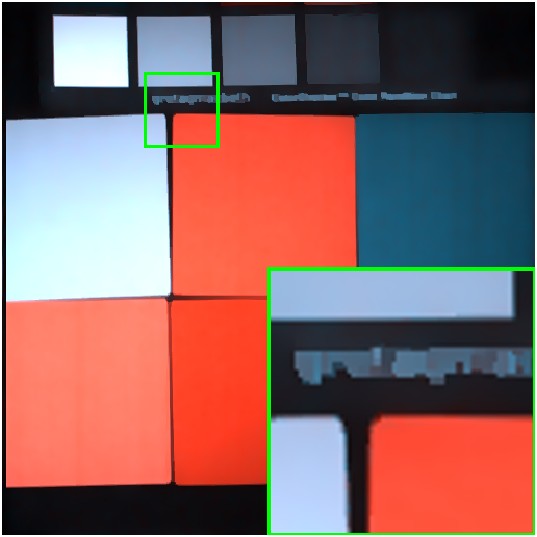} \\
			(25.01, 0.9121)
		\end{minipage}\hspace{\minivs}
		\begin{minipage}[b]{\mysize}
			\centering
			N 
			\includegraphics[width=\mysize]{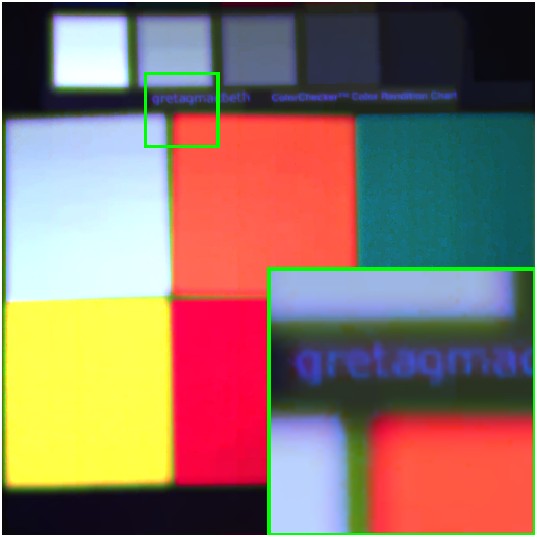} \\
			(26.07, 0.8612)
		\end{minipage}\hspace{\minivs}
		\begin{minipage}[b]{\mysize}
			\centering
			T 
			\includegraphics[width=\mysize]{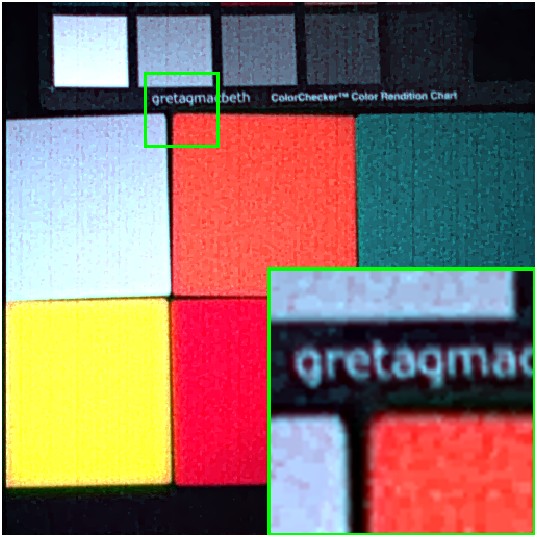} \\
			(27.93, 0.7923)
		\end{minipage} \vspace{6pt} \\
		\begin{minipage}[b]{\mysize}
			\centering
			TS 
			\includegraphics[width=\mysize]{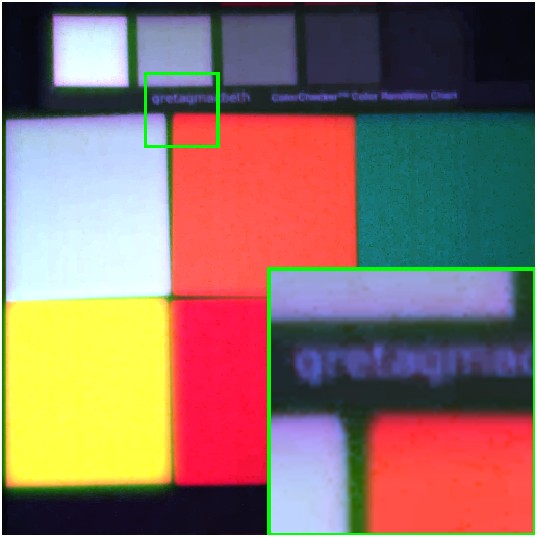} \\
			(26.42, 0.9019)
		\end{minipage}\hspace{\minivs}
		\begin{minipage}[b]{\mysize}
			\centering
			N+T 
			\includegraphics[width=\mysize]{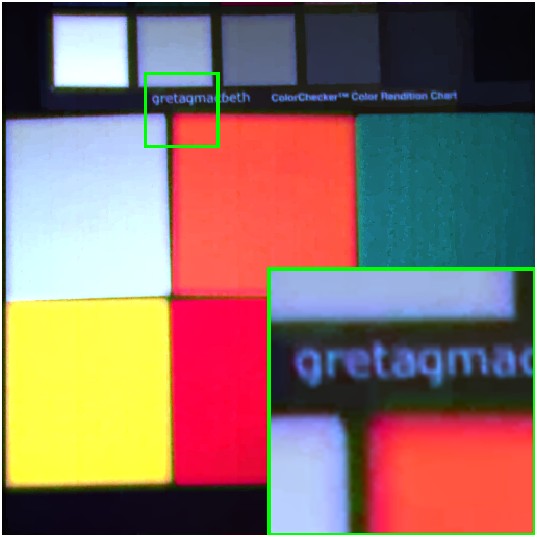} \\
			(27.08, 0.8928)
		\end{minipage}\hspace{\minivs}
		\begin{minipage}[b]{\mysize}
			\centering
			N+TS 
			\includegraphics[width=\mysize]{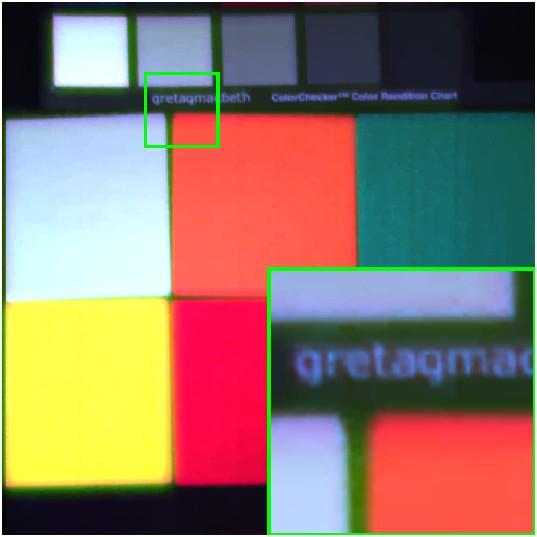} \\
			(27.01, 0.9079)
		\end{minipage}\hspace{\minivs}
		\begin{minipage}[b]{\mysize}
			\centering
			T+TS 
			\includegraphics[width=\mysize]{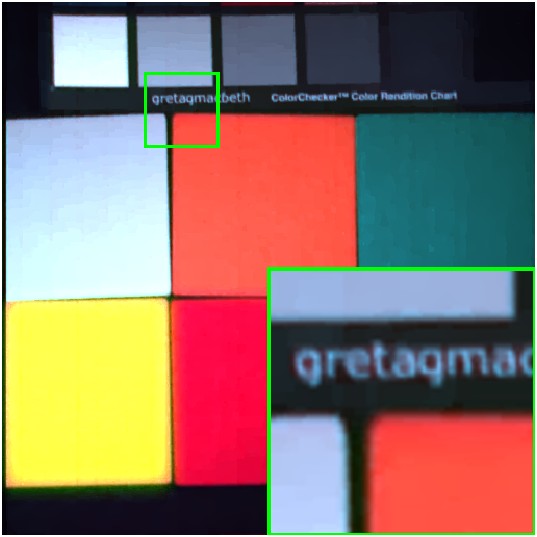} \\
			(\tb{28.36}, \tb{0.9223})
		\end{minipage}\hspace{\minivs}
		\begin{minipage}[b]{\mysize}
			\centering
			N+T+TS 
			\includegraphics[width=\mysize]{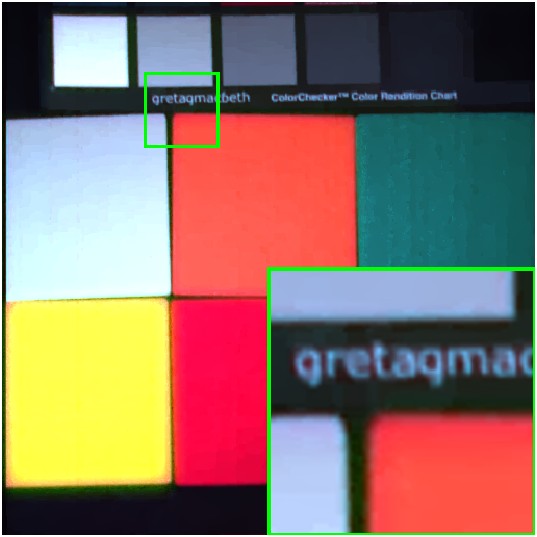} \\
			(\ul{28.17}, \ul{0.9181})
		\end{minipage}\hspace{\minivs}
	\end{minipage}
	\caption{Denoising results (pesudo-color image) of LRTFDFR and their corresponding ``DLW-" models on image ``sponges" of CAVE dataset. The noisy type is ``Gaussian+deadline" (Case 3).}
	\label{fig-append-add-lrtfdfr}
\end{figure}

\begin{figure*}[ht]
	\fontsize{8.5}{9.5}\selectfont
	\renewcommand{\arraystretch}{1.25}
	\newcommand{\mysizeone}{3cm}
	\newcommand{\mysizetwo}{2.05cm}
	\newcommand{\minivs}{-2pt}
	\newcommand{\ttt}{1.7cm}
	\centering
	\begin{tabular}{M{\ttt} M{\ttt} M{\ttt} M{\ttt} M{\ttt} M{\ttt} M{\ttt}}
		\multicolumn{7}{c}{
			\begin{minipage}[t]{\mysizeone}
				\centering
				\includegraphics[width=\mysizeone]{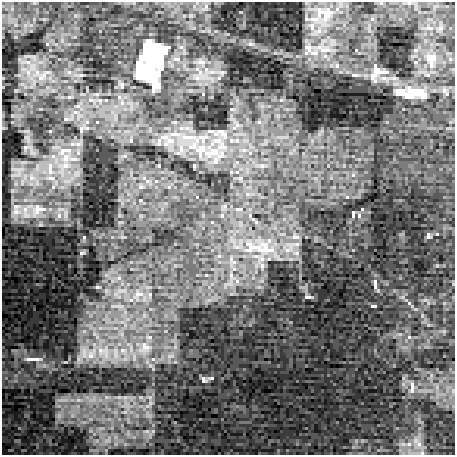}\\
				Band 2
			\end{minipage}
			\begin{minipage}[t]{\mysizeone}
				\centering
				\includegraphics[width=\mysizeone]{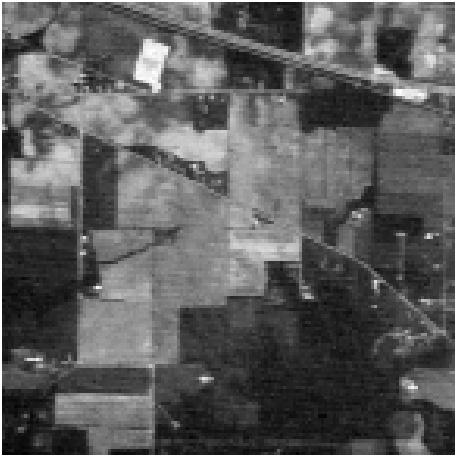}\\
				LRTV
			\end{minipage}
			\begin{minipage}[t]{\mysizeone}
				\centering
				\includegraphics[width=\mysizeone]{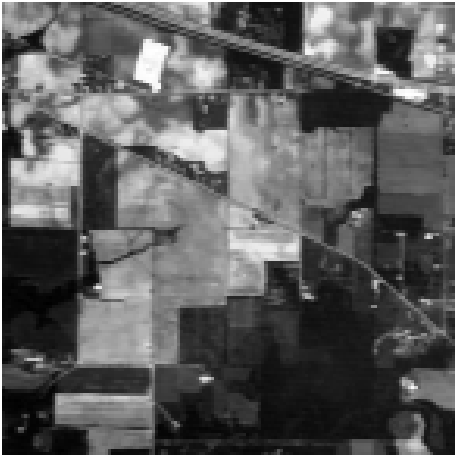}\\
				E3DTV
			\end{minipage}
			\begin{minipage}[t]{\mysizeone}
				\centering
				\includegraphics[width=\mysizeone]{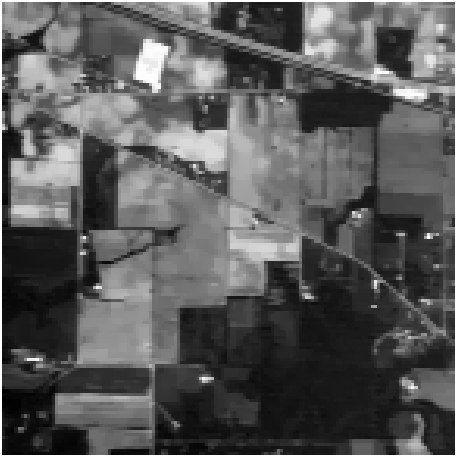}\\
				LRTFDFR
			\end{minipage}
		}\vspace{2pt}\\
		
		\Xhline{0.8pt}
		N & T & TS & N+T & N+TS & T+TS & N+T+TS \\
		
		\hline
		\multicolumn{7}{c}{target model: DLW-LRTV} \\
		\multicolumn{7}{c}{
			\begin{minipage}[t]{\mysizetwo}
				\centering
				\includegraphics[width=\mysizetwo]{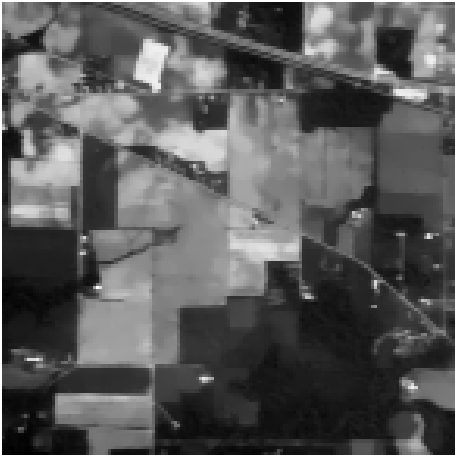} \\
			\end{minipage} 
			\begin{minipage}[t]{\mysizetwo}
				\centering
				\includegraphics[width=\mysizetwo]{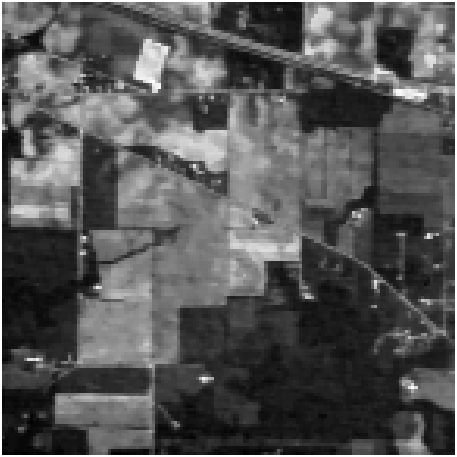} \\
			\end{minipage} 
			\begin{minipage}[t]{\mysizetwo}
				\centering
				\includegraphics[width=\mysizetwo]{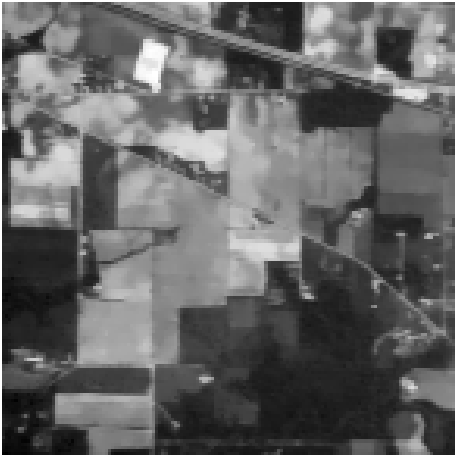} \\
			\end{minipage}  
			\begin{minipage}[t]{\mysizetwo}
				\centering
				\includegraphics[width=\mysizetwo]{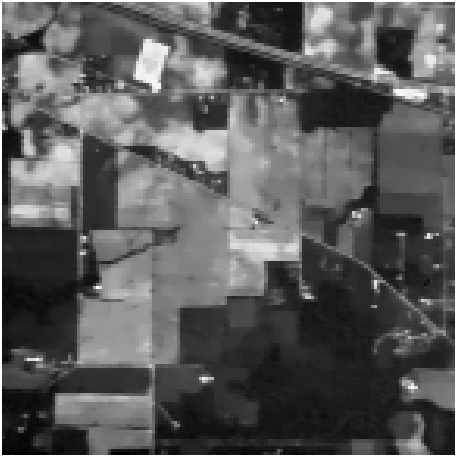} \\
			\end{minipage}  
			\begin{minipage}[t]{\mysizetwo}
				\centering
				\includegraphics[width=\mysizetwo]{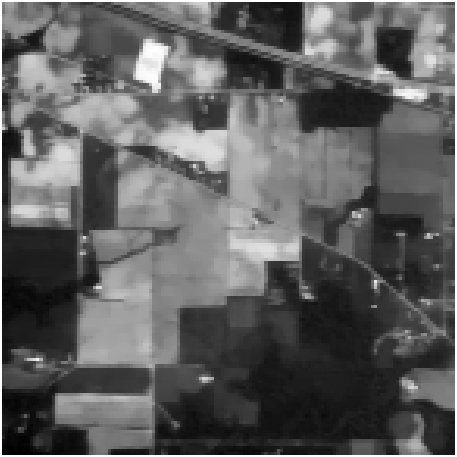} \\
			\end{minipage}  
			\begin{minipage}[t]{\mysizetwo}
				\centering
				\includegraphics[width=\mysizetwo]{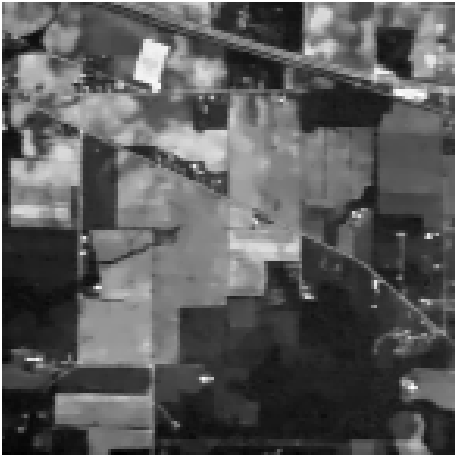} \\
			\end{minipage} 
			\begin{minipage}[t]{\mysizetwo}
				\centering
				\includegraphics[width=\mysizetwo]{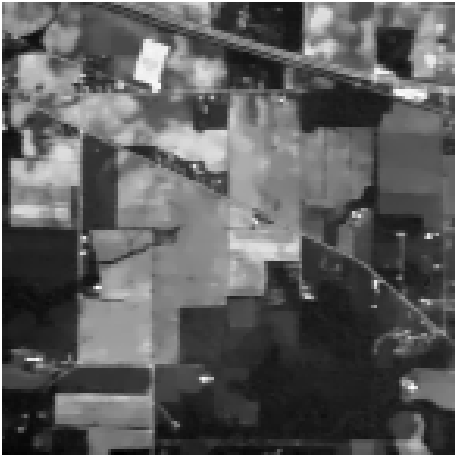} \\
		\end{minipage}} \\
		
		\hline
		\multicolumn{7}{c}{target model: DLW-E3DTV} \\
		\multicolumn{7}{c}{
			\begin{minipage}[t]{\mysizetwo}
				\centering
				\includegraphics[width=\mysizetwo]{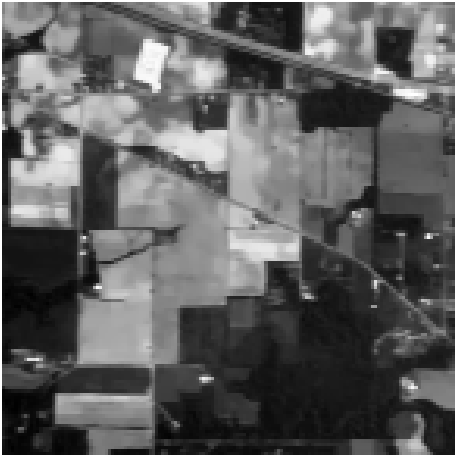} \\
			\end{minipage}
			\begin{minipage}[t]{\mysizetwo}
				\centering
				\includegraphics[width=\mysizetwo]{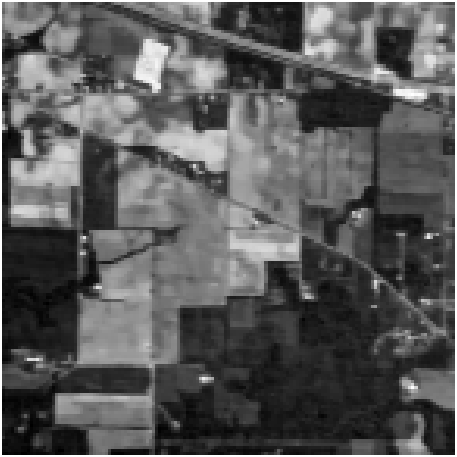} \\
			\end{minipage}
			\begin{minipage}[t]{\mysizetwo}
				\centering
				\includegraphics[width=\mysizetwo]{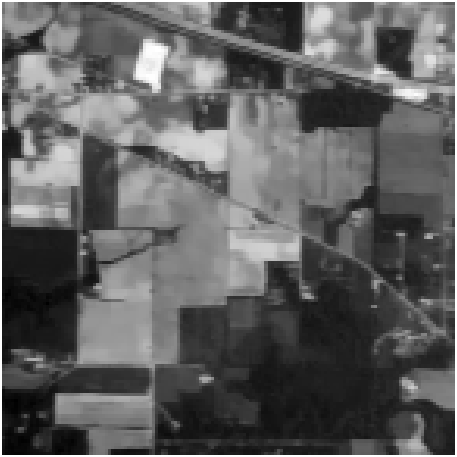} \\
			\end{minipage}
			\begin{minipage}[t]{\mysizetwo}
				\centering
				\includegraphics[width=\mysizetwo]{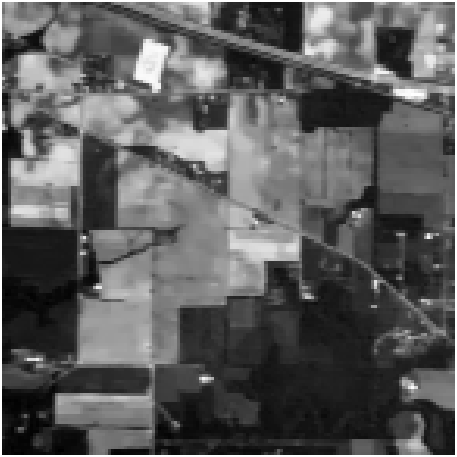} \\
			\end{minipage}
			\begin{minipage}[t]{\mysizetwo}
				\centering
				\includegraphics[width=\mysizetwo]{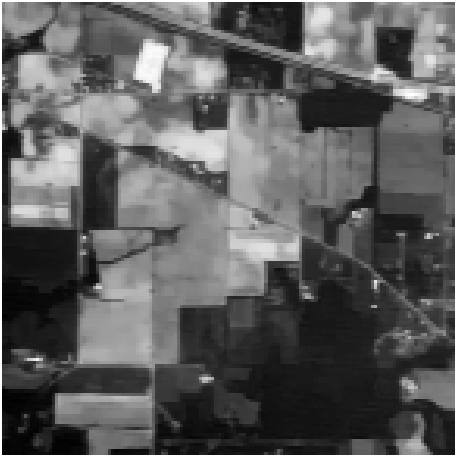} \\
			\end{minipage}
			\begin{minipage}[t]{\mysizetwo}
				\centering
				\includegraphics[width=\mysizetwo]{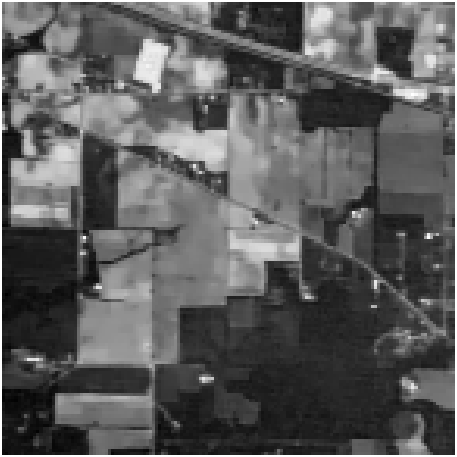} \\
			\end{minipage}
			\begin{minipage}[t]{\mysizetwo}
				\centering
				\includegraphics[width=\mysizetwo]{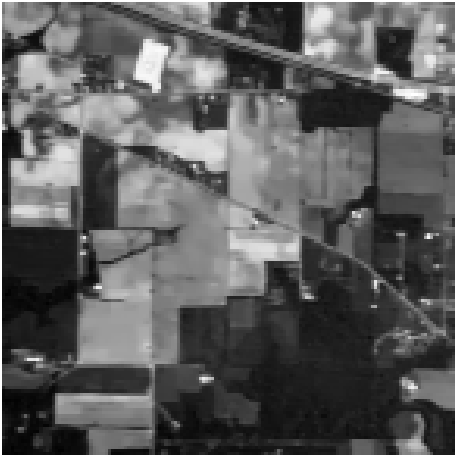} \\
		\end{minipage}} \\
		
		\hline
		\multicolumn{7}{c}{target model: DLW-LRTFDFR} \\
		\multicolumn{7}{c}{
			\begin{minipage}[t]{\mysizetwo}
				\centering
				\includegraphics[width=\mysizetwo]{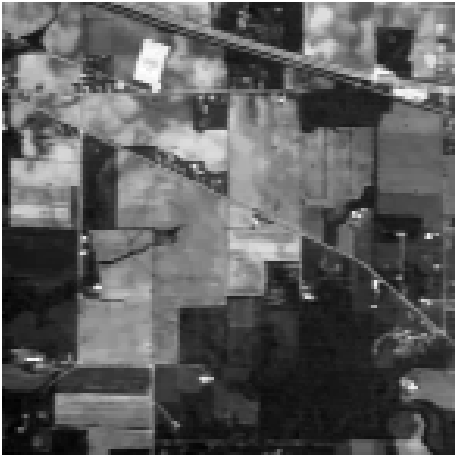} \\
			\end{minipage}
			\begin{minipage}[t]{\mysizetwo}
				\centering
				\includegraphics[width=\mysizetwo]{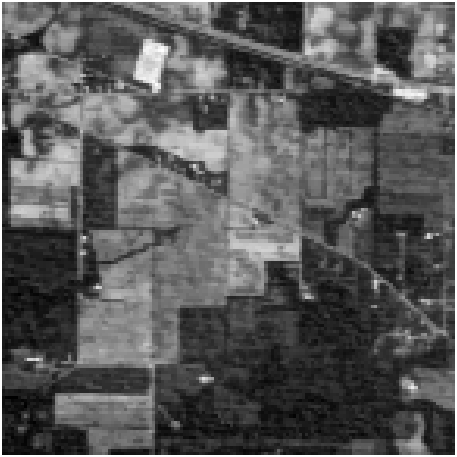} \\
			\end{minipage}
			\begin{minipage}[t]{\mysizetwo}
				\centering
				\includegraphics[width=\mysizetwo]{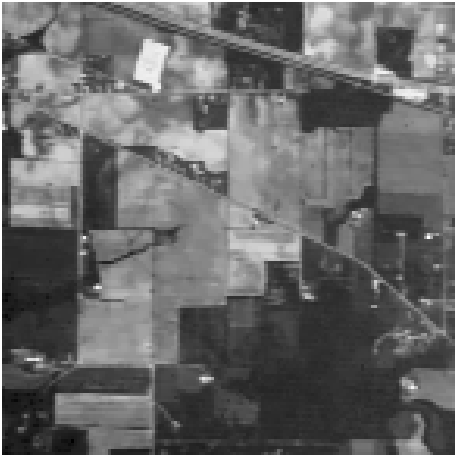} \\
			\end{minipage}
			\begin{minipage}[t]{\mysizetwo}
				\centering
				\includegraphics[width=\mysizetwo]{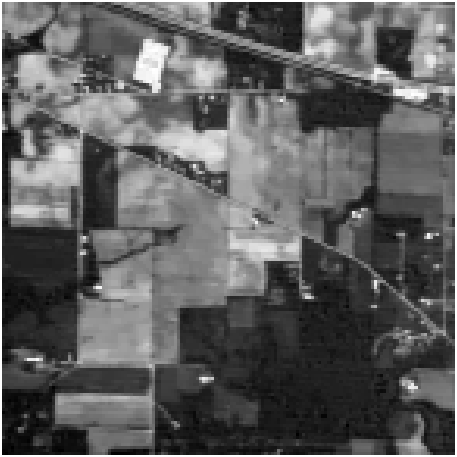} \\
			\end{minipage}
			\begin{minipage}[t]{\mysizetwo}
				\centering
				\includegraphics[width=\mysizetwo]{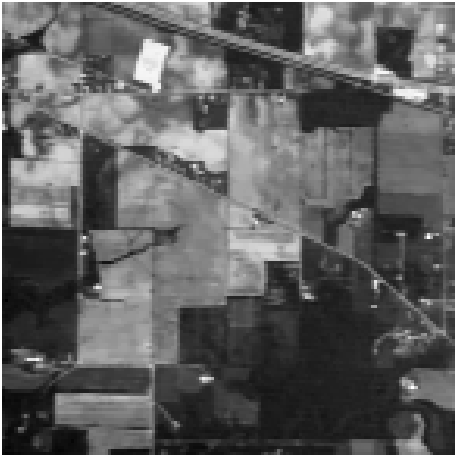} \\
			\end{minipage}
			\begin{minipage}[t]{\mysizetwo}
				\centering
				\includegraphics[width=\mysizetwo]{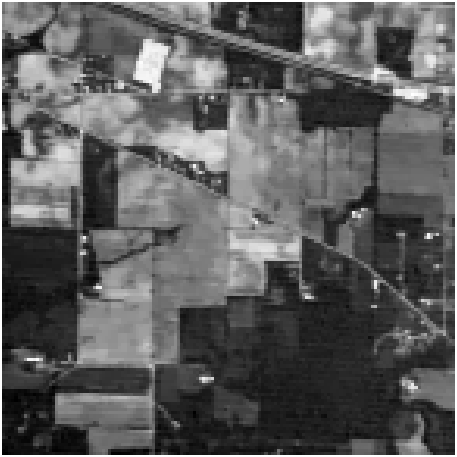} \\
			\end{minipage}
			\begin{minipage}[t]{\mysizetwo}
				\centering
				\includegraphics[width=\mysizetwo]{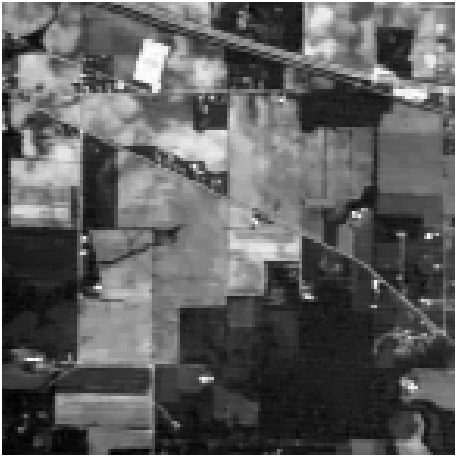} \\
		\end{minipage}} \\
		
		\Xhline{0.8pt}
		
	\end{tabular}
	\caption{Visual comparison of the denoising performance on Indian Pine dataset obtained by LRTV, E3DTV, LRTFDFR and their corresponding DLW-models.}
	\label{fig-tran-indian}
\end{figure*}

\begin{figure*}[ht]
	\fontsize{8.5}{9.5}\selectfont
	\renewcommand{\arraystretch}{1.25}
	\newcommand{\mysizeone}{3cm}
	\newcommand{\mysizetwo}{2.05cm}
	\newcommand{\minivs}{-2pt}
	\newcommand{\ttt}{1.7cm}
	\centering
	\begin{tabular}{M{\ttt} M{\ttt} M{\ttt} M{\ttt} M{\ttt} M{\ttt} M{\ttt}}
		\multicolumn{7}{c}{
			\begin{minipage}[t]{\mysizeone}
				\centering
				\includegraphics[width=\mysizeone]{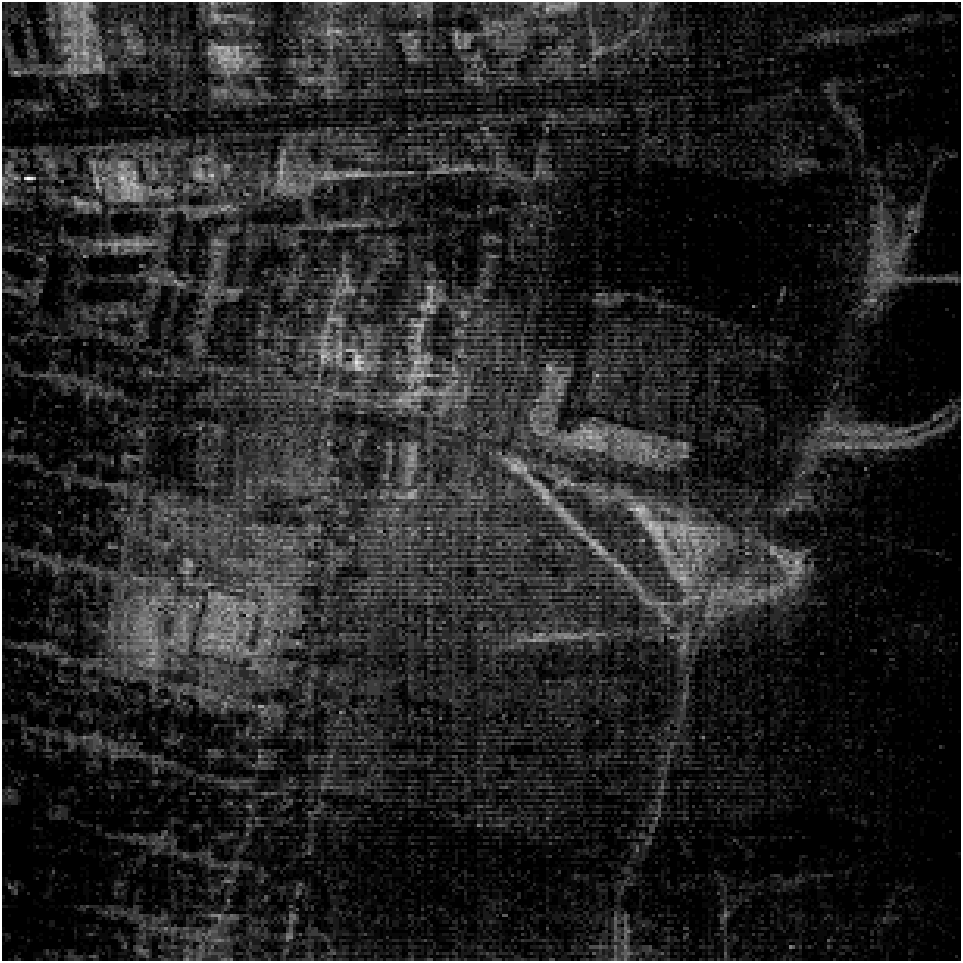}\\
				Band 2
			\end{minipage}
			\begin{minipage}[t]{\mysizeone}
				\centering
				\includegraphics[width=\mysizeone]{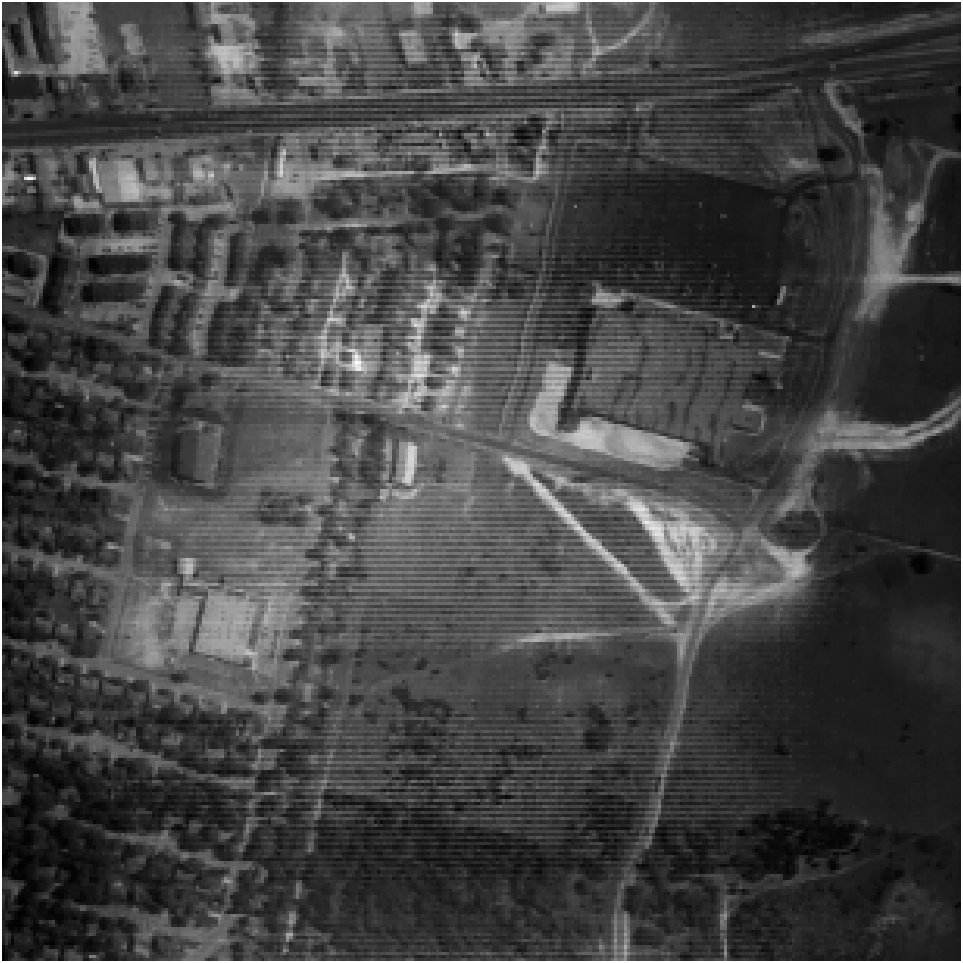}\\
				LRTV
			\end{minipage}
			\begin{minipage}[t]{\mysizeone}
				\centering
				\includegraphics[width=\mysizeone]{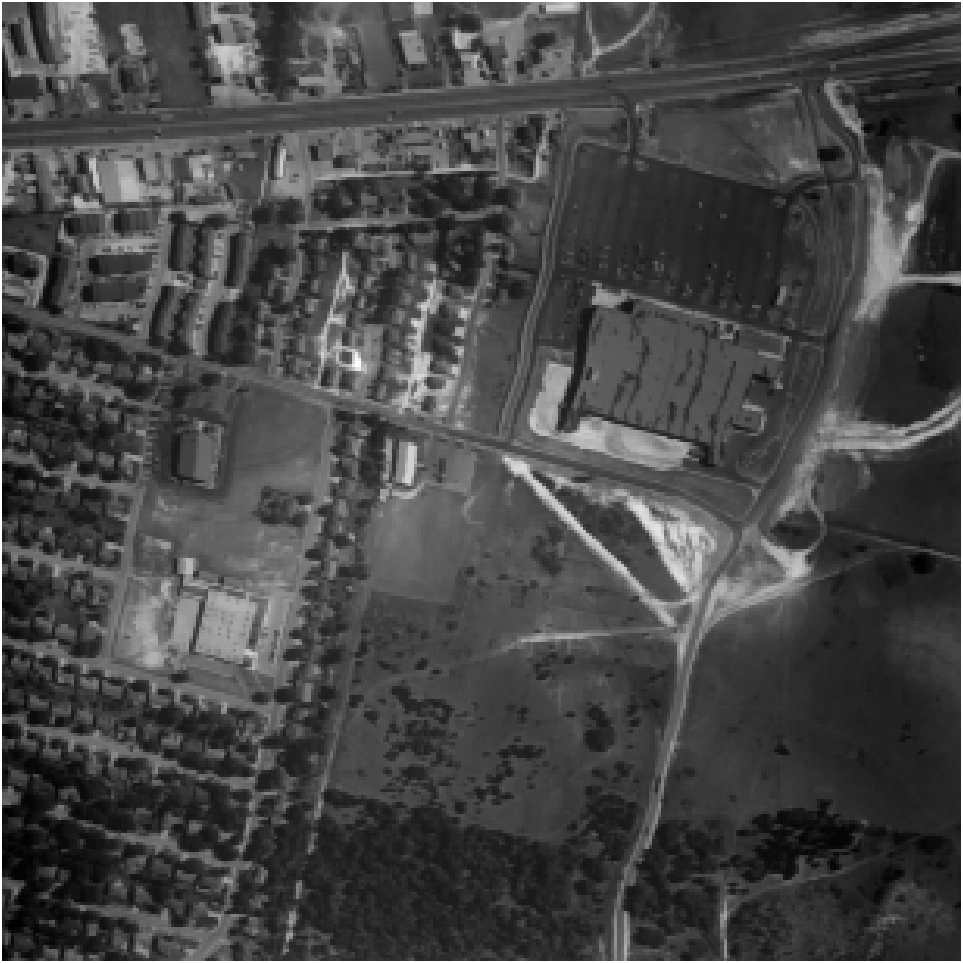}\\
				E3DTV
			\end{minipage}
			\begin{minipage}[t]{\mysizeone}
				\centering
				\includegraphics[width=\mysizeone]{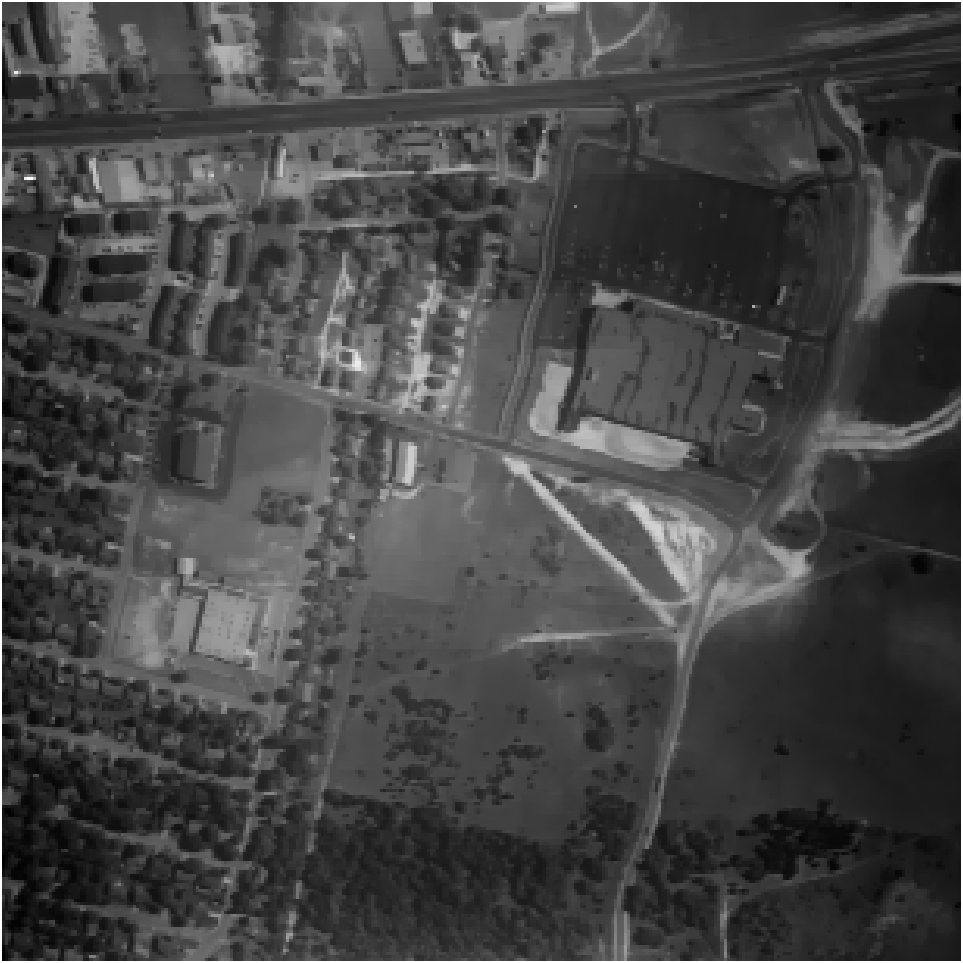}\\
				LRTFDFR
			\end{minipage}
		}\vspace{2pt}\\
		
		\Xhline{0.8pt}
		N & T & TS & N+T & N+TS & T+TS & N+T+TS \\
		
		\hline
		\multicolumn{7}{c}{target model: DLW-LRTV} \\
		\multicolumn{7}{c}{
			\begin{minipage}[t]{\mysizetwo}
				\centering
				\includegraphics[width=\mysizetwo]{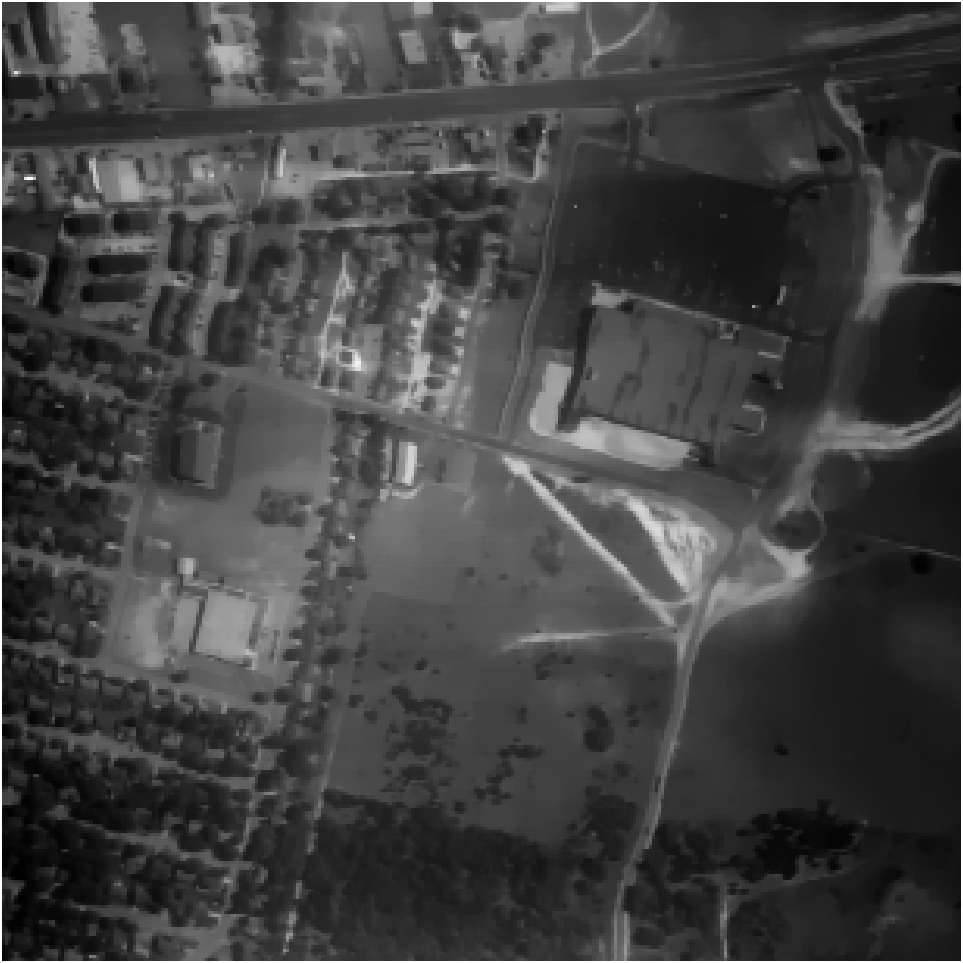} \\
			\end{minipage}
			\begin{minipage}[t]{\mysizetwo}
				\centering
				\includegraphics[width=\mysizetwo]{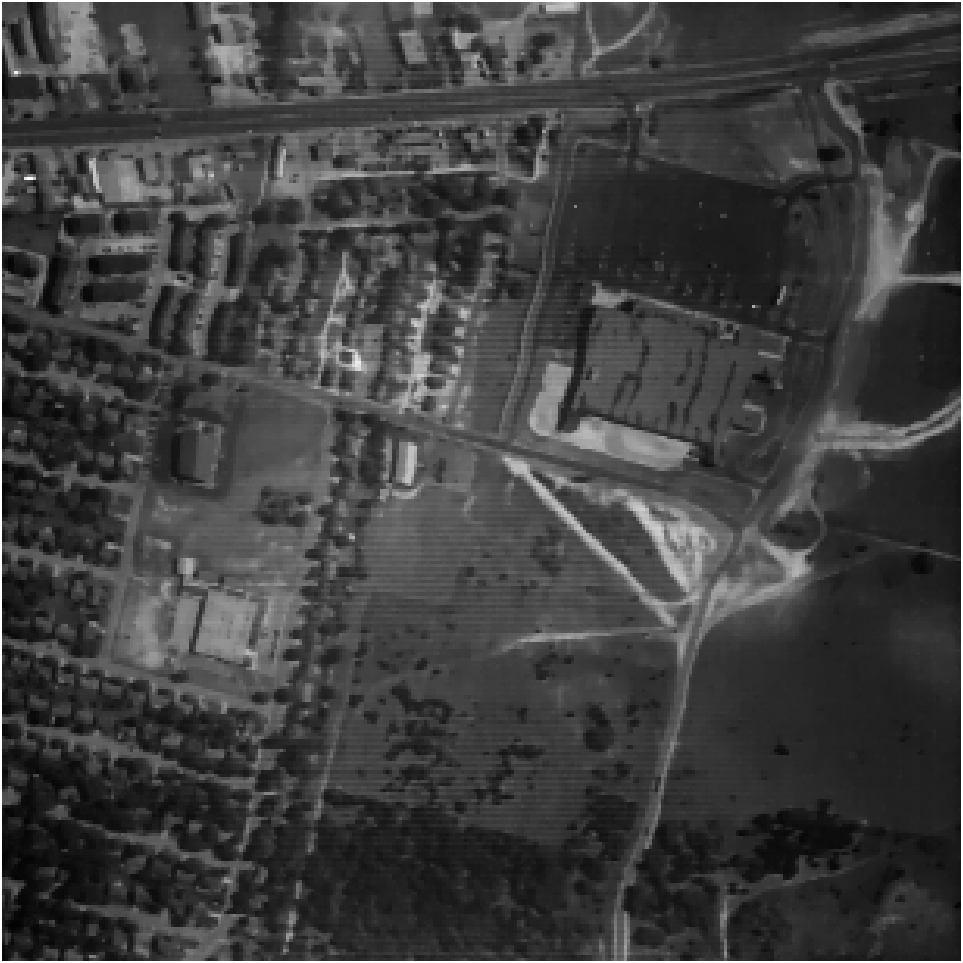} \\
			\end{minipage}
			\begin{minipage}[t]{\mysizetwo}
				\centering
				\includegraphics[width=\mysizetwo]{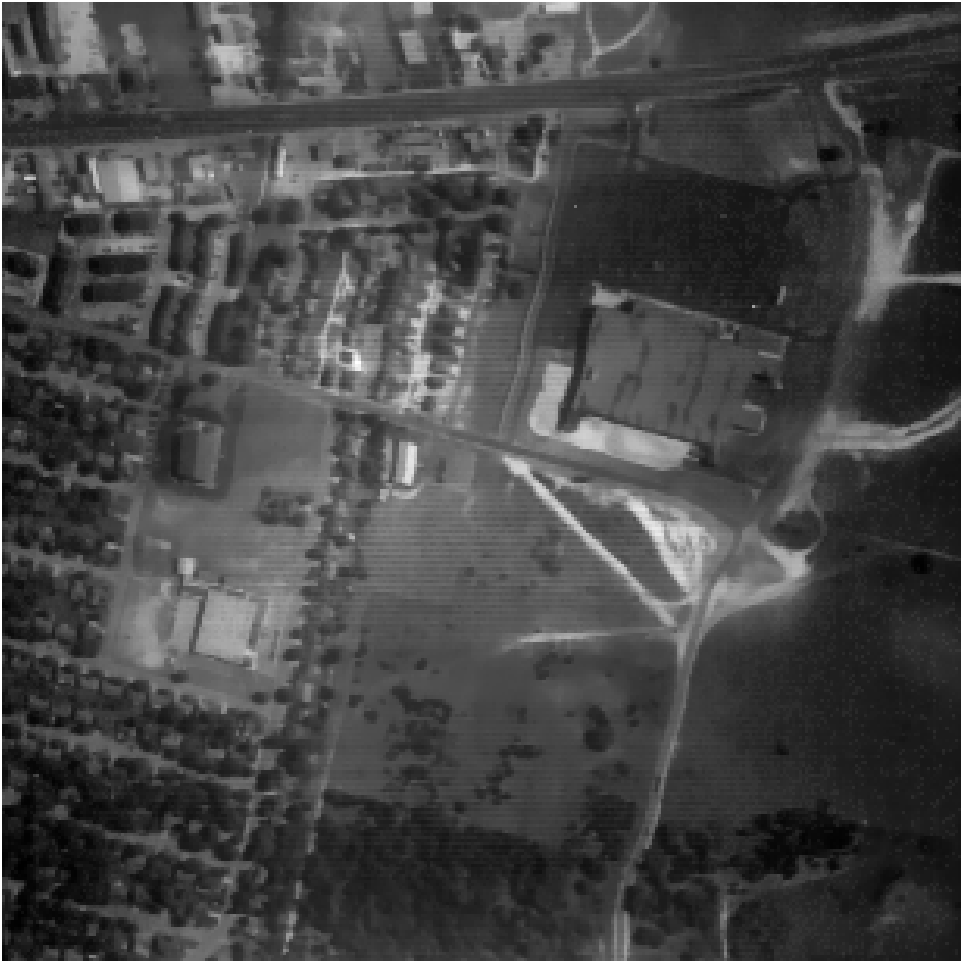} \\
			\end{minipage}
			\begin{minipage}[t]{\mysizetwo}
				\centering
				\includegraphics[width=\mysizetwo]{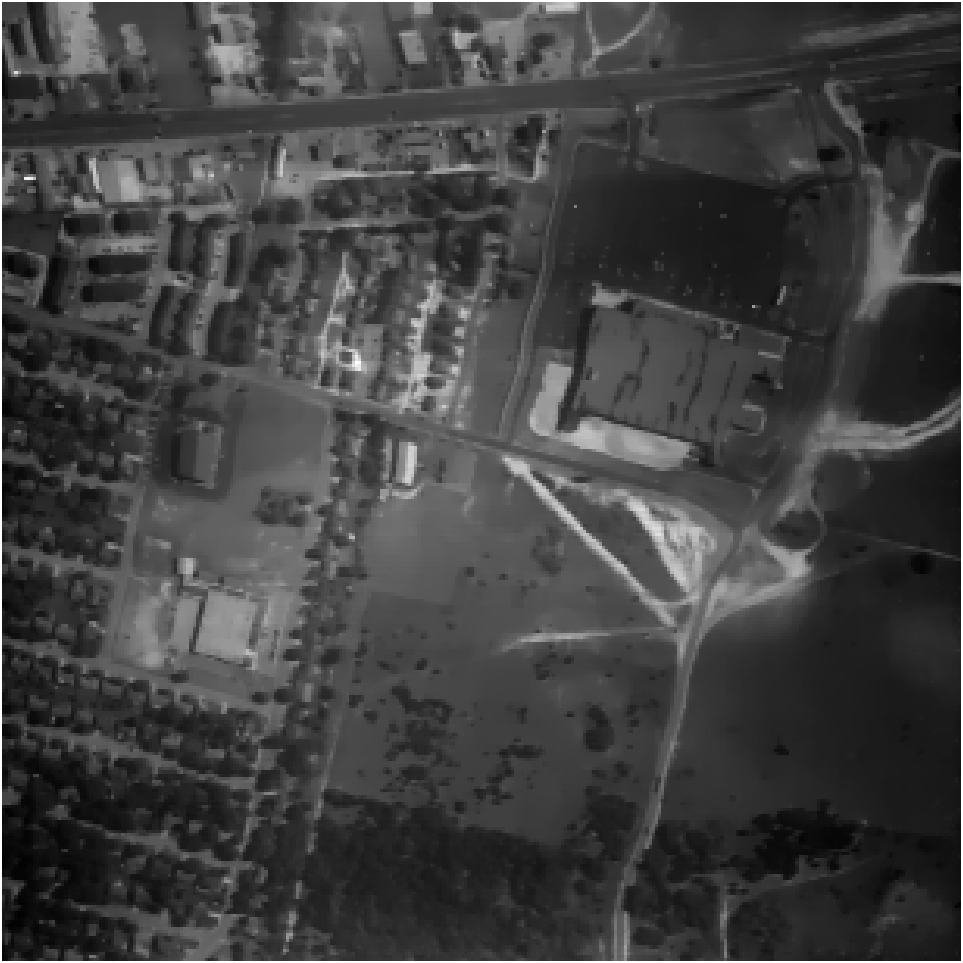} \\
			\end{minipage}
			\begin{minipage}[t]{\mysizetwo}
				\centering
				\includegraphics[width=\mysizetwo]{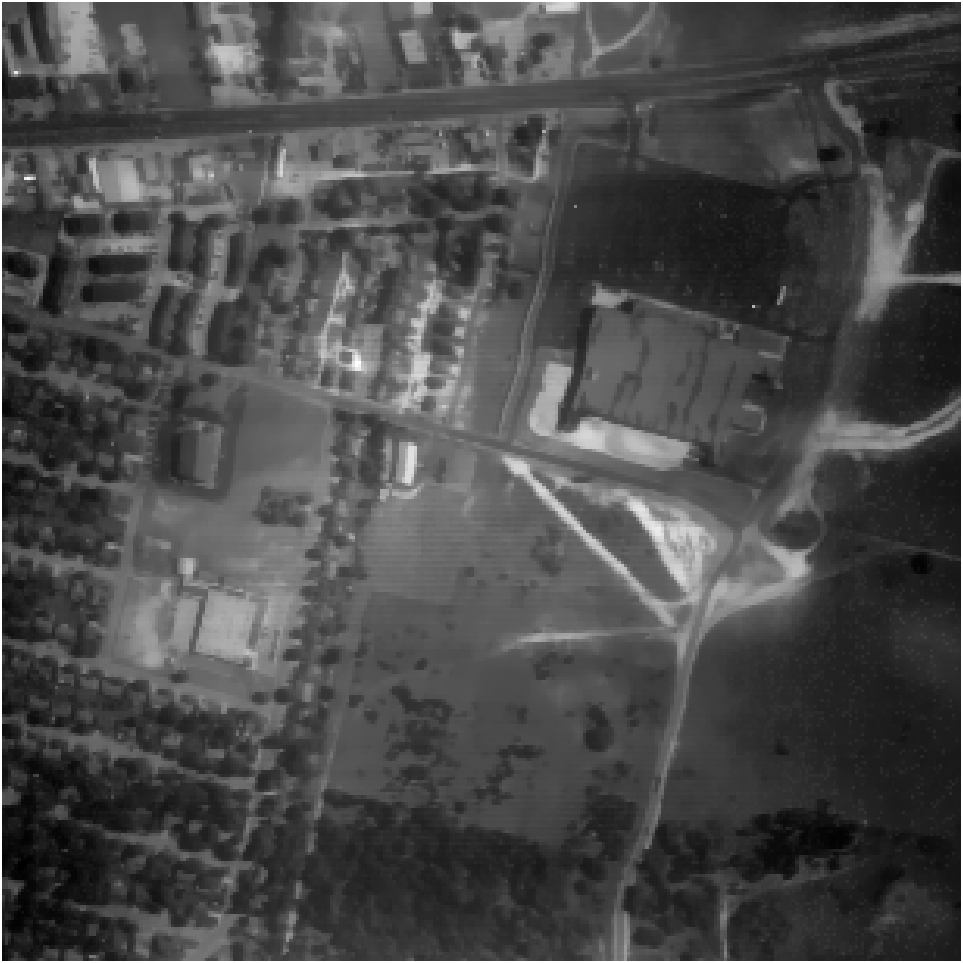} \\
			\end{minipage}
			\begin{minipage}[t]{\mysizetwo}
				\centering
				\includegraphics[width=\mysizetwo]{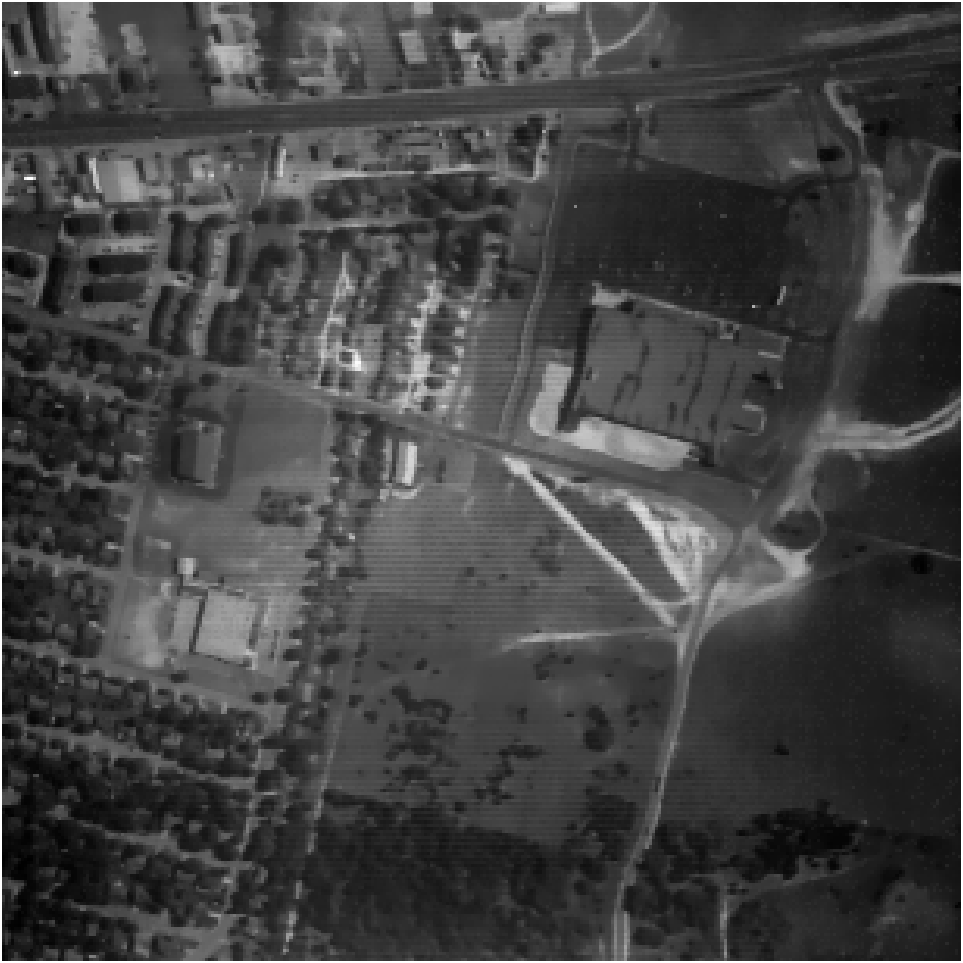} \\
			\end{minipage}
			\begin{minipage}[t]{\mysizetwo}
				\centering
				\includegraphics[width=\mysizetwo]{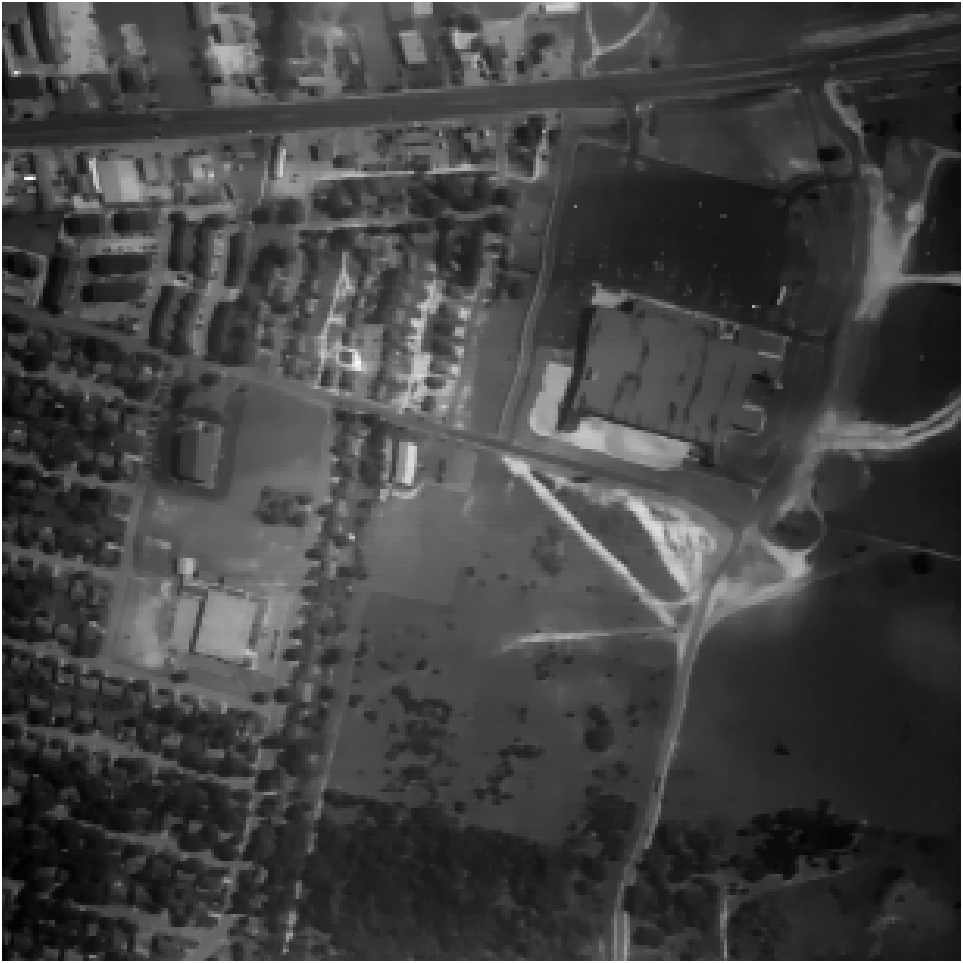} \\
		\end{minipage}} \\
		
		\hline
		\multicolumn{7}{c}{target model: DLW-E3DTV} \\
		\multicolumn{7}{c}{
			\begin{minipage}[t]{\mysizetwo}
				\centering
				\includegraphics[width=\mysizetwo]{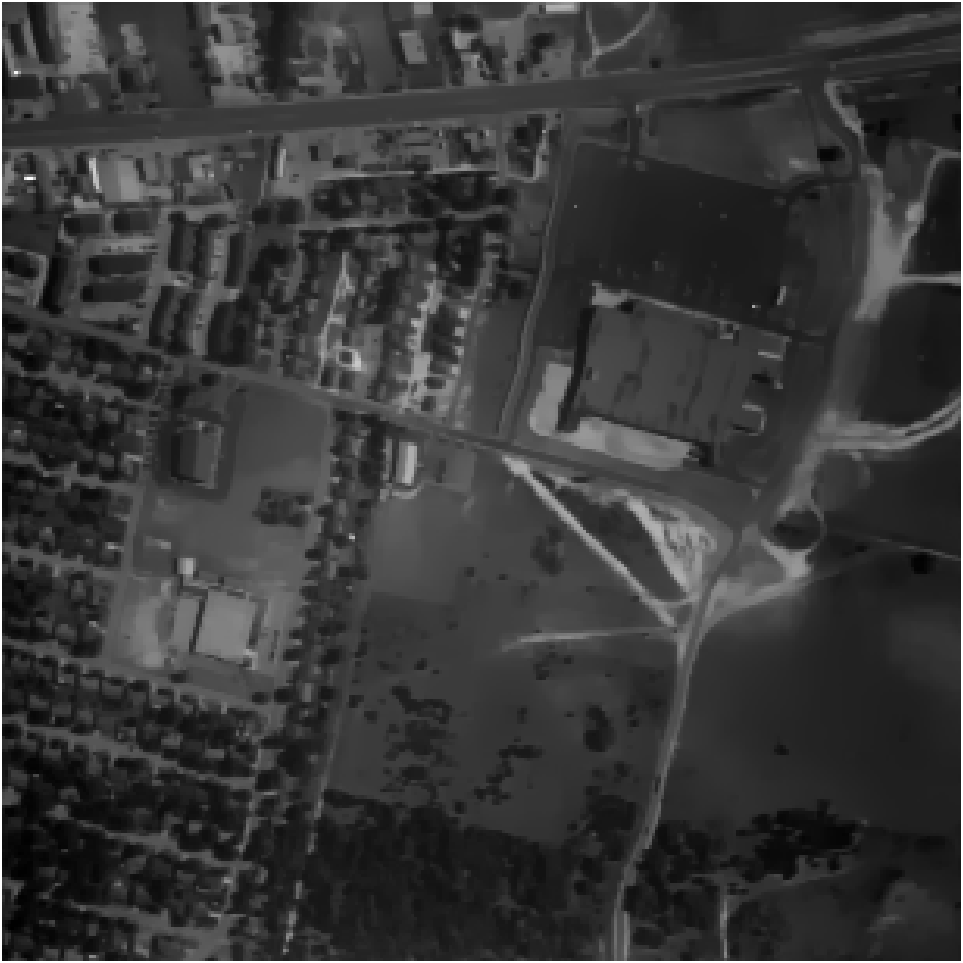} \\
			\end{minipage}
			\begin{minipage}[t]{\mysizetwo}
				\centering
				\includegraphics[width=\mysizetwo]{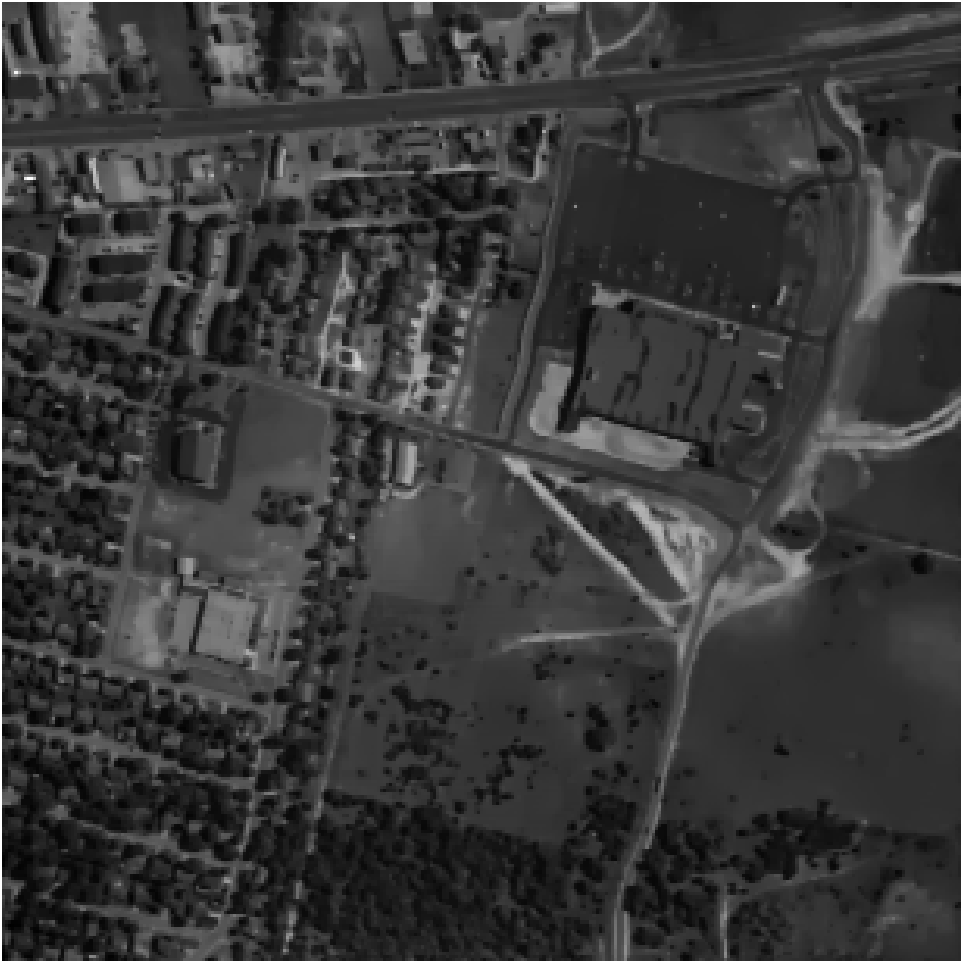} \\
			\end{minipage}
			\begin{minipage}[t]{\mysizetwo}
				\centering
				\includegraphics[width=\mysizetwo]{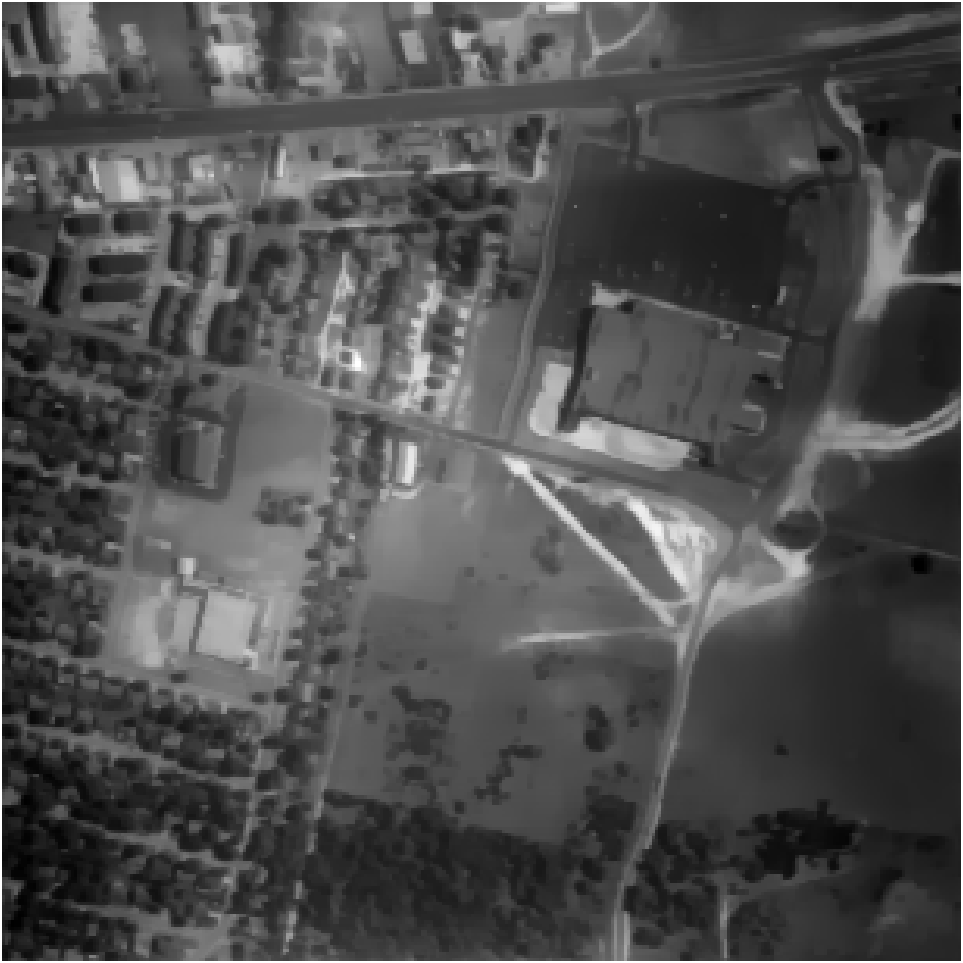} \\
			\end{minipage}
			\begin{minipage}[t]{\mysizetwo}
				\centering
				\includegraphics[width=\mysizetwo]{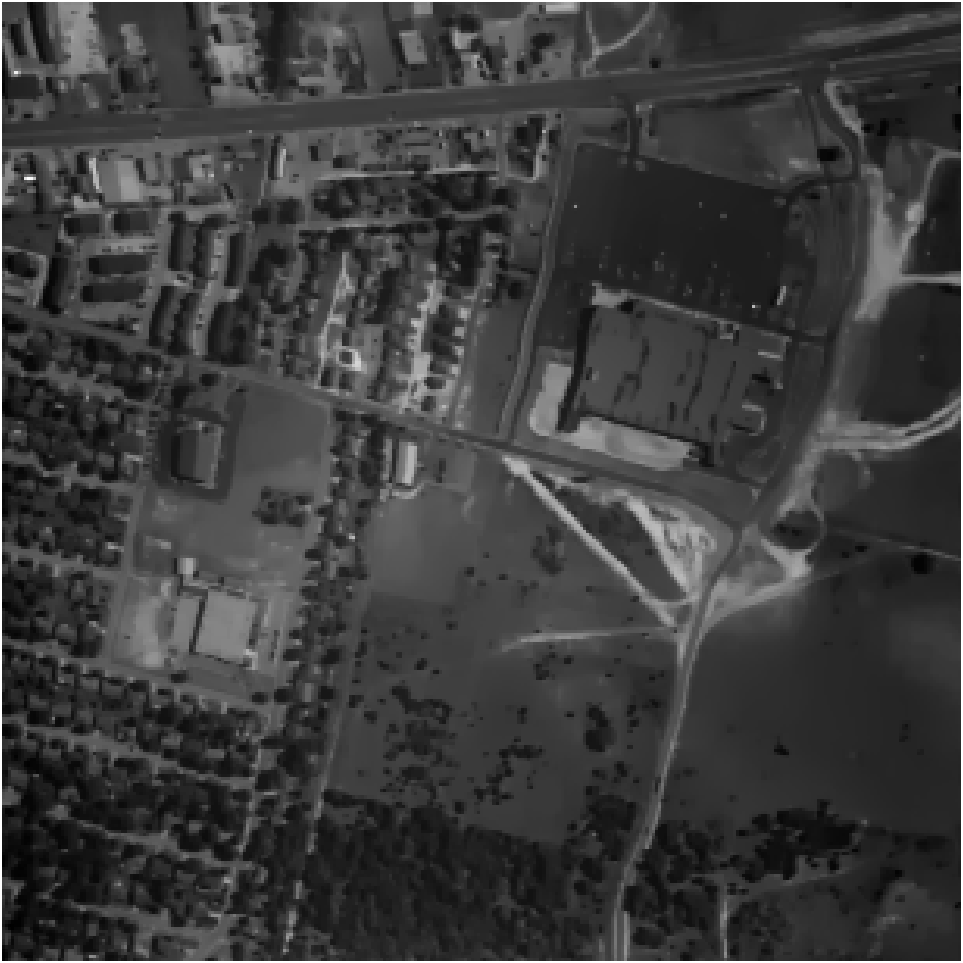} \\
			\end{minipage}
			\begin{minipage}[t]{\mysizetwo}
				\centering
				\includegraphics[width=\mysizetwo]{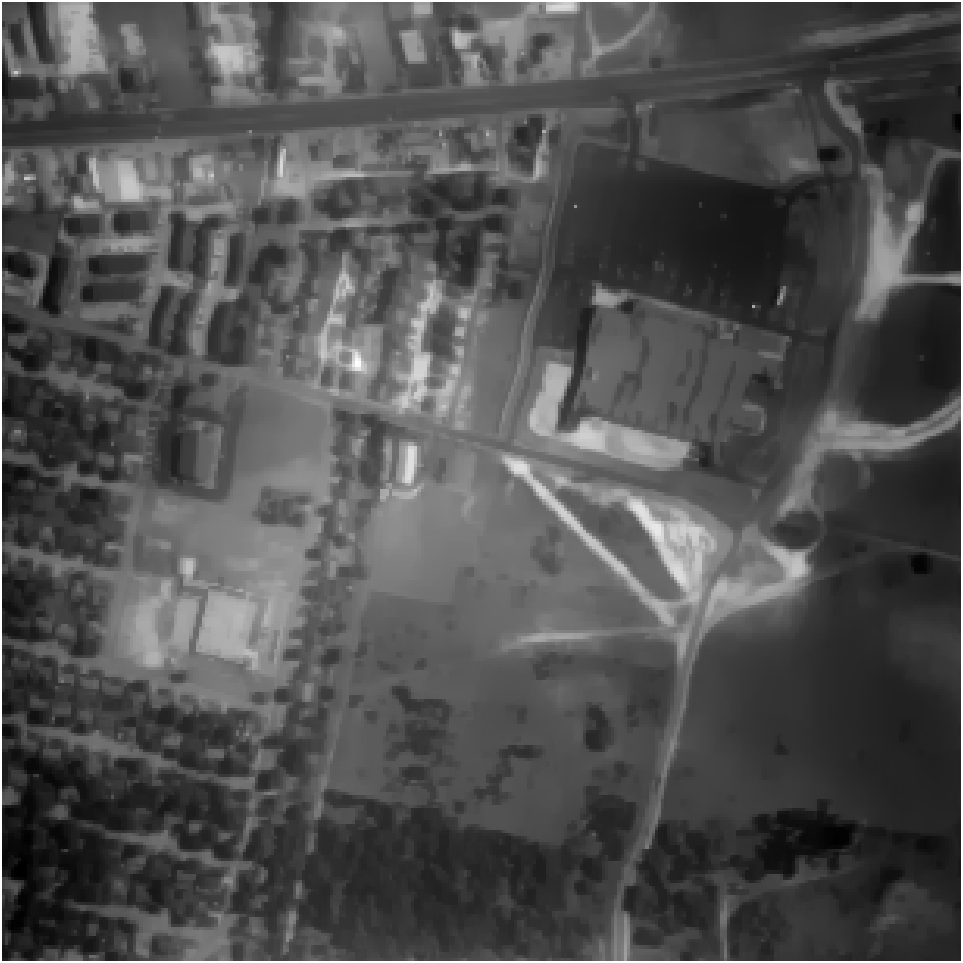} \\
			\end{minipage}
			\begin{minipage}[t]{\mysizetwo}
				\centering
				\includegraphics[width=\mysizetwo]{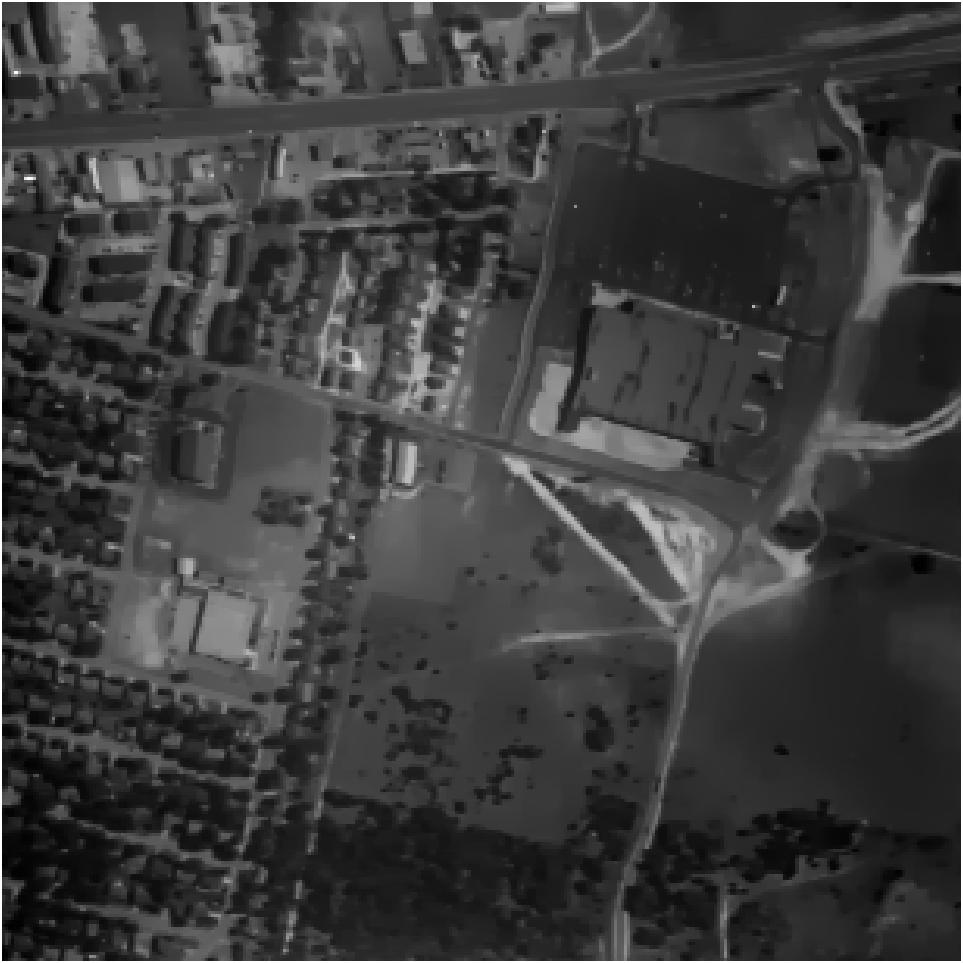} \\
			\end{minipage}
			\begin{minipage}[t]{\mysizetwo}
				\centering
				\includegraphics[width=\mysizetwo]{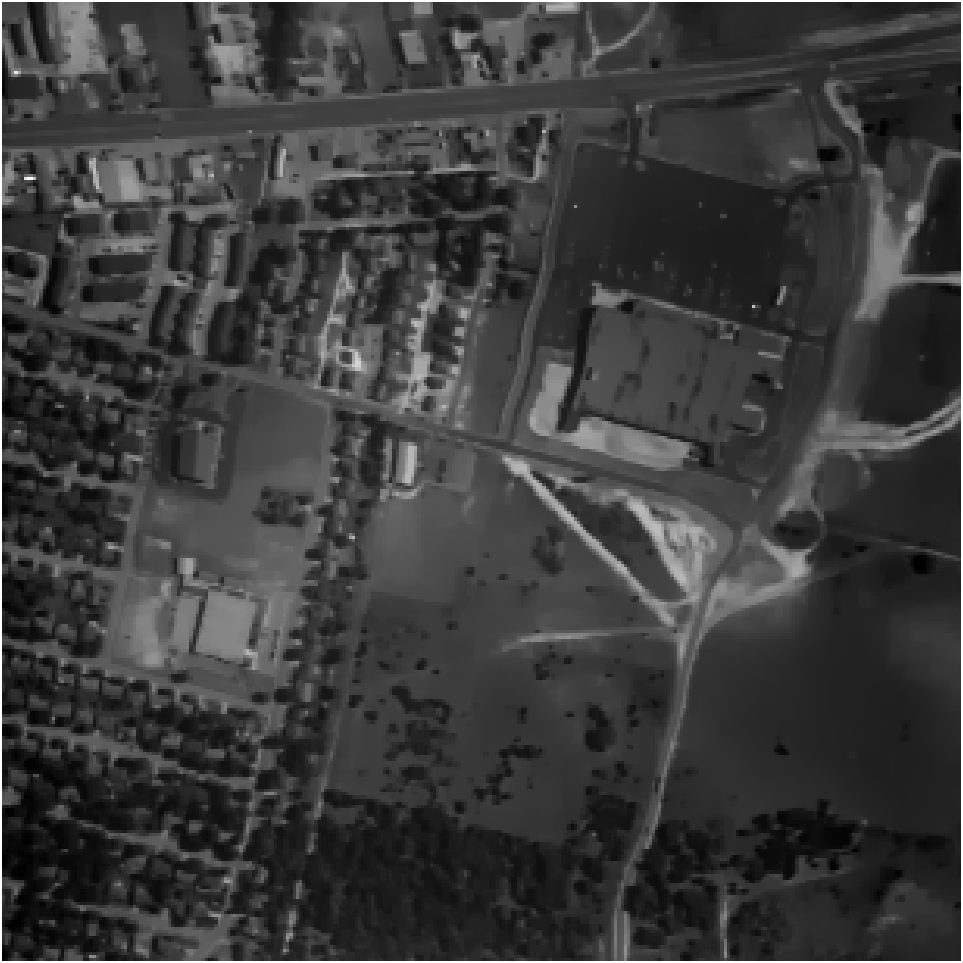} \\
		\end{minipage}} \\
		
		\hline
		\multicolumn{7}{c}{target model: DLW-LRTFDFR} \\
		\multicolumn{7}{c}{
			\begin{minipage}[t]{\mysizetwo}
				\centering
				\includegraphics[width=\mysizetwo]{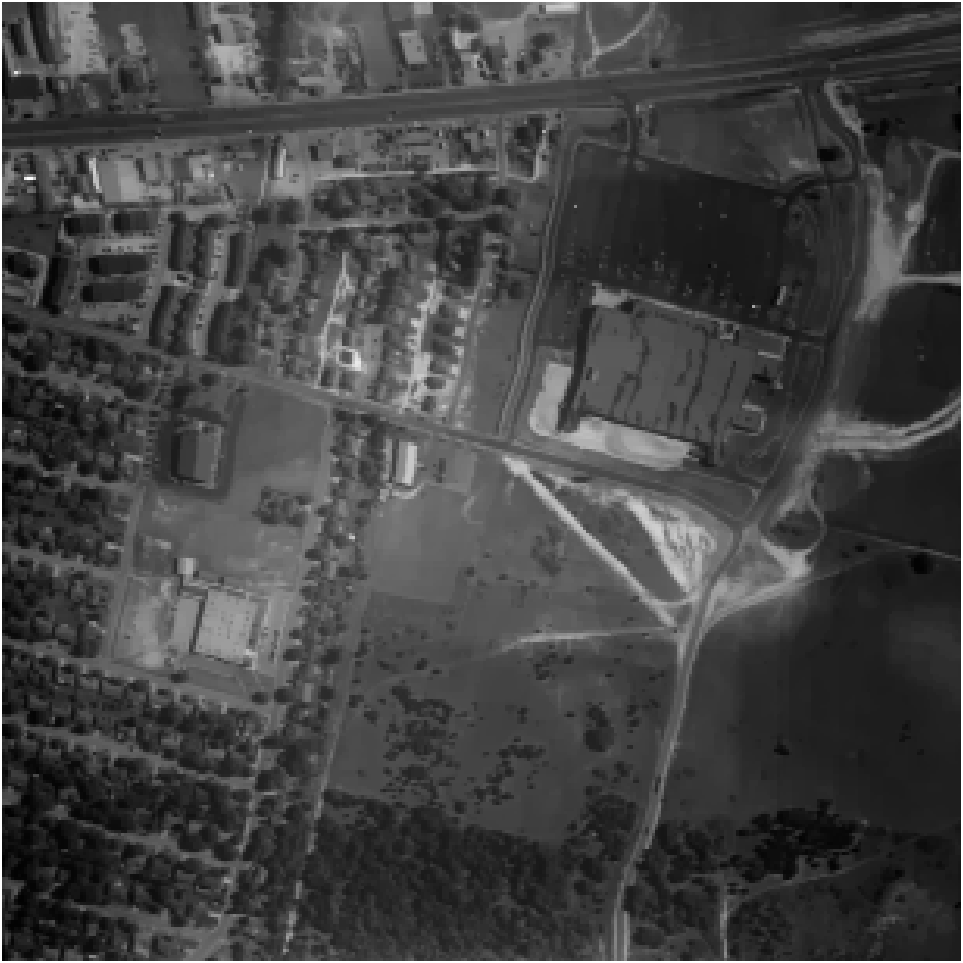} \\
			\end{minipage}
			\begin{minipage}[t]{\mysizetwo}
				\centering
				\includegraphics[width=\mysizetwo]{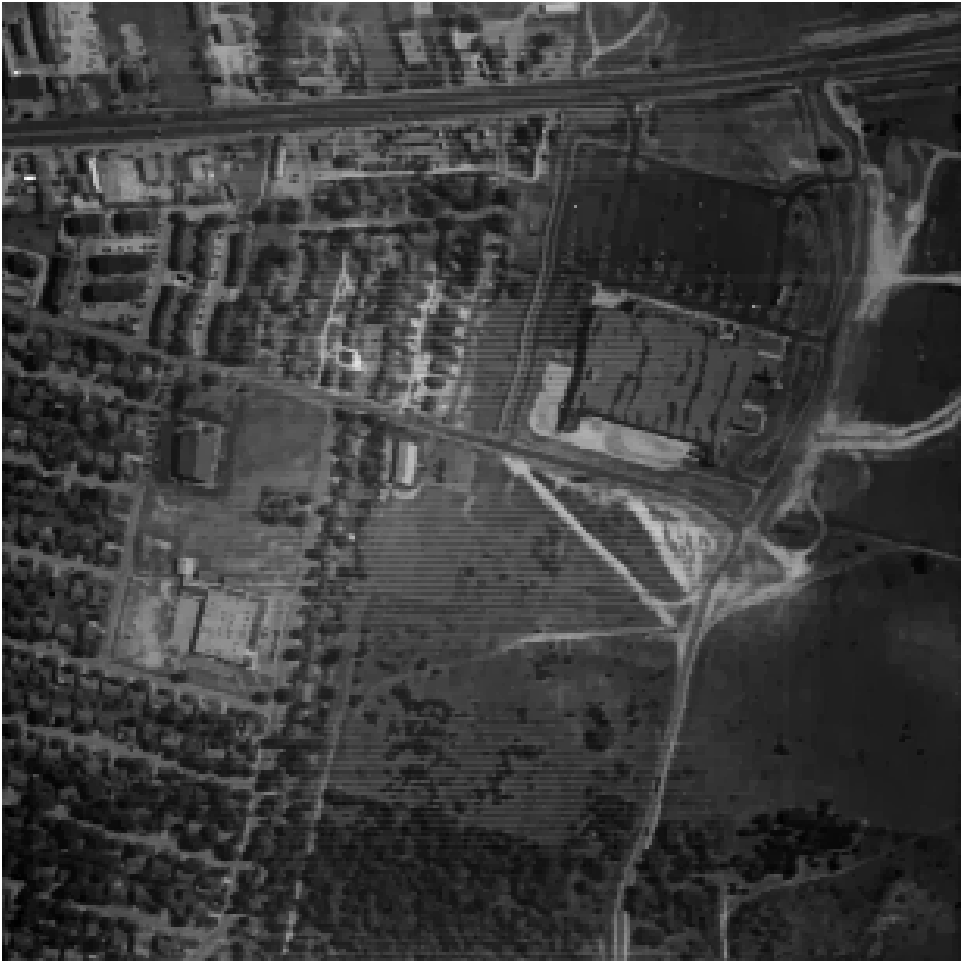} \\
			\end{minipage}
			\begin{minipage}[t]{\mysizetwo}
				\centering
				\includegraphics[width=\mysizetwo]{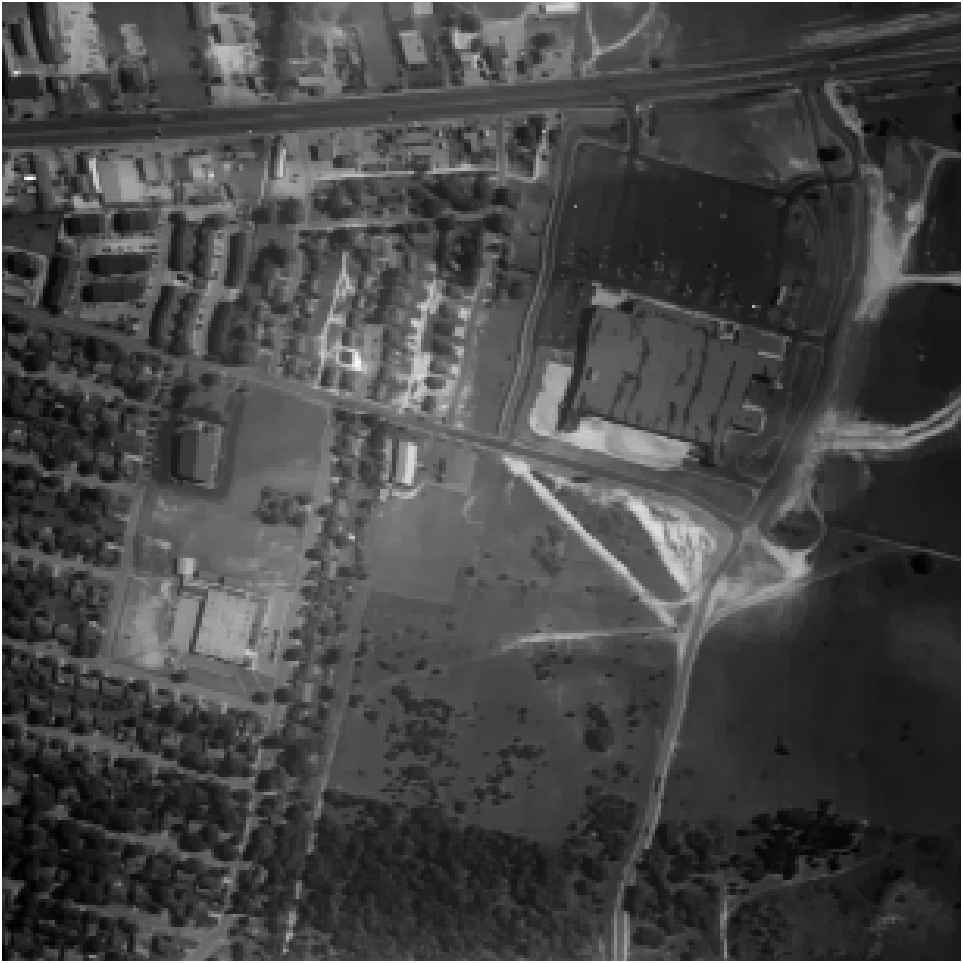} \\
			\end{minipage}
			\begin{minipage}[t]{\mysizetwo}
				\centering
				\includegraphics[width=\mysizetwo]{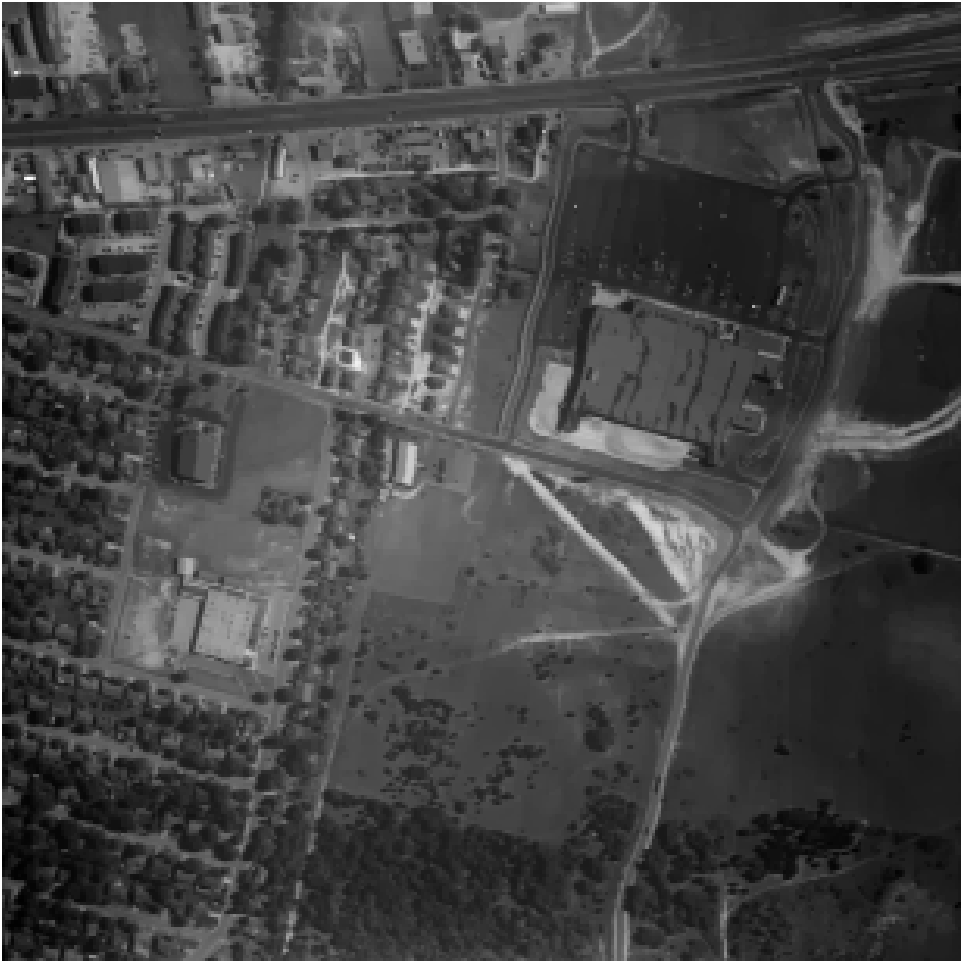} \\
			\end{minipage}
			\begin{minipage}[t]{\mysizetwo}
				\centering
				\includegraphics[width=\mysizetwo]{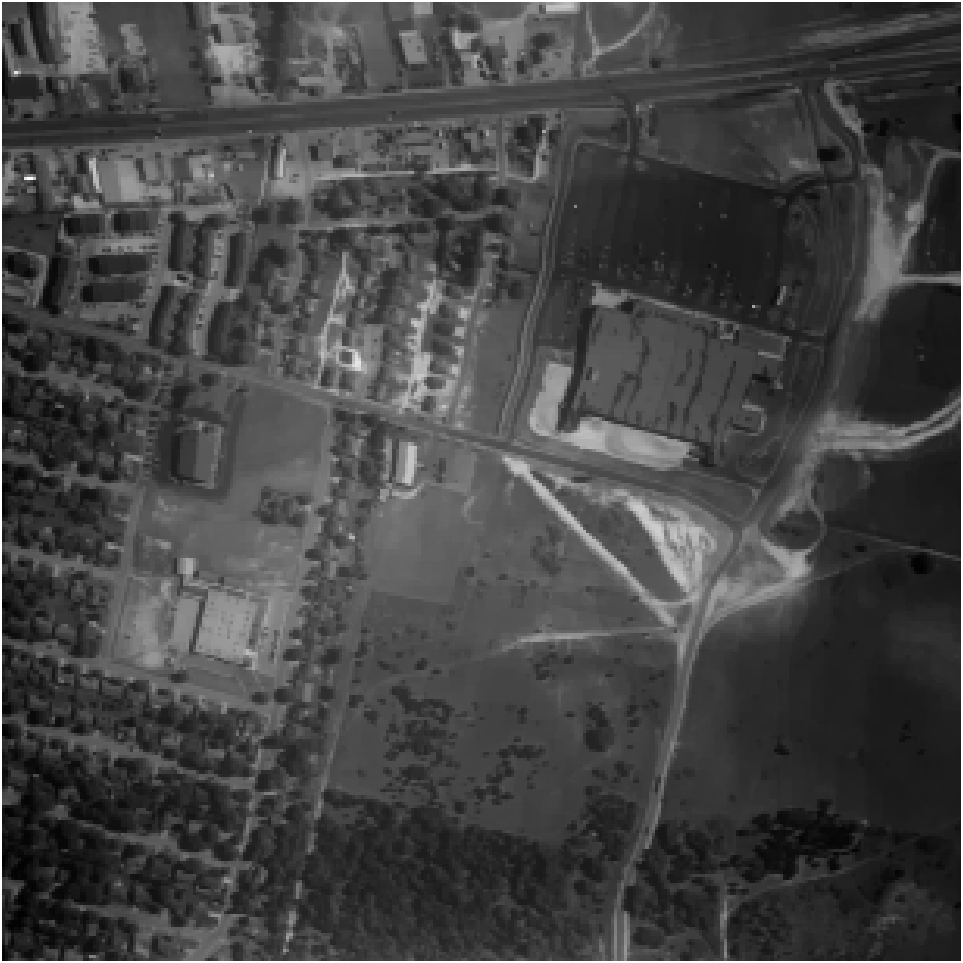} \\
			\end{minipage}
			\begin{minipage}[t]{\mysizetwo}
				\centering
				\includegraphics[width=\mysizetwo]{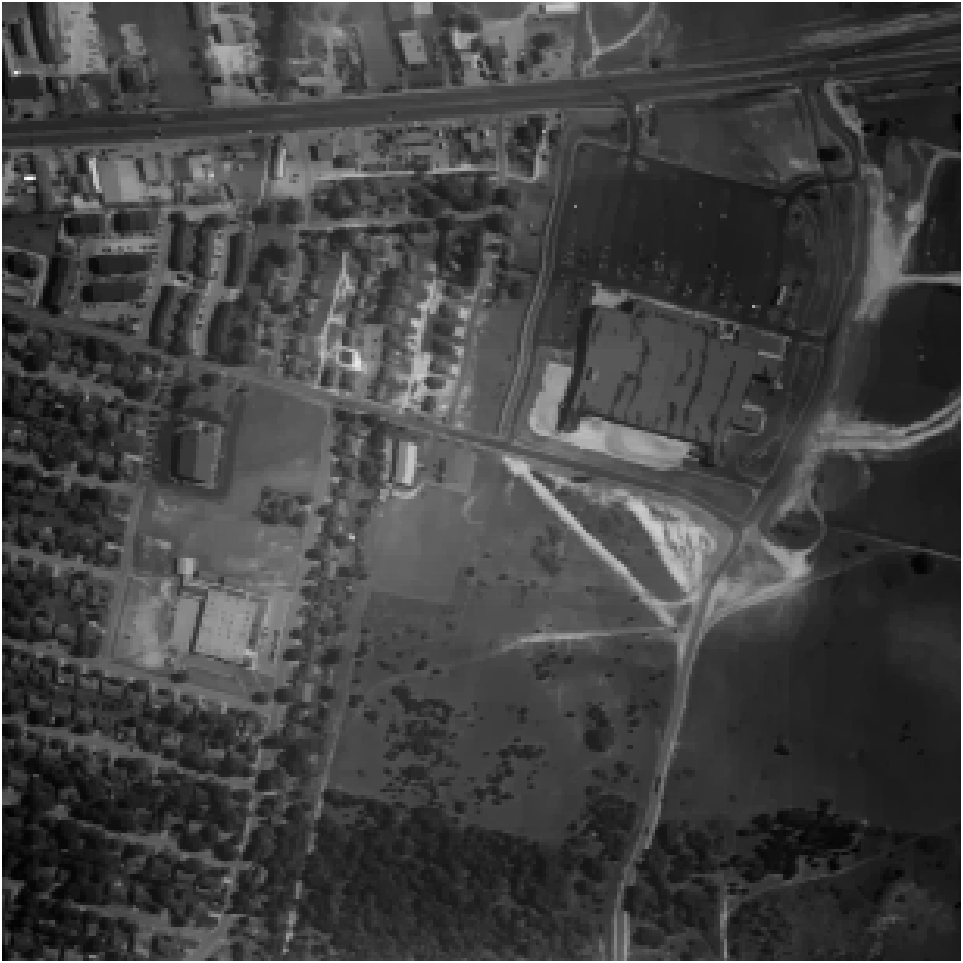} \\
			\end{minipage}
			\begin{minipage}[t]{\mysizetwo}
				\centering
				\includegraphics[width=\mysizetwo]{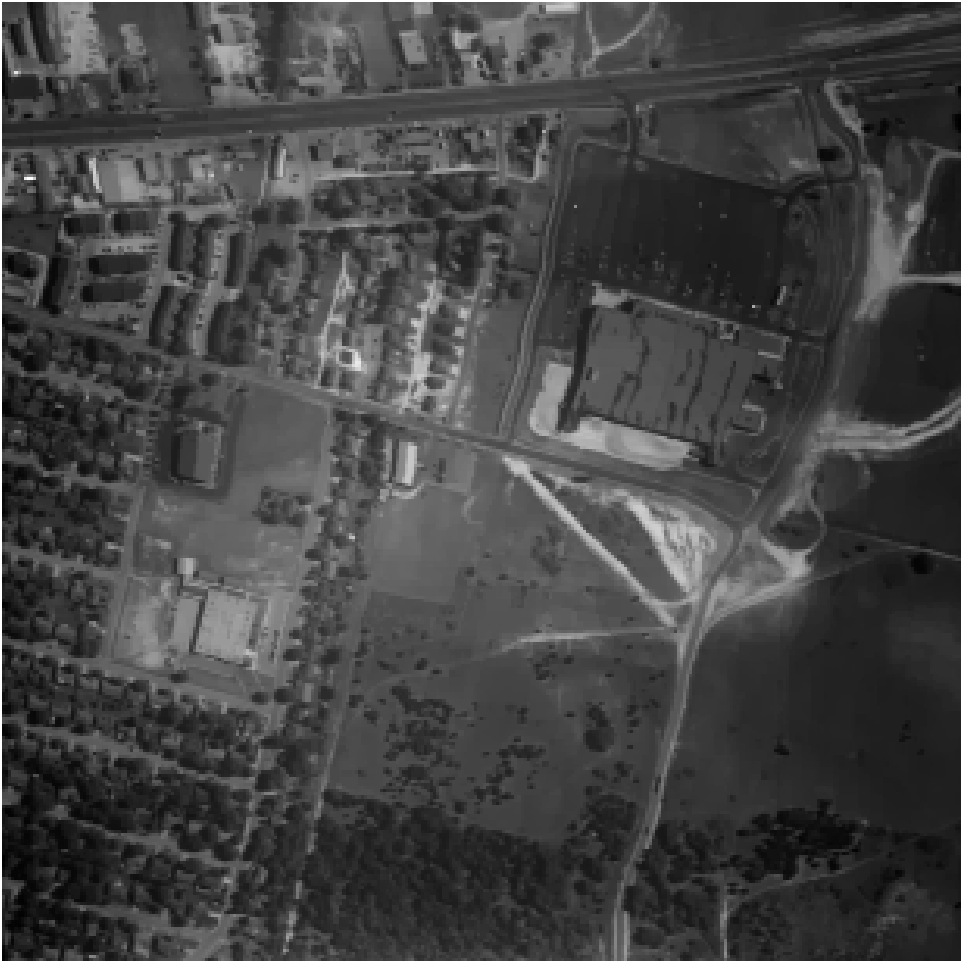} \\
		\end{minipage}} \\
		
		\Xhline{0.8pt}
		
	\end{tabular}
	\caption{Visual comparison of the denoising performance on Urban dataset obtained by LRTV, E3DTV, LRTFDFR and their corresponding DLW-models.}
	\label{fig-tran-urban}
\end{figure*}

\begin{figure}[t]
	\renewcommand{\arraystretch}{1.15}
	\newcommand{\mysize}{3cm}
	\fontsize{8.5}{9.5}\selectfont
	\newcommand{\minivs}{5pt}
	\newcommand{\vs}{2pt}
	\centering
	\begin{minipage}[t]{\mysize*4+2cm}
		\centering
		\begin{minipage}[b]{\mysize}
			\centering
			clean 
			\includegraphics[width=\mysize]{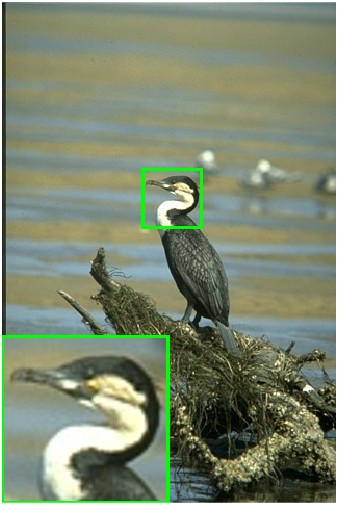} \\
			(PSNR, SSIM)
		\end{minipage}\hspace{\minivs}
		\begin{minipage}[b]{\mysize}
			\centering
			noisy 
			\includegraphics[width=\mysize]{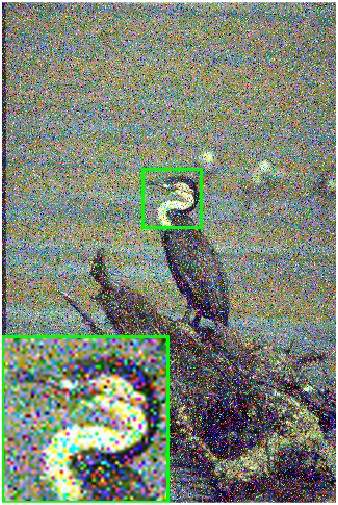} \\
			(11.69, 0.0892)
		\end{minipage}\hspace{\minivs}
		\begin{minipage}[b]{\mysize}
			\centering
			SVTV 
			\includegraphics[width=\mysize]{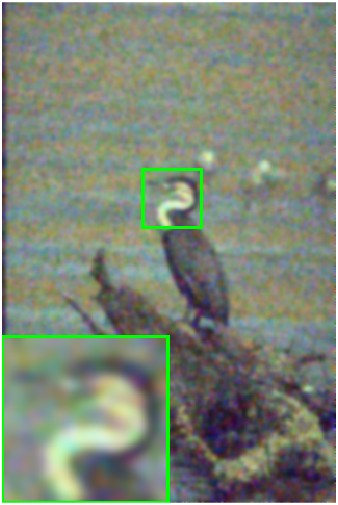} \\
			(\ul{21.89}, \ul{0.5355})
		\end{minipage}\hspace{\minivs}
		\begin{minipage}[b]{\mysize}
			\centering
			T 
			\includegraphics[width=\mysize]{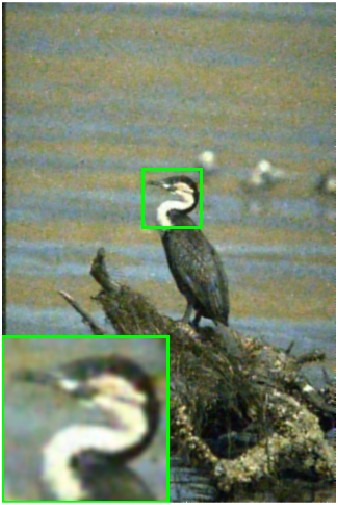} \\
			(\tb{26.78}, \tb{0.7024})
		\end{minipage}\hspace{\minivs}
	\end{minipage}
	\caption{Denoising results of SVTV and its corresponding ``DLW-" model on image ``268002" of BSDS dataset. The noisy type is ``Gaussian+impulse" (Case 1).}
	\label{fig-append-svtv}
\end{figure}

\bibliographystyle{siamplain}
\bibliography{references}

\end{document}